\documentclass[journal]{IEEEtran}
\usepackage{tikz}
\usetikzlibrary{arrows, calc, decorations.markings, positioning,trees}
\newenvironment{timeline}[6]{%

    \newcommand{\startyear}{#1}
    \newcommand{\tlendyear}{#2}

    \newcommand{\yearcolumnwidth}{#3}
    \newcommand{\rulecolumnwidth}{#4}
    \newcommand{\entrycolumnwidth}{#5}
    \newcommand{\timelineheight}{#6}

    \newcommand{\templength}{}

    \newcommand{\entrycounter}{0}

    \long\def\ifnodedefined##1##2##3{%
        \@ifundefined{pgf@sh@ns@##1}{##3}{##2}%
    }

    \newcommand{\ifnodeundefined}[2]{%
        \ifnodedefined{##1}{}{##2}
    }

    \newcommand{\drawtimeline}{%
        \draw[timelinerule] (\yearcolumnwidth+5pt, 0pt) -- (\yearcolumnwidth+5pt, -\timelineheight);
        \draw (\yearcolumnwidth+0pt, -10pt) -- (\yearcolumnwidth+10pt, -10pt);
        \draw (\yearcolumnwidth+0pt, -\timelineheight+15pt) -- (\yearcolumnwidth+10pt, -\timelineheight+15pt);

        \pgfmathsetlengthmacro{\templength}{neg(add(multiply(subtract(\startyear, \startyear), divide(subtract(\timelineheight, 25), subtract(\tlendyear, \startyear))), 10))}
        \node[year] (year-\startyear) at (\yearcolumnwidth, \templength) {\startyear};

        \pgfmathsetlengthmacro{\templength}{neg(add(multiply(subtract(\tlendyear, \startyear), divide(subtract(\timelineheight, 25), subtract(\tlendyear, \startyear))), 10))}
        \node[year] (year-\tlendyear) at (\yearcolumnwidth, \templength) {\tlendyear};
    }

    \newcommand{\entry}[2]{%

        \pgfmathtruncatemacro{\lastentrycount}{\entrycounter}
        \pgfmathtruncatemacro{\entrycounter}{\entrycounter + 1}

        \ifdim \lastentrycount pt > 0 pt%
            \node[entry] (entry-\entrycounter) [below of=entry-\lastentrycount] {##2};
        \else%
            \pgfmathsetlengthmacro{\templength}{neg(add(multiply(subtract(\startyear, \startyear), divide(subtract(\timelineheight, 25), subtract(\tlendyear, \startyear))), 10))}
            \node[entry] (entry-\entrycounter) at (\yearcolumnwidth+\rulecolumnwidth+10pt, \templength) {##2};
        \fi

        \ifnodeundefined{year-##1}{%
            \pgfmathsetlengthmacro{\templength}{neg(add(multiply(subtract(##1, \startyear), divide(subtract(\timelineheight, 25), subtract(\tlendyear, \startyear))), 10))}
            \draw (\yearcolumnwidth+2.5pt, \templength) -- (\yearcolumnwidth+7.5pt, \templength);
            \node[year] (year-##1) at (\yearcolumnwidth, \templength) {##1};
        }

        \draw ($(year-##1.east)+(2.5pt, 0pt)$) -- ($(year-##1.east)+(7.5pt, 0pt)$) -- ($(entry-\entrycounter.west)-(5pt,0)$) -- (entry-\entrycounter.west);
    }

    \newcommand{\plainentry}[2]{

        \pgfmathtruncatemacro{\lastentrycount}{\entrycounter}
        \pgfmathtruncatemacro{\entrycounter}{\entrycounter + 1}

        \ifdim \lastentrycount pt > 0 pt%
            \node[entry] (entry-\entrycounter) [below of=entry-\lastentrycount] {##2};
        \else%
            \pgfmathsetlengthmacro{\templength}{neg(add(multiply(subtract(\startyear, \startyear), divide(subtract(\timelineheight, 25), subtract(\tlendyear, \startyear))), 10))}
            \node[entry] (entry-\entrycounter) at (\yearcolumnwidth+\rulecolumnwidth+10pt, \templength) {##2};
        \fi

        \ifnodeundefined{invisible-year-##1}{%
            \pgfmathsetlengthmacro{\templength}{neg(add(multiply(subtract(##1, \startyear), divide(subtract(\timelineheight, 25), subtract(\tlendyear, \startyear))), 10))}
            \draw (\yearcolumnwidth+2.5pt, \templength) -- (\yearcolumnwidth+7.5pt, \templength);
            \node[year] (invisible-year-##1) at (\yearcolumnwidth, \templength) {};
        }

        \draw ($(invisible-year-##1.east)+(2.5pt, 0pt)$) -- ($(invisible-year-##1.east)+(7.5pt, 0pt)$) -- ($(entry-\entrycounter.west)-(5pt,0)$) -- (entry-\entrycounter.west);
    }

    \begin{tikzpicture}
        \tikzstyle{entry} = [%
            align=left,%
            text width=\entrycolumnwidth,%
            node distance=10mm,%
            anchor=west]
        \tikzstyle{year} = [anchor=east]
        \tikzstyle{timelinerule} = [%
            draw,%
            decoration={markings, mark=at position 1 with {\arrow[scale=1.5]{latex'}}},%
            postaction={decorate},%
            shorten >=0.4pt]

        \drawtimeline
}
{
    \end{tikzpicture}
}
\makeatother
\usepackage{colortbl} 
\usepackage[utf8]{inputenc}
\usepackage{longtable}
\usepackage{amsthm}
\usepackage{amsmath}
\usepackage{amssymb}
\usepackage{amsfonts}
\usepackage{cite}

\usepackage{booktabs}
\usepackage{bbm}
\usepackage{mathrsfs}

\newcommand{\Rmnum}[1]{\expandafter\@slowromancap\romannumeral #1@}
\usepackage{graphicx}
\usepackage{subfigure}
\usepackage{epsfig}
\usepackage[numbers,sort&compress]{natbib}
\usepackage{xcolor}
\usepackage{color}
\usepackage{enumerate}
\usepackage{extarrows}
\usepackage[ruled,linesnumbered,lined]{algorithm2e}  
\usepackage{algorithmic}
\usepackage{tensor}
\usepackage[colorlinks,linkcolor=blue,anchorcolor=blue,citecolor=blue,urlcolor=black]{hyperref}
\usepackage{url}

\newcommand{\fref}[1]{Fig.~\ref{#1}}
\newcommand{\ket}[1]{| #1 \rangle}
\newcommand{\eqr}[1]{Eq.~(\ref{#1})}
\hypersetup{%
    ,urlcolor=black
    ,citecolor=black
    ,linkcolor=black
    }

\usepackage{makecell}
\usepackage{multirow}
\usepackage{diagbox}
\usepackage{subfigure}
\DeclareMathAlphabet\mathbfcal{OMS}{cmsy}{b}{n} 

\newsavebox\ltmcbox
\newenvironment{fakelongtable}
        {\setbox\ltmcbox\vbox\bgroup
        \csname @twocolumnfalse\endcsname
        \csname col@number\endcsname\csname @ne\endcsname}
        {\unskip\unpenalty\unpenalty\egroup\unvbox\ltmcbox}

\usepackage{physics}

\begin{document}

%
\title{Quantum Information Processing, Sensing and Communications:\\
  Their Myths, Realities and Futures}
\author{L. Hanzo\thanks{L. Hanzo would like to acknowledge the financial support of the Engineering and Physical Sciences Research Council (EPSRC) projects under grant EP/Y037243/1, EP/W016605/1, EP/X01228X/1, EP/Y026721/1, EP/W032635/1, EP/Y037243/1 and EP/X04047X/1 as well as of the European Research Council's Advanced Fellow Grant QuantCom (Grant No. 789028).}, Z. Babar, ZY. Cai\thanks{ZY. Cai acknowledges support from the EPSRC QCS Hub EP/T001062/1, EPSRC projects Robust and Reliable Quantum Computing (RoaRQ, EP/W032635/1), Software Enabling Early Quantum Advantage (SEEQA, EP/Y004655/1) and the Junior Research Fellowship from St John’s College, Oxford.}, D. Chandra, I. Djordjevic\thanks{I. Djordjevic's research was supported in part by the Science and Technology Center New Frontiers of Sound through National Science Foundation (NSF) Grant No. 2242925 and NSF under Grant 2244365.}, B. Koczor\thanks{B. Koczor thanks the University of Oxford for a Glasstone Research Fellowship, Lady Margaret Hall Oxford for a Research Fellowship, UKRI for the Future Leaders Fellowship project titled Theory to Enable Practical Quantum Advantage (MR/Y015843/1) and the EPSRC projects Robust and Reliable Quantum Computing (RoaRQ, EP/W032635/1) and Software Enabling Early Quantum Advantage (SEEQA, EP/Y004655/1).}, S-X. Ng, M. Razavi, O. Simeone\thanks{The work of O. Simeone is supported by the European Union’s Horizon Europe project CENTRIC (101096379), by an Open Fellowship of the EPSRC (EP/W024101/1), and by the EPSRC project (EP/X011852/1).}} 
\maketitle


\IEEEpeerreviewmaketitle

\begin{abstract}
  The recent advances in quantum information processing, sensing and
  communications are surveyed with the objective of identifying the
  associated knowledge gaps and formulating a roadmap for their future
  evolution. Since the operation of quantum systems is prone to the
  deleterious effects of decoherence, which manifests itself in terms
  of bit-flips, phase-flips or both, the pivotal subject of quantum
  error mitigation is reviewed both in the presence and absence of
  quantum coding. The state-of-the-art, knowledge gaps and future
  evolution of quantum machine learning are also discussed, followed
  by a discourse on quantum radar systems and briefly hypothesizing
  about the feasibility of integrated sensing and communications in
  the quantum domain. Finally, we conclude with a set of promising
  future research ideas in the field of ultimately secure quantum
  communications with the objective of harnessing ideas from the
  classical communications field.
  
\end{abstract}

\section*{Acronyms}
\begin{fakelongtable}
\begin{longtable}{l l}
AO & Adaptive Optics\\
AWGn & Additive White Gaussian Noise\\
BBS & Balanced Beam Splitter\\
CD & Classical Domain \\
CDMA & Code Division Multiple Access\\
CF & Cost Function \\
CI & Configuration Interaction\\
COMP & Cooperative Multicell Processing\\
CPU & Central Processing Unit \\
CSS & Calderbank-Shor-Steane\\
CV & Continuous-Variable\\
DFG & Difference Frequency Generation\\
DFRC & Dual-Function Radar Communication\\
DV & Discrete-Variable\\
EA & Entanglement-Assisted \\
EV & Echo Verification\\
EXIT & EXtrinsic Information Transfer \\
FPQTD & Fully-Parallel Quantum Turbo Decoder \\
FSO & Free Space Optical \\
GRAND & Guessing Random Additive Noise Decoding\\
HPC & Hypergraph Product Code \\
ISAC & Integrated Sensing And Communications\\
LADAR & LAser Detection And Ranging\\
LDGM & Low Density Generator Matrix \\
LEO & Low-Earth-Orbit \\
LPC & Lifted-Product Code \\
LTE & Long-Term Evolution \\
MDI & Measurement-Device-Independent\\
MIMO & Multiple-Input Multiple-Output \\
MMD & Maximum Mean Discrepancy\\
MUD & Multi-User Detection\\
MU-MIMO & Multi-User MIMO\\
MUT & Multi-User Transmission \\
NISC & Nearterm Intermediate-Scale Computer \\
NISQ & Noisy Intermediate Scale Quantum\\
NTN & Non-Terrestrial Networking \\
OFDM & Orthogonal Frequency Division Multiplexing\\
OPA & Optical Parametric Amplifier\\
OPC & Optical Phase-Conjugation\\
OSD & Ordered-Statistic Decoding\\
PC & Phase-Conjugated\\
PCM & Parity Check Matrix\\
PEC & Probabilistic Error Cancellation\\
PPLN & Periodically Poled LiNbO3\\
PSCC & Phase-Sensitive Cross-Correlation\\
PQC & Parameterized Quantum Circuit\\
PQC & Post-Quantum Cryptography\\
QAOA & Quantum Approximate Optimization Algorithm\\
QBCH & Quantum Bose-Chaudhuri-Hocquenghem \\
QBER & Quantum Bit Error Ratio\\
QC & Quasi-Cylic \\
QCC & Quantum Convolutional Code\\
QCNN & Quantum Convolutional Neural Networks\\
QD & Quantum Domain \\
QEC & Quantum Error Correction \\
QECC & Quantum Error Correction Code\\
QEM & Quantum Error Mitigation \\
QGAN & Quantum Generative Adversarial Network\\
QI & Quantum Illumination \\
QIrCC & Quantum Irregular Convolutional Code\\
QKD & Quantum Key Distribution \\
QLDPC & Quantum Low-Density Parity-Check \\
QM & Quantum Memory \\
QML & Quantum Machine Learning \\
QNC & Quantum Network Code \\
QNN & Quantum Neural Network\\
QPC & Quantum Polar Code\\
QPU & Quantum Processor Unit\\
QRM & Quantum Reed-Muller \\
QRS & Quantum Reed-Solomon \\
QSBC & Quantum Short-Block Code\\
QSC & Quantum Stabilizer Code \\
QSDC & Quantum-Secured Direct Communications\\
QTC & Quantum Turbo Code\\
QTECC & Quantum Topological Error-Correction Code\\
QURC & Quantum Unity-Rate Code\\
RF & Radio-Frequency\\
SC & Spatially-Coupled \\
SM & Spatial Multiplexing\\
SPDC & Spontaneous Parametric Down Conversion\\
SNR & Signal-to-Noise Ratio\\
STC & Space-Time code\\
SV & Symmetry Verification\\
TMSV & Two-Mode Squeezed Vacuum\\
URLLC & Ultra-Reliable Low-Latency Communication\\
VL & Visible Light\\
WDM & Wavelength Division Multiplexing\\
VQE & Variational Quantum Eigensolver\\
ZNE & Zero Noise Extrapolation \\
\end{longtable}
\end{fakelongtable}
\section{Introduction}

Back in 1965 Gordon Moore hypothesized that the integration density of micro-electronics chips would be doubled every 18 months or so and - perhaps somewhat surprisingly - this prediction has remained valid ever since. As a result, at the time of writing the integration density of chips has reached nano-meter scales, hence the quantum-effects may no longer be ignored by chip designers. As a further advance, Richard Feynman suggested that the ubiquitous bits conveying digital information could in fact be mapped not only to a pair of distinct voltage levels, but also either to a pair of different electron-charges or to the up- and down-oriented spin of an electron for quantum-domain information processing. Naturally, this quantum-domain representation has its pros and cons, because the resultant quantum bits abbreviated as qubits no longer obey the laws of classical physics - instead they are governed by the laws of quantum  physics. What are these?

To elaborate briefly, a qubit may be in the so-called superposition of a logical one and a logical zero. We may be able to interpret this by referring to a coin spinning in a closed box, which might be deemed to be in the equi-probable superposition of head and tail, representing a logical zero and one. Based on these qubits we may construct arbitrarily long strings of qubits as the operands of quantum information processing. However, when this hypothetical coin stops spinning and we lift the lid of the box, we can reveal/observe the qubit, which will be either in a state of logical zero or logical one. We might say that upon its 'observation' the qubits fall back into the classical domain and they may no longer be 'processed or manipulated' in the quantum domain. Another salient quantum-domain feature is that the qubits must not be copied, which is formulated in terms of the so-called no-cloning theorem of quantum physics. The qubits may be processed by unitary operators or quantum gates, which are the quantum-domain counterparts of classical gates.

\subsection{Applications of Quantum-Domain Information Processing}

At the time of writing the most prominent quantum-domain applications are in the field of quantum communications, since quantum key distribution (QKD) is already a commercial off-the-shelf reality. By contrast, quantum computing is still in its infancy, because the most capable quantum computer commercialized by DWave only handles 2048 qubits.

\subsubsection{Quantum Information Processing and its Applications}

However, even the powerful quantum computers of the future are not expected to outperform classical computers in all tasks - they are more suitable for carrying out rather specific tasks at a high speed by exploiting their true parallel processing capability. This fact also motivates the design of so-called near-term intermediate-scale computers (NISCs), where bespoke quantum circuits are harnessed for evaluating a particularly demanding cost function (CF) and the results are then fed into a classical computer for further processing. This might be deemed reminiscent of having a powerful external quantum processor. One of the main impediments of quantum computers is their very limited so-called coherence time, which limits the number of operations/actions that may be carried out before avalanche-like error proliferation sets in. The above-mentioned NISC philosophy mitigates this decoherence problem by forwarding the CF value to a classical computer before catastrophic decoherence occurs. Another potent quantum-domain error mitigation technique is constituted by the family of quantum error correction codes (QEC). The design of these QECs may be deemed to be reminiscent  of classical-domain error correction codes, provided that the so-called simplectic conditions exploited in~\cite{babar2018duality,babar2019polar,chandra2019near,chandra2023exit} are satisfied.

When quantum computers capable of running large-scale quantum search algorithms become widely available, numerous large-scale search problems of wireless communications that have hitherto been deemed to have excessive complexity may be solved more efficiently than ever before, as detailed for example in~\cite{botsinis14tcom,botsinis15tcom,botsinis15acc}. These may be exemplified by multiple-symbol-based differential detection~\cite{8540839}, multi-user detection~\cite{8540839}, multi-objective Pareto optimization of large-scale routing problems~\cite{7268791,7745885}, localization problems~\cite{7997701} as well as network coding solutions~\cite{8019800}.
\begin{figure*}
\center
\includegraphics[width=\linewidth]{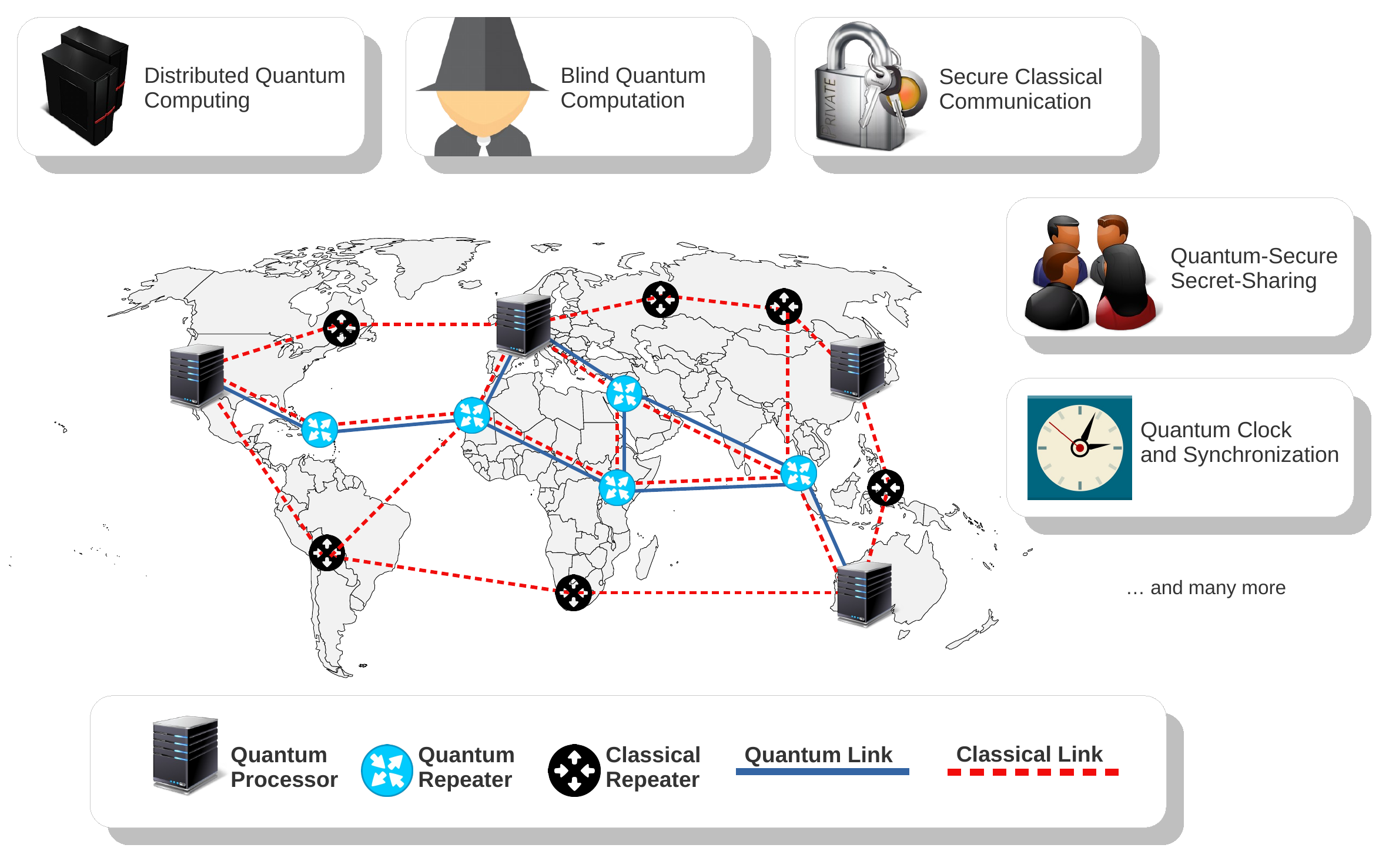}
\caption{Stylized vision of the Quantum Internet of the near future, which will rely on a combination of both classical and quantum devices \copyright Chandra {\em et al.}~\cite{hanzo-hte}.}
\label{fig:internet}
\end{figure*}

\subsubsection{Quantum Key Distribution}

In the 5G Advanced and 6G era there is more emphasis on information security than ever before. Hence QKD networks are proliferating at a fast pace, which are capable of providing an extremely high level of information security, because if an eavesdropper tampers with the confidential key negotiation protocol, it may be detected with a near-unity probability. Furthermore, as mentioned above, observation of the qubits by an eavesdropper results in destroying  the confidential quantum state and as a result, the qubits collapse back into the classical domain, as detailed  in great depth with the aid of tutorial examples in~\cite{Hosseinidehaj_SatQKD2017,Hanzo:QInternet}. Once the secret key has been negotiated and agreed by the communicating parties, it can  be readily appied in a similar way to the classic crypto systems, where in its simplest form the modulo two connection of the information bits and the key bits are transmitted over the channel. Suffice to say however that a severe limitation  of this concept is that the secret key has to be at least as long as the data sequence, which represents a 100 percent security overhead. Furthermore, in the interest of high security, the key has to be changed rather frequently to avoid its potential long-term observation by eavesdroppers. But the most severe limitation of QKD networks is that as the transmission distance is increased, the maximum attainable key-rate is reduced due to the attenuation of the channel, because the quantum signal must not be amplified. This phenomenon is reminiscent of the reduced bitrate {\em vs.} distance trend of classical systems. As a remedy, similarly to classical systems, so-called trusted relaying may be used for extending the distance, but this trusted relay must be accommodated in  protected premises to avoid tampering with them. We note at this stage that given the similarity of the QKD-based and classical-domain encryption procedures, it is becoming realistic to incorporate QKD-based key negotiation in next-generation wireless systems.

\subsubsection{A Vision of the Quantum Internet}  
At the time of writing the research of the interdisciplinary subject
of quantum science and engineering is attracting substantial
investment right across the globe. Many of the IEEE societies -
including the Computer, the Communications, Signal Processing,
Antennas and Propagation Society, just to name a few - have joined
forces in creating the IEEE Quantum Engineering Journal and also
support the recently created conference referred to as the IEEE
Quantum Week. Their broad vision is to make the creation of the {\bf
  Quantum Internet} (Qinternet)~\cite{caleffi2020rise} a reality,
which is portrayed in the stylized illustration of
Fig.~\ref{fig:internet}. It is expected to support both ultimate
information security as well eavesdropping detection, neither of which
is feasible in the classical Internet.  Numerous hitherto non-existent
services might be created~\cite{van2016path, fitzsimons2017private},
but 'quantum-leaps' are required in quantum engineering to make this
vision a reality.

\begin{figure*}
\center
\includegraphics[width=\linewidth]{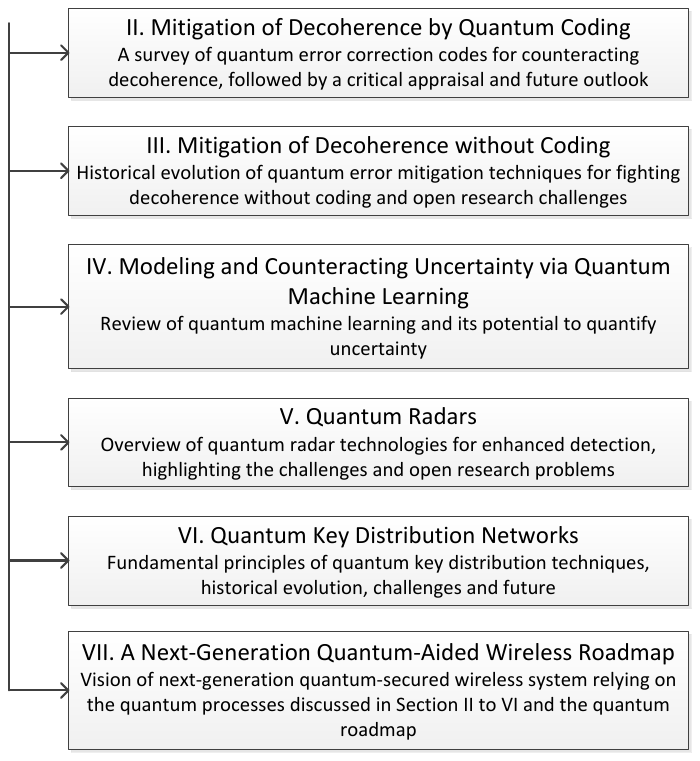}
\vspace*{-4cm}
\caption{Paper structure.}
\label{fig:outline}
\end{figure*}

\fref{fig:outline} provides an overview of the paper. In each of the Sections~\ref{quantum-coding} -- \ref{qkd-mohsen} we
embark on critically appraising the state-of-the-art in the critical
components of quantum coding, quantum error mitigation, quantum
machine learning, quantum radar and QKD along the evolutionary road of
creating the Qinternet. Based on this analysis, we then identify the
knowledge gaps and speculate about the future roadmap of filling these
gaps.

Section~\ref{quantum-coding} is dedicated to the central topic of
mitigating quantum-domain impairments by quantum coding, while
Section~\ref{qem-lh} explores quantum error mitigation without
coding. Section~\ref{qml-osvaldo} puts quantum machine learning under
the magnifier glass, followed by Section~\ref{quantum-radar} devoted
to the radically new quantum radar subject area. Although QKD
solutions are now off-the-shelf commercial reality, numerous open
problems exist in architecting the global Qinternet, as discussed in
Section~\ref{qkd-mohsen}. Finally, in Section~\ref{final} we
hypothesize about the next steps along the road leading to the
construction of an ultimately secure quantum communications system.

Let us now delve into deeper technical details concerning the
relationship between classical as well as quantum error correction coding.

\section{Mitigation of Decoherence by Quantum Coding}
\label{quantum-coding}
      {\bf The Myth:} Quantum coding is readily capable of eliminating
      the deleterious effects of quantum decoherence.

      {\bf The Reality:} Only short, low-complexity quantum codes --
      such as for example topological codes -- may be implemented at
      the current state-of-the-art, which are unable to approach the
      ultimate performance limit of the hashing bound.

      {\bf The Future:} The hardware of quantum codes also suffers
      from decoherence effects, but this is typically assumed to be
      flawless at this early stage. Future investigations have to
      incorporate the imperfections of the coding hardware as well.
      
{\bf Abstract:} {\em Quantum error-correction codes (QECCs) can be constructed from the known classical coding paradigm by exploiting the inherent isomorphism between the classical and quantum regimes, while also addressing the challenges imposed by the strange laws of quantum physics. In this spirit, we provide insights into the duality of quantum and classical coding theory, hence aiming to bridge the gap between them. Explicitly, we survey the rich history of both classical as well as quantum codes, followed by a critical appraisal of QECCs, as exemplified by the family of dual-containing and non-dual-containing Calderbank-Shor-Steane (CSS) codes, non-CSS codes, and entanglement-assisted codes. Finally, we provide an outlook on the potential evolution of QECCs.}

\subsection{State-of-the-Art}
\label{quantum-coding-state-of-the-art}

The deleterious effects of quantum decoherence can be completely characterized by amplitude-damping and phase-damping channel models. Explicitly, amplitude damping models the loss of qubit energy, for example, when the excited state of a qubit decays due to the spontaneous emission of a photon or when photons are lost during their transmission through a free-space optical (FSO) channel or optical fibers~\cite{chuang1997bosonic}. By contrast, phase damping (or dephasing) models the contamination of quantum information without any loss of energy. This may occur, for example, due to the scattering of photons or owing to the perturbation of electronic states induced by stray electrical charges. This combined amplitude-damping and phase-damping model of quantum decoherence may be further reduced to a so-called Pauli channel, which merely inflicts bit-flip and/or phase-flip errors upon the qubit. Specifically, a Pauli channel $\mathcal{N}_{\text{P}}$ may inflict a phase-flip ($\mathbf{Z}$), a bit-flip ($\mathbf{X}$) or a bit-and-phase-flip ($\mathbf{Y}$) error with probabilities of $p_z$, $p_x$ and $p_y$, respectively, on a qubit having density matrix $\rho$,  which may be formulated as~\cite{sarvepalli2009asymmetric}:
\begin{equation}
    \mathcal{N}_{\text{P}} (\rho) = (1 - p_z - p_x - p_y)\rho  + p_z \mathbf{Z}\rho\mathbf{Z} + p_x \mathbf{X}\rho\mathbf{X} + p_y \mathbf{Y}\rho\mathbf{Y}.
    \label{eq:pauli:channel}
\end{equation}
The error probabilities $p_z$, $p_x$ and $p_y$ depend on the qubit relaxation time $T_1$ and the qubit dephasing time $T_2$, as given below:
\begin{align}
    p_x = p_y &= \frac{1}{4} \left( 1 - e^{-t/T_1} \right) \nonumber \\
    p_z &= \frac{1}{4} \left( 1 + e^{-t/T_1} - 2 e^{-t/T_2} \right).
    \label{eq:pxpz}
\end{align}
$T_1$ is the time it takes for the excited state to decay to ground state, while $T_2$ is the time for which the superimposed quantum state is maintained. Both $T_1$
and $T_2$ times vary with the qubit implementation technology (e.g. superconducting, trapped ion, quantum dot) because they depend on the properties of the material and the associated environmental interaction mechanisms.

This Pauli channel may also be viewed as the quantum-domain counterpart of a quaternary classical channel.  The Pauli channel is also often represented as a pair of independent channels, namely the bit-flip and the phase-flip channels, inflicting errors with probabilities $(p_x + p_y)$ and $(p_z + p_y)$, respectively. These individual bit-flip and phase-flip channels are analogous to classical binary symmetric channels. Furthermore, a widely used class of Pauli channels - the so-called \emph{depolarizing channel} - assumes that all three types of errors are equally likely, i.e., ($p_z = p_x = p_y $).

\begin{figure*}[tb]
    \centering
    \includegraphics[width=0.9\linewidth]{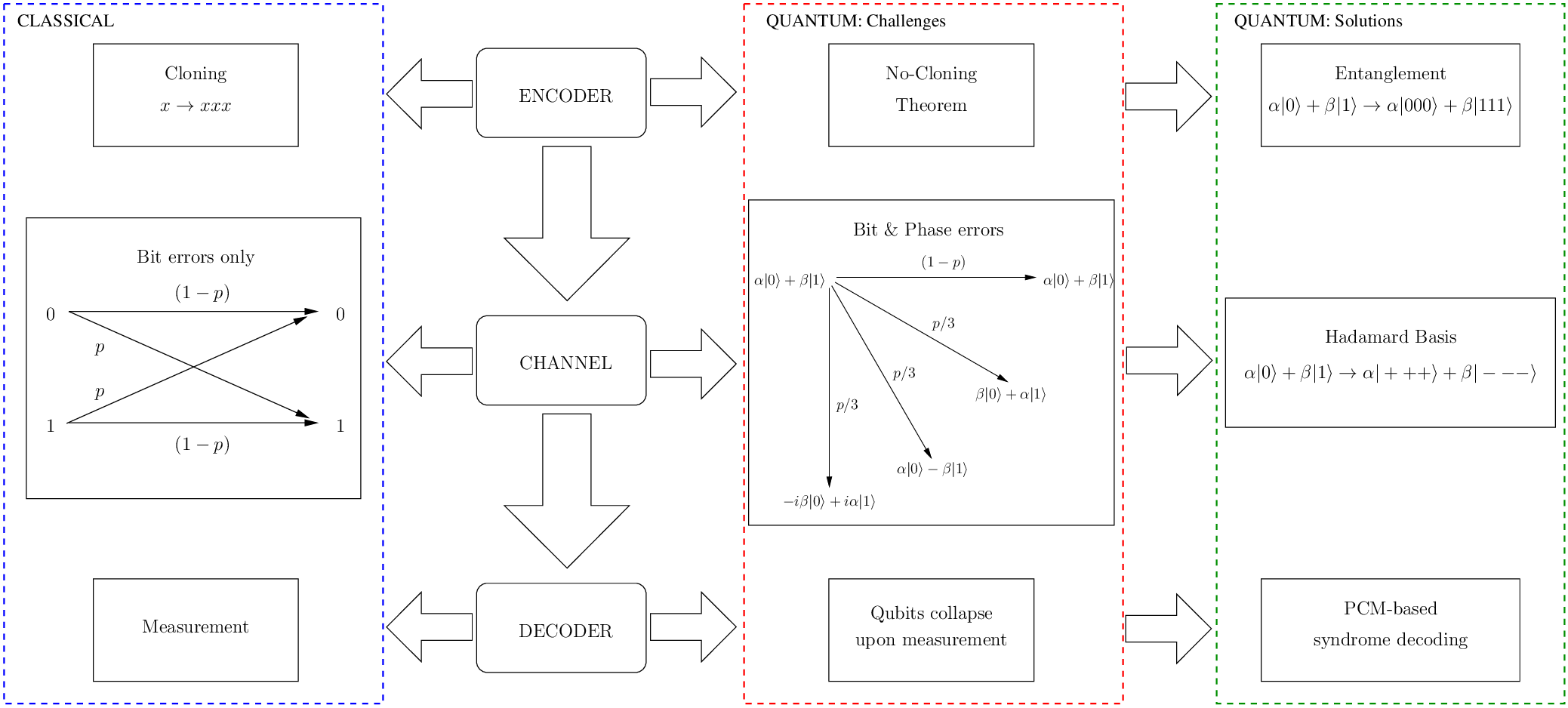}
    \caption{Transition of error correction codes from the classical to the quantum domain~\cite{babar2015road}). \textbf{Encoder}: Classical encoders replicate the information bits. Since qubits cannot be cloned, quantum encoders exploit entanglement to clone the information in the basis states. \textbf{Channel}: While classical channel incurs only bit errors, qubits may experience bit-flip as well as phase-flip errors. Hadamard basis $\{\ket{+},\ket{-}\}$ is used to protect against phase errors. \textbf{Decoder}: Classical decoders rely on measuring the received bits. Since qubits collapse upon measurement, quantum decoders invoke parity-check matrix (PCM)-based syndrome decoding for estimating the channel errors without measuring the received qubits.}
    \label{fig:comp}
\end{figure*}

Despite the existence of the above-mentioned inherent isomorphism between the quantum and classical channels, the conception of quantum codes was deemed infeasible in the early quantum era. This was partly because of the so-called no-cloning theorem~\cite{wootters1982single}, which does not allow the qubits to be copied -- they can only be processed by so-called unitary operations to avoid their \emph(collapse) to the classical domain. Hence, it was hard to envision how redundancy attached to or imposed on the original information qubits by quantum error correction coding (QECC) could be exploited by the QECC decoder for correcting quantum-domain (QD) errors caused by decoherence. More explicitly, since the qubits collapse to the classical domain (CD) upon their observation or measurement, QECC decoding poses a challenge. To elaborate, classical decoding processes rely on observing/measuring the received bits. Therefore, correcting qubits without perturbing their coherent superimposed state was considered an infeasible task. However, Shor dispelled these notions in 1995 by conceiving his seminal $9$-qubit QECC~\cite{shor1995scheme} that is capable of correcting a single-qubit error. This $9$-qubit QECC having a coding rate of $1/9$ is reminiscent of the classical $1/3$-rate $3$-bit repetition code, as exemplified in~\fref{fig:comp}, where one of the $1/3$-rate codes is used for correcting a single bit-flip error and the other a phase-flip error.

Inspired by this significant contribution, Calderbank and Shor~\cite{calderbank1996good}, as well as Steane~\cite{steane1996multiple, steane1996error}, separately developed a generalized framework for designing quantum codes from binary linear classical codes. This framework laid the foundation of the well-known family of Calderbank-Shor-Steane (CSS) codes, which allows constructing an $[n,k_1-k_2]$ CSS code from a pair of classical linear block codes $C_1(n,k_1)$ and $C_2(n,k_2)$, provided that $C_2 \subset C_1$. The code $C_1$ is exploited for bit-flip error correction, while the dual of code $C_2$, denoted as $C_2^\bot,$ is used for phase-flip error correction. Hence, the resultant CSS code is capable of correcting $t$ bit-flips and $t$ phase-flips, if both $C_1$ and $C_2^\bot$ can correct $t$ errors in the classical domain. Furthermore, a specific category of CSS codes known as {\em dual-containing} CSS codes was introduced, which originates from dual-containing binary codes. In essence, dual-containing CSS codes constitute a distinct subset of CSS codes, where $C_2 = C_1^\bot$. Following these principles, Steane~\cite{steane1996error} used the classical $(7,4)$ Hamming code to design a single-error correcting $[[7,1]]$ dual-containing CSS code.

Both dual-containing and non-dual-containing CSS codes can be viewed as an amalgam of two independent quantum codes, one for bit-flip and the other for phase-flip correction. Since the bit-flip and phase-flip errors are corrected independently, CSS codes do not fully exploit the redundant qubits. This led to the development of non-CSS codes, such as the optimal $5$-qubit quantum code by Laflamme {\em et al.}~\cite{laflamme1996perfect} and Bennett {\em et al.}~\cite{bennett1996mixed}, representing the shortest possible codeword required for single-qubit error correction.

As a further development, Gottesman formalized the design of quantum codes from the known classical binary and quaternary codes during his Ph.D~\cite{gottesman1997stabilizer} by presenting quantum stabilizer codes (QSC)~\cite{gottesman1996class}. Recall that an $(n,k)$ classical linear block code uses an $(n-k)$ by $n$ parity-check matrix (PCM) for computing the error syndrome of length $(n - k)$\footnote{Let $y = \overline{x} + e$ be the received codeword, $\overline{x}$ the transmitted codeword and $e$ the channel error. The classical syndrome of length $(n-k)$ is computed by using: $s = yH^T = \overline{x}H^T + eH^T= eH^T$. This syndrome is then harnessed for either detecting the expected codeword (codeword decoding) or the expected error pattern (error decoding).}. Similarly, an $[[n,k]]$ QSC invokes a set of $(n - k)$ \emph{commuting} Pauli generators $g_i \in \mathcal{G}_n$\footnote{The Pauli group $\mathcal{G}_n$ is an $n$-fold tensor product of single-qubit Pauli operators, i.e. $\mathbf{I}$ (identity), $\mathbf{X}$ (bit-flip), $\mathbf{Y}$ (simultaneous bit-and-phase-flip) and $\mathbf{Z}$ (phase-flip).}, called the stabilizers, for computing syndromes. The stabilizers can be linked to a classical PCM by mapping each constituent Pauli operator, i.e. $\mathbf{I}$, $\mathbf{X}$, $\mathbf{Y}$ and $\mathbf{Z}$, to a pair of classical bits as follows~\cite{cleve1997quantum, mackay2004sparse}:
\begin{align}
    \mathbf{I} \rightarrow (0,0), \; 
    \mathbf{X} \rightarrow (0,1), \; 
    \mathbf{Y} \rightarrow (1,1), \; 
    \mathbf{Z} \rightarrow (1,0).
    \label{eq:mapping}
\end{align}

Based on this Pauli-to-binary isomorphism, the $(n - k)$ stabilizers of an $[[n,k]]$ QSC can be mapped to an $(n - k)$ by $2n$ PCM $\mathbf{H}$, which consists of two concatenated $(n - k) \times n$ binary matrices $\mathbf{H}_z$ and $\mathbf{H}_x$, as given below:
\begin{equation}
    \mathbf{H} = \left( \mathbf{H}_z | \mathbf{H}_x \right). 
    \label{eq:eq-classical}
\end{equation}
The matrix $\mathbf{H}_z$ is used for bit-flip error correction, while the matrix $\mathbf{H}_x$ is used for phase-flip error correction. Furthermore, the commutative constraint of stabilizers is translated into the orthogonality of rows of $\mathbf{H}$ with respect to the symplectic product (or the twisted product), which is satisfied for all the rows of $\mathbf{H}$ if and only if we have:
\begin{equation}
    \mathbf{H}_z \mathbf{H}_x^T + \mathbf{H}_x \mathbf{H}_z^T = 0.
    \label{eq:twist}
\end{equation}
Hence, any classical PCM which meets the symplectic product criterion of~\eqr{eq:twist} may be used for designing a QSC. It is pertinent to point out here that when the equivalent classical PCM assumes the following structure:
\begin{equation}
    \mathbf{H} = \begin{pmatrix} 
    \mathbf{H}_z^\prime & \mathbf{0} \\
    \mathbf{0} & \mathbf{H}_x^\prime
\end{pmatrix},
\label{eq:H_Shor}
\end{equation}
where we have $\mathbf{H}_z = \begin{pmatrix}
                                \mathbf{H}_z^\prime \\
                                \mathbf{0}
                            \end{pmatrix}$, 
$\mathbf{H}_x = \begin{pmatrix}
	\mathbf{0} \\
    \mathbf{H}_x^\prime 
\end{pmatrix}$, then it reduces to a CSS code. 

The conception of the above stabilizer formalism marked a major breakthrough in the realm of QECCs, serving as a universal panacea for converting classical codes into their quantum counterparts. Research efforts have remained more focused on the conversion of algebraic codes in the initial years, aiming for maximizing the minimum Hamming distance of legitimate codewords by leveraging finite Galois-field arithmetic. This led to the development of quantum Bose-Chaudhuri-Hocquenghem (QBCH) codes~\cite{steane1996simple, grassl1997codes, calderbank1998quantum, grassl1999quantum1, steane1999enlargement, xiaoyan2004quantum}, quantum Reed-Muller (QRM) codes~\cite{steane1999quantum}, and quantum Reed-Solomon (QRS) codes~\cite{grassl1999quantum2}. While leveraging the classical-to-quantum isomorphism has proven valuable in terms of importing classical codes into the quantum domain, it also became clear that they require a high number of qubits, and hence their practical employment will only become realistic in years to come.

This realization has inspired expedited research in the field of short-block QECCs, placing the design of QSCs based on the intricacies of quantum topology and homology in the limelight. This includes for example toric codes~\cite{kitaev1997quantum, kitaev2003fault, fujii2015quantum}, surface codes~\cite{bravyi1998quantum, horsman2012surface}, color codes~\cite{bombin2006topological}, cubic codes~\cite{haah2011local}, hyperbolic codes~\cite{zemor2009cayley, delfosse2013tradeoffs}, homological product codes~\cite{bravyi2014homological}.

While algebraic coding provides a powerful mechanism for designing codes having a high minimum distance, it does not guarantee a near-capacity design, which requires probabilistic codes. Explicitly, the probabilistic coding avenue is inspired by Shannon's random coding theory~\cite{shannon1948mathematical} and strives for striking a beneficial performance {\em vs.}  complexity trade-off. The quantum-domain counterpart of the famous Shannon capacity is the so-called hashing bound~\cite{lloyd1997capacity, shor2003capacities, devetak2005private}, which sets a lower limit on the capacity of Pauli channels. More specifically, analogous to the Shannon capacity of classical binary symmetric channels, the hashing bound specifies the capacity of quantum channels based on the coherent information output of the channel. In case of a depolarizing channel associated with the probability $p$, the hashing bound is given by~\cite{bennett1996mixed, wilde2013entanglement}:
\begin{equation}
    C_Q(p) = 1 - H_2(p) - p \log_2(3),
    \label{eq:capacity_q}
\end{equation}
where $H_2(p)$ denotes the binary entropy function. It must be pointed out here that the actual quantum channel capacity is higher than the hashing bound due to the so-called quantum {\em degeneracy}~\cite{divincenzo1998quantum, smith2007degenerate, fuentes2021degeneracy}. This is a unique quantum phenomenon, which is inherently present in quantum stabilizer codes but does not exist in the classical world. To elaborate, quantum degeneracy implies that different channel errors may have the same impact on the encoded quantum state and hence can be corrected by the same error correction operation\footnote{Let us consider a two-qubit state $\ket{\psi} = \ket{00} + \ket{11}$. The error patterns $\mathbf{IZ}$ as well as $\mathbf{ZI}$ yield the same corrupted state $(\ket{00} - \ket{11})$, hence are termed as degenerate errors. Similarly, the error pattern $\mathbf{ZZ}$ does not perturb the original state $\ket{\psi}$. Hence, the identity operation $\mathbf{II}$ and the error pattern $\mathbf{ZZ}$ also constitute a degenerate pair.}. This in turn enhances the channel capacity.

\begin{figure}[tb]
\begin{center}
    \includegraphics[width=\linewidth]{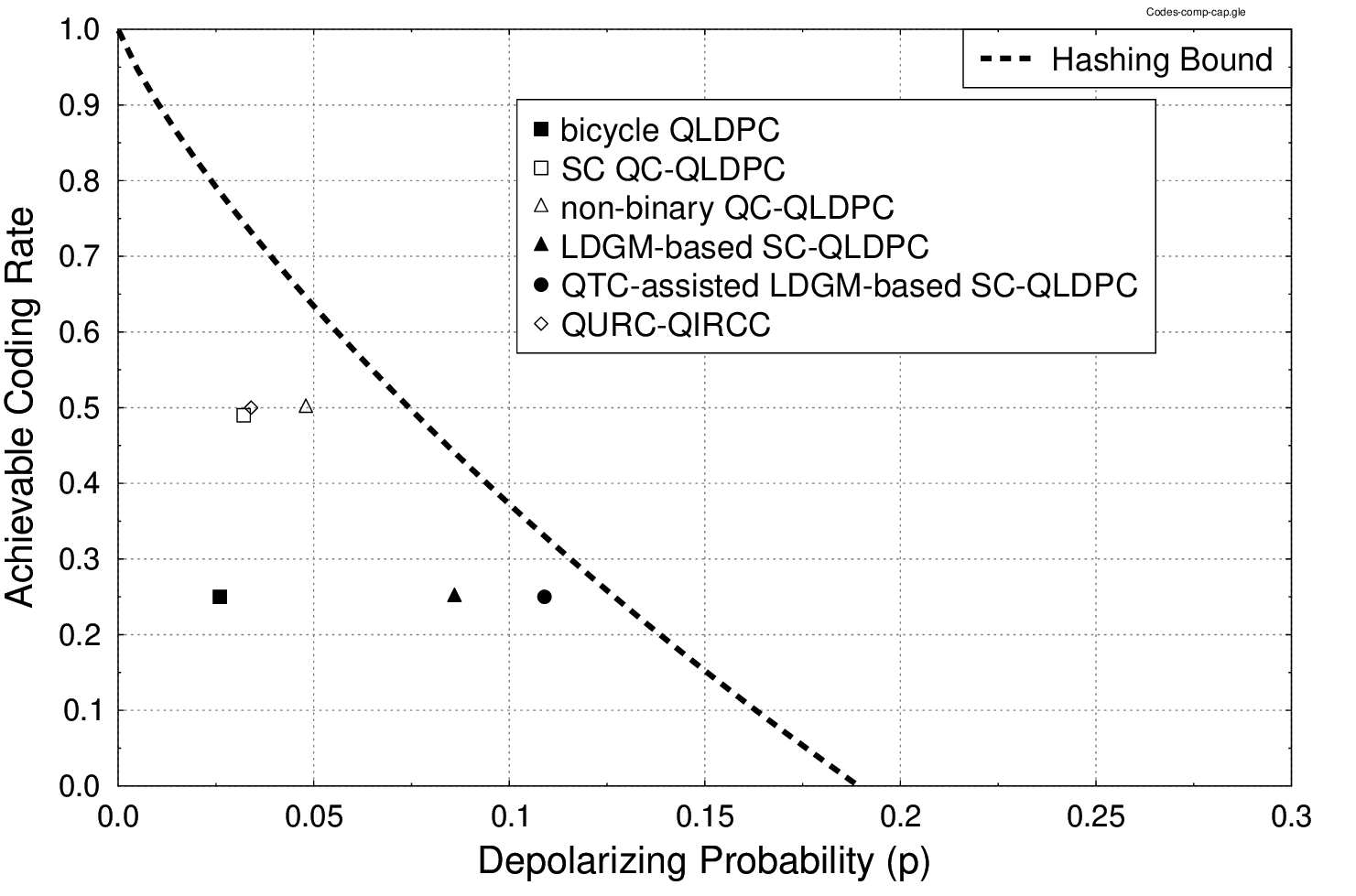}
    \caption{Achievable performance at a word error rate (or frame error rate) of $10^{-3}$ benchmarked against the hashing bound~\cite{babar2018duality} for the `bicycle QLDPC' code ($R = 0.25$, $n = 19,014$) of~\cite{mackay2004sparse}, `spatially-coupled (SC) quasi-cylic (QC)-QLDPC' code ($R = 0.49$, $n = 181,000$) of~\cite{hagiwara2011spatially}, `non-binary QC-QLDPC' code ($R = 0.5$, $n = 20,560$, GF($2^{10}$)) of~\cite{kasai2011nonbinary, kasai2011quantum}, `low-density generator matrix (LDGM)-based SC-QLDPC' code ($R = 0.25$, $n = 76,800$) of~\cite{andriyanova2012spatially}, `QTC-assisted LDGM-based SC-QLDPC' code ($R = 0.25$, $n = 821,760$) of~\cite{maurice2013family} and `quantum-unity rate code (QURC) concatenated with quantum irregular convolutional code (QIrCC)' ($R = 0.5$, $n = 2,000$) of~\cite{babar2016serially}.}
    \label{fig:codes-comp-cap}
\end{center}
\end{figure}

In duality to the classical probabilistic coding theory, a random quantum code $\mathcal{C}$ having a sufficiently long codeword is expected to incur an arbitrarily low quantum bit error ratio (QBER) at a depolarizing probability of $p$, provided that its coding rate is below the hashing limit $C_Q(p)$ of~\eqr{eq:capacity_q}. Although having long codewords is not desirable in the quantum domain due to stringent latency requirements arising from the short qubit relaxation and dephasing times, the urge to approach the capacity triggered interest in probabilistic code designs. This led to the development of quantum low-density parity-check (QLDPC) codes~\cite{postol2001proposed, mackay2004sparse, camara2005constructions, camara2007class}, quantum convolutional codes (QCC)~\cite{ollivier2003description, ollivier2004quantum, forney2005simple, forney2007convolutional}, quantum turbo codes (QTC)~\cite{poulin2008quantum, poulin2009quantum}, quantum irregular convolutional codes (QIrCC)~\cite{babar2015road}, quantum polar codes (QPC)~\cite{renes2015efficient, babar2019polar}, and quantum unity-rate codes (QURC)~\cite{babar2016serially}. The performance of some of these codes is benchmarked against the hashing bound in~\fref{fig:codes-comp-cap}.

\begin{figure}[tb]
\begin{center}
    \includegraphics[width=\linewidth]{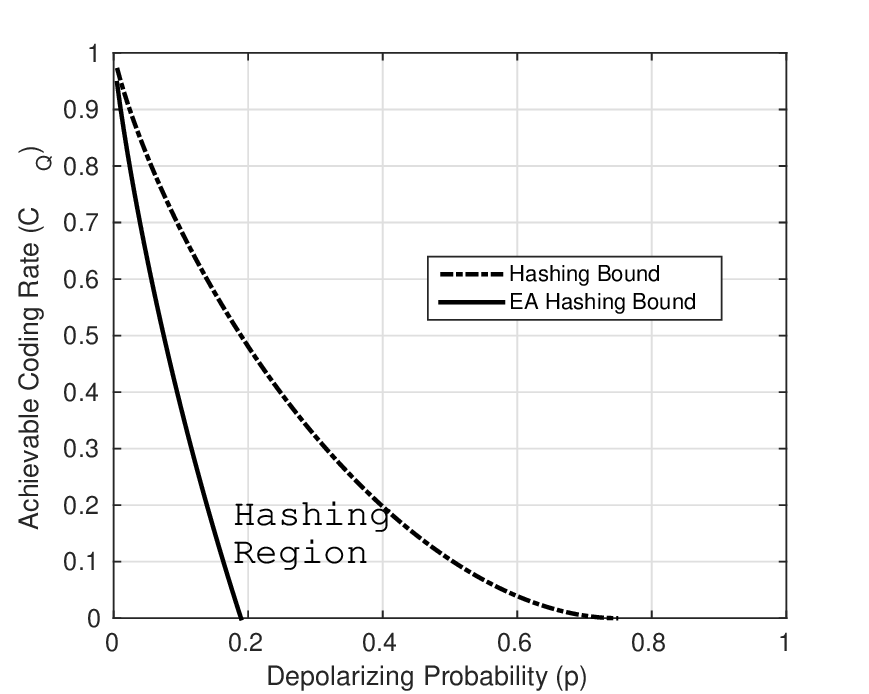}
    \caption{Hashing bounds for the unassisted and maximally-entangled quantum codes. The enclosed region, labeled the `hashing region', quantifies the capacity of EA codes with the varying number of pre-shared entangled qubits, i.e. $0 < c < (n - k)$~\cite{babar2018duality}.}
    \label{fig:cap-H}
\end{center}
\end{figure}

Recall that classical codes used for designing QSCs must satisfy the stringent symplectic product criterion. This symplectic criterion may sometimes lead to undesirable code properties, such as having short cycles in QLDPC codes, which degrade the decoding performance. Furthermore, not every classical code can be transformed into a quantum code owing to the symplectic criterion. For example, there is no quantum counterpart for a family of recursive non-catastrophic convolutional codes used for QTCs. Hence QTCs tend to exhibit a bounded minimum distance\footnote{If the minimum distance of a concatenated code with an interleaver increases almost linearly with the interleaver length, then it has unbounded minimum distance.}. To circumvent the symplectic product criterion of QSCs, entanglement-assisted (EA) quantum codes~\cite{bowen2002entanglement, brun2006correcting, brun2007general, hsieh2007general} were conceived, which rely on the availability of pre-shared entangled qubits. Since these pre-shared entangled qubits are generally assumed to be transmitted reliably before actual transmission, EA codes are capable of achieving a higher capacity. \fref{fig:cap-H} compares the hashing bound of the maximally-entangled\footnote{An EA code may have $0 \leq c \leq (n - k)$ pre-shared qubits. It reduces to an unassisted code when $c = 0$, while it is called a maximally-entangled code when $c = n - k$.} EA codes to that of their unassisted counterparts.

An EA code can operate anywhere in the hashing region, which is bounded by the maximally-entangled and the unassisted hashing limits. The notion of entanglement-assistance has been extended to nearly all quantum coding families, including for example EA-QLDPC codes~\cite{hsieh2009entanglement}, EA-QCCs~\cite{wilde2010entanglement}, EA-QTCs~\cite{wilde2011entanglement, wilde2013entanglement}, and EA-QPCs~\cite{wilde2012polar, wilde2013polar, renes2012efficient}.

\begin{figure}[tb]
    \centering
    \includegraphics[width=0.95\linewidth]{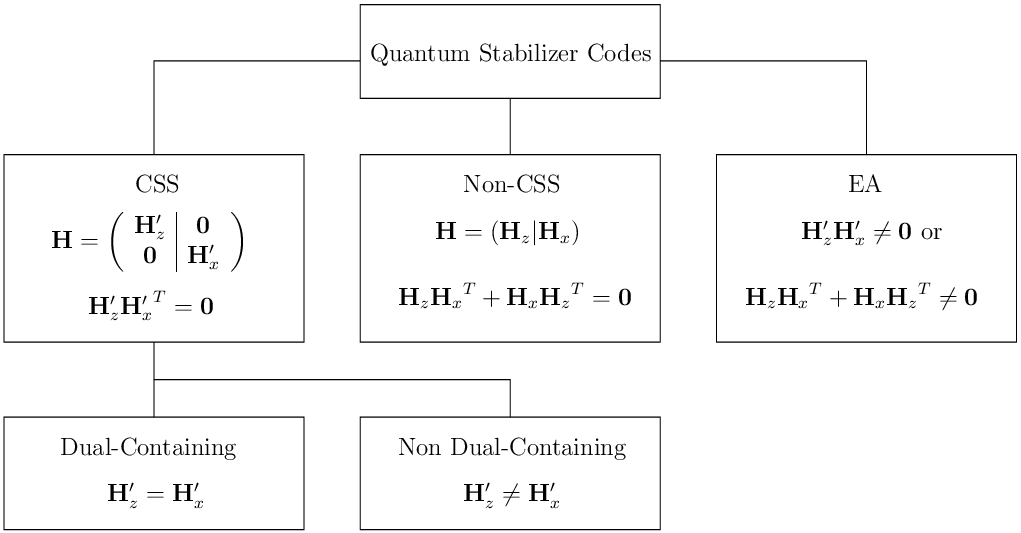}
    \caption{The taxonomy of quantum stabilizer codes based on their binary parity-check matrix representations~\cite{chandra2017quantum}.}
    \label{fig:classification}
\end{figure}

\begin{figure*}[!th]
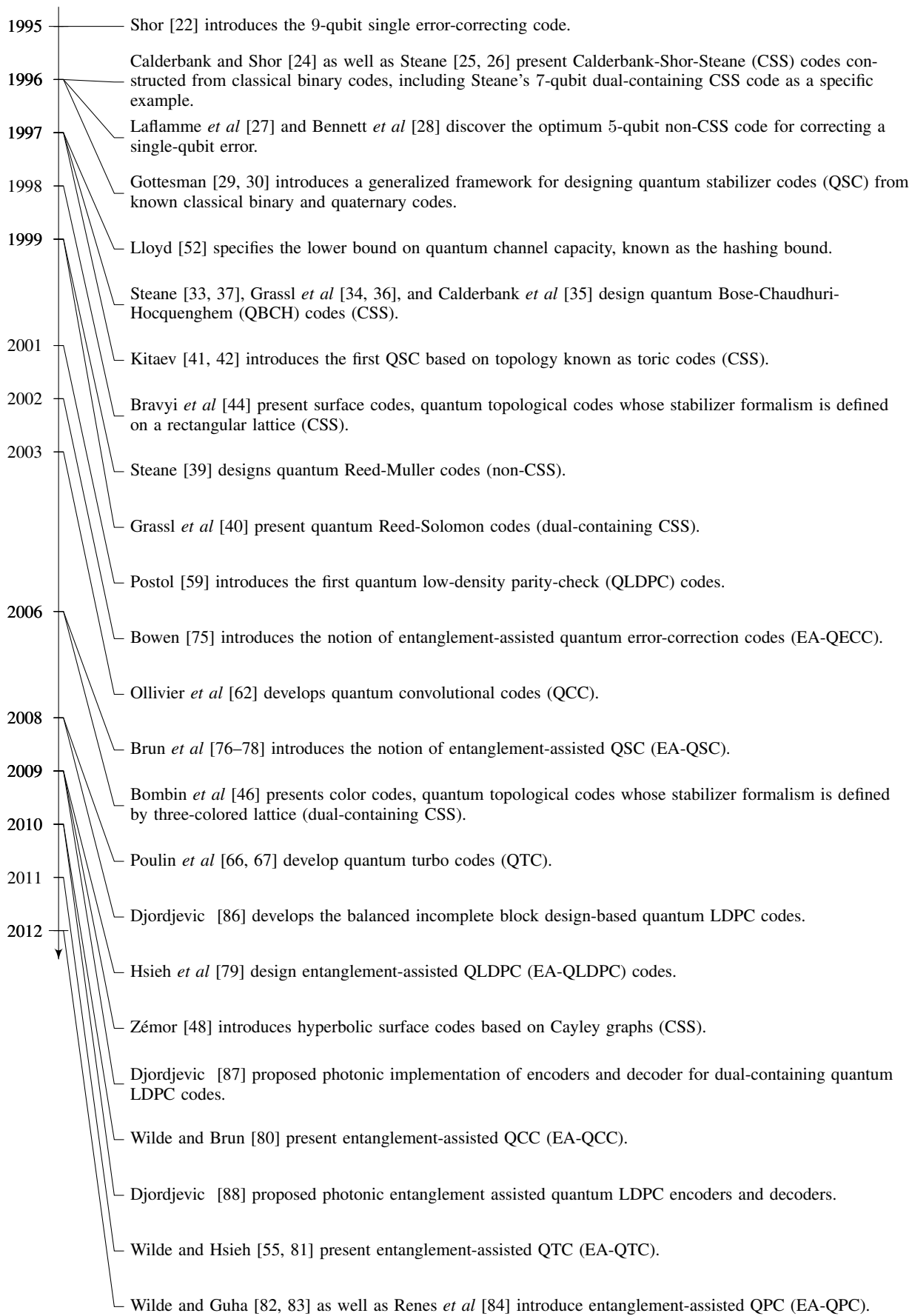

\begin{small}
\begin{timeline}{1995}{2012}{1cm}{1cm}{14cm}{0.7\textheight}
\entry{1995}{Shor~\cite{shor1995scheme} introduces the $9$-qubit single error-correcting code.}
\entry{1996}{Calderbank and Shor~\cite{calderbank1996good} as well as Steane~\cite{steane1996multiple, steane1996error} present Calderbank-Shor-Steane (CSS) codes constructed from classical binary codes, including Steane's $7$-qubit dual-containing CSS code as a specific example.}
\entry{1996}{Laflamme~\textit{et al}~\cite{laflamme1996perfect} and Bennett~\textit{et al}~\cite{bennett1996mixed} discover the optimum $5$-qubit non-CSS code for correcting a single-qubit error.}
\entry{1996}{Gottesman~\cite{gottesman1996class, gottesman1997stabilizer} introduces a generalized framework for designing quantum stabilizer codes (QSC) from known classical binary and quaternary codes.}
\entry{1997}{Lloyd~\cite{lloyd1997capacity} specifies the lower bound on quantum channel capacity, known as the hashing bound.}
\entry{1997}{Steane~\cite{steane1996simple, steane1999enlargement}, Grassl~\textit{et al}~\cite{grassl1997codes, grassl1999quantum1}, and Calderbank~\textit{et al}~\cite{calderbank1998quantum} design quantum Bose-Chaudhuri-Hocquenghem (QBCH) codes (CSS).}
\entry{1997}{Kitaev~\cite{kitaev1997quantum, kitaev2003fault} introduces the first QSC based on topology known as toric codes (CSS).}
\entry{1998}{Bravyi~\textit{et al}~\cite{bravyi1998quantum} present surface codes, quantum topological codes whose stabilizer formalism is defined on a rectangular lattice (CSS).}
\entry{1999}{Steane~\cite{steane1999quantum} designs quantum Reed-Muller codes (non-CSS).}
\entry{1999}{Grassl~\textit{et al}~\cite{grassl1999quantum2} present quantum Reed-Solomon codes (dual-containing CSS).}
\entry{2001}{Postol~\cite{postol2001proposed} introduces the first quantum low-density parity-check (QLDPC) codes.}
\entry{2002}{Bowen~\cite{bowen2002entanglement} introduces the notion of entanglement-assisted quantum error-correction codes (EA-QECC).}
\entry{2003}{Ollivier~\textit{et al}~\cite{ollivier2003description} develops quantum convolutional codes (QCC).}
\entry{2006}{Brun~\textit{et al}~\cite{brun2006correcting, brun2007general, hsieh2007general} introduces the notion of entanglement-assisted QSC (EA-QSC).}
\entry{2006}{Bombin~\textit{et al}~\cite{bombin2006topological} presents color codes, quantum topological codes whose stabilizer formalism is defined by three-colored lattice (dual-containing CSS).}
\entry{2008}{Poulin~\textit{et al}~\cite{poulin2008quantum, poulin2009quantum} develop quantum turbo codes (QTC).}
\entry{2008}{Djordjevic ~\cite{BIBD2008} develops the balanced incomplete block design-based quantum LDPC codes.}
\entry{2009}{Hsieh~\textit{et al}~\cite{hsieh2009entanglement} design entanglement-assisted QLDPC (EA-QLDPC) codes.}
\entry{2009}{Z{\'e}mor~\cite{zemor2009cayley} introduces hyperbolic surface codes
based on Cayley graphs (CSS).}
\entry{2009}{Djordjevic ~\cite{Photonic_QLDPC2009} proposed photonic implementation of encoders and decoder for dual-containing quantum LDPC codes.}
\entry{2010}{Wilde and Brun~\cite{wilde2010entanglement} present entanglement-assisted QCC (EA-QCC).}
\entry{2010}{Djordjevic ~\cite{Photonic_EA_QLDPC2010} proposed photonic entanglement assisted quantum LDPC encoders and decoders.}
\entry{2011}{Wilde and Hsieh~\cite{wilde2011entanglement, wilde2013entanglement} present entanglement-assisted QTC (EA-QTC).}
\entry{2012}{Wilde and Guha~\cite{wilde2012polar, wilde2013polar} as well as Renes~\textit{et al}~\cite{renes2012efficient} introduce entanglement-assisted QPC (EA-QPC).}
\end{timeline}
\end{small}
\caption{Timeline of quantum error-correction codes milestones (continued).} 
\label{fig:timeline:QECC-1}
\end{figure*}

To conclude this section, \fref{fig:classification} depicts the taxonomy of QSCs based on their binary PCM representations, and \fref{fig:timeline:QECC-1} as well as \fref{fig:timeline:QECC-2} portray the timeline of historical milestones in the field of QECC.

\begin{figure*}[!th]
\begin{small}
\begin{timeline}{2013}{2023}{1cm}{1cm}{14cm}{0.7\textheight}
\entry{2013}{Pelchat and Poulin~\cite{pelchat2013degenerate} introduce degenerate Viterbi decoding for QCC.}
\entry{2013}{Babar~\textit{et al}~\cite{babar2013near} present near-capacity code designs for entanglement-assisted classical communications.}
\entry{2013}{Delfosse~\cite{delfosse2013tradeoffs} designs hyperbolic color codes (dual-containing CSS).}
\entry{2013}{Tillich and Z{\'e}mor~\cite{tillich2013quantum} introduce hypergraph product codes (HPC) that always satisfy the symplectic product by taking two classical LDPC codes (CSS).}
\entry{2014}{Babar~\textit{et al}~\cite{babar2014exit} introduces the utilization of extrinsic information transfer (EXIT) charts for designing QTCs.}
\entry{2014}{Bravyi and Hastings~\cite{bravyi2014homological} design homology-inspired QLDPC codes known as homological product codes.}
\entry{2015}{Babar~\textit{et al}~\cite{babar2015road} develop quantum irregular convolutional codes (QIrCC) for concatenated quantum codes operating close to the hashing bound.}
\entry{2015}{Renes~\textit{et al}~\cite{renes2015efficient} introduce unassisted QPC.}
\entry{2015}{Leverrier~\textit{et al}~\cite{leverrier2015quantum} present the quantum counterparts of the classical expander codes.}
\entry{2016}{Babar~\textit{et al}~\cite{babar2016serially} develop quantum unity-rate codes (QURC) for facilitating the design of concatenated quantum codes.}
\entry{2016}{Babar~\textit{et al}~\cite{babar2016fully} conceive a fully-parallel quantum turbo decoder (FPQTD) with reduced decoding latency.}
\entry{2016}{Nguyen~\textit{et al}~\cite{nguyen2016exit} present the design of QTCs tailored for asymmetric Pauli channels.}
\entry{2019}{Panteleev and Kalachev~\cite{panteleev2021degenerate, panteleev2021quantum} introduce lifted-product code (LPC) the generalization of HPC (CSS).}
\entry{2019}{Chandra~\textit{et al}~\cite{chandra2019near, chandra2023exit} design multiple-rate QTC.}
\entry{2022}{Leverrier and Z{\'e}mor~\cite{leverrier2022quantum} conceive the quantum counterparts of the classical Tanner codes.}
\entry{2022}{Vuillot and Breuckmann~\cite{vuillot2022quantum} introduce a generalization of quantum color codes and Reed-Muller codes known as quantum pin codes.}
\entry{2022}{Roffe~\textit{et al}~\cite{roffe2023bias} present the design of QLDPC codes tailored for asymmetric Pauli channels.}
\entry{2022}{Christandl and M{\"u}ller-Hermes~\cite{christandl2022fault} introduce the notion of fault-tolerant QECC for quantum communications.}
\entry{2023}{Chandra~\textit{et al}~\cite{chandra2023universal} as well as Cruz~\textit{et al}~\cite{cruz2023quantum} propose a universal decoding scheme for QSCs inspired by guessing random additive noise decoding (GRAND).}
\end{timeline}
\end{small}
\caption{Timeline of quantum error-correction codes milestones.} 
\label{fig:timeline:QECC-2}
\end{figure*}

\subsection{Knowledge Gaps, Challenges}
\label{quantum-coding-knowledge-gaps-challenges}

The near-hashing-limit performance can be achieved by utilizing QSCs having a very long codeword. However, with state-of-the-art quantum technologies, the number of qubits that can be used to perform error correction is limited. Thus, finding a family of QSCs that exhibit excellent error-correction capability at short to moderate codeword length and are also efficiently decodable has become a more pivotal problem than ever. In contrast to the classical domain -- where error correction can be implemented using error-free encoding and decoding circuits -- the implementation of QSCs must contend with imperfect quantum gates for processing both the information and redundancy qubits. Quantum gates -- the fundamental building blocks for quantum circuits -- are not yet flawless and their imperfections tend to introduce more errors during the encoding and decoding process, potentially corrupting the information they are meant to safeguard. Therefore, the design and the implementation of QSCs have to correct qubit errors imposed both by external gates to be protected by the QECC and also by the gates used in the QECC encoding and decoding process. This requires the employment of a so-called fault-tolerant approach to QSCs~\cite{christandl2022fault}, where the encoding and decoding steps are intricately designed to limit the error proliferation within the quantum circuits.

To compound the issue further, qubits inherently suffer from short coherence times, which is the duration of maintaining their uncontaminated quantum state. This imposes a stringent time constraint during which all encoding and decoding algorithms must be completed. In light of this race against time, the fault-tolerant implementation of QSCs must operate within this limited coherence time frame to prevent the degradation of quantum information. If the qubits begin to decohere before the completion of the error correction cycle -- encoding, stabilizer measurements, decoding --, the information may become irretrievably lost, undermining the very purpose of quantum coding. This necessitates not only \emph{good} QSC constructions for short to moderate codeword lengths but also the development of reliable encoding and stabilizer measurements -– a method to measure the syndrome values of QSCs without collapsing the encoded quantum state -– as well as decoding algorithms. These decoders must be capable of high-fidelity operation even in the face of imperfect encoding and stabilizer measurements in order to make correct decoding decisions, thereby preserving the quantum states essential for reliable quantum communications.

Significant advances have been made in the field of code constructions, particularly in developing near-hashing-bound performances using only moderate codeword lengths. A notable development relies on harnessing a quantum-domain unity-rate code (QURC) in conjunction with quantum irregular convolutional codes (QIrCCs) to create near-hashing-limit QTCs. This approach, detailed in~\cite{babar2016serially}, achieves a remarkable error-correction performance without compromising the coding rate or the number of information qubits. A similar strategy that also achieves near-hashing-limit error correction performance is presented in~\cite{chandra2019near, chandra2023exit} upon substituting the QIrCCs by quantum short-block codes (QSBCs) as the outer codes of this serially concatenated arrangement.

Impressive progress has also been observed in the field of QLDPC codes. For instance, quasi-cyclic QLDPC code constructions having outstanding error-correction performance were proposed in~\cite{panteleev2021degenerate}, while spatially-coupled QLDPC codes were designed in~\cite{yang2023quantum}. Both these treatises demonstrate near-hashing-limit performance, despite using significantly shorter codewords than the earlier studies in~\cite{hagiwara2007quantum, hagiwara2011spatially, andriyanova2012spatially}. Efforts to develop unassisted QPCs began to unfold from~\cite{renes2015efficient}, which were then formalized in~\cite{babar2019polar}. However, these initial findings suggest that QPCs may not perform well for short to moderate codeword lengths~\cite{yi2022quantum, chandra2023universal}.

The standard encoding for QSCs as described in~\cite{cleve1997efficient} and the more specific encoding rules proposed for dual-containing CSS codes and detailed in~\cite{mackay2004sparse} are inherently not fault-tolerant. This means that a single quantum gate error in the encoder may proliferate the number of errors, which the QSCs are unable to correct. A fault-tolerant approach to encode information qubits has been demonstrated for the family of quantum topological error-correction codes (QTECCs), which involves performing a round of stabilizer measurements to encode the information qubits, as seen in~\cite{fowler2012surface}. However, it remains to be investigated, whether the same approach can result in fault-tolerant encoding for the popular families of probabilistic codes, such as QTCs, QPCs, and QLDPC codes.

In the realm of stabilizer measurements, substantial efforts have been dedicated to designing circuits having inherent capabilities for error detection and, in some cases, error correction, as detailed in~\cite{shor1996fault, steane1997active, steane2002fast}. However, these methods necessitate that the redundant qubits are \emph{encoded}, which in turn introduces additional overhead for preparing these fault-tolerant redundant qubits. Consequently, significant efforts have been channeled into minimizing the overhead associated with redundant qubits, while preserving their error detection capabilities~\cite{chao2018quantum} or into striking a balance between the overhead and the error-correction/error-detection capabilities of the encoded redundant qubits~\cite{huang2021between}.

Finally, since stabilizer measurements can provide additional error information, it is crucial for the decoder to integrate this information into making an informed correction decision. For QTECCs, the reliability of stabilizer measurements often requires multiple repetitions to enhance confidence in the measured syndrome values, as discussed in~\cite{dennis2002topological}. However, it also has been shown that some QLDPC codes inherently possess a \emph{single-shot} property, as described in~\cite{bombin2015single,campbell2019theory}. Notable results in QLDPC code construction and decoder pairs that exhibit high single-shot error-correction performance are highlighted in~\cite{breuckmann2021single, quintavalle2021single, grospellier2021combining, higgott2023improved, gu2023single}.

\subsection{Research Roadmap}
\label{quantum-coding-reseach-roadmap}

The field of fault-tolerant quantum communications encompasses diverse active research areas, each addressing critical challenges to achieve reliable and scalable quantum communications. Key areas of ongoing research include practical quantum code constructions and real-time decoding.

Creating or discovering a specific set of QSC constructions that are guaranteed to deliver good error-correction performance for various codeword lengths and coding rates is of the essence. At the time of writing, only QTCs have been able to offer this assurance, although promising methods of single-shot decoding remain largely unexplored. The preliminary findings pertain only to the application of single-shot decoding on QCCs~\cite{zeng2019quantum}. Conversely, while some QLDPC codes possess single-shot error correction capabilities, efficient techniques of generating QLDPC codes that maintain good performance across various codeword lengths and coding rates are still unknown.

Another significant factor in the design of QSCs for quantum communications is the time-varying nature of quantum channels~\cite{etxezarreta2021time}. Given that the limited number of qubits in state-of-the-art quantum technology must be frugally used, taking into account the noise levels or depolarizing probabilities of the quantum channels is crucial. Therefore, the development of adaptive QSCs, which allow their error correction capabilities to be adjusted in response to noise levels may be expected to lead to more efficient exploitation of the valuable qubits. Preliminary studies indicate the feasibility of designing such adaptive codes within the realms of QTCs~\cite{chandra2019near, chandra2023exit} and QLDPC codes~\cite{wang2022construction}.

To meet the demand for a fast and reliable decoder, enhancements to the speed of cutting-edge decoders can be achieved through parallelization and intelligent scheduling within the state-of-the-art decoding algorithm~\cite{babar2016fully, skoric2023parallel, battistel2023real}. Additionally, adopting relatively low-complexity decoding algorithms for short to moderate codeword lengths is another avenue that can be explored. This path has been recently considered in classical ultra-reliable low-latency communication (URLLC), as demonstrated in~\cite{yue2023efficient}, by investigating the potential of ordered-statistic decoding (OSD)~\cite{fossorier1995soft, yue2021probability} and guessing random additive noise decoding (GRAND)~\cite{duffy2019capacity}. In the quantum domain, the use of OSD has been explored primarily as a post-processing tool in~\cite{panteleev2021degenerate}, rather than as a standalone decoder. On the other hand, the first studies of GRAND as a standalone decoder have been disseminated in~\cite{chandra2023universal, cruz2023quantum, roque2023efficient}. GRAND is often referred to as a universal decoder, because it may be readily harnessed for the entire family of the above-mentioned CSS codes. Hence GRAND decoders may be expected to gain popularity and attract further research. 

QECCs may also be specifically designed for the amplitude damping
channel, which were investigated by harnessing the popular stabilizer
formalism in~\cite{4675715}, where improved error recovery operations
were conceived as a function of the amplitude damping probability.

Since the family of QECC schemes has to correct both bit-flips as well
as phase-flips, its members tend to have a low coding rate owing to
requiring numerous redundant bits. An attractive design alternative is
to harness sophisticated error mitigation techniques dispensing with
QECCs, as detailed in the next section.


\section{Mitigation of Decoherence Without Coding}
\label{qem-lh}
{\bf The Myth:} The impressive hardware developments of recent years
that led to quantum computers in excess of a hundred controllable
qubits are already enabling useful applications.

{\bf The Reality:} Unfortunately current devices are incapable of
quantum error correction and uncontrolled errors severely limit the
practical applicability of early quantum hardware.  Quantum error
mitigation is a major enabler in utilising noisy quantum devices but
practical quantum advantage, i.e., the point at which a quantum
computer solves a practical problem faster than any classical
supercomputer, remains to be demonstrated.

{\bf The Future:} Further theoretical as well as hardware developments
are required but it is anticipated that some form of early practical
quantum advantage may be achieved in the near term and the techniques
we are about to describe will be absolutely instrumental in this
endeavour.

{\bf Abstract:} {\em Quantum computers are becoming a reality at the time of
writing. However, hardware imperfections still overwhelm these devices
and fault-tolerant quantum devices relying on quantum error correction
codes require substantial hardware developments. Explicitly, a single
{\em logical qubit} may have to be encoded into a potentially large
number of {\em physical qubits} and one has to harness additional
complex measures such as {\em magic-state distillation} or the
constant monitoring of stabilisers -- which may be deemed prohibitive
with the aid of existing and near-term technology.

The limited number of qubits available in the near-term means that we
cannot use powerful QECCs for cleaning up all qubit errors. Hence, it
is an exciting challenge to find applications without full
fault-tolerance using noisy intermediate-scale quantum (NISQ) devices
in the near term or early fault-tolerant devices further down the
road. To circumvent the limitation of having a low number of qubits,
we have to resort to quantum error mitigation (QEM) techniques.  In
this section we highlight their basic concepts, historic evolution and
open research challenges.  }

\subsection{State-of-the-Art}
\subsubsection{Scope of QEM}

QEM may be viewed as a lower-complexity design alternative to the
QECCs discussed in Section~\ref{quantum-coding}, which does not
require significant qubit overhead and thus can be implemented already
at the time of writing relying on existing quantum devices.  However,
the significant difference is that, while QECCs spot errors and
correct them on the fly to recover the pure quantum states, QEM is
unable to correct errors in quantum states.  Rather, QEM mitigates the
effect of errors on average by measuring the associated expected
values through many consecutive activations of a quantum circuit.  In
particular, in near-term applications, such as in certain QML
applications discussed in Section~\ref{qem-lh}, one has to repeatedly
activate a quantum circuit many times to obtain an accurate estimate
of the particular cost function, which is often chosen to be the
expectation value $\tr [\rho O]$ of some observable $O$.

\begin{figure}[t]
	\centering
	\includegraphics[width=0.45\textwidth]{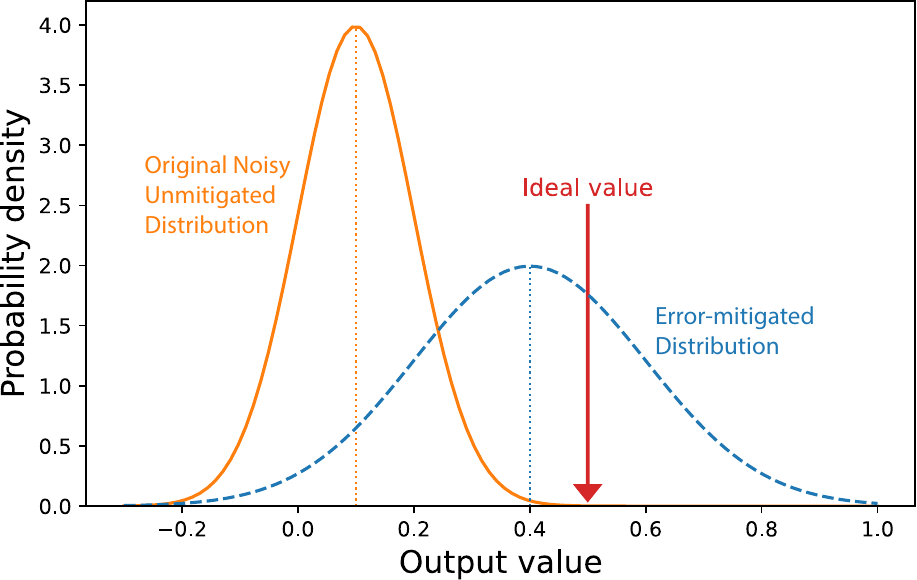}
	\caption{ Quantum error mitigation (QEM) techniques require
          drastically fewer quantum resources than Quantum Error
          Correction (QEC) but are limited to mitigating errors in
          expected value measurements: even with noise free qubits,
          one needs to repeat a quantum circuit many times to obtain
          an accurate expected value of an observable (red
          arrow). (Orange distribution) Due to errors affecting the
          qubits, the expected-value measurement yields a biased
          distribution (shifted mean) while QEM techniques generally
          aim to mitigate the effect of errors on average by reducing
          the bias, i.e., mean of blue distribution is closer to ideal
          value. QEM techniques come at the cost of an increased
          measurement count, i.e., width of blue distribution is
          larger.  }
	\label{fig:qem}
\end{figure}

We can illustrate this through an example whereby we run a circuit
once and measure a qubit which indeed yields an outcome of either $+1$
or $-1$.  Then the expected value of the Pauli operator $\langle Z
\rangle = \tr [\rho_{id} Z] = p_+ - p_-$ is estimated by repeated
activations of the same circuit through the relative frequency $p_\pm$
of the $\pm 1$ outcomes.  Even with noise-free qubits one obtains a
distribution for $\langle Z \rangle_{id}$ whose width is determined by
the number of samples $N_s$ as $N_s^{-1/2}$.  In a realistic noisy
quantum device, however, the centre of the distribution is biassed
$\langle Z \rangle$ as illustrated in Fig~\ref{fig:qem} (red arrow
vs. orange distribution).  Again, quantum error mitigation does not
correct errors in the individual circuit runs but rather introduces a
new distribution whose bias is reduced, see Fig~\ref{fig:qem} (blue
distribution).  However, this comes at the cost that the width of the
distribution following error-mitigation is increased. Hence one
requires an increased number of samples to get a sufficiently accurate
estimate of $\langle Z \rangle_{id}$.  As we discuss below, this
overhead quantified in terms of the number of samples increases
exponentially with the expected number of circuit errors. Nonetheless,
this is a tradeoff that is worth making in many practical scenarios:
error mitigation techniques may deliver accurate results at reasonable
measurement overheads, provided that the expected number of errors in
a quantum circuits is not significantly higher than 1.

\begin{figure*}[t]
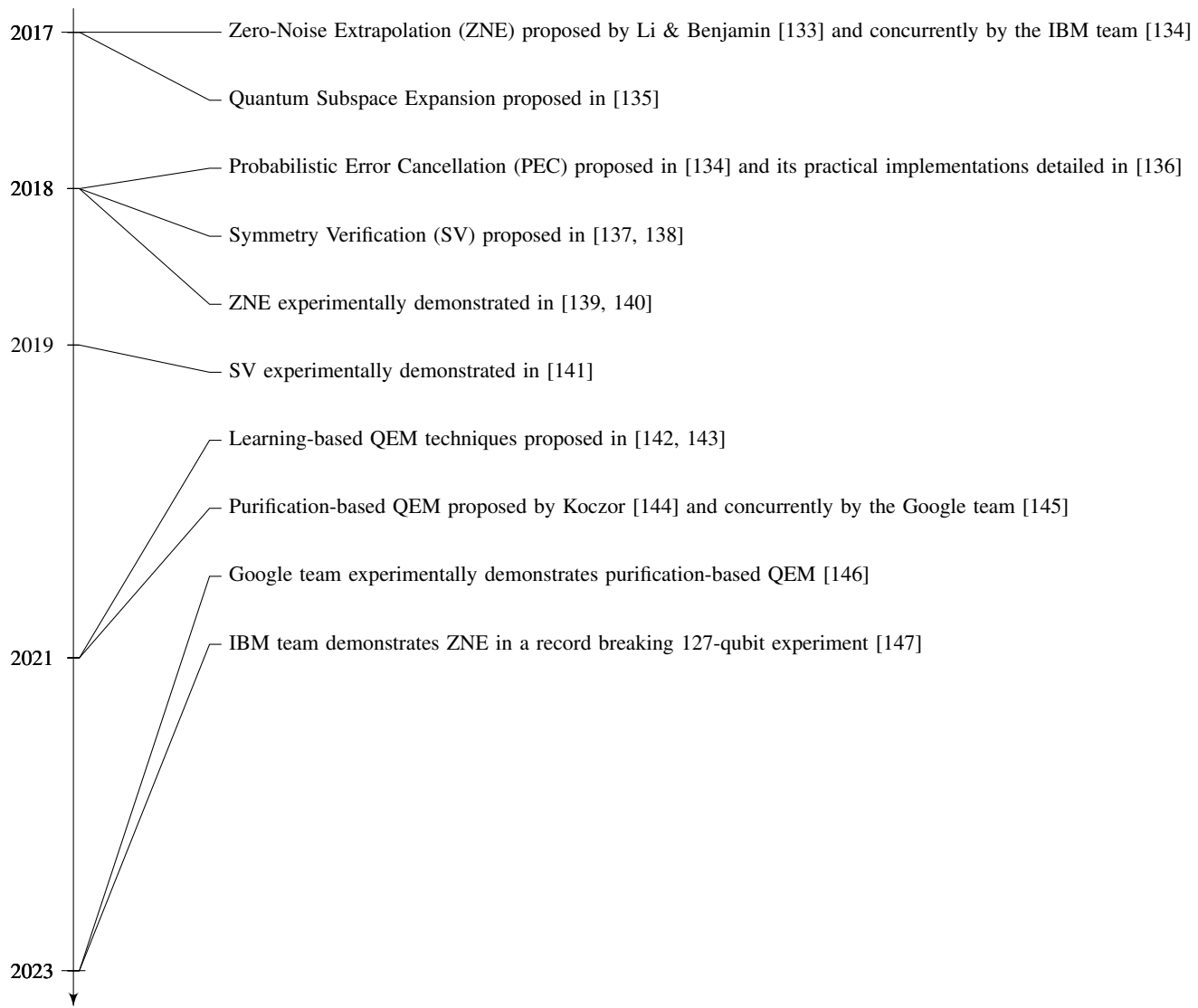

	\begin{small}
		\begin{timeline}{2017}{2023}{2cm}{2cm}{15cm}{0.6\textheight}
			\entry{2017}{Zero-Noise Extrapolation (ZNE) proposed by Li \& Benjamin~\cite{liEfficientVariationalQuantum2017}
				 and concurrently by the IBM team \cite{temmeErrorMitigationShortDepth2017}}
			\entry{2017}{Quantum Subspace Expansion proposed in \cite{mccleanHybridQuantumclassicalHierarchy2017}}
			\entry{2018}{Probabilistic Error Cancellation (PEC) proposed in \cite{temmeErrorMitigationShortDepth2017} and its
			practical implementations detailed in \cite{endoPracticalQuantumError2018}}
			\entry{2018}{Symmetry Verification (SV) proposed in \cite{mcardleErrorMitigatedDigitalQuantum2019,bonet-monroigLowcostErrorMitigation2018}}
			\entry{2018}{ZNE experimentally demonstrated in~\cite{dumitrescuCloudQuantumComputing2018,kandalaErrorMitigationExtends2019} }
			\entry{2019}{SV experimentally demonstrated in~\cite{sagastizabalExperimentalErrorMitigation2019} }
			\entry{2021}{Learning-based QEM techniques proposed in~\cite{czarnikErrorMitigationClifford2021,strikis2021learning}}
			\entry{2021}{Purification-based QEM proposed by Koczor~\cite{koczorExponentialErrorSuppression2021}
			and concurrently by the Google team~\cite{hugginsVirtualDistillationQuantum2021}}
			\entry{2023}{Google team experimentally demonstrates purification-based QEM~\cite{obrienPurificationbasedQuantumError2023}}
			\entry{2023}{IBM team demonstrates ZNE in a record breaking 127-qubit experiment~\cite{kimEvidenceUtilityQuantum2023}}
		\end{timeline}
	\end{small}
	\caption{Timeline of quantum error mitigation milestones.}\label{fig:timeline-qem}
\end{figure*}

Indeed the advantage of QEM is that at the current state-of-the-art it
requires drastically reduced quantum resources compared to QECC.  Due
to the high complexity of QECC, experimental demonstrations have been
limited to correcting a single logical qubit in a short codeword,
whereas QEM techniques have already been routinely applied in
experiments relying on hundred plus qubits.  QEM may be viewed as a
family of numerous diverse techniques and `tricks' and in the following
we review some of the most popular ones.

\subsubsection{Zero Noise Extrapolation}
The seminal contributions
\cite{liEfficientVariationalQuantum2017,temmeErrorMitigationShortDepth2017}
made the observation that by artificially increasing the error burden
on the qubits one can learn how the quantum-domain noise affects the
measured expected values and reduce the bias by extrapolating to zero
error rates. These constitute the family of Zero Noise
Extrapolation (ZNE) techniques.  In particular, by denoting the
average number of errors in each circuit run as $\lambda$, the output
state will be denoted by the density matrix $\rho_\lambda$. One then
aims for measuring a set of expected values $\langle O_k \rangle =
\Tr[\rho_{\lambda_k} O]$ at increased error rates
$\lambda_k$. Fitting a model---most typically a linear or exponential
function~\cite{endoPracticalQuantumError2018,caiMultiexponentialErrorExtrapolation2021}---to
the graph $\langle O_k \rangle$ {\em vs.} $\{\lambda_k\}$
allows us to estimate the expected value $\Tr[\rho_{\lambda
\rightarrow 0} O]$ at zero noise.

Several methods have been developed for artificially boosting the
error rates in a controllable fashion. The first of such methods increases
the pulse time used for implementing the gates in the
circuit~\cite{temmeErrorMitigationShortDepth2017}, which
simultaneously exposes the qubits to the decoherence mechanisms for a
longer period.  This led to the first successful demonstration of
quantum error mitigation in an experiment that achieved accurate
measurements of expected values in small-scale
systems~\cite{kandalaErrorMitigationExtends2019} using linear
extrapolation model functions.

Further methods for boosting errors have also been explored, such as
inserting sequences of gates and their inverses, which would ideally
combine into identity operations, but introduce additional errors due
to realistic imperfections~\cite{dumitrescuCloudQuantumComputing2018}.
Given that the ZNE techniques is very simple and they do not require
any advanced pre- or post-processing steps, ZNE has become one of the
most widely used QEM techniques. In a recent experiment ZNE was used
for demonstrating that it is indeed possible to arrive at accurate
expectation values in a 127-qubit quantum computer, with an accuracy
comparable to those of the most advanced classical approximate
simulation methods~\cite{kimEvidenceUtilityQuantum2023}.

\subsubsection{Probabilistic Error Cancellation}
Another family of error mitigation techniques relies on a
sophisticated Monte-Carlo scheme that probabilistically inserts
additional gates into each activation of the quantum circuit, so that
on average these additional gates can cancel out the damage inflicted
by the errors in the expectation values.  These techniques belong to
the family of Probabilistic Error Cancellation (PEC) solutions.

This approach was originally introduced in
\cite{temmeErrorMitigationShortDepth2017}. We can consider the simple
example of trying to prepare the ideal state $\rho_0$, but we end up
with the noisy state $\rho = \mathcal{N}_X(\rho_0) = (1-p) \rho_0 + p
X \rho_0 X$ which was corrupted by the bit-flip channel
introduced in Section~\ref{quantum-coding}. We can also
express the inverse of this noise process as:
\begin{equation}\label{eq:quasiprob}
	\rho_0 = \mathcal{N}_X^{-1} \rho = \gamma_1 \rho + \gamma_2 X \rho X,
\end{equation}
where we have $\gamma_1 = \frac{1-p}{1-2 p}$ and $\gamma_2 = - \frac{p}{1-
  2p}$.  While $\mathcal{N}_X^{-1} $ does not represent a channel that
can be physically implemented, we are actually interested in the
expectation value of some observable $O$ as
\begin{align*}
	\Tr(O\rho_0) = \gamma_1 \Tr(O\rho) + \gamma_2 \Tr(OX \rho X).
\end{align*}
The ideal expected value $\Tr(O\rho_0)$ is thus simply a linear
combination of the expectation value measured in the noisy state
$\rho$ and the expectation value measured in the noisy state $X\rho X$
to which we apply an additional $X$ gate.

The corresponding Monte-Carlo sampling scheme may be formulated by re-writing
the above linear combination as
\begin{align*}
	\Tr(O\rho_0) &=  p_1 (\abs{\gamma_1} + \abs{\gamma_2} )  \text{sgn}(\gamma_1) \Tr(O\rho) \\
	 &+p_2 (\abs{\gamma_1} + \abs{\gamma_2} ) \text{sgn}(\gamma_2) \Tr(O X \rho X)
	.
\end{align*}
The above may be interpreted as obtaining the ideal expectation value
by activating the original circuit with the probability of $p_1 =
\frac{\abs{\gamma_1}}{\abs{\gamma_1} + \abs{\gamma_2}}$ and modifying
the output with the aid of both the sign $\text{sgn}(\gamma_1)$ and
the scaling factor $(\abs{\gamma_1} + \abs{\gamma_2} )$, while the
circuit associated with the additional $X$ gate is chosen with the
probability of $p_2 = \frac{\abs{\gamma_2}}{\abs{\gamma_1} +
  \abs{\gamma_2}}$ and its output is modified both by the sign
$\text{sgn}(\gamma_2)$ and the scaling factor $(\abs{\gamma_1} +
\abs{\gamma_2} )$.  Hence, we simulate the effect of
$\mathcal{N}_X^{-1}$ by probabilistically applying an additional $X$
gate to the circuit while also adjusting the sign of the output
accordingly.

The above arguments may be readily extended beyond bit-flip
impairments, and they still hold even if the additional gates applied
are imperfect~\cite{endoPracticalQuantumError2018}, provided that we
have accurate knowledge of the nature of errors.  However, PEC
requires full knowledge of the specific form of the channel imposing
qubit errors, which is classically intractable for circuits that are
applied to a large number of qubits.  Hence, in practice, PEC is often
directly applied to individual gates in the circuit, whose noise
models can be accurately characterised. Instead of completely removing
all noise, this approach is only capable of mitigating the deleterious
effects of noise contributions.  This philosophy may be harnessed for
obtaining data points of reduced noise for zero-noise extrapolation,
which can be more effective than extrapolating with boosted noise
levels~\cite{caiMultiexponentialErrorExtrapolation2021}.

Measurement error mitigation is closely connected to probabilistic
error cancellation. At the measurement stage, both the ideal and the
noisy output states are transformed into the corresponding noisy and
ideal probabilistic distribution of bit strings, respectively, denoted
as $\vec{p}_{0}$ and $\vec{p}_{\mathrm{noi}}$. Hence, the channel
representing the impairments simply becomes a stochastic matrix
(transition matrix) that transforms the ideal distribution
$\vec{p}_{0}$ into the noisy distribution
$\vec{p}_{\mathrm{noi}}$~\cite{chowDetectingHighlyEntangled2010} as
\begin{align}\label{eqn:inverse_prob}
    \vec{p}_{\mathrm{noi}} = A \vec{p}_{0}.
\end{align}
The effect of noise can thus be removed through inverting the
transition matrix as $\vec{p}_0 = A^{-1}
\vec{p}_{\mathrm{noi}}$~\cite{kandalaHardwareefficientVariationalQuantum2017}.
Similarly to PEC, fully characterising the noise matrix $A$ becomes
intractable, as the circuits are scaled up and one thus has to make
assumptions concerning the noise model to simplify the form of $A$.
For example, we can assume that the measurement noise corrupting the
individual qubits is uncorrelated, which means that the global noise
matrix $A$ is simply a tensor product of single-qubit measurement
noise matrices.

\begin{figure}[t]
	\centering
	\includegraphics[width=0.45\textwidth]{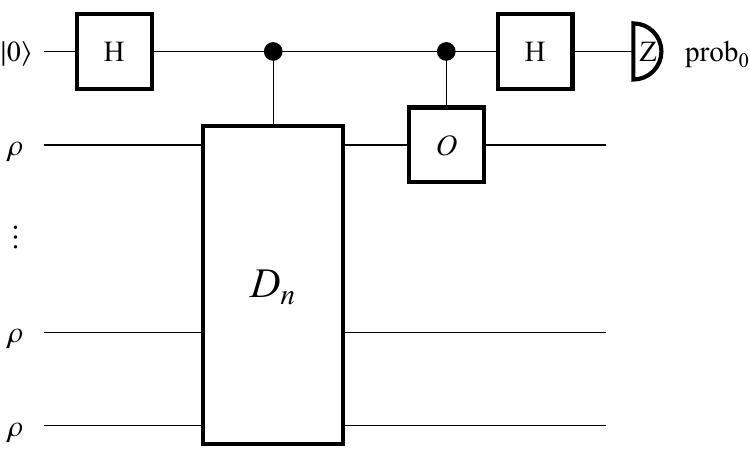}
	\caption{Purification-based techniques prepare $n$ copies 
	of a noisy quantum state $\rho$ either by splitting an array of qubits into
	batches~\cite{obrienPurificationbasedQuantumError2023} or by using multiple quantum cores for state preparation~\cite{PhysRevApplied.18.044064}.
	A derangement $D_n$ is a generalisation of the SWAP gate which applies a permutation to the qubits
	controlled on an ancilla qubit (topmost qubit).
	Estimating the probability of the $|0\rangle$ outcome allows one to suppress
	errors in the expected value measurement of the operator $\sigma$ exponentially in $n$.
	Reproduced from~\cite{koczorExponentialErrorSuppression2021} (CC BY 4.0).
	}
	\label{fig:derangement}
\end{figure}

\subsubsection{Purification-based Error Mitigation}
The QEM techniques we detailed so far were clearly distinct from QECCs
in that they did not introduce any redundant qubits, but rather made
use of noisy outputs that were linearly combined in a post-processing
stage. By contrast, purification-based techniques in a way share more
similarities with QECCs, since they also introduce redundancy --
but as we will describe now, the approach only requires a handful of
copies of the qubits as opposed to potentially thousands in the case
of QECCs.

Recall that a noisy quantum state $\rho$, which we describe by a
density matrix, can be diagonalised to obtain probabilities $p_k$ as
eigenvalues and pure states $| \psi_k \rangle$ as eigenvectors.
Purification-based techniques commonly exploit that, somewhat
surprisingly, the dominant eigenvector $| \psi_1 \rangle$ of the
density matrix is a good approximation of the noise-free quantum
state~\cite{koczor2021dominant}.  Thus, one proceeds by preparing
multiple copies of the state as $\rho^{\otimes
  n}$~\cite{koczorExponentialErrorSuppression2021,hugginsVirtualDistillationQuantum2021}.
By performing a cyclic permutation operation, or more generally by
creating a derangement
operation~\cite{koczorExponentialErrorSuppression2021}, one can
guarantee that only permutation-symmetric combinations such as $| \psi_k
\rangle^{\otimes n}$ contribute to the measurement process, which
happen with probabilities of $(p_k)^n$. Thus, low probabilities are
exponentially suppressed. Hence this technique acts similarly to a
high-pass filter that passes the dominant component $p_1 \gg p_k$ but
drastically attenuates low-probability ``error events'' associated
with $k>1$~\cite{9638483}.

More specifically, this approach takes $n$ independently prepared
copies of the computational state and applies to it a derangement
circuit that is controlled by an ancilla qubit, as illustrated in
Fig~\ref{fig:derangement}.  Measuring the probability of the
$|0\rangle$ outcome of the ancilla qubit allows one to formally
estimate the expected value~\cite{9638483}:
\begin{equation*}
	\frac{\tr[\rho^n \sigma]}{\tr[\rho^n]}
= \frac{1}{\sum_{k=1}^{2^N} (p_k)^n}    \sum_{k=1}^{2^N} (p_k)^n  \langle\psi_k |\sigma | \psi_k \rangle,	
\end{equation*}
which indeed gets exponentially close to the desired, ``noise-free'' expected value $\langle\psi_1 |\sigma | \psi_1 \rangle$ as we increase $n$.

This approach has the significant advantages that one does not have to
know the noise model of the quantum device and that there are various
different ways of physically implementing the derangement operations.
As such, many variants have been proposed that explore various
tradeoffs.  First, the approach was demonstrated in an experiment
whereby an array of qubits in Google's Sycamore quantum processor was
divided into two
halves~\cite{obrienPurificationbasedQuantumError2023}; the same noisy
state preparation circuit was applied to both halves of the qubit
array and finally the qubits were measured in the Bell basis -- this
variant only applies to $n=2$ copies but forgoes the challenging
controlled operations on an ancilla qubit.  A similar approach --
termed as Echo Verification (EV) -- was also conceived, whereby one
does not introduce any redundant qubits, but rather the redundancy is
introduced in the time-domain by consecutively activating the state
preparation circuit and its
inverse~\cite{huoDualstatePurificationPractical2022,obrienPurificationbasedQuantumError2023}.
Further generalisations have been explored whereby multiple copies are
used both in the space- and
time-domain~\cite{caiResourceefficientPurificationbasedQuantum2021}.

It was also pointed out that bespoke architectures should be designed
for optimally accommodating purification-based
techniques~\cite{PhysRevApplied.18.044064}. Semiconductor based qubit
technology constitutes a promising platform for cost-effectively
manufacturing a multitude of quantum cores on a single chip. Each of
these quantum cores could then be used for implementing the same
quantum circuits in parallel. One would then apply the derangement
operator in a distributed mode through a linking of the separate
quantum cores via quantum communication channels. Ion trap quantum
computers are similarly promising and indeed entangling operations
across the quantum cores have been demonstrated experimentally
in~\cite{PhysRevLett.124.110501}. For hardware platforms relying on
semiconductor spins and trapped ions, one can also use the pipeline
architecture of~\cite{caiLoopedPipelinesEnabling2023} which allows
both for native parallel processing of multiple copies and for
transversal operations between the copies, thus dispensing with the
space-time overhead associated with the purification-based QEM.

\subsubsection{Symmetry Verification}

In many quantum algorithms, such as in QML applications of
Section~\ref{qem-lh}, the problem considered exhibits inherent
symmetries, which forces symmetry constraints on the target output
state.  We can exploit these symmetries by performing measurements
that verify the symmetry of the output state and thus project the
output state to the subspace of correct
symmetry~\cite{mcardleErrorMitigatedDigitalQuantum2019,bonet-monroigLowcostErrorMitigation2018,10048485}.
These techniques are quite powerful, since they allow us to mitigate
all noise sources that break the symmetry of the output state. A
simple example is represented by the family of variational algorithms
that are used for simulating physical systems. In this context the
Jordan-Wigner encoding simply states that the number of particles
corresponds to the Hamming weight of bitstrings modulo a known
constant shift.  Exploiting this allows us to discard all measurement
outcomes of incorrect Hamming weight, which hence mitigates the effect
of quantum impairments.

In practice, Pauli symmetry operators are straightforward to implement,
whereby the target output state $\ket{\psi}$ is in the $+1$ subspace
of the symmetry operator $S$ via the eigenvalue equation
\begin{align*}
	S \ket{\psi} = \ket{\psi}.
\end{align*}
Non-Pauli symmetry operators can often be transformed into a related
Pauli symmetry. For example, instead of measuring the number of
particles, we can measure the odd or even parity of this number, which
indeed corresponds to the action of a Pauli operator.  Then, by
performing the measurement of the symmetry operator in a
non-destructive way and post-selecting the $+1$ outcome, one
effectively projects the noisy output state to the subspace of correct
symmetry as
\begin{align}\label{eqn:sym_state}
	\rho_{\mathrm{sym}} = \frac{\Pi_S \rho \Pi_S}{\Tr(\Pi_S \rho)}
\end{align}
where $\Pi_S = \frac{1 + S}{2}$ is the projector to the $+1$ subspace of $S$.

The symmetry operator $S$ is often a global property of the state and
thus may be challenging to measure, since the measurements in quantum
devices are typically related to single qubits.  A potential
workaround is to transform $S$ into a single-qubit measurement through
a basis-change (global Clifford circuit) which, however, may be too
costly to implement in near-term devices.  Another design alternative
exploits the fact that $S$ is a tensor product of single-qubit Pauli
operators $S = \bigotimes_{n = 1}^N S_n$ that we can directly measure;
one then multiplies the individual outcomes to obtain the desired
output in post-processing.  Of course, the output state in the
individual runs will differ from Eq~\ref{eqn:sym_state}.

Furthermore, one often only cares about obtaining the expectation
value of a Pauli observable $O = \bigotimes_{n = 1}^N O_n$ with
respect to the correct symmetries $\Tr(O\rho_{\mathrm{sym}})$, which
allows us to further simplify the problem.  If $O$ commutes qubit-wise 
with $S$ for each qubit, i.e. we have $[S_n, O_n] = 0$ for all $n$, then we
can simply measure $O$ on the post-selected state by measuring all of
its constituent single-qubit Paulis $\{O_n\}$ and then multiply the
individual outcomes together.  As such, performing only local
measurements to obtain $S$ and $O$ via post-processing is effectively
equivalent to post-selecting the $+1$ outcome of $S$ as long as $S$
and $O$ commute qubit-wise and the aim is to estimate the expectation
value $\Tr(O\rho_{\mathrm{sym}})$.  Given that post-selection implies
that only a fraction $\Tr(\Pi_S \rho)$ of the circuit runs is retained,
i.e., the ``useful'' circuit runs that pass the test, the sampling
overhead of this method depends on $\Tr(\Pi_S \rho)^{-1}$.

On the other hand, if $S$ and $O$ do not commute qubit-wise, we
rewrite the 'symmetrised' expectation value in Eq~\ref{eqn:sym_state}
as
\begin{equation*}
	\Tr(O\rho_{\mathrm{sym}}) = \frac{\Tr(O \rho) + \Tr(SO\rho) + \Tr(OS\rho ) + \Tr(SOS\rho )}{2 + 2\Tr(S\rho)}
\end{equation*}
which can be obtained by post-processing the expectation values of $O$,
$S$, $SO$, $OS$ and $SOS$.  If both $S$ and $O$ are Pauli strings, then
their products are Pauli strings too, thus the desired expected values
can be obtained through single-qubit Pauli measurements and 
post-processing, as detailed above.  The measurement overhead of this
approach then depends on $\Tr(\Pi_S
\rho)^{-2}$~\cite{hugginsEfficientNoiseResilient2021,caiQuantumErrorMitigation2021}.

Beyond exploiting the inherent symmetries of the problem, one can
similarly exploit symmetries that arise from encoding particles, such
as fermions into
qubits~\cite{bravyiFermionicQuantumComputation2002,setiaBravyiKitaevSuperfastSimulation2018,derbyCompactFermionQubit2021,jiangMajoranaLoopStabilizer2019},
or artificially constructed symmetries like parity checks in
QECC~\cite{mccleanDecodingQuantumErrors2020}.


\subsubsection{Quantum Subspace Expansion}
The most typical problem in QML -- which was the subject of
Section~\ref{qem-lh} -- is to prepare the eigenvector $\ket{\psi_0}$
of a Hamiltonian $H$ matrix of smallest eigenvalue that is often
termed as the ground state.  Due to circuit-depth limitations,
however, one can only prepare an approximation $\ket{\phi_0}$.  In
Quantum Subspace
Expansion~\cite{mccleanHybridQuantumclassicalHierarchy2017} one
proceeds by applying a set of operators $\{G_m\}$ to the approximate
state. The resultant collection of vectors $\{G_m\ket{\phi_0}\}$ spans
a subspace of the full Hilbert space and allows one to efficiently
find an improved approximation within this subspace.

A general state within this subspace can be written as
$\ket{\psi_{\vec{w}}} = \sum_{m = 1}^{M} w_m G_m \ket{\phi_0}$ with
the weights $\vec{w}$ chosen for ensuring that we have
$\braket{\psi_{\vec{w}}} = 1$. In this way, we can identify the
optimal ground state within the subspace by solving
\begin{align}\label{eqn:pure_state_expansion}
    \vec{w}^* = \text{arg\,min}_{\vec{w}} \bra{\psi_{\vec{w}}} H \ket{\psi_{\vec{w}}} \quad\quad \text{s.t. \ } \braket{\psi_{\vec{w}}}  = 1.
\end{align}
The above optimisation problem can be solved efficiently through
defining the matrices
\begin{align}
    \overline{H}_{ij} = \bra{\phi_i} H \ket{\phi_j}, \quad\quad \overline{S}_{ij} = \braket{\phi_i}{\phi_j},
\end{align}
and solving $\overline{H} \, W = \overline{S} \, W E$ which is often
referred to as the generalised eigenvalue equation.  The eigenvector
in $W$ that corresponds to the lowest eigenvalue in $E$ is the optimal
of coefficient vector $\vec{w}^*$ that solves the problem in
Eq~\ref{eqn:pure_state_expansion}.  Given that the approximation prepared
by the quantum device $\ket{\phi_0}$ is within the subspace, the
optimal ground state $\ket{\psi_{\vec{w}^*}}$ is guaranteed to be an
approximation no worse than $\ket{\phi_0}$.

In practice, one proceeds by preparing $\ket{\phi_0}$ with a quantum device and estimates
the matrix entries
\begin{equation}\label{eqn:expand_matrix}
    \begin{aligned}
        \overline{H}_{ij} &= \bra{\phi_0} G_i^\dagger H G_j \ket{\phi_0} = \Tr[G_i^\dagger H G_j\rho ] \\
        \overline{S}_{ij} &= \bra{\phi_0} G_i^\dagger G_j \ket{\phi_0} = \Tr[G_i^\dagger G_j\rho ]
    \end{aligned}
\end{equation}
for all $i$ and $j$.  The right-hand sides above define the matrix
elements in terms of measurements applied to a density matrix $\rho =
\ketbra{\phi_0}$ and this allows us to generalise the approach to
mixed states $\rho$, since the subspace expansion technique may be
applied to circuits undergoing stochastic
noise~\cite{mccleanHybridQuantumclassicalHierarchy2017,collessComputationMolecularSpectra2018,sagastizabalExperimentalErrorMitigation2019,urbanekChemistryQuantumComputers2020}.

Alluding briefly to the context of quantum chemistry, a possible
choice of the expansion operators $\{G_m\}$ are constituted by
products of fermionic raising and lowering
operators~\cite{mccleanHybridQuantumclassicalHierarchy2017} in line
with classical configuration interaction (CI) techniques.
Improvements to the ground state can also be achieved by applying
powers of $H$~\cite{mottaDeterminingEigenstatesThermal2020} or
$\rho$~\cite{yoshiokaGeneralizedQuantumSubspace2022} to the probe
state.  Furthermore, symmetry verification may actually be viewed as a
specific instance of subspace expansion using symmetry
operators~\cite{mccleanDecodingQuantumErrors2020,caiQuantumErrorMitigation2021}
and this can also be further linked to purification-based
methods~\cite{caiQuantumErrorMitigation2021}.

\subsubsection{Learning-Based Error Mitigation}
As we briefly alluded to in Section~\ref{qem-lh}, classical
machine-learning techniques may be employed for improving error
mitigation
techniques~\cite{czarnikErrorMitigationClifford2021,strikis2021learning}.  In particular, one is
typically interested in the expected value at the output
$E(\mathbf{C}_{0})$ of an ideal circuit $\mathbf{C}_{0}$, but only has
access to the noisy circuit $\mathbf{C}_{\mathrm{noi}}$ and its noisy
output expectation $E(\mathbf{C}_{\mathrm{noi}})$.  Then our aim is to
estimate a more accurate expected value
$f_{\vec{\alpha}}(\mathbf{C}_{\mathrm{noi}})$ through an error
mitigation strategy that results in reduced distance from the ideal
value formulated as $\abs{f_{\vec{\alpha}}(\mathbf{C}_{\mathrm{noi}})
  - E(\mathbf{C}_{0})} < \abs{\mathbf{C}_{\mathrm{noi}} -
  E(\mathbf{C}_{0})} $.

However, our model depends on a set of unknown parameters
$\vec{\alpha}$, which we can find by training.  Thus, one aims for
constructing training circuits $\mathbf{T}_{0}$, which resemble to
$\mathbf{C}_{0}$ in terms of their circuit structure, but can be
efficiently simulated classically, i.e. we can readily calculate
$E(\mathbf{T}_{0})$.  We then activate the circuit (and any of its
variants) on the quantum machine---which individually result in noisy
expected values---and feed them to our model. The model is then
trained on these noisy expected values in order to find the optimal
parameters $\vec{\alpha}^*$, which minimise the distance from the
ideal output formulated as
$\abs{f_{\vec{\alpha}}(\mathbf{T}_{\mathrm{noi}}) -
  E(\mathbf{T}_{0})}$.  Upon applying this optimised model to our
circuit of interest we obtain the desired error-mitigated estimates
$f_{\vec{\alpha}^*}(\mathbf{C}_{\mathrm{noi}})$.

The simplest such error-mitigation model employs a scaling and
shifting of the noisy expectated value, which is expressed as
$f_{\vec{\alpha}}(\mathbf{C}_{\mathrm{noi}}) = \alpha_0 + \alpha_1
E(\mathbf{C}_{\mathrm{noi}})$~\cite{czarnikErrorMitigationClifford2021,foldager2023can}. The
potential training circuits to use include the family of Clifford gate
variants of the original
circuit~\cite{czarnikErrorMitigationClifford2021,strikis2021learning}
and the so-called free fermion circuits
of~\cite{montanaro2021error}. One can also introduce parametrised
models into other QEM methods and apply learning-based
techniques. This has been demonstrated for PEC in order to account for
the presence of correlated quantum noise~\cite{strikis2021learning}.

The historic evolution of QEM is depicted at a glance in
Figure~\ref{fig:timeline-qem}, which assists us in identifying the
knowledge gaps in the next subsection.

\subsection{Knowledge Gaps and Challenges}
We now continue by highlighting some of the open challenges and main
limitations of the above-mentioned QEM techniques and we also identify
a number of knowledge gaps.

\subsubsection{Sampling overhead}
Let us now detail the main limitation of QEM, which is the increased
variance of the expected
values~\cite{caiMultiexponentialErrorExtrapolation2021,caiPracticalFrameworkQuantum2021,quek2022exponentially,9294106,9684862}.
The average number of errors occurring in the quantum
circuit $\lambda$ -- which may also be termed as the circuit fault
rate -- is roughly the sum of the error rates of all the gates in the
circuit. For example, if we have a circuit constructed from $N$ gates
and each gate has an error rate of $p$, then the circuit fault rate
will be $\lambda = Np$. For stochastic errors, the number of errors in
the circuit follows a Poisson distribution, for which the probability
of $\ell$ errors occurring is $e^{-\lambda}
\lambda^{\ell}/\ell!$. Hence the fraction of circuit activations that
are noiseless ($\ell = 0$) is
$e^{-\lambda}$~\cite{caiMultiexponentialErrorExtrapolation2021,caiPracticalFrameworkQuantum2021}.
We may surmise from this simplified picture that to obtain the same
amount of information that is contained in a single noise-free circuit
activation we need approximately $\sim e^{\lambda}$ noisy circuit
executions. Clearly, this represents the sampling overhead associated
with QEM.  However, QEM techniques typically fail to remove all noise
contributions, thus the above argument only applies to the specific
fraction of error contributions that we succeed in removing: with
$\lambda_{\mathrm{rm}}$ representing the average number of removable
errors in the circuit for a given QEM technique, the associated
sampling overhead is $\sim e^{\lambda_{\mathrm{rm}}}$, while the
effective noise level of the resultant circuit become $\lambda -
\lambda_{\mathrm{rm}}$.

This illustrates that indeed in general \emph{the sampling overhead
  imposed by QEM grows exponentially with the number of errors in the
  circuit that are removed by a given QEM technique}.  These overheads
are usually specified as upper bounds, and thus account for worst-case
scenarios. However, the actual overhead of some of the most potent
error mitigation techniques can be
lower~\cite{kimEvidenceUtilityQuantum2023}.  On the other hand, given
that the sampling overhead grows exponentially with the circuit size,
in contrast to QECC, QEM is not scalable to arbitrary system sizes.
Thus in practice, QEM techniques are applicable in the regime, where
the number of removable errors is on the order of $\lambda \approx 1$
or lower. This ensures that the associated sampling overhead remains
reasonable.

\subsubsection{QEM Architectures}
Advanced QEM techniques, such as purification-based ones, require
multiple copies of the quantum state.  As we detailed above, these can
be prepared either by splitting a large qubit array into multiple
regions or by actually assigning the same quantum circuit to multiple,
physically separate quantum processors that are interlinked in the
quantum domain~\cite{PhysRevApplied.18.044064,PhysRevLett.124.110501}.

While fascinating progress has been made, improving the quantum
communication channels -- such as the depolarizing channel
characterizing the circuit imperfections, rather than for example
free-space optical satellite channels -- is an immense engineering
challenge.  This may require for example Bell-pair purification
synchronisation across multiple quantum
cores~\cite{PhysRevApplied.18.044064}.  However, multicore
architectures are likely to deliver value even beyond QEM and may be a
substrate for the so-called SWAP-test~\cite{PhysRevApplied.18.044064}
which is an important subroutine for many of the QML algorithms
mentioned in Section~\ref{qem-lh}.  Furthermore, multicore
architectures will enable noise-resilient implementations of advanced,
gradient-based training algorithms for conceived for QML, such as the
quantum natural gradient techniques
of~\cite{koczor2022quantum,stokes2020quantum}.

\subsubsection{Learning and controlling noise models}

The above QEM techniques tend to rely on rather different amount of
knowledge concerning the particular noise contaminating the
qubits. For example, PEC requires that one explicitly learns the noise
model. By contrast, purification-based techniques are -- at least to a
first approximation -- oblivious to the error model, albeit the
efficacy of the derangement circuits depends on the particular noise
statistics.  This has motivated the community to conceive
sophisticated techniques for accurately learning the noise
models~\cite{berg2022probabilistic} -- some of these techniques were
touched upon above.  However as the experiments are scaled up in size
and are improved in accuracy, the importance of efficiently, reliably
and accurately learning and controlling the error statistics is
becoming ever more crucial. The same is true for the design of
potent QECC techniques.

\subsubsection{Classical communication bottleneck}

QEM protocols often require extensive classical communication whereby
the circuits are frequently recalibrated. For example, PEC requires a
new circuit for each activation.  As such, at the time of writing
classical communication is one of the main bottlenecks. For example, the
actual circuit execution time on some platforms is on the order of microseconds, while
updating the circuit description through classical communication
requires orders of magnitude more
time~\cite{kimEvidenceUtilityQuantum2023}.  Particularly relevant is
the intrinsic drift in the error statistics over time, which
necessitates occasional re-calibration.


\subsection{Research Roadmap}

In this subsection we identify a suite of promising future research
directions that may be of interest to readers with a background in the
broad field of signal processing and machine learning in support of
improved quantum computing and communications.

\subsubsection{Integration with randomised measurement protocols}
The above QEM techniques assumed that the expectated values are
directly measured and then used in classical post processing. In
recent years, randomised measurement protocols, such as classical
shadows, have become an area of active
research~\cite{elben2023randomized}.  Rather than directly estimating
the expected values, these techniques apply random measurements to the
qubits and store the individual outcomes classically. The collection
of these individual measurement outcomes has the fond conotation of a
classical shadow and can be used for simultaneously predicting the
expected values of many observables in classical post-processing.
Promising early works have shown the benefit of combining these
powerful randomised protocols both with
QEM~\cite{jnaneQuantumErrorMitigated2023,seifShadowDistillationQuantum2023}
and QML~\cite{PhysRevX.12.041022} techniques.

\subsubsection{Error-resilient algorithms and protocols}

QEM constitutes a pivotal subroutine in near-term quantum algorithms
in terms of estimating accurate expected values. However, for certain
applications one can construct bespoke protocols that are by
construction resilient to noise.

For example, 'quantum supremacy'
experiments~\cite{aruteQuantumSupremacyUsing2019} based on
random-circuit sampling techniques are by construction
noise-resilient.  The reason for this is that the deep random circuits
employed result in local qubit noise reminiscent of white noise, which
only trivially affects the expected value measurements.  Although
practical quantum circuits may exhibit noise characteristics that are
significantly different from the global white
noise~\cite{foldager2023can}, randomised compiling techniques have
been shown to be effective in combination with the above linear,
learning-based QEM
techniques~\cite{czarnikErrorMitigationClifford2021,foldager2023can}.

Another example is shadow spectroscopy, whereby one aims for
estimating eigenvalue differences in a Hamiltonian matrix by encoding
the eigenvalue differences as periodic signals that are estimated as
time-dependent excepted values $S(t) = \tr[O
  \rho(t)]$~\cite{chan2022algorithmic}.  One can then use established
signal processing techniques for estimatiing the frequencies of the
signals in a way that the approach is provably immune to stochastic
noise occuring in the corresponding circuits.

\subsubsection{Exploring tradeoffs  and combinations of QEM/QECC techniques}
For some application areas, such as training variational circuits, it
may be more beneficial to only remove a fraction of noise
contributions using QEM in exchange for a reduced measurement
overhead.  Furthermore, it has been shown that the locality of the
observable may crucially affect the measurement
overhead~\cite{kimEvidenceUtilityQuantum2023}.  It is thus an exciting
challenge to further explore non-trivial techniques that might assist
in reducing the measurement costs. QEM techniques constitute a
suite of diverse 'tricks of the trade' and each variant has its own
pros and cons, thus there is immense potential for further improvements by combining different techniques~\cite{caiMultiexponentialErrorExtrapolation2021,caiPracticalFrameworkQuantum2021,loweUnifiedApproachDatadriven2021,bultriniUnifyingBenchmarkingStateoftheart2023}.

Going beyond pure QEM, even when QECC can be successfully implemented
in experiments, there will still be an extended period of time when
the effects of quantum impairments cannot be sufficiently mitigated
owing to the limited number of qubits. This period is often referred
to as the early fault-tolerant era. In this era, harnessing QEM
techniques is still essential for mitigating the residual errors for
demonstrating quantum
advantage~\cite{suzukiQuantumErrorMitigation2022,piveteauErrorMitigationUniversal2021}. There
is also a range of ideas for combining QEM and QECC. These ideas tend
to be centered around sufficiently mitigating the qubit error ratio by
QECC techniques, so that the residual errors may be 'cleaned up' by
QECCs without encountering avalanche-like error proloferation, as
intimated in~\cite{9294106}.  In this context {\em
    near-hashing-baound multiple-rate QECCs} may be beneficial in
  terms of adjusting the coding-rate of QECC based on the estimated
  error rate in liaison with QEM
  techniques~\cite{chandra2019near}. Another attractive proposition is
  to harness the {\em universal decoding} concept, which allows the
  employment of the same decoder for different types linear codes,
  such as polar codes, BCH codes etc. used in combination with QEM
  techniques~\cite{chandra2023universal}.

Having surveyed the family of error mitigation solutions operating
both with and without the aid of QECCs, let us now focus our attention
on the pros and cons of quantum machine learning.


\section{Modeling and Counteracting Uncertainty via Quantum Machine Learning}
\label{qml-osvaldo}
{\bf The Myth:} Quantum computers can speed up machine learning on
large-scale data by exploiting the exponentially large vector space of
quantum states.

{\bf The Reality:} The benefits of quantum machine learning applied to
classical data are practically limited by the need to map the data onto
quantum states. While classical processing units excel at handling
large volumes of data by following deterministic computational graphs,
quantum computers may serve as co-processors for the implementation of
probabilistic tasks, such as generating discrete data.

{\bf The Future:} Fundamental research is required for developing
theoretical insights into the potential benefits of quantum machine
learning for the processing of both classical- and quantum-domain
data. The algorithms conceived for training and inference are likely
to be informed by different principles than those of the classical
machine learning workflow based on over-parameterization and
stochastic gradient descent.

\begin{figure}[t]
    \centering
    \includegraphics[width=3.2in]{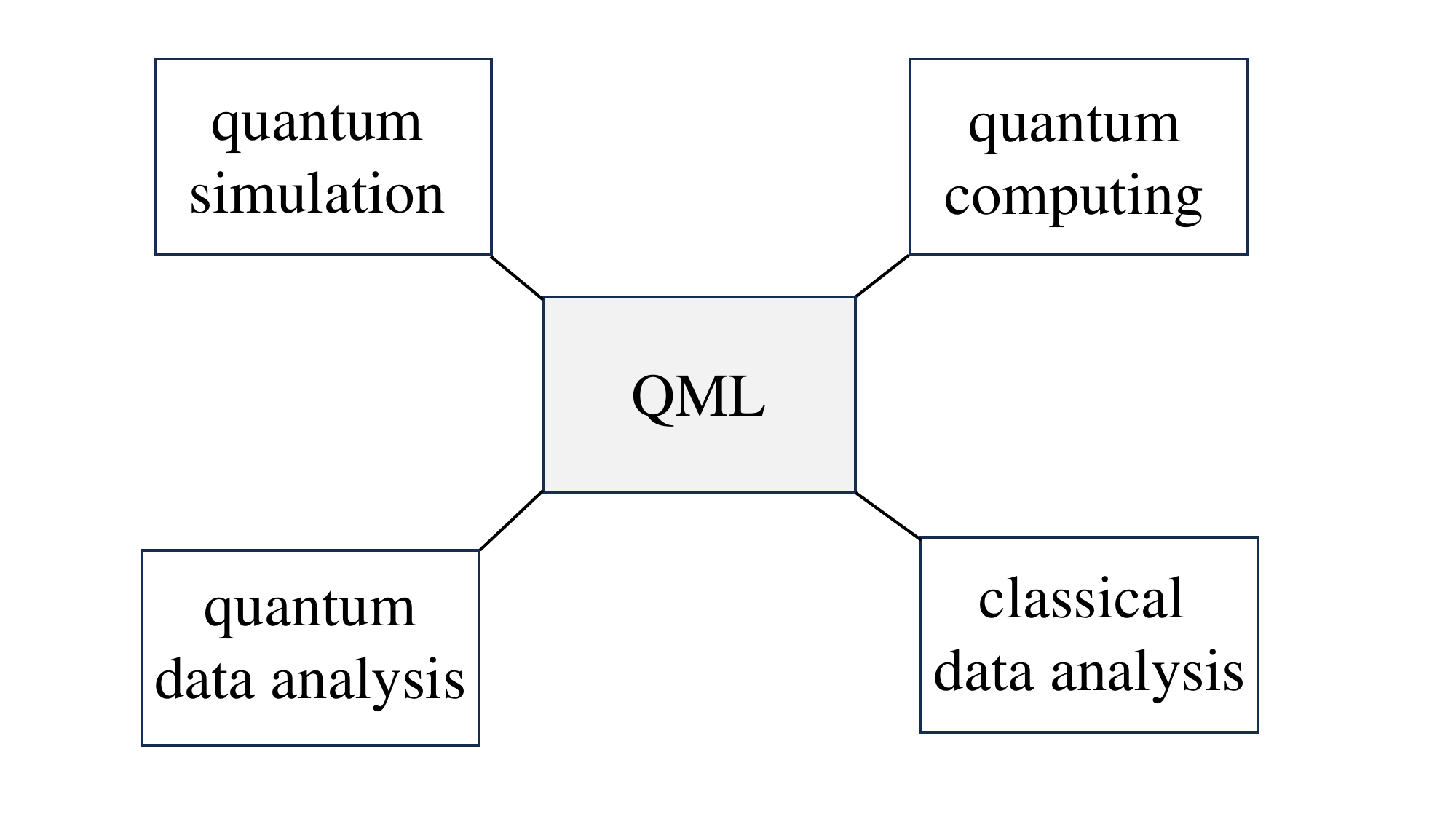}   
    \caption{Possible applications of QML (adapted from~\cite{cerezo2022challenges}).} 
    \label{fig:QMLapp}
\end{figure}

{\bf Abstract:} {\em In the previous sections, we have considered the
popular design paradigm of quantum algorithms, which is based on a
handcrafted selection of quantum gates and routines implementing
functionalities such as quantum error mitigation and error
correction. In practice, optimizing gate placement and the circuit
architecture is a complex task, particularly in the noisy
intermediate-scale quantum (NISQ) computers relying on a limited
number of qubits. Fro these the efficient exploitation of quantum
resources is essential. In this context, quantum machine learning
(QML) is emerging as a promising paradigm to program gate-based
quantum computers using a methodology that borrows insights from
classical machine learning~\cite{schuld2021machine,
  simeone2022introduction}. Applications of QML to the design of
improved quantum computing algorithms, such as quantum error
correction and quantum error mitigation, are under intense
investigation~\cite{valenti2019hamiltonian, nautrup2019optimizing,
  kim2020quantum, strikis2021learning, locher2023quantum}, along with
a host of other bespoke applications. As illustrated in
Fig.~\ref{fig:QMLapp}, such applications include also quantum
simulation, quantum data analysis, and classical data
analysis~\cite{nawaz2019quantum, tabi2021evaluation, cui2022quantum,
  chittoor2023quantum}.

At a technical level, QML is potentially capable of addressing
combinatorial optimization problems, implementing probabilistic
generative models, and carrying out inference tasks such as
classification as well as regression. They can be instantiated within
actual quantum computers via cloud-based interfaces accessible through
several software libraries -- such as IBM’s Qiskit, Google’s Cirq, and
Xanadu’s PennyLane. In this section, we provide a short overview of
QML, and we point to the potential of QML as a tool to quantify and
represent uncertainty.}

\subsection{State-of-the-Art}
\label{quantum-machine-learning-state-of-the-art}

\subsubsection{Introducing Quantum Machine Learning (QML)} 
\label{quantum-machine-learning-introducing-qml}

\begin{figure}[t] 
    \centering
    \includegraphics[width=3.2in]{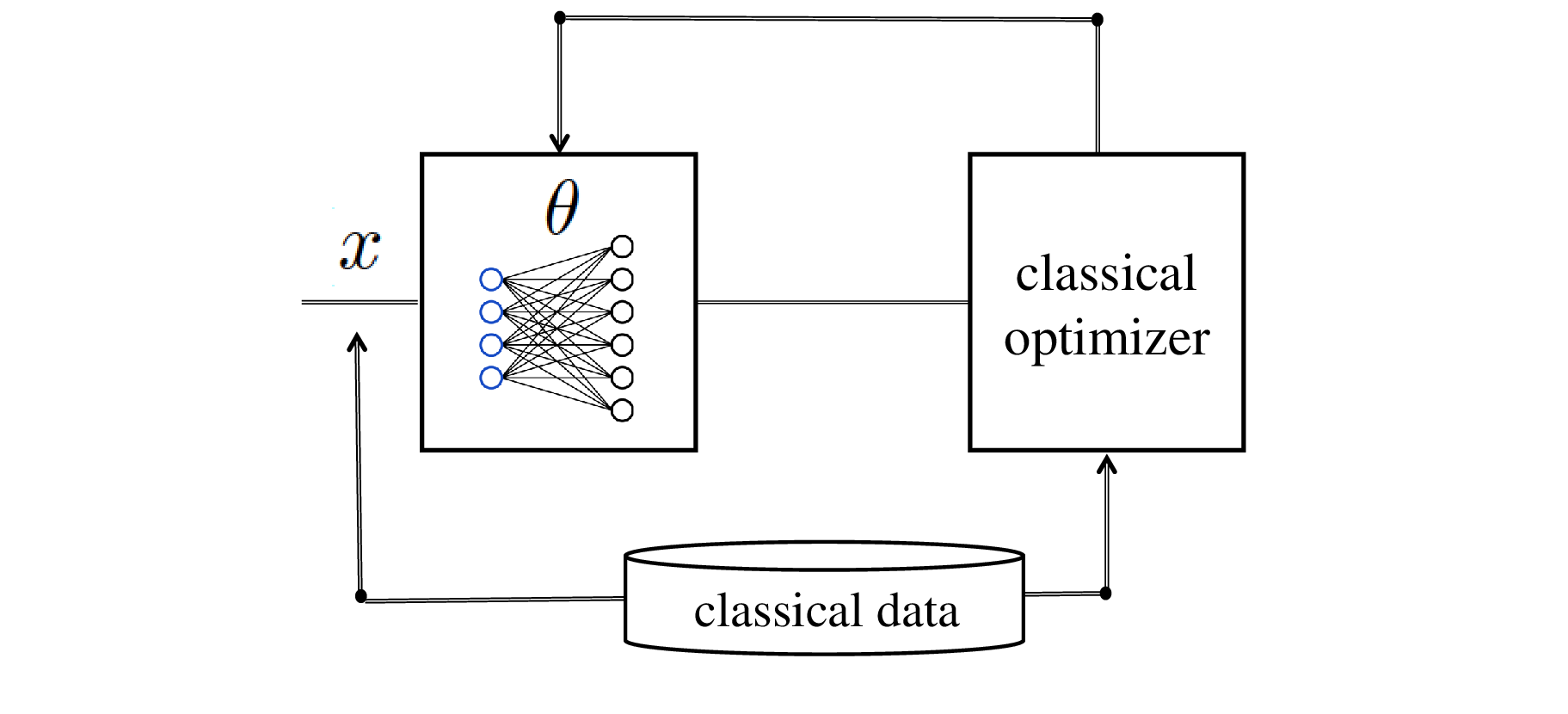}
    \centering
    \includegraphics[width=3.2in]{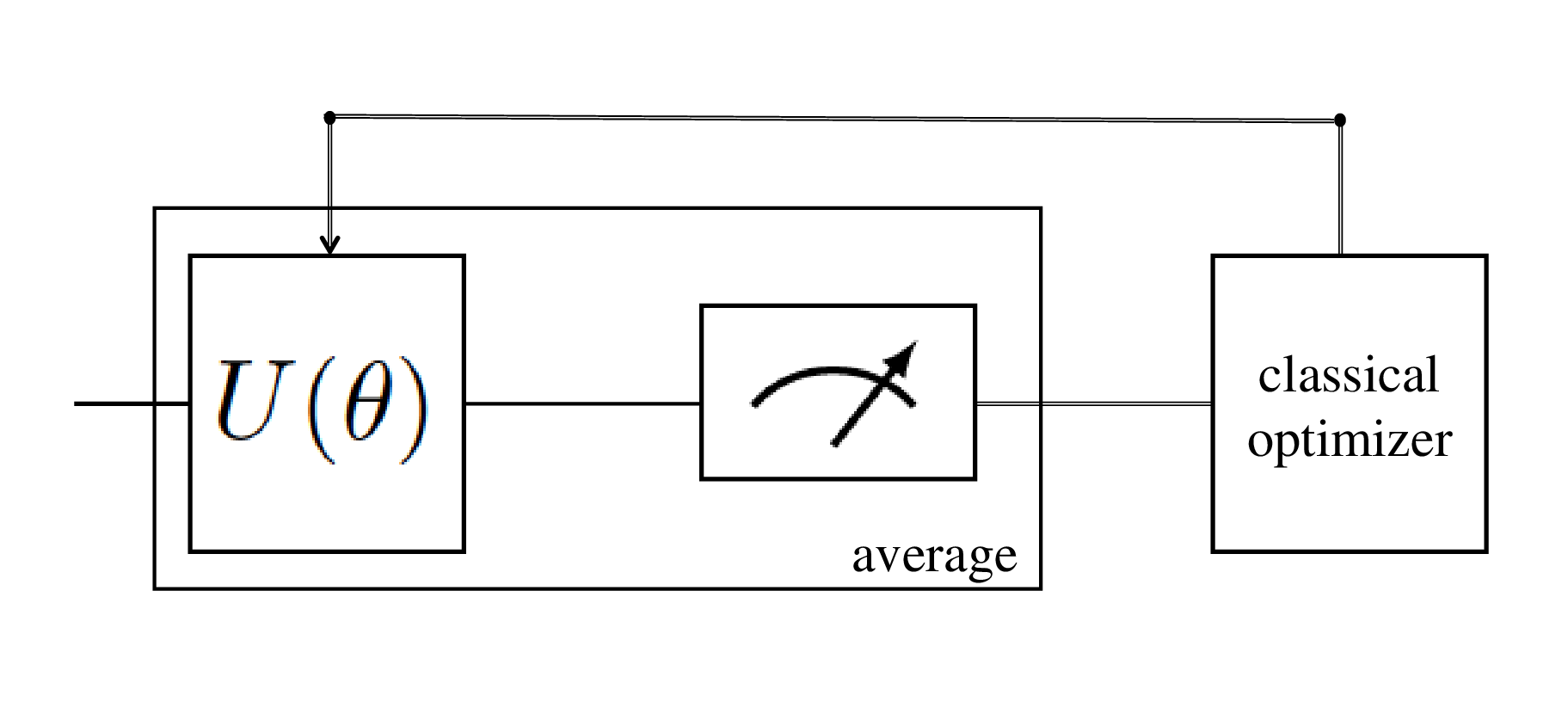}
    \caption{(\textbf{Top}:) Classical machine learning versus (\textbf{Bottom}:) quantum machine learning (QML).} 
    \label{fig:QML1}
\end{figure}

As illustrated in Fig.~\ref{fig:QML1}-(\textbf{Top}), in classical machine learning a {parameterized function} $f(x|\theta)$, e.g., a neural network, is optimized by adjusting the free parameters $\theta$ based on an available training data set. This is typically done by comparing the model outputs $f(x|\theta)$ with desired outputs extracted from the training data set, and making local adjustments to the free parameters $\theta$ to reduce the discrepancy between the two outputs.

As discussed in earlier sections, a {quantum algorithm} is specified by a quantum circuit operating on a set of $n$ qubits. Furthermore, a {quantum circuit} consists of a sequence of {quantum gates} that are applied sequentially and {in place} to the $n$ qubits, followed by {measurements} that convert the state of the $n$ qubits into $n$ classical bits. Quantum measurements are inherently \emph{random}, producing a jointly distributed vector of $n$ classical bits, and they cause the \emph{collapse} of the quantum state.

As shown in Fig.~\ref{fig:QML1}-(\textbf{Bottom}), in QML, the gates of a quantum circuit depend on the free parameters $\theta$, defining a \emph{parameterized quantum circuit} (PQC) that implements a parameterized unitary matrix $U(\theta)$. PQCs are also known as \emph{quantum neural networks}. As we will see in this section, in a manner similar to classical machine learning, the parameters $\theta$ are tuned via classical optimization based on data and measurements of the outputs of the circuit. An important advantage of the QML framework is that, by keeping the quantum computer in the loop, the classical optimizer can account for the non-idealities and limitations of quantum operations on NISQ computers.

\subsubsection{Taxonomy of QML Solutions}
\label{quantum-machine-learning-taxonomy-of-qml-solutions}

\begin{table}
    \caption{Taxonomy of QML strategies (``C'' stands for classical and ``Q'' for quantum).}
    \label{table:QMLtable}
    \begin{center}
    \begin{tabular}{cccc}
     &  & \textbf{data} & \textbf{processing} \tabularnewline
     &  & C & Q\tabularnewline
    \cmidrule{3-4} \cmidrule{4-4} 
    \textbf{data} & C & CC & CQ \tabularnewline
    \cmidrule{3-4} \cmidrule{4-4} 
    \textbf{generation} & Q & QC & QQ \tabularnewline
    \cmidrule{3-4} \cmidrule{4-4} 
    \end{tabular}
    \par
    \end{center}
\end{table}

As illustrated in Table~\ref{table:QMLtable}, data and processing can be generally of a quantum or classical nature. Quantum data refers to quantum states -- which may be encoded by physical systems produced by quantum sensors~\cite{banchi2023statistical} -- while quantum processing refers to the use of quantum computers. Classical machine learning corresponds to the ``CC'' corner in the table, with classical data and processing. The other three cases, with data and/or processing being quantum are the domain of QML.

\begin{figure}[t]
    \centering
    \includegraphics[width=3.2in]{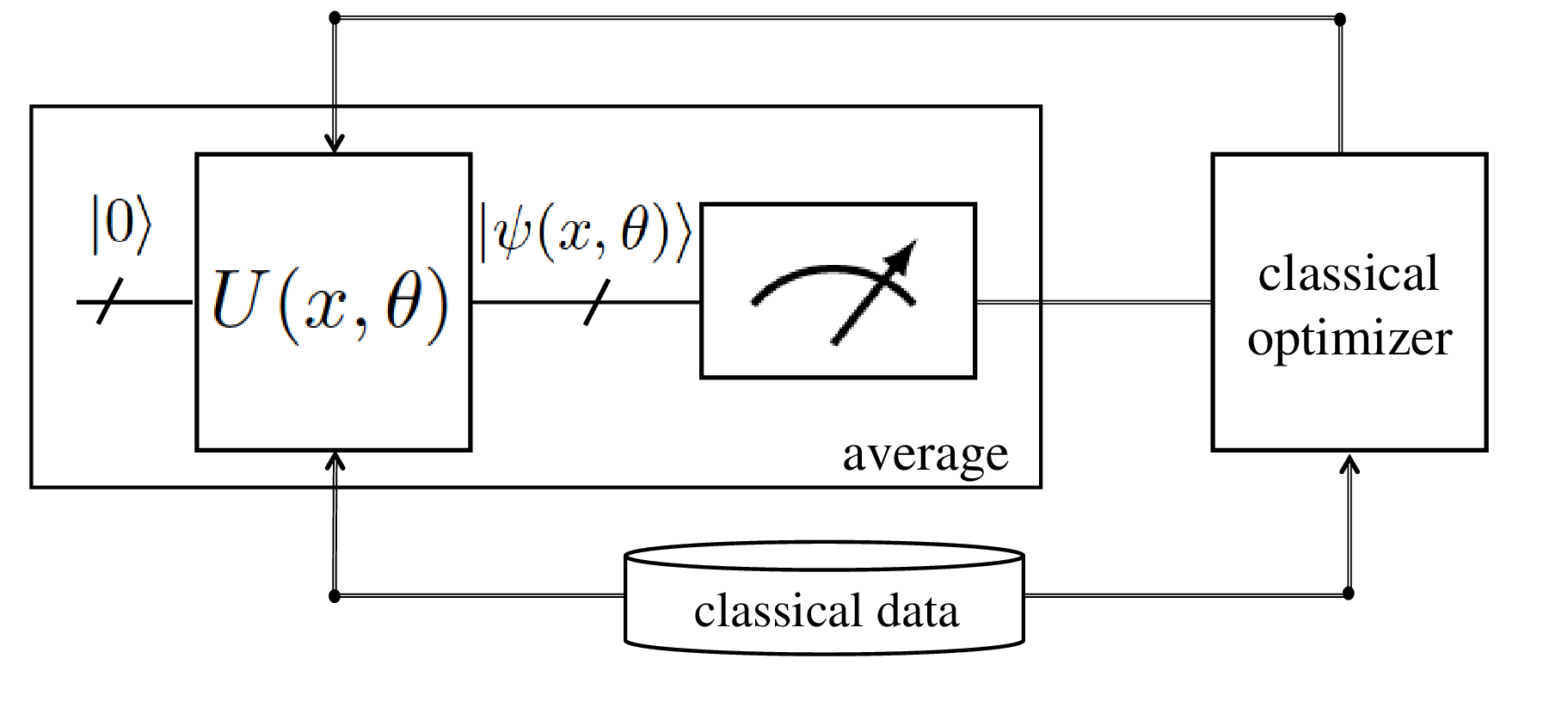}   
    \centering
    \includegraphics[width=3.2in]{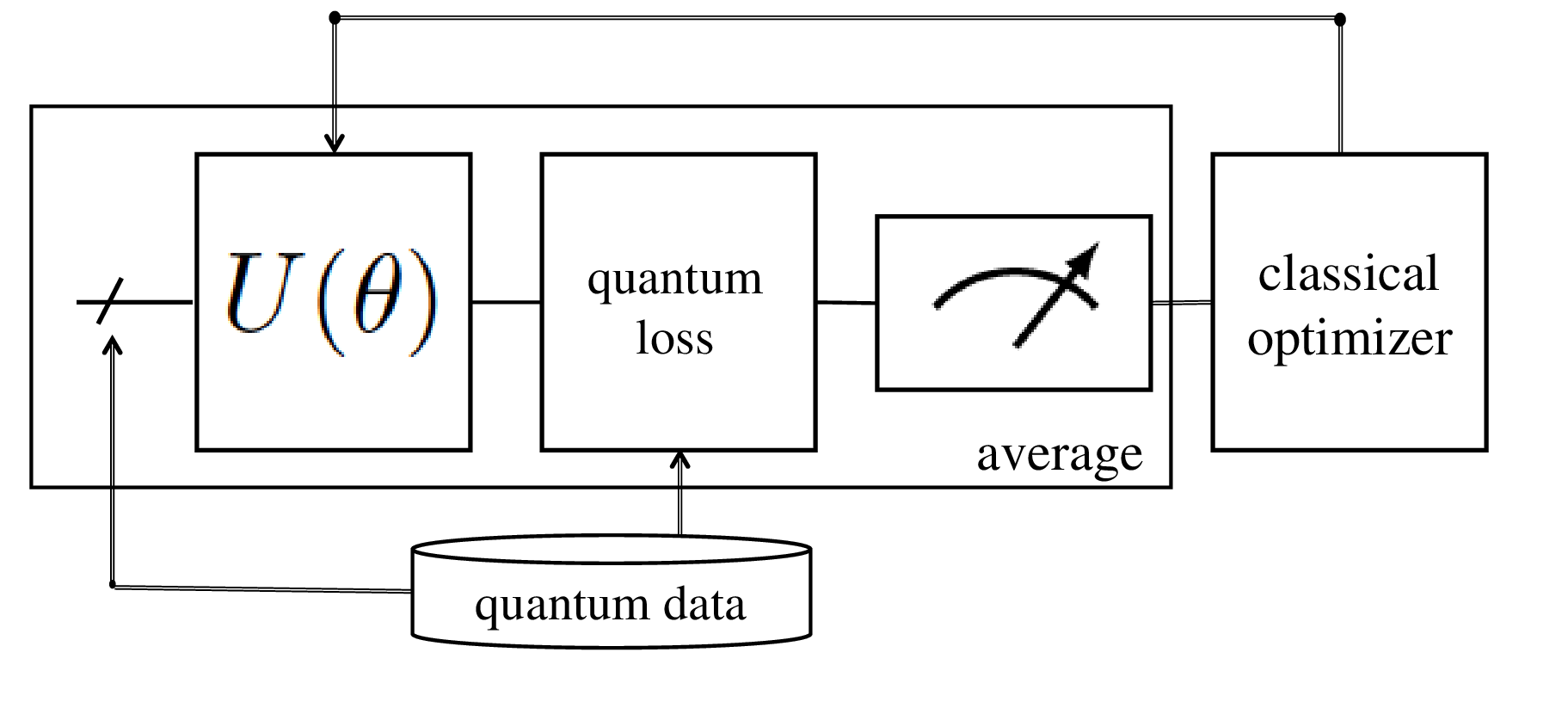}  
    \centering
    \caption{QML architectures for (\textbf{Top}:) classical data, quantum computing; and (\textbf{Bottom}:) quantum data and quantum computing.} 
    \label{fig:QMLarch}
\end{figure}

While processing quantum data for, e.g., chemistry and biology, is widely considered to be most promising in the long run, currently, the most common QML case is CQ: data are classical, while processing is quantum. The popularity of this setting is also due to the availability of many classical data sets. As seen in Fig.~\ref{fig:QMLarch}-(\textbf{Top}), in the CQ case, the measurement outputs from a PQC are compared to classical data within a classical computer, which then adjusts the local parameters $\theta$ to minimize the discrepancy between PQC outputs and targets. In principle, QC-based QML can implement any classical machine learning task. The largely open question at this stage is whether there are useful tasks for which QML can provide gains in terms of performance or efficiency.

As seen in Fig.~\ref{fig:QMLarch}-(\textbf{Top}), in the QQ case, the quantum state produced by the PQC is directly compared with a quantum data target to optimize $\theta.$ This comparison is done within the quantum computer, evaluating some loss metric that can then be measured by a classical optimizer for the update of parameters $\theta$. Two examples of QQ models are quantum autoencoders~\cite{romero2017quantum} and quantum generative adversarial networks (GANs)~\cite{dallaire2018quantum}.

Finally, in the QC case, there is no PQC, and the outputs of measurements of a quantum state are processed by a classical machine learning model. Examples include quantum tomography, in which the goal is that of estimating or representing an unknown quantum state~\cite{banchi2023statistical}.

\subsubsection{Parameterized Quantum Circuits}
\label{quantum-machine-learning-parameterized-quantum-circuits}

As mentioned, a PQC is defined by a fixed sequence of quantum gates that can depend on a vector of classical parameters $\theta$, defining a parameterized unitary matrix $U(\theta)$. The choice of the architecture of the PQC is akin to the selection of the {model class} in classical machine learning, which typically involves validating several different neural network architectures and hyperparameters.  In QML, the architecture of the PQC $U(\theta)$ is referred to as the \emph{ansatz} (from the German term for ``approach'' or ``attempt''). As for the model class in machine learning, one should choose the ansatz, if possible, based on domain knowledge. For instance, in quantum chemistry, some ansatz can closely represent the physics of the system.

\begin{figure}[t] 
    \centering
    \includegraphics[width=3.2in]{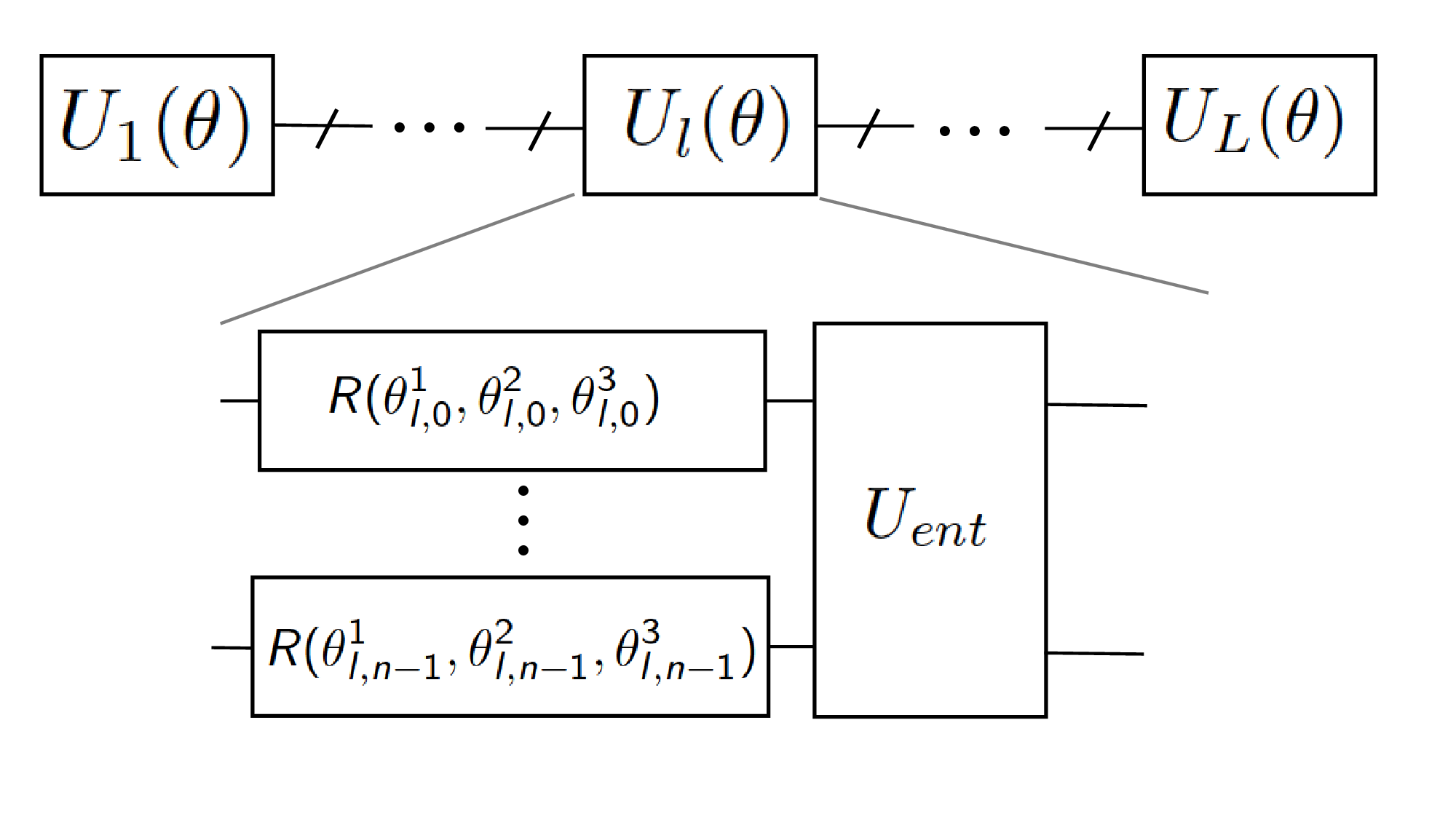}   
    \centering
    \includegraphics[width=3.2in]{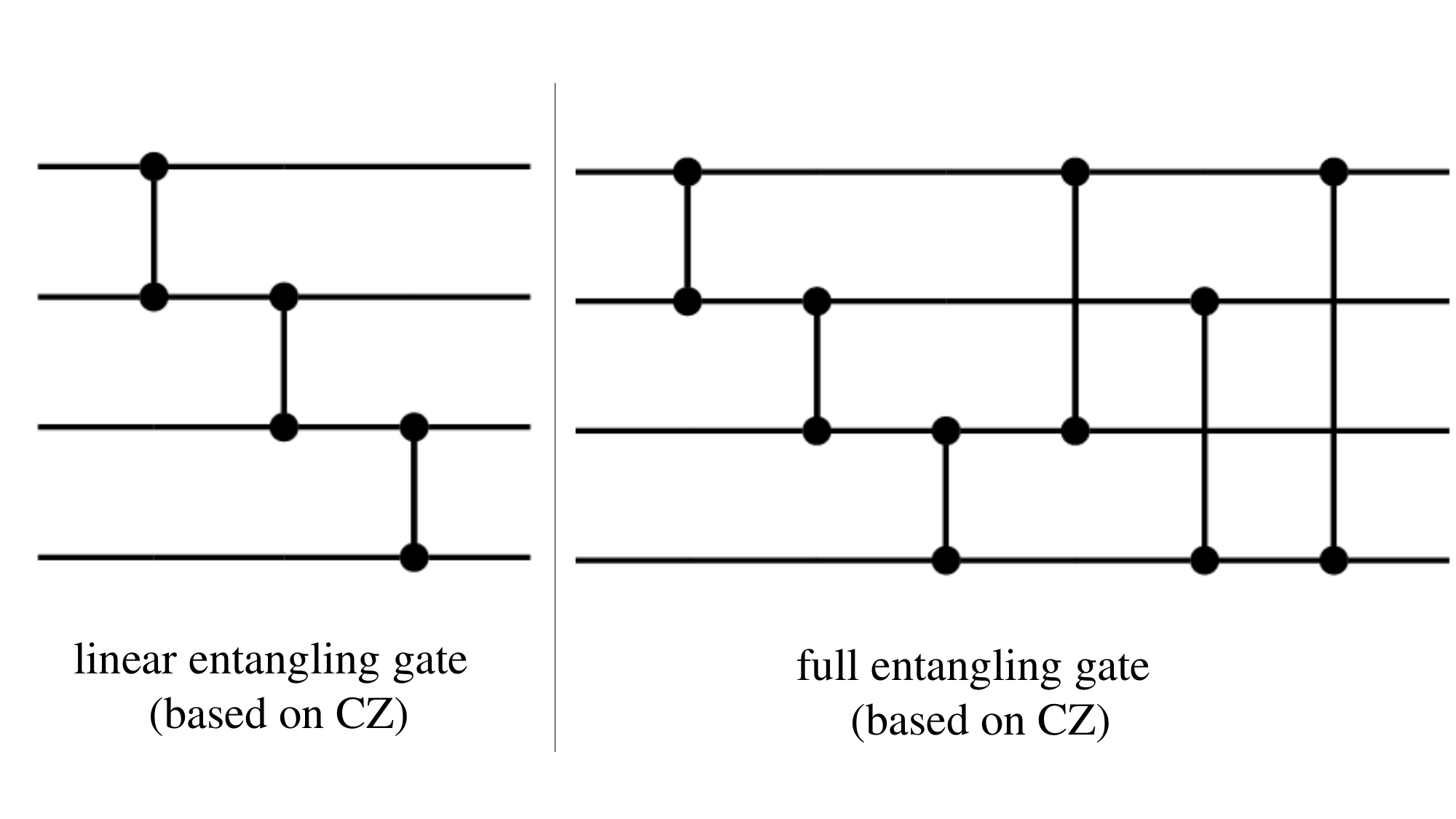}    
    \caption{(\textbf{Top}:) General form of a hardware-aware ansatz; and (\textbf{Bottom}:) example of an entangling gate.} 
    \label{fig:QMLarch1}
\end{figure}

When lacking insights from the physics of the problem, one can select generic ansatzes, with the important constraint that they can be efficiently implementable on the given hardware. As shown in Fig.~\ref{fig:QMLarch1}, the \emph{hardware-efficient ansatz} applies layers of separate rotations on each qubit -- which are typically available on all gate-based quantum computers -- as well as fixed multi-qubit entangling gates. The multi-qubit entangling gate $U_{ent}$ consists of a fixed cascade of two-qubit gates, whose connectivity structure abides by the locality properties of the architecture of the quantum computer.

The hardware-efficient ansatz is generic, and it is often viewed as playing a similar role to fully connected classical neural networks. In this regard, it is important to stress that a hardware-efficient ansatz does not have similar properties to fully connected classical neural networks in terms of dependence on model parameters. Notably, in fully connected classical neural networks, one has significant freedom in optimizing the connectivity among neurons by designing the individual synaptic weights as biases. By contrast, in the hardware-efficient ansatz, one can control only the rotations applied to each individual qubit, and the inter-dependencies among qubits are dictated by fixed entangling circuits.

It is finally noted that there are more complex alternatives to the hardware-aware ansatz, such as circuits that include parameterized two-qubit gates and ansatzes associated with increasing/decreasing number of qubits along the layers of the PQC~\cite{cerezo2022challenges}.

\subsubsection{Unsupervised Generative Learning}
\label{quantum-machine-learning-unsupervised-generative-learning}

The most direct QML application with classical data is generative modeling -- a key task in applications requiring the modeling of uncertainty. In particular, a \emph{Born machine} is a generative model constructed for binary strings $x$ that is implemented via a PQC~\cite{coyle2020born}. In a Born machine, a measurement of the output of the PQC on $n$ qubits, which produces a {random} $n$-bit string $x\sim p(x|\theta)$, is considered to be the generated data. The distribution $p(x|\theta)$ of the data generated by a Born machine can be controlled via the PQC parameters $\theta$ by Born's rule. Note that \emph{shot noise}, i.e., the inherent randomness of quantum measurements, is the key feature leveraged by Born machines to produce random samples.

Many current claims of {quantum supremacy/advantage} rest on the capability of quantum circuits to generate samples from {joint discrete distributions} in a more efficient manner than classical devices, although theoretical conclusions in this regard are conflicting~\cite{gao2022enhancing, pirnay2023superpolynomial, hinsche2023one}. It is an open question, in particular, whether Born machines can be efficiently learned~\cite{pirnay2023superpolynomial, hinsche2023one}.

\begin{figure}[t] 
    \centering
    \includegraphics[width=3.4in]{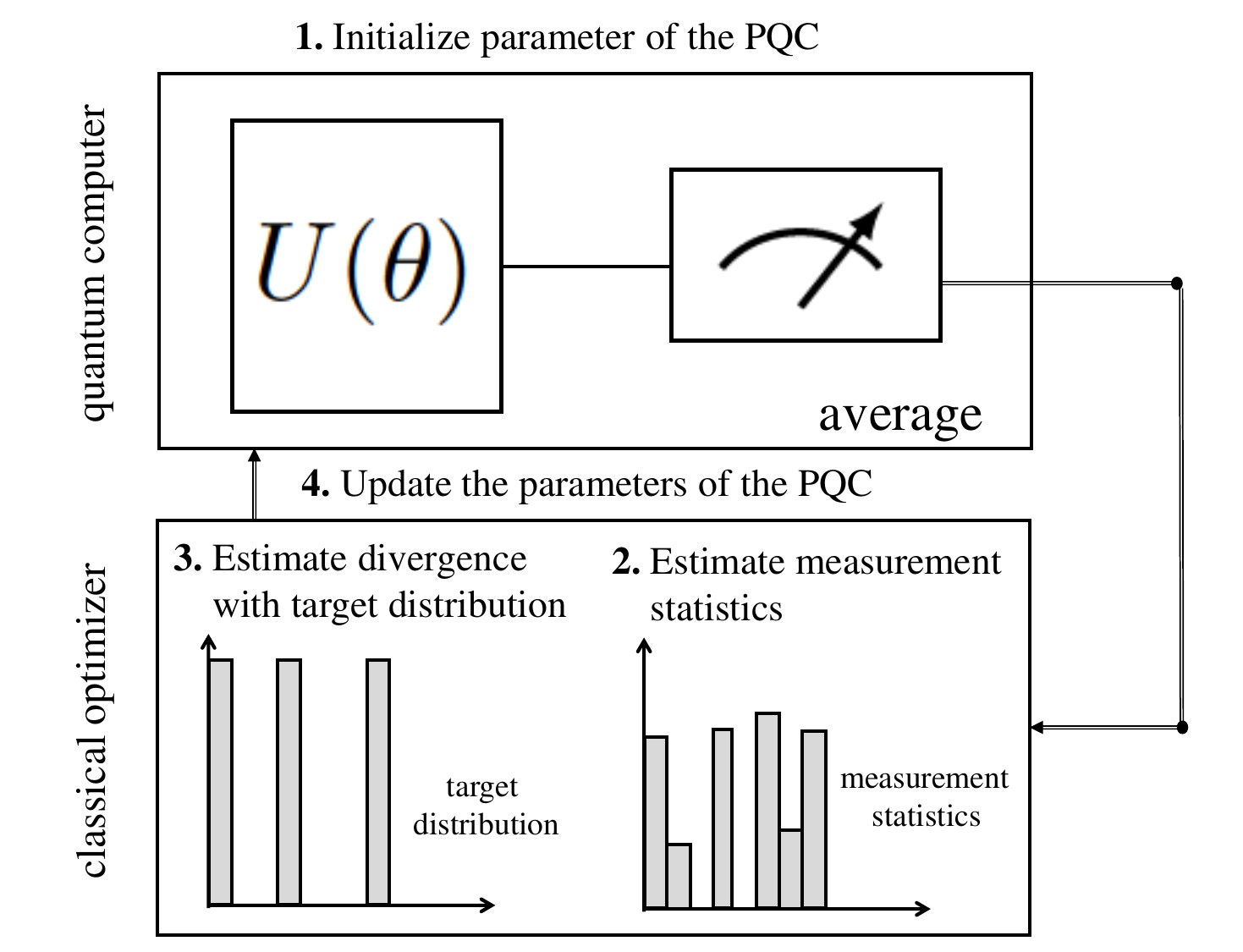}   
    \caption{Illustration of the process of training a PQC.}
    \label{fig:QMLtrain}
\end{figure}

In practice, as illustrated in Fig.~\ref{fig:QMLtrain}, training of a quantum generative model for discrete classical data is based on comparing the statistics of samples $x\sim p(x|\theta)$ generated by the model with the desired distribution inferred from training data. This is done by minimizing a specific measure of \emph{divergence} between the target data distribution and $p(x|\theta)$. This estimate may entail evaluating the {expected value of a cost observable} or applying {kernel-based} divergence measures that directly depend on all samples in the data set $\mathcal{D}$ and on measurements from the circuit, such as the {maximum mean discrepancy} (MMD)~\cite{du2022theory}. Note that, in contrast to sample generation, evaluating the loss requires repeating the measurement of the observable multiple times. 

Following standard practice in classical machine learning, optimization is typically carried out via {gradient descent}. However, in contrast to classical machine learning models, the {backpropagation} algorithm is not applicable, since we do not have access to the internal operations of the PQC. Rather, the gradient is typically estimated via a {perturbation-based}, {zero-th order} method known as a \emph{parameter shift rule}~\cite{simeone2022introduction}. This entails a complexity that scales with the square of the number of parameters, rather than being a constant multiplicative factor of the complexity for the forward pass as is the case for classical models~\cite{abbas2023quantum}.

\subsubsection{Supervised Learning}
\label{quantum-machine-learning-supervised-learning}
 
PQCs can also be used for supervised learning tasks with classical data. In this case, the classical input $x$ is typically {encoded} in the operation of the PQC $U(x,\theta)$ in a manner similar to the model parameters $\theta$, i.e., in the angle of single-qubit {rotations}. In the case of angle encoding, the PQC is obtained by alternating unitary operators dependent on $x$ and dependent on $\theta$. In this regard, it is advantageous to encode input $x$ multiple times -- a process known as \emph{data reuploading} -- to improve the expressiveness of the function encoded by the PQC. However, there are other ways of embedding classical information into a quantum state, such as amplitude encoding and basis encoding~\cite{schuld2021machine, simeone2022introduction}.

Given a classical input $x$, {probabilistic models} obtain a {randomized decision} $y$ through a single measurement of the PQC output. Given that the output is discrete, such models are suitable for classification. By contrast, {deterministic models} rely on a parametric function of the input $x$, which may be used for regression or classification. This is achieved by estimating the expectations of one or more observables. In contrast to probabilistic modes, shot noise averaging is required also for inference, not only for learning. Deterministic quantum models are akin to classical \emph{kernel methods} in that they operate over a large feature space -- the Hilbert space of dimension $2^{n}$ -- via linear operations~\cite{schuld2021machine}. Supervised learning follows the same type of optimization strategies, as described above for unsupervised generative modeling.

\begin{figure*}[!th]
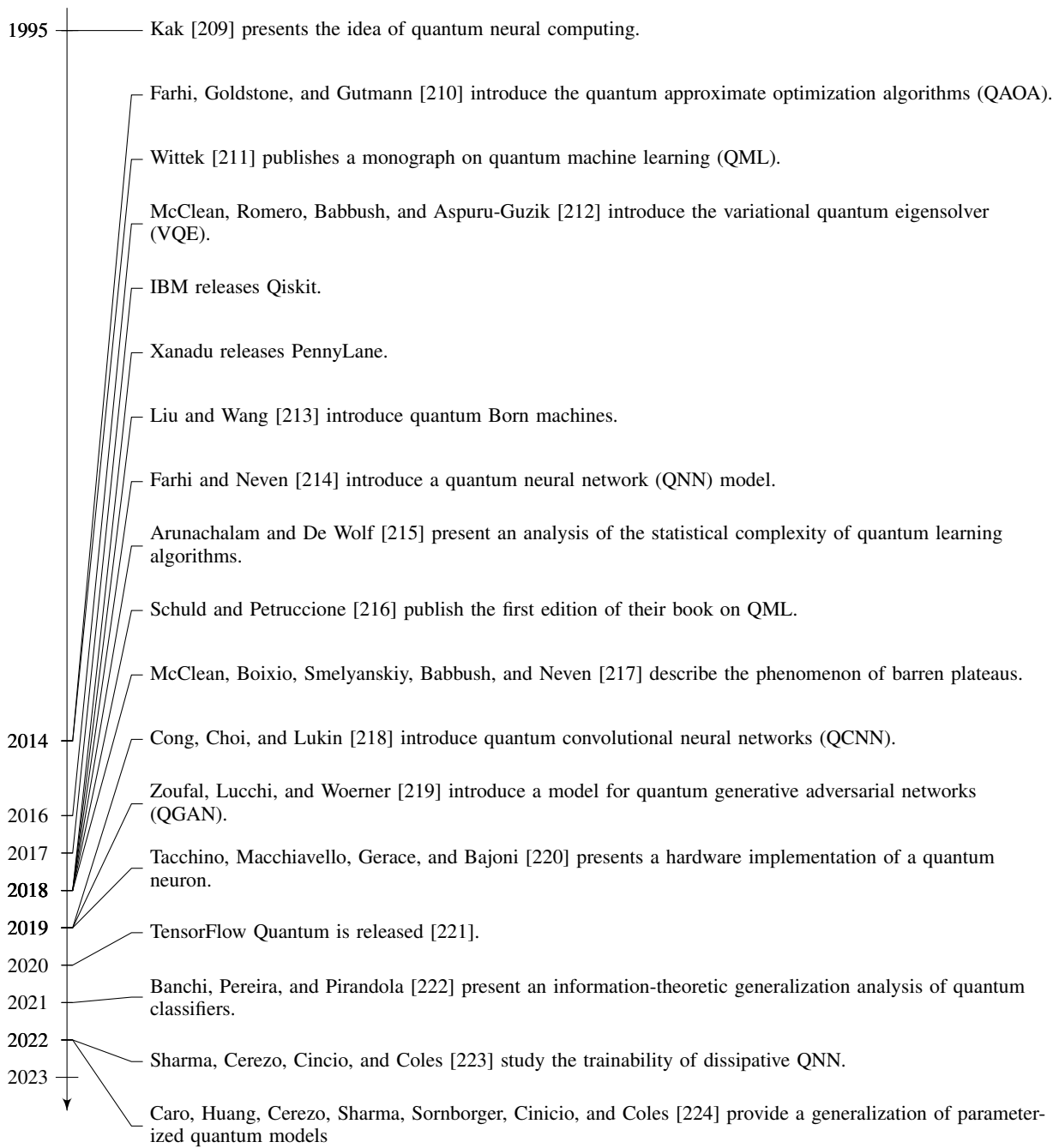

\begin{small}
\begin{timeline}{1995}{2023}{1cm}{1cm}{14cm}{0.7\textheight}
\entry{1995}{Kak~\cite{kak1995quantum} presents the idea of quantum neural computing.}
\entry{2014}{Farhi, Goldstone, and Gutmann~\cite{farhi2014quantum} introduce the quantum approximate optimization algorithms (QAOA).}
\entry{2014}{Wittek~\cite{wittek2014quantum} publishes a monograph on quantum machine learning (QML).} 
\entry{2016}{McClean, Romero, Babbush, and Aspuru-Guzik~\cite{mcclean2016theory} introduce the variational quantum eigensolver (VQE).}
\entry{2017}{IBM releases Qiskit.}
\entry{2018}{Xanadu releases PennyLane.}
\entry{2018}{Liu and Wang~\cite{liu2018differentiable} introduce quantum Born machines.}
\entry{2018}{Farhi and Neven~\cite{farhi2018classification} introduce a quantum neural network (QNN) model.}
\entry{2018}{Arunachalam and De Wolf~\cite{arunachalam2018optimal} present an analysis of the statistical complexity of quantum learning algorithms.}
\entry{2018}{Schuld and Petruccione~\cite{schuld2018supervised} publish the first edition of their book on QML.}
\entry{2018}{McClean, Boixio, Smelyanskiy, Babbush, and Neven~\cite{mcclean2018barren} describe the phenomenon of barren plateaus.}
\entry{2019}{Cong, Choi, and Lukin~\cite{cong2019quantum} introduce quantum convolutional neural networks (QCNN).}
\entry{2019}{Zoufal, Lucchi, and Woerner~\cite{zoufal2019quantum} introduce a model for quantum generative adversarial networks (QGAN).}
\entry{2019}{Tacchino, Macchiavello, Gerace, and Bajoni~\cite{tacchino2019artificial} presents a hardware implementation of a quantum neuron.}
\entry{2020}{TensorFlow Quantum is released~\cite{broughton2020tensorflow}.}
\entry{2021}{Banchi, Pereira, and Pirandola~\cite{banchi2021generalization} present an information-theoretic generalization analysis of quantum classifiers.}
\entry{2022}{Sharma, Cerezo, Cincio, and Coles~\cite{sharma2022trainability} study the trainability of dissipative QNN.}
\entry{2022}{Caro, Huang, Cerezo, Sharma, Sornborger, Cinicio, and Coles~\cite{caro2022generalization} provide a generalization of parameterized quantum models}
\end{timeline}
\end{small}
\caption{Timeline of quantum machine learning milestones (limiting the survey to variational quantum algorithms based on parameterized quantum circuits).} 
\label{fig:timeline-qml}
\end{figure*}

Next we embark on identifying the associated knowledge gaps in the light of the
timeline seen in Figure~\ref{fig:timeline-qml}.

\subsection{Knowledge Gaps and Challenges}
\label{quantum-machine-learning-knowledge-gaps-and-challenges}

We now briefly review a number of knowledge gaps and challenges in the QML field, which will be further addressed in the next subsection. We divide the discussion into three main areas of research: architecture, optimization, and generalization theory.

\subsubsection{Architecture}
\label{quantum-machine-learning-architecture}

As discussed in the previous subsection, the state-of-the-art tends to rely on generic ansatzes that are suitable for implementation on existing quantum computers in the absence of insights from the physics of the problem. Even using such ansatzes, one may ask whether there are useful tasks that may be addressed in ways that would be infeasible using classical means. Generative modeling provides a useful benchmark in this regard, and some initial studies have provided mixed conclusions~\cite{pirnay2023superpolynomial, hinsche2023one, nietner2023average}. Results about a possible separation between classical and quantum computers in their capacity to address special classes of learning problems are provided in~\cite{liu2021rigorous}.

Going beyond hardware-tailored ansatzes, it is important to identify alternative architectures satisfying the following requirements:
\begin{enumerate}
    \item They are expressive enough to capture solutions to problems of interest for the analysis of classical or quantum data.
    \item They can be efficiently implemented on hardware, being robust to impairments such as quantum noise, as well as to shot noise.
    \item They can be efficiently learned using a limited amount of training data. In this regard, one is interested in assessing the scaling of the data requirements and of the time complexity with respect to the number of qubits, in the hope of finding solutions, whose complexity escalates slower than exponentially with the number of qubits.
\end{enumerate}

Another interesting direction, espoused by the programs of companies like IBM, is the integration of quantum processor units (QPUs) with classical central processing units (CPUs) and graphics processing units (GPUs). We will return to this point below in the context of a research roadmap.

\subsubsection{Optimization}
\label{quantum-machine-learning-optimization}

At the core of QML lies the optimization of the free parameter $\theta$ via a classical computer. As covered in the previous subsection, QML cannot rely on the scaling efficiency of backpropagation, calling into question the adoption of gradient-based methods~\cite{abbas2023quantum}. Furthermore, implementing gradient descent is practically made complicated by the fact that the {loss landscape} associated with {generic ansatzes} is not well behaved as the number of qubits increases~\cite{ragone2023unified}.

In particular, the loss function for a randomly selected model parameter tends to have exponentially vanishing variance as the dimension of the circuit increases. This indicates that distinguishing different values of the loss functions requires an exponentially large number of measurements in order to overcome short noise. This behavior -- known as \emph{barren plateaus} -- is verified under general conditions, which are exacerbated by the presence of entanglement noise and measurements that involve multiple qubits~\cite{ragone2023unified}. This points once more to the importance of choosing well-structured bespoke ansatzes~\cite{pesah2021absence}.

\subsubsection{Generalization theory}
\label{quantum-machine-learning-generalization-theory}

The performance of both classical and quantum machine learning must be quantified by applying the model to test data that were not used during training and validation. In other words, the goal of machine learning is generalization. Accordingly, the analysis of QML strategies calls for the development of theoretical tools that can correctly describe the dependence of the generalization performance on the amount of data and time required for the training of the QML model of interest. As in classical machine learning, several approaches exist for this purpose, from combinatorial arguments to information-theoretic methods~\cite{caro2022generalization, banchi2023statistical}. Open research questions concern aspects such as the impact of overfitting and quantum noise~\cite{peters2022generalization}.

\subsection{Research Roadmap}
\label{quantum-machine-learning-roadmap}

In this final subsection, we point to several directions for future research in QML by concentrating on aspects that may be of particular interest to researchers with a background in signal processing, information theory, and communications.

\subsubsection{Architecture}
\label{quantum-machine-learning-roadmap-architecture}

A principled way to identify useful ansatzes revolves around the idea of encoding the geometric properties of the data of interest into the architecture of a PQC. Geometric properties are defined by symmetries of the data. Classical examples include the translational invariance leveraged by convolutional neural networks and the permutation invariance encoded by graph neural networks~\cite{bronstein2021geometric}. The identification of relevant symmetries can lead to the design of efficient ansatzes that encapsulate useful domain knowledge about the problem, preserving invariance or equivariance to the transformations underlying the symmetry~\cite{perrier2020quantum}. An example is given by the recent study~\cite{nikoloska2023time}, which introduced a class of quantum recurrent neural networks by encoding the specific property that the model class should be able to represent all classical time series that are related by a time-warping transformation.

\begin{figure}
    \centering
    \includegraphics[width=3.6in]{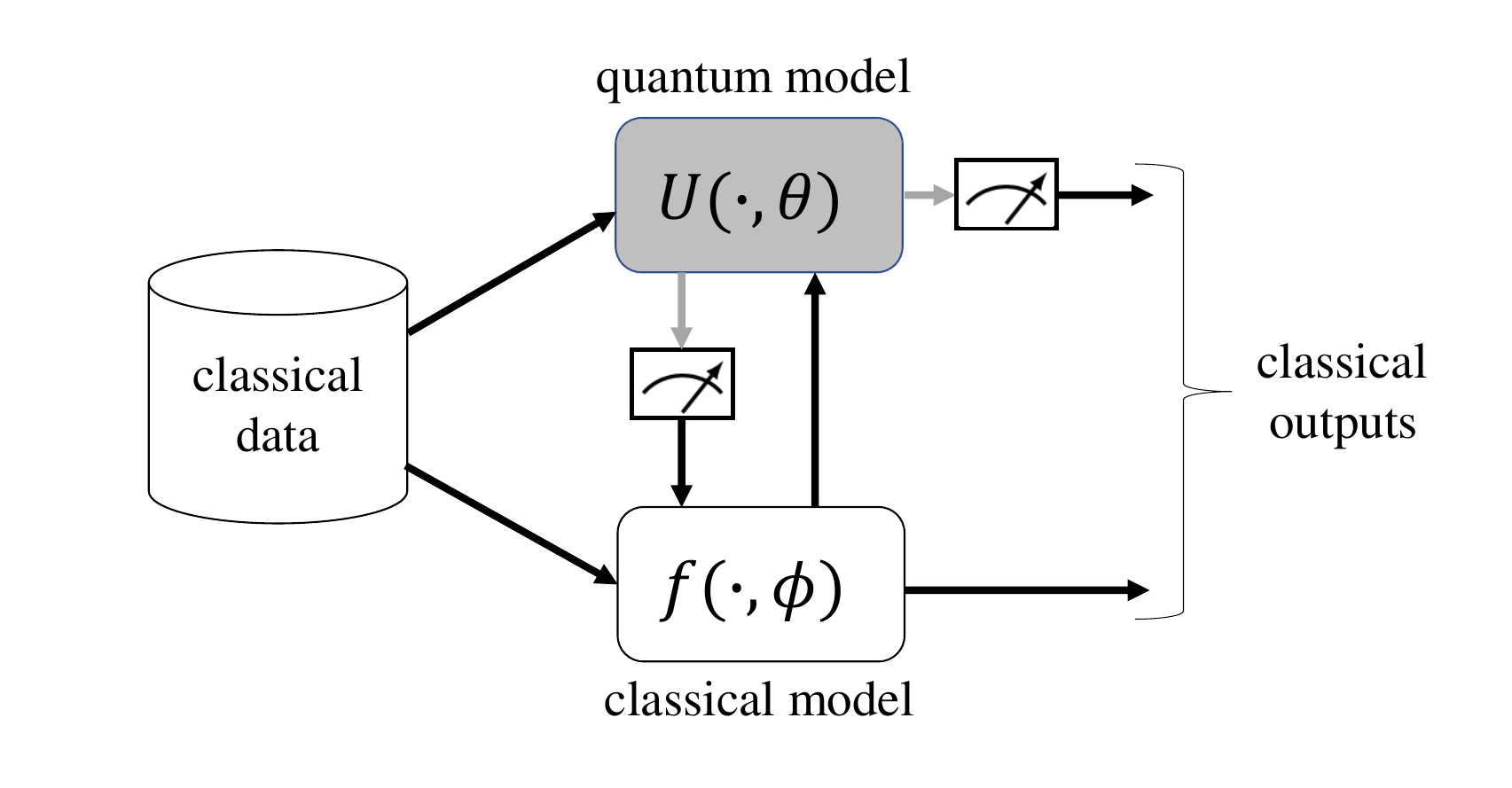}
    \caption{The general architecture of a hybrid classical-quantum model for QML.}
    \label{fig:QMLhybrid}
\end{figure}

Beyond purely quantum models, there is an interest, both commercial and academic, in finding effective and efficient ways of combining classical and quantum machine learning models. As illustrated in Fig.~\ref{fig:QMLhybrid} for the case of classical data, one may envision the combination of a classical model defined, e.g., by a neural network, and of a quantum model, implemented as a PQC. Such an architecture may benefit from the strengths of both classes of models. For instance, classical neural networks may be leveraged to process high-dimensional data, while quantum models could be tasked with generative modeling routines for discrete data.

For example, in~\cite{nikoloska2022quantum}, a Born machine is leveraged to encode the distribution over the binary weights of a classical Bayesian neural network (see also~\cite{carrasquilla2023quantum} for a related study). Bayesian neural networks maintain a distribution over the model parameters, which allows them to account for \emph{epistemic uncertainty} via ensembling. Representing complex distributions in the model parameter space can enhance the capability of Bayesian neural networks to quantify and represent uncertainty, but this is a complex task for classical models, particularly in the case of discrete parameters. Therefore, adopting a PQC for the purpose of encoding model parameter uncertainty holds the promise of enhancing the reliability of classical machine learning models.

Parameterized quantum circuits can also be implemented in the form of \emph{measurement-based quantum computation}, whereby the sequence of gates is encoded by the sequence of measurement settings applied to the qubits of a cluster state. This type of circuit has the additional source of uncertainty caused by the outcome of the intermediate measurements, which are typically compensated for in order to realize a deterministic unitary transformation~\cite{raussendorf2003measurement}. Recent work~\cite{majumder2023variational} proposed to exploit this randomness for the purpose of improving generative modeling.

\subsubsection{Optimization}
\label{quantum-machine-learning-roadmap-optimization}

In the context of QML-aided optimization, the open questions include, but are not limited to: How to improve the performance of gradient descent in the presence of barren plateaus? How to account for \emph{quantum decoherence}?

With regards to the first question, in light of the limitations of gradient-based optimization, it may be desirable to use {global optimization strategies}. A well-established approach for global optimization is to leverage {surrogate objective functions} that extrapolate the value of the loss across different values of the model parameter vector $\theta$. The surrogate function is typically updated sequentially based on previous measurements. One such method is \emph{Bayesian optimization}~\cite{tibaldi2023bayesian}.

As for the second question, it is important to evaluate the impact of quantum decoherence on optimization routines relying on QML. For example, quantum decoherence makes the gradients estimated from the measurements of a PQC {biased}. As seen earlier in this paper, {quantum error mitigation} trades chip area (qubits) against time, by running multiple noisy quantum circuits to emulate a noiseless one. Quantum error mitigation is capable of reducing the bias imposed by quantum gate noise on the estimate of gradients, but it increases the variance. An analysis of the impact of quantum error mitigation on gradient descent-based training of PQCs can be found in~\cite{jose2022error}.

\subsubsection{Theory}
\label{quantum-machine-learning-roadmap-theory}

Generalization analysis formulates bounds -- either average or probabilistic bounds -- for the {generalization error} of QML algorithms, with the main goal of understanding the associated scaling laws with respect to the size of the data set. So far, the focus has been almost exclusively on supervised learning, leaving the important task of unsupervised generative learning largely unexplored by the existing theory~\cite{banchi2023statistical}.

By its very nature, generalization analysis does not provide operational \emph{error bounds} for the performance of quantum models, focusing instead on scaling laws. A recent piece of work~\cite{park2023quantum} has initiated the investigation of statistical tools that can offer provable guarantees concerning the reliability of quantum models for test inputs. It is envisaged that this can be a fruitful direction for further research.

Following the above discourse on quantum machine learning, let us now
focus our attention on the radical frontier research of quantum radar
systems.
\section{Quantum Radars}
\label{quantum-radar}
{\bf The Myth:} With the aid of quantum radars stealth aircrafts can
be tracked up to several thousand km without being detected by the
aircraft's radars.

{\bf The Reality:} The microwave quantum radar demonstrations so far
have provided only a marginal improvement over their classical
counterpart within rather limited ranges.

{\bf The Future:} Significant research efforts are required for
developing high-fidelity entanglement generation sources and detectors
at microwave frequencies. Entanglement assisted quantum radars
operated at optical frequencies are more mature, but optical
frequencies have limited propagation through both clouds as well as
fog, hence further research is needed to solve for this problem.

{\bf Abstract:} {\em Quantum radars offer alternatives to classical radars
and are relevant in low-brightness and high-attenuation scenarios. The
key promise of quantum radars is to outperform the quantum limit of
classical sensors, and thus improve the detection probability at very
low signal-to-noise ratios. Even though some progress has been made in
terms of quantum radars at microwave frequencies, further significant
efforts are required for developing improved entangled sources and the
corresponding detectors. In this section we provide an overview of
various quantum radar technologies, including both quantum
interferometry and quantum illumination based radars. The focus will
be on a particular version of quantum illumination based radars,
namely on entanglement assisted (EA) radars.  The EA radars operated
in the C- and L-bands are much more mature than their microwave
counterparts and lend themselves to supporting both monostatic and
multistatic solutions, which are reminiscent of single- and multi-cell
communications scenarios. The EA multistatic radar concept also offers
the possibility of dual-function radar and communication (DFRC)
operation, which is also often referred to as integrated sensing and
communications (ISAC).}

\subsection{State-of-the-Art}
\label{state-of-the-art}

Quantum radar is a particular quantum sensing technique that exploits
quantum-mechanical features of the electromagnetic fields to attain
improvements in target detection compared to the classical
scenario~\cite{1,2,3,4,5}. The key motivation behind the quantum radar
studies is to outperform the quantum limit of classical
sensors~\cite{1}. The potential advantages of quantum radars compared
to the classical radars can be summarized as follows~\cite{1,2,3,4,5}:
improved receiver sensitivity, enhanced target detection probability
at low signal-to-noise ratios (SNR), improved penetration through
clouds and fog when microwave photons are used, up-graded resilience
to jamming and improved synthetic-aperture radar imaging
quality. Furthermore, the quantum radar signals are harder to detect
than their classical counterparts, and quantum radars have higher
cross-section (as shown in~\cite{1}), just to mention a few for their
promises. However, unfortunately, they are significantly more
challenging to implement, in particular at microwave frequencies. A
pair of popular quantum radar designs are: (i) the quantum radar
employing the quantum illumination sensing
concept~\cite{1,2,3,4,5,6,7,8,9,10,11,12,13} and (ii) interferometric
quantum radar~\cite{1,2,14,15,16,17} relying on a concept reminiscent
of quantum interferometry.  Depending on the specific underlying
quantum phenomenon employed, the quantum sensors can be categorized
into three types~\cite{1,14}:

{\em Type~1:} a quantum sensor transmits the quantum states of light that are
not entangled with the receiver;

{\em Type~2:} the sensor transmits the classical states of lights, but employs
the quantum detectors for improving the performance;

{\em Type~3:} the quantum transmitter emits quantum states that are
entangled with the reference states available at the receiver.

\begin{figure}
    \centering
    \includegraphics[width=4.5in]{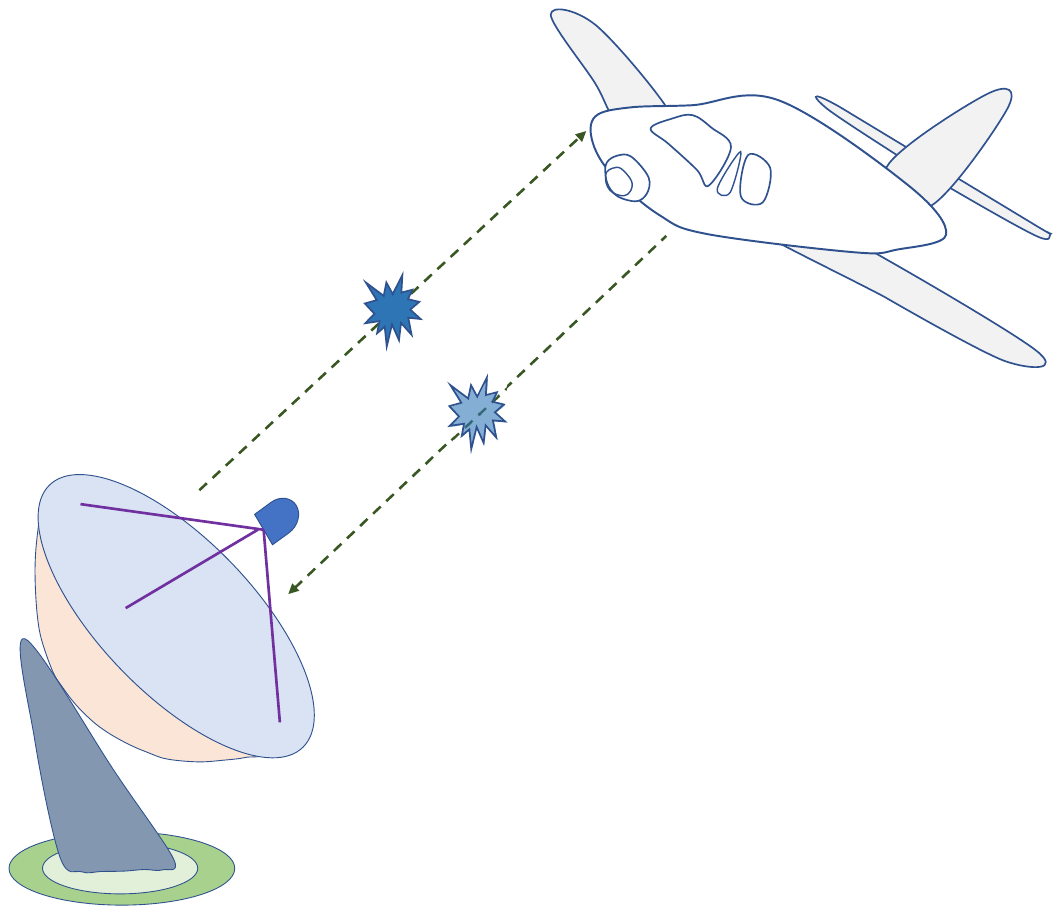}
    \caption{Illustration of the single photon monostatic quantum radar concept.}
    \label{fig:1}
\end{figure}

{\em The monostatic single photon radar, illustrated in Fig.~\ref{fig:1},
belongs to the Type 1 quantum sensors}, and its operating principle is
similar to that of the classical radars, but a single photon pulse is
used instead of an RF pulse. The single-photon quantum radar has
larger radar cross-section than classical radars~\cite{1,18},
but requires a large number of single-photon pulses to be transmitted.

{\em The quantum laser detection and ranging (LADAR) solution belongs to
the Type-2 quantum sensors} and employs principles similar to that of
the light detection and ranging (LIDAR)
technology~\cite{1,14,15,16,17}. Because the LADARs operate in
the visible and near-infrared wavelength regions, the laser beam does
not propagate well through clouds and fog, in particular in the
visible-light band around 800~THz. Nevertheless, given that the
operating wavelengths are much lower compared to the classical radars,
the LADARs have much better spectral resolution.

\begin{figure}
    \centering
    \includegraphics[width=3.8in]{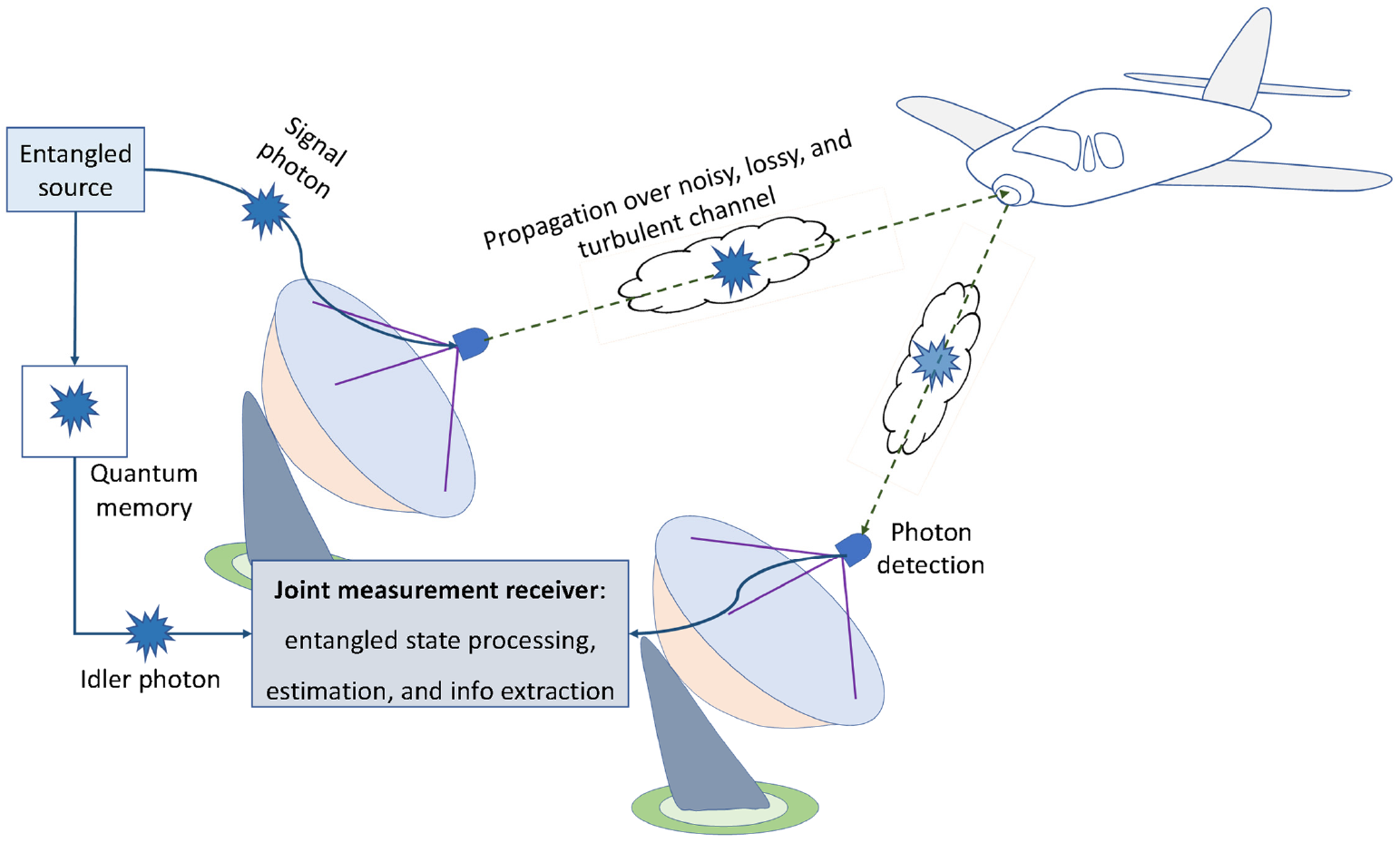}
    \caption{Illustration of the entanglement based bistatic quantum radar concept.}
    \label{fig:2}
\end{figure}

{\em Entanglement assisted radars belong to the Type-3 quantum
  sensors}, and the corresponding operational principle of the
bistatic radar is illustrated in Fig.~\ref{fig:2}. The entanglement
generation source orchestrates an entangled pair of photons, namely
the signal and idler photons. The idler photon is kept in the quantum
memory of the receiver, but we hasten to add that at the time of
writing there is no quantum memeory having a long retention duration.
Nonetheless, for optical frequencies this quantum memory may rely on
optical delay lines. By contrast, the signal photon is transmitted
over a realistic noisy, lossy, and atmospheric turbulent channel
towards the target. The reflected signal photon is detected by the
radar and the quantum-domain correlation between the signal and idler
photons is exploited at receiver side for improving both the receiver
sensitivity and target detection probability.

In the next subsections we describe interferometric quantum radars,
quantum illumination based radars, and entanglement assisted radars.

\subsubsection{Interferometric Quantum Radars}
\label{IQR}

\begin{figure}
    \centering
    \includegraphics[width=4.5in]{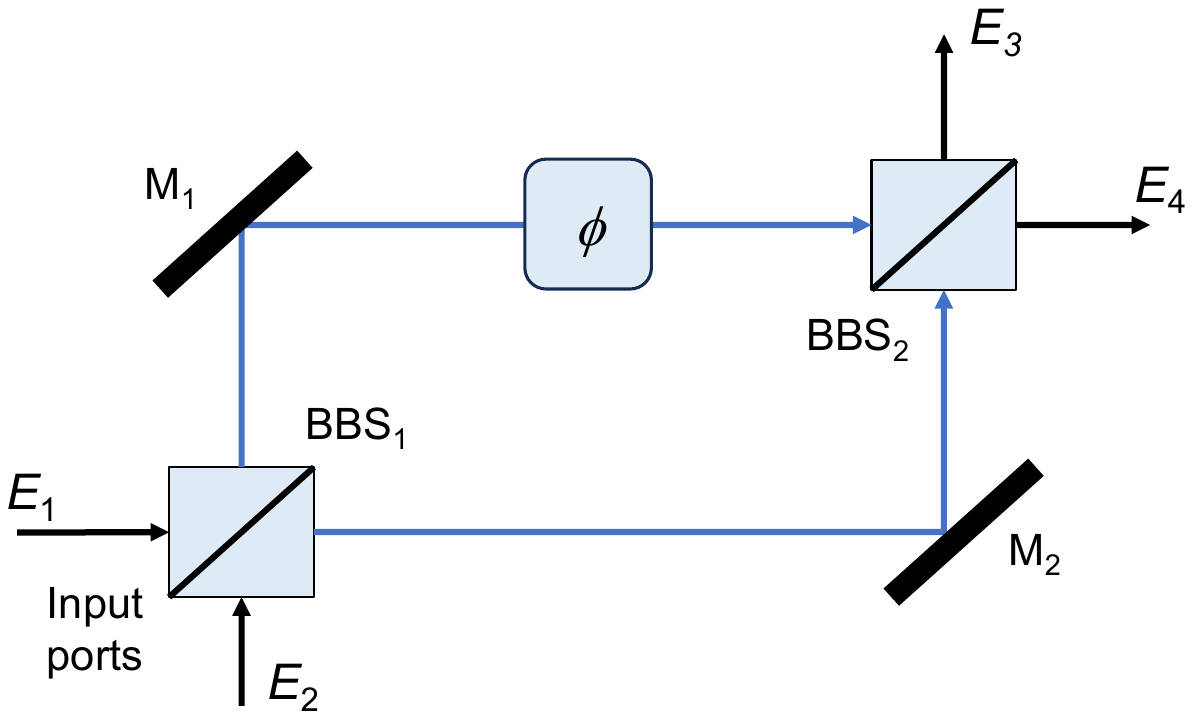}
    \caption{Illustration of the interferometric quantum radar concept. BBS: balanced beam splitter. M: mirror.}
    \label{fig:3}
\end{figure}

\begin{figure}
    \centering
    \includegraphics[width=4in]{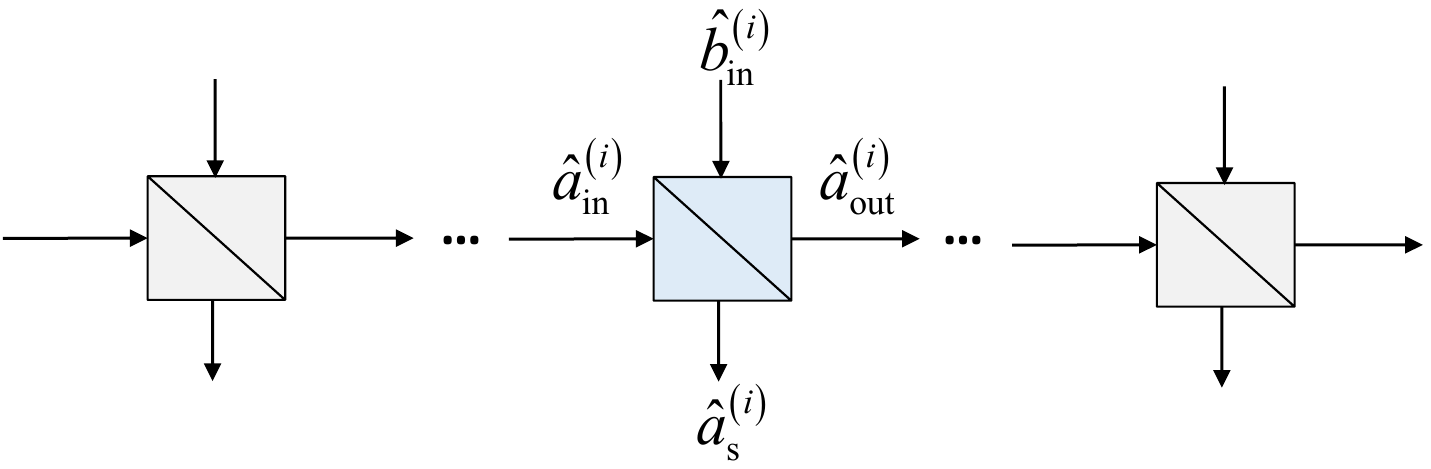}
    \caption{The light beam attenuation effects can be modeled by a cascade of lossy thermal Bosonic channel models.}
    \label{fig:4}
\end{figure}

The interferometric quantum radars rely on the concept introduced in
Fig.~\ref{fig:3}, where a single photon is present at the input port
$E_1$ of the balanced beam splitter 1 (BBS1), while no photon is at
the input $E_2$. The upper branch corresponds to the target branch,
where the phase shift introduced by the presence of the target is
denoted by $\phi$. The expected value at the output port E4 in the
absence of any background radiation is proportional to
$R_t$cos$^2(\phi/2)$, where $R_t$ is the reflectivity of the
target. If we repeat the target interrogation problem $N$ times, the
uncertainty of the phase estimate will become~\cite{1,2}
$N^{-1/2}$, which is commonly referred to as the standard quantum
limit (SQL). If we apply entangled states at the input ports of the
BBS1 in Fig.~\ref{fig:3}, that is $|\psi>=2^{-1/2}(|N0>+|0N>)$, the
uncertainty for the phase estimate will be $1/N$, which is known as
the Heisenberg limit. Naturally, in practice we have to take the
realistic propagation effects of absorption, scattering, diffraction,
and turbulence into account together with the background radiation
into account, which may be modelled by a cascade of thermal Bosonic
channels, as shown in Fig.~\ref{fig:4}. These are popularly harnessed
for studying the efficiency of quantum radar techniques. The
atmospheric turbulence~\cite{19,20} is caused by the refractive index
variations imposed on the beam propagation path by temperature and
pressure fluctuations, introducing wavefront distortions. This
indicates that the transmissivities $T_i$ of the different Bosonic
stages of Fig.~\ref{fig:4} are random variables. In the face of these
hostile propagation effects we can use the adaptive optics
(AO)~\cite{21,22,23,24} for enhancing the quantum radars sensitivity
attained.

An AO system is composed of: (1) a wavefront sensor used for detecting
the atmospheric distortions, (2) a wavefront corrector harnessed for
compensating the turbulence effects, and (3) a control processor
employed for monitoring the wavefront sensor information and for
updating the wavefront corrector~\cite{21,22,23,24}. At the time of
writing the AO based correction is capable of maintaining
super-sensitivity over ranges of up 5000 km~\cite{21}. The key idea
behind Smith's proposal~\cite{21} is to introduce the phase shift
$\phi_{AO}$ in the upper branch of Fig.~\ref{fig:3} for interrogating
the target, which is controlled by the radar operator. In the quantum
radar scenario of Fig.~\ref{fig:3}, let us assume that the entangled
source, generating the the so-called NOON states~\footnote{In quantum
  information science, NOON states typically refer to quantum states
  where a certain number of particles are in one state and the rest
  are in another, usually entangled in a specific way. These states
  have beneficial properties for precision measurements.}, is located
at the transmitter side of the radar system. Hence the distortions
inflicted by turbulence and attenuation of the lower branch spanning
from the transmitter side to the receiver side of the radar system can
be neglected. When $\phi_{AO}$ is appropriately chosen, the corrected
phase is also a function of the corrected transmissivity of the
optical medium and hence it will never exactly reach the Heisenberg
limit. Unfortunately, this method requires the accurate knowledge of
the target range. Alternatively, we can use the AO to compensate for
the aberrations in the wavefront of a single photon imposed by
turbulence to improve the transmissivity of the channel in a similar
fashion to that in~\cite{23,24}. As a benefit, the latter
approach does not require the knowledge of the target range.

\subsubsection{Quantum Illumination Based Radars}
\label{QI_R}

The quantum illumination technique proposed in~\cite{6} is capable of
improving the sensitivity, when operating in a hostile noisy
regime. Since it is not restricted to any specific wavelength, it may
also be readily applied to both LADAR and microwave quantum
radars. This approach does not require any Mach-Zehnder interferometer
-- simple photon counters are adequate for detection.  They are also
readily applicable to both entanglement based and non-entangled
quantum radars, and the corresponding principle was already
highlighted in Fig.~\ref{fig:2}. Assuming that the target reflectivity
is ${R_t}$ and the average number of background photons is $N_b$, when
entangled photons are used and the system dimensionality is $D$, the
corresponding $SNR$ will be~\cite{1,2}:
\begin{equation}
    SNR_{ent}=\frac{R_t+(1-R_t)\frac{N_b}{D}}{\frac{N_b}{D}}
    \label{eq:SNR_ent}
\end{equation}
 and the improvement in the SNR when entangled photons are used over non-entangled  case, assuming that the target reflectivity is close to 1 will be:
\begin{equation}
     \frac{SNR_{ent}}{SNR}=\frac{1-R_t+{\frac{R_t D}{N_b}}}{1-R_t+\frac{R_t}{N_b}}\approx D
    \label{eq:ratio}
\end{equation}

Clearly, for a typical system dimensionality of $D=2$, the improvement
in the SNR originating from entanglement is up to 3 dB.  Tan {\em et
  al.}~\cite{Tan2008} proposed to use a spontaneous parametric down
conversion (SPDC) source as an entanglement generation source, but the
corresponding detector has not been proposed. The authors of~\cite{26}
used the quantum Chernoff bound under the assumption that the target
is present with a probability of 0.5 and that the system is operated
in a hostile noisy scenario in order to obtain the following bound for
the asymptotic error probability:
\begin{equation}
     P{_{e}}^{(QCB)}\sim 0.5  e^{-MTN_s/N_b}
    \label{eq:bound}
\end{equation}
where $T$ is the overall transmissivity, $M$ is the number of
signal-idler photon pairs being utilized, and the mean photon count
of the background radiation noise is given by $Nb/(1-T)$. This bound
is the same as the quantum Bhattacharyya bound, and it has been shown
in~\cite{3} to be valid only for high SNR values. The
corresponding classical counterpart relying on a coherent state at the
transmitter side and on homodyne detection at receiver side has the
following bound~\cite{Tan2008}:
\begin{equation}
     P{_{e}}^{(CB)}\sim 0.5  e^{-MTN_s/(c_f N_b)}
    \label{eq:cl_bound}
\end{equation}
where $c_f=4$ for the upper bound and $c_f=2$ for the lower
bound. Therefore, the expected entanglement advantage based
on~\cite{Tan2008} is between 3 and 6 dB.

\begin{figure}
    \centering
    \includegraphics[width=3.6in]{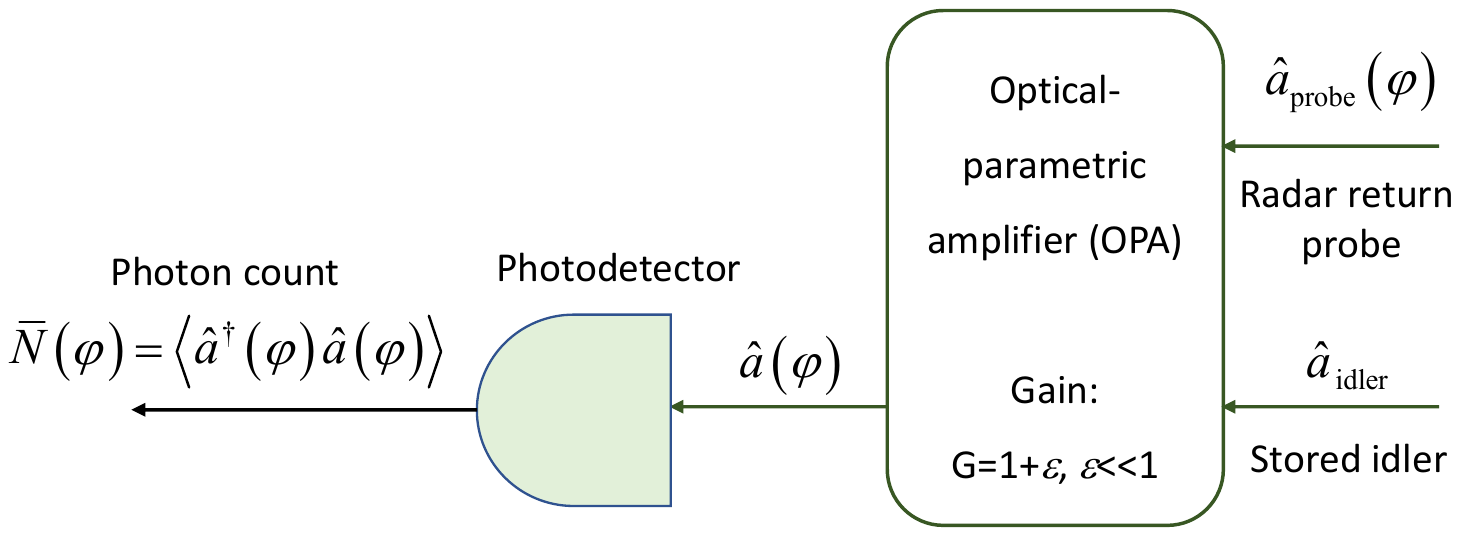}
    \caption{The optical parametric amplifier (OPA)-based receiver.}
    \label{fig:5}
\end{figure}

To perform the joint measurement required we can arrange for the
interaction of the radar return probe and of the stored idler by
harnesssing an optical parametric amplifier (OPA), as illustrated in
Fig.~\ref{fig:5}~\cite{2}. The bottom output port's annihilation
operator $\hat{a}(\varphi )$ is related to the radar return probe
$\hat{a}_{\textup{probe}}(\varphi )$, where $\phi$ is the phase shift
introduced by the target. The associated idler annihilation operator
is denoted by $\hat{a}_{\textup{idler}}$. The variable $G$ in
Fig.~\ref{fig:5} denotes the OPA's gain, which is in this case chosen
to be low, namely $G=1+\varepsilon, \varepsilon <<1$. The concatenated
transmitter-target-radar receiver channel is modeled as a lossy
thermal Bosonic channel according to Fig.~\ref{fig:4}. It has been
shown in~\cite{27} that the maximum possible SNR improvement is 3 dB,
which is consistent with~\eqr{eq:ratio}. However, once all realistic
experimental imperfections have been included, the quantum
illumination outperforms the classical limit by only about
1~dB~\cite{27}. This experiment carried out at optical frequencies was
performed in a hostile noisy environment. Nevertheless, a quantum
advantage was demonstrated, as an explicit benefit of entanglement. To
improve the performance of the joint measurement based detection
scheme, the authors of~\cite{28} proposed to the employment of
sum-frequency generation, but the complexity of this detection scheme
was found excessive.
\subsubsection{Entanglement Assisted  Radars}
\label{EA_R}
\begin{figure}
    \centering
    \includegraphics[width=3.6in]{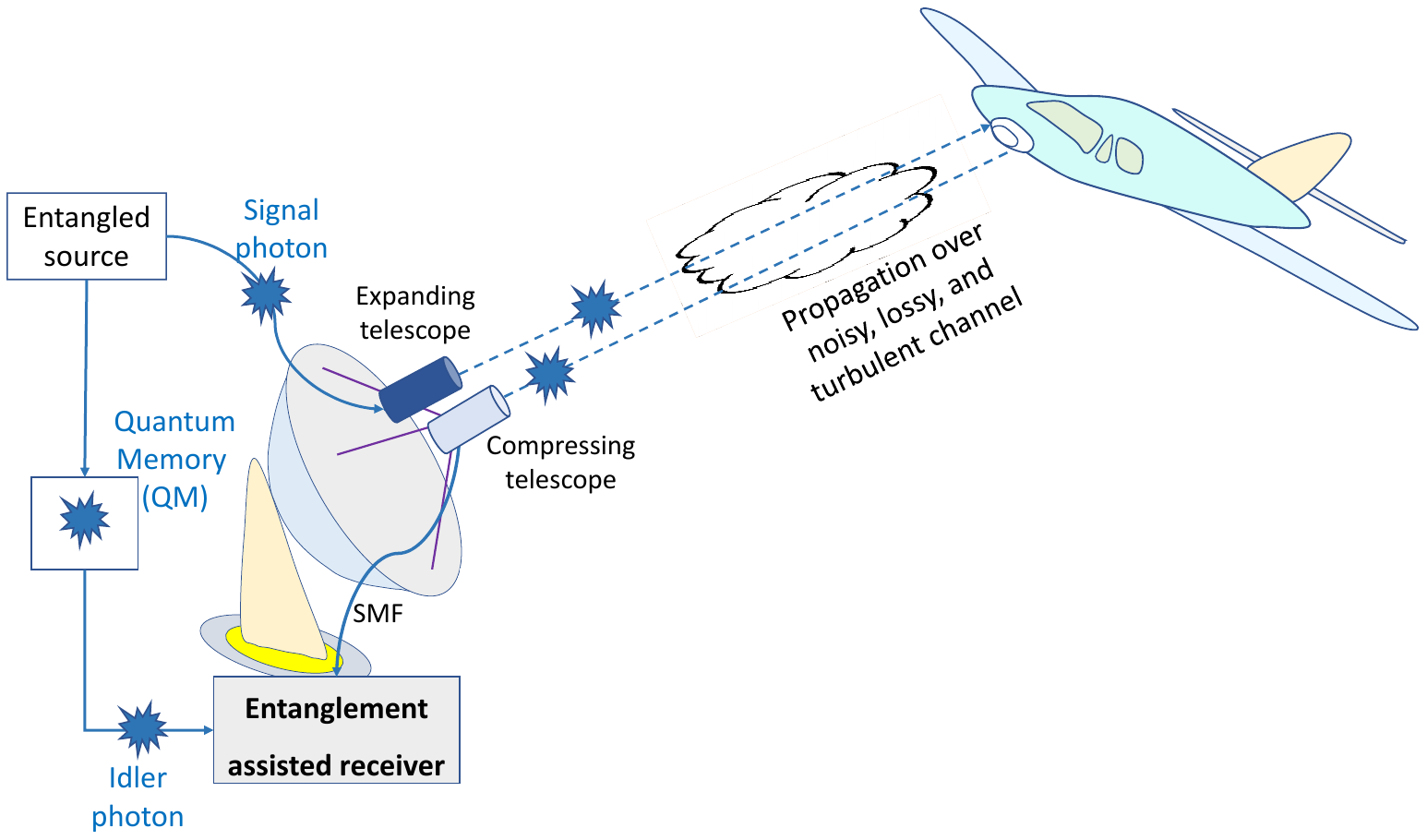}
    \caption{The EA monostatic quantum radar (modified from ref. ~\cite{4}).}
    \label{fig:6}
\end{figure}

Now we shift our focus to the particular version of the quantum
illumination relying on entangled sources. We describe several
entanglement-assisted (EA) radar schemes: 1) EA monostatic~\cite{4}
(depicted in Fig.~\ref{fig:6}), 2) EA joint
monostatic-bistatic~\cite{4}, and EA multistatic~\cite{5} radar
schemes. The EA bistatic scheme~\cite{3} was already described in
Fig.~\ref{fig:2}. All of these schemes employ the Gaussian states
generated through the continuous-wave SPDC process. The SPDC-based
entangled source represents a broadband source having $M=T_m W$
i.i.d. signal-idler photon pairs, where $T_m$ is the measurement
interval and $W$ is the phase-matching SPDC bandwidth. Each
signal-idler photons pair, which are represented by blue photons in
Fig.~\ref{fig:6}, is in fact a two-mode squeezed vacuum (TMSV)
state~\cite{2}. The signal-idler entanglement is characterized by the
phase-sensitive cross-correlation (PSCC) coefficient, defined as
${C_{si}}=\left< {\hat{a}_{s} \hat{a}_i}\right>=\sqrt{N_s(N_s+1)}$,
which can be considered as the quantum limit. Clearly, in the
low-brightness regime of $N_s<<1$, the PSCC is $\sqrt{N_s}$ and it is
much larger than the corresponding classical limit $N_s$. By going
back to Fig.~\ref{fig:6}, an entangled source is used at transmitter
side to generate a quantum correlated signal photon (probe) and an
idler photon, where the latter serves as a local reference. The signal
photon is transmitted over noisy, lossy, and atmospheric turbulent
channels towards the target with the aid of an expanding
telescope. The reflected photon representing the radar return is then
collected by the compressing telescope and detected by the radar's
receiver. In this context the quantum correlation between the radar
return and retained reference represented by the idler photon is
exploited at receive side for improving the receiver's sensitivity. The
interaction between the probe (signal) photon and the target can be
described by a beam splitter of transmissivity $T^{(r)}$. Therefore,
we can model the concatenated transmitter-target-receiver (directly
reflected mode) link (direct return channel) as a lossy thermal
Bosonic channel, similar to Fig.~\ref{fig:4}, where the target is
assumed to introduce the phase shift.

\begin{figure}
    \centering
    \includegraphics[width=5in]{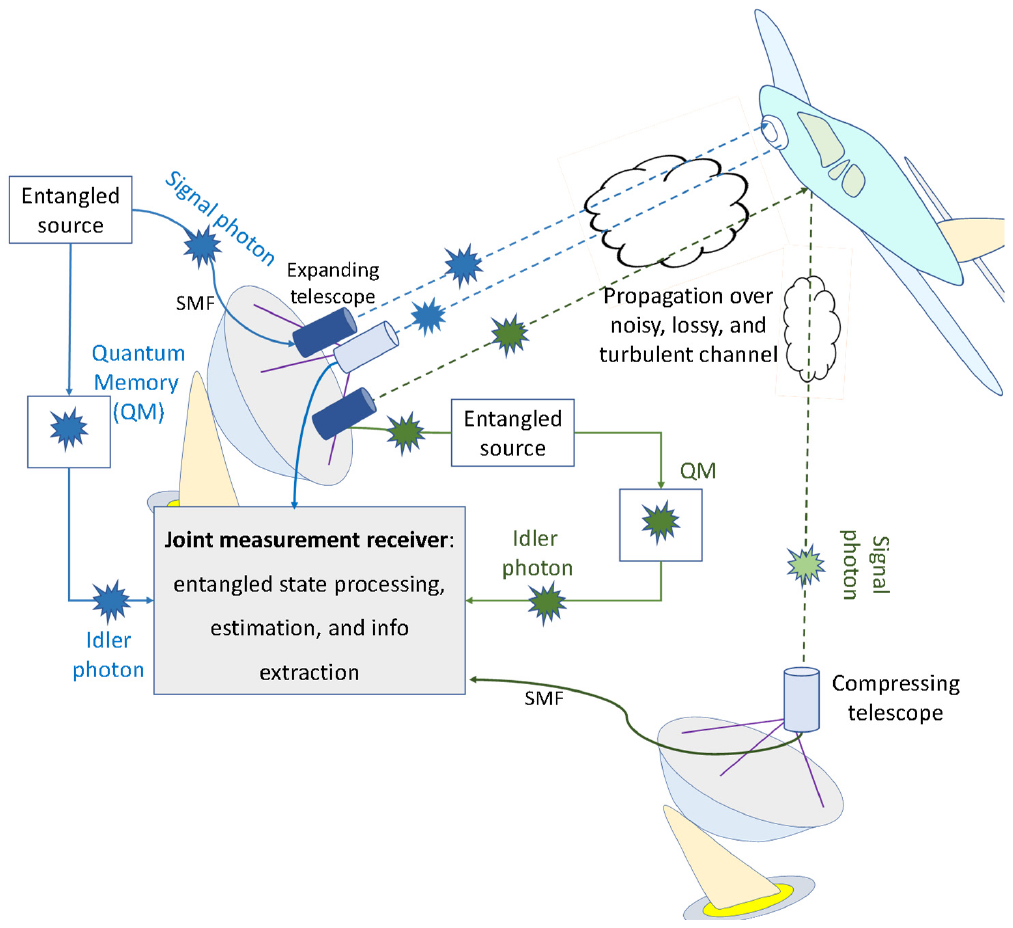}
    \caption{The EA joint monostatic-bistatic quantum radar scheme (modified from ref. ~\cite{4}).}
    \label{fig:7}
\end{figure}

The operational principle of the EA joint monostatic-bistatic quantum
radar detection scheme proposed in~\cite{4} is portrayed in
Fig.~\ref{fig:7}. The wideband entangled source generates two
entangled photon pair, where each pair contains a signal and an idler
photon. The idler photons are stored in the quantum memories of the
receivers. Both signal photons are then transmitted with the aid of
the corresponding expanding telescopes over noisy, lossy, and
atmospheric turbulent channels to the target. The directly reflected
photon is collected by the compressing telescope and it is detected by
the first radar receiver, while the forward scattered photon is
collected by the second compressing telescope and it is detected by
the second radar receiver. The quantum correlation is then exploited
at the receive sides for improving the overall target detection
probability. The inherent spatial diversity is also exploited for
improving the overall SNR.

\begin{figure}
    \centering
    \includegraphics[width=4.5in]{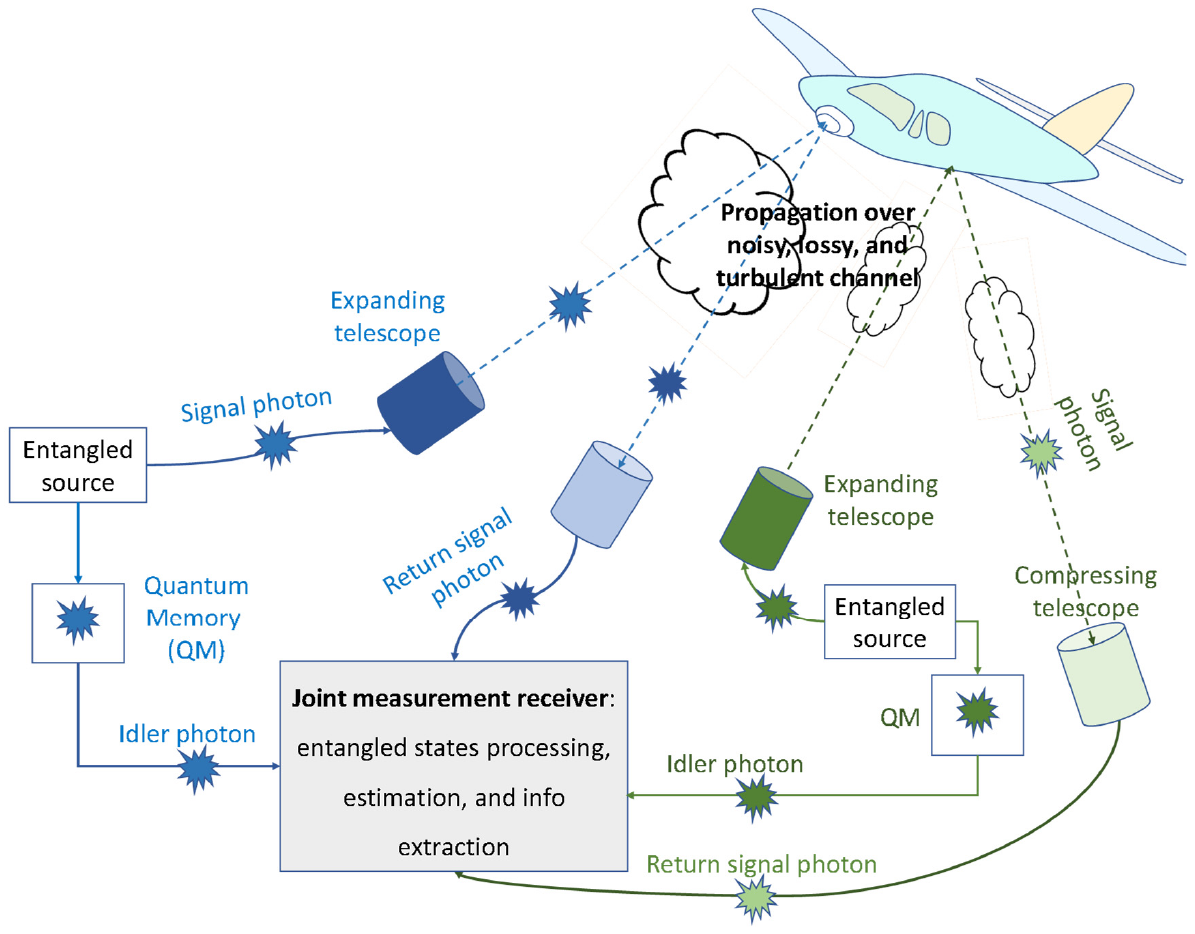}
    \caption{The 2x2 EA multistatic (MIMO) quantum radar scheme (modified from ref.~\cite{5}).}
    \label{fig:8}
\end{figure}

The operational principle of the EA multistatic quantum radar
detection technique of~\cite{5} is depicted in
Fig.~\ref{fig:8}. Compared to the joint monostatic-bistatic scheme
from Fig.~\ref{fig:7}, the EA multistatic radar scheme has slightly
higher complexity, but much better performance and improved
flexibility. The multistatic radar scheme employs multiple entangled
transmitters and multiple coherent detection-based receivers, as shown
in Fig.~\ref{fig:8}, while the previous scheme only relied on a single
transmitter and monostatic as well as bistatic receivers. In this
scheme, the phase-sensitive quantum correlation is exploited at the
receivers' sides with the objective of improving the overall detection
probability of the target. Moreover, to increase the overall SNR, the
spatial MIMO concept is harnessed.  Compared to the reflected and
forward scattered components available in joint monostatic and
bistatic radars, which are correlated and as such do not provide full
spatial diversity, by using multiple transmitters that are
sufficiently fasr apart in space we can ensure statistical
independence of the different optical paths, thus achieving the
maximum attainable diversity order of the multistatic radar
concept. This leads to improvements both in terms of the diversity
order and the array gain compared to joint bistatic-monostatic
schemes. An alternative terminology for this scheme is the EA MIMO
radar concept~\cite{32}.

\begin{figure}   \centering
    \includegraphics[width=3.4in, height=3.4in]{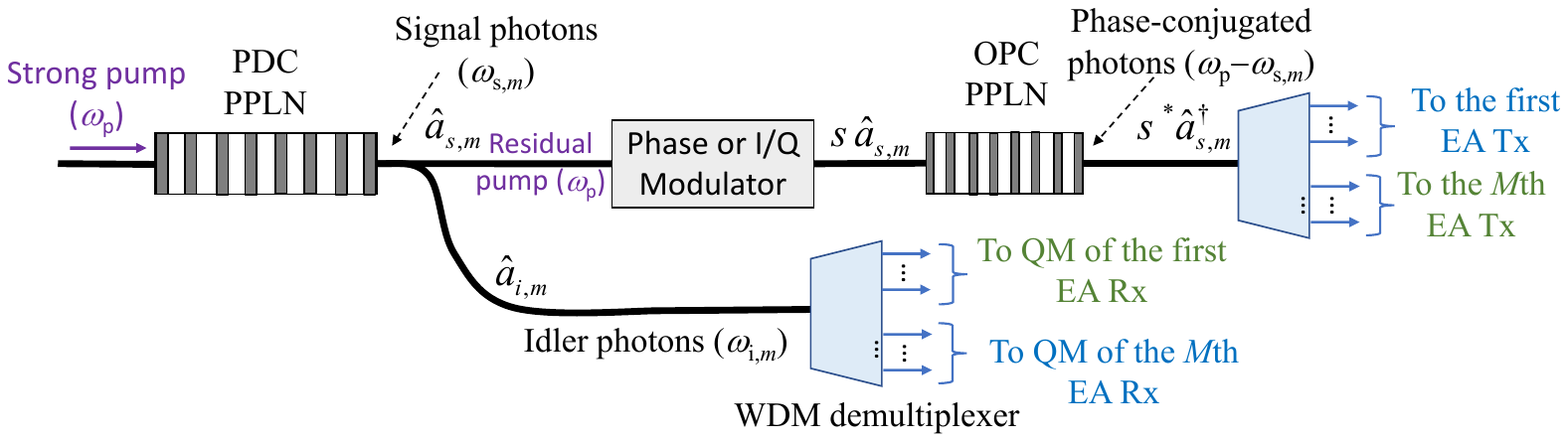}
    \caption{The integrated multistatic (MIMO) EA transmitter with transmit side OPC. We use s to denote a phase (or I/Q) modulator induced signal constellation point. QM: quantum memory, PPLN: periodically poled LiNbO3 waveguide, PDC: parametric down conversion, OPC: optical phase-conjugation, and WDM: wavelength-division multiplexing. (Modified from ref.~\cite{5}).}
    \label{fig:9}
\end{figure}

To simplify the transceivers' design and reduce the system's cost, the
transmit sides' optical phase-conjugation (OPC) -- which is required
before the detection takes place -- has been performed at transmitter
side so that classical balanced coherent detectors can be utilized as
the EA detectors~\cite{5}. Furthermore, the employment of a single
broadband entangled source combined with a WDM demultiplexer has been
proposed as the common source for all transmitters, which is
illustrated in Fig.~\ref{fig:9}. First a periodically poled LiNbO3
(PPLN) waveguide serves as the SPDC source, which generates a large
number of signal-photon pairs, where only the $m$-th signal-photon
pair is illustrated in Fig.~\ref{fig:9}. The signal and idler photons
become separated by an appropriately designed Y-junction. The idler
photons become further separated by the WDM demultiplexer, whose
outputs are directed towards the quantum memories (QMs) of the
corresponding EA receivers. On the other hand, all signal photons get
simultaneously modulated by a training sequence known to all EA
receivers, imposed by a Q-ary PSK modulator.  This sequence is then
used for estimating the phase shift introduced by the target and the
channel. The common sequence is also used for determining the target's
range more precisely by harnessing the cross-correlation method. The
second PPLN waveguide is then used for carrying out the OPC by
employing the popular difference frequency generation (DFG) process,
in which the $m$-th signal photon at angular frequency $\omega _{s,
  m}$ interacts with the pump photon $\omega _{p}$ to get the
phase-conjugated (PC) photon at the radial frequency $\omega
_{p}-\omega _{s, m}$. We then use the WDM demultiplexer for
demultiplexing the signal photons to be used in the multistatic
transmitters, as depicted in Fig.~\ref{fig:9}.

\begin{figure}
    \centering
    \includegraphics[width=3.2in]{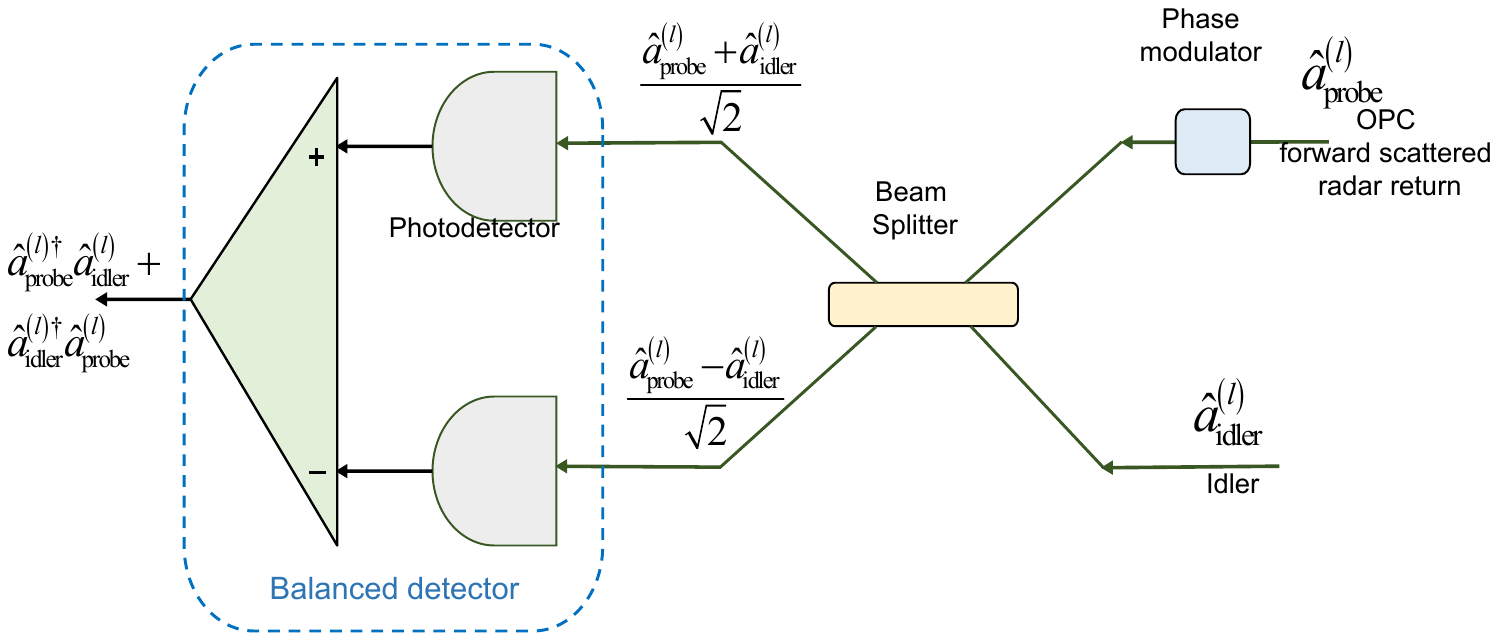}
    \caption{The EA receiver corresponding to the $l$-th forward scattered component. The receive side phase modulator is used to select either in-phase or quadrature component of the corresponding phase-conjugated signal. The photodiode responsivity is set to 1 A/W. (Modified from ref. ~\cite{5}).}
    \label{fig:10}
\end{figure}

Given that the OPC is carried out at the transmiter, we do not have to
use OPC-based EA receivers -- instead, commercially available
classical balanced coherent detectors may be used as the EA receivers,
such as the one shown in Fig.~\ref{fig:10}. This substantially reduces
the overall system cost and complexity. For the associated simulation
based results interested readers might like to refer to~\cite{5}. For
experimental characterization of the EA radar concept in the context
of realistic turbulent free-space optical channels motivated readers
might like to consult~\cite{24}.

\subsection{Knowledge Gaps and Challenges}
\label{Knowledge_Gaps_and_Challenges}

As a step towards quantum illumination absed radar operating at
microwave frequencies, authors of~\cite{10} proposed an
optical-microwave transduction scheme for generating the entangled
microwave and optical idler photons. The microwave photon return probe
is converted by the same type of device into the optical domain so
that the joint measurement can be carried out in the optical domain.'
For example, the OPA-based detection scheme of Fig.~\ref{fig:5} may be
harnessed in this context. Unfortunately, the current
optical-microwave transduction devices~\cite{29,30} still have a low
transduction efficiency to be of practical importance. The Josephson
parametric converter -- which serves a similar role to the SPDC device
-- has been used in~\cite{11} for generating the entangled microwave
photons. However, the quantum advantage of the experimental
demonstration remained rather limited. In a recent
experiment~\cite{31}, the Josephson ring modulator has been employed
for coupling a pair of microwave resonators and thus for producing the
entangled microwave photons. This experiment employed the OPA-based
receiver and suceeded in achieving an approximately 20 percent
advantage over its corresponding classical counterpart.

The first experimental demonstration of quantum illumination based on
entangled photons at optical frequencies over a free-space optical
link length of 750 m was documented in~\cite{24} in the face of strong
atmospheric turbulence.  The system applied the optical
phase-conjugation to the idler photons before the balanced homodyne
detector by relying on the concept introduced in~\cite{23}. In order
to improve the tolerance to turbulence effects adaptive optics was
used.

\begin{figure}
    \centering
    \includegraphics[width=3.2in]{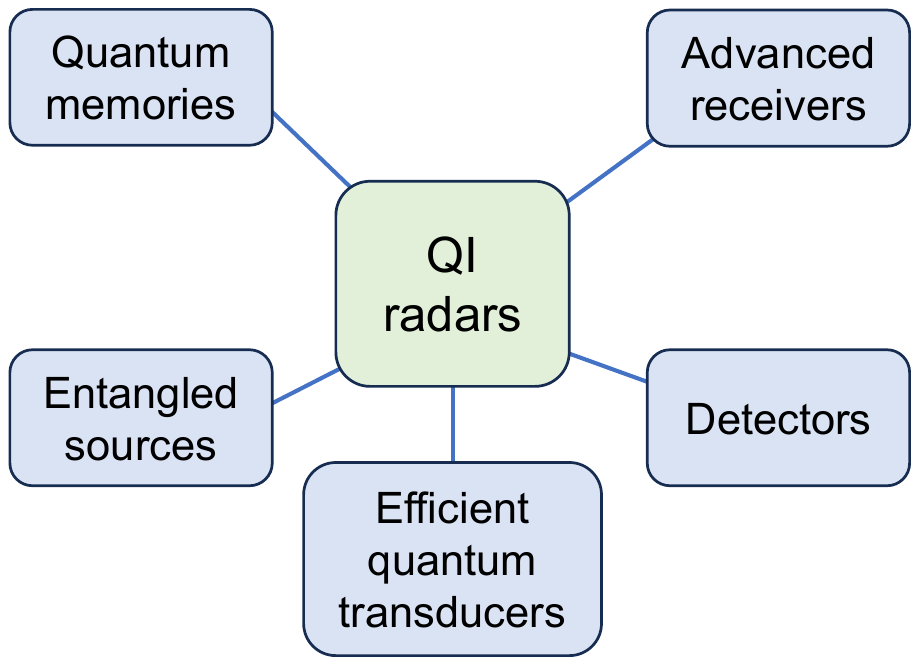}
    \caption{The quantum illumination-based radar enabling devices.}
    \label{fig:11}
\end{figure}

The devices facilitating quantum illumination based radar services at
microwave frequencies are summarized in Fig.~\ref{fig:11}. One of the
important issues for quantum illumination is the need for quantum
memory, which will store the idler photons until they are needed for
balanced detection. In experimental demonstration~\cite{24} an optical
delay line is used instead of real quantum memory. For quantum radar
applications, variable delay lines can be used. However, the
commercially available ones are bulky, slow, and expensive. The
research of quantum memory having improved retention duration is
currently ongoing~\cite{40,41,42,43,44,45,46,47}. However, these
solutions are still far from commercialization. Another challenging
open problem is is the design of detectors having near-unity quantum
efficiency~\cite{48,49,50}. As a further advance, for the OPA-based
receiver of Fig.~\ref{fig:5} photon-number-resolving detectors are
required~\cite{51,52}. As for microwave radar illumination, efficient
quantum transducers should be developed~\cite{30,53,54}. Furthermore,
new types of receivers suitable for quantum illumination should be
developed~\cite{55, 56}. Finally, for quantum multistatic radars
associated with numerous transmitters, large-scale
entanglement sources are needed~\cite{57,58,59}.

\subsection{Research Road Map}
\label{Research_Road_Map}

\begin{figure*}[!th]
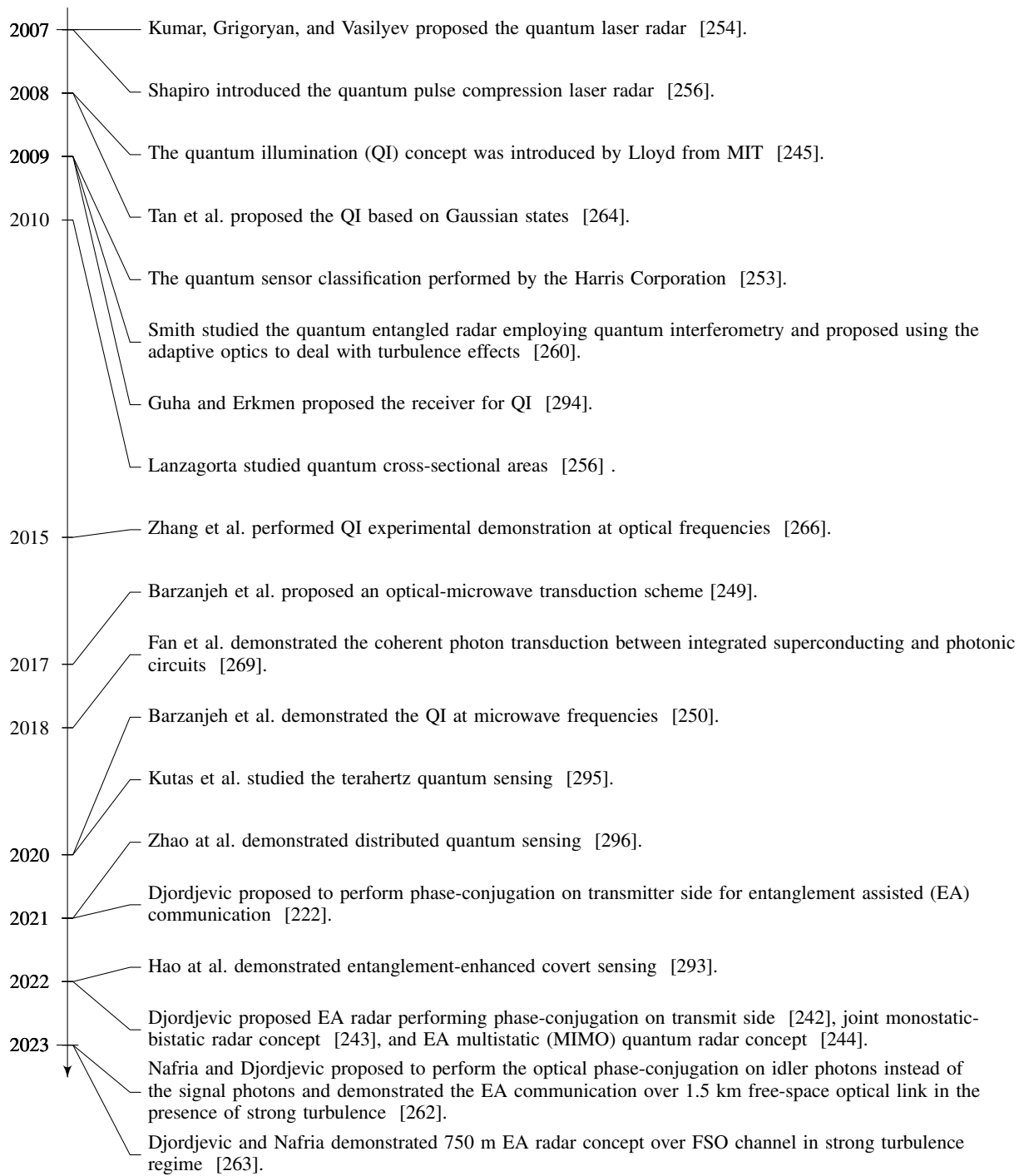

\begin{small}
\begin{timeline}{2007}{2023}{1cm}{1cm}{14cm}{0.7\textheight}
\entry{2007}{Kumar, Grigoryan, and  Vasilyev proposed the quantum laser radar ~\cite{15}. }
\entry{2007}{Shapiro introduced the quantum pulse compression laser radar ~\cite{17}.}
\entry{2008}{The quantum illumination (QI) concept was introduced by Lloyd from MIT ~\cite{6}.} 
\entry{2008}{Tan et al. proposed the QI based on Gaussian states ~\cite{Tan2008}.}
\entry{2009}{The quantum sensor classification performed by the Harris Corporation ~\cite{14}.}
\entry{2009}{Smith studied the quantum entangled radar employing quantum interferometry and proposed using the adaptive optics to deal with turbulence effects ~\cite{21}.}
\entry{2009}{Guha and Erkmen proposed the receiver for QI ~\cite{35}.}
\entry{2010}{Lanzagorta studied quantum cross-sectional areas ~\cite{17} .}
\entry{2015}{Zhang et al. performed QI experimental demonstration at optical frequencies  ~\cite{27}.}
\entry{2017}{Barzanjeh et al. proposed an optical-microwave transduction scheme~\cite{10}.}
\entry{2018}{Fan et al. demonstrated the coherent photon transduction between integrated superconducting and photonic circuits ~\cite{29}.}
\entry{2020}{Barzanjeh et al. demonstrated the QI at microwave frequencies ~\cite{11}.}
\entry{2020}{Kutas et al. studied the terahertz quantum sensing ~\cite{36}.}
\entry{2021}{Zhao at al. demonstrated distributed quantum sensing ~\cite{37}.}
\entry{2021}{Djordjevic proposed to perform phase-conjugation on transmitter side for entanglement assisted (EA) communication ~\cite{banchi2021generalization}.}
\entry{2022}{Hao at al. demonstrated entanglement-enhanced covert sensing ~\cite{39}.}
\entry{2022}{Djordjevic proposed EA radar performing phase-conjugation on transmit side ~\cite{3}, joint monostatic-bistatic radar concept ~\cite{4}, and EA multistatic (MIMO) quantum radar concept ~\cite{5}.}
\entry{2023}{Nafria and Djordjevic proposed to perform the optical phase-conjugation  on idler photons instead of the signal photons and demonstrated the EA communication over 1.5 km free-space optical link in the presence of strong turbulence  ~\cite{23}.}
\entry{2023}{Djordjevic and Nafria demonstrated 750 m EA radar concept over FSO channel in strong turbulence regime  ~\cite{24}.}
\end{timeline}
\end{small}
\caption{The timeline describing the quantum radar research activities.} 
\label{fig:timeline3}
\end{figure*}

The quantum radar relies on a relatively new concept compared to its
classical counterparts. In Fig.~\ref{fig:timeline3} we provide the
timeline describing the quantum radar research activities. The quantum
illumination radars appear to be more practical than the quantum
interferometry based radars. Among various quantum illumination radar
schemes the entanglement assisted radars have received the most attention
at the time of writing.

For quantum illumination at microwave frequencies, the quantum
advantage over classical radar has remained limited so
far~\cite{11,31}. Substantial community effort is required for
developing improved entangled sources and detectors operating at
microwave frequencies. Again, the optical-microwave transduction
devices have limited efficiency~\cite{29,30,38,39} and significant
research efforts have to be invested in this field as well.

The quantum illumination based on entangled states at optical
frequencies~\cite{2,3,4,5} appears to be much more mature compared to
its microwave counterparts. There are some relevant EA quantum
demonstrations at these frequencies~\cite{24,27}, but this field is
very much in its infancy and it may be expected to attract scientists
looking for exciting open challenges. Clearly, the EA radars operating
at optical frequencies suffer from substantial performance erosion in
the face of clouds and fog, hence additional research efforts are likely
to be invested in this field as well.

Another research area along this evolutionary quantum radar research
path is related to the covert sensing concept, with the first
experimental demonstration reported in~\cite{37}. The quantum covert
radar concept can be considered as a specific quantum illumination
scheme in which the transmission of the probe is controlled by an
agent, who tries perceive the presence of any sensing. By
ensuring that the probe is partially masked by the background
noise this agent will not be able to detect the sensing
attempt owing to the statistical fluctuations in his measurement
attempts. The quantum illumination concept has also become relevant
for LIDAR applications~\cite{60,61} and it may be expected to attract
substantial research attention.

As part of the roadmap, it is necessary to develop the quantum
illumination radars operating at microwave frequencies closer to the
commercialization, which requires the development of new type of
receivers and devices~\cite{62,63}. As for implementations at optical
frequencies, the photonic integration research is making
progress~\cite{2,3,4,5}, while on theoretical side, tighter bounds are
expected to be developed for quantum illumination~\cite{64}.

Another interesting, but hitherto largely unexplored area in the
quantum illumination literature will be to combine quantum radars and
communications. This research problem is known in the classical
radio-frequency (RF) and optical wireless communication literature as
dual-function radar communication (DFRC)~\cite{66,67,68,69}. To
elaborate briefly, the integrated multistatic entanglement source of
Fig.~\ref{fig:9} is also suitable for simultaneous radar and
communication services.  Briefly, similar to EA
communications~\cite{23}, we organize the data to be transmitted into
packets, with the packet-header known to the receiver. The data
sequence may be protected by a variety of error correction codes, such
as an LDPC code and mapped to the payload.  In both radar and
communication applications, we use the packet-header to determine the
beginning of the packet by the popular cross-correlation method. In radar
systems we only need the header for estimating the target detection
probability and the target range, while in the EA communication
system we further process the payload and carry out LDPC decoding for
recovering the transmitted sequence.

The original quantum illumination concept was conceived for target
detection. In this context it is also relevant to demonstrate quantum
advantage in terms of the range~\cite{2,3,4,5},~\cite{65}, angle, and
speed of the target, which is an active research topic. Harnessing
quantum illumination at microwave frequencies is very challenging,
because the background radiation noise is significantly higher than at
optical frequencies. Hence significant research efforts are needed to
make efficient quantum illumination at microwave frequencies a
reality.

Whilst the quantum radar systems of this section are still very much
in their infancy, the QKD systems of the next section are now
commercially available. However, there is a pressing need for further
research to architect the global Qinternet relying on large-scale
multi-protocol relay-aided networking solutions providing end-to-end
security, as discussed in the next section.

\section{Quantum Key Distribution Networks}
\label{qkd-mohsen}

{\bf The Myth:} There is no need for quantum key distribution (QKD);
post-quantum cryptography (PQC) would eliminate all security risks
that future quantum computers may cause.

{\bf The Reality:} PQC could certainly be part of the solution for
providing data security in future telecommunications networks, but it
could still be susceptible to 'harvest now, decrypt later' type of
attacks. QKD solutions are already an off-the-shelf reality and they
constitute the most reliable solution that offers long-term security,
provided that it is implemented appropriately with the aid of
high-quality devices avoiding information leakage.

{\bf The Future:} Having said that, the economy of scale has to reduce
the cost to make it commercially viable for private subscribers {\em
  and} to offer end-to-end security for all relevant applications. A
seven-step evolutionary pathway was constructed
in~\cite{long2022evolutionary} for outlining the associated roadmap.
This would require further advances ib both the relevant quantum and
classical technologies, and a global research effort is required for
integrating quantum and classical communications networks.

{\bf Abstract:} {\em QKD has been one of the most successful
  applications of quantum technologies, which crucially addresses some
  security gaps in our current communications
  systems~\cite{Pirandola:AQCrypt, Hanzo:QInternet, RazaviQCbook}. In
  particular, the threat from quantum computers becoming able to crack
  some of the widely employed public-key cryptosystems has required
  developing new methodologies for sharing secret cryptographic keys
  among legitimate users. QKD offers a solution, based on the
  previously discussed laws of quantum mechanics, that offers
  confidentiality even for the future, when large-scale quantum
  computers become available. Examples of such scenarios include the
  exchange of medical records over the Internet, which, for privacy
  reasons, may have to remain confidential during the lifetime of an
  individual and even beyond that. For such personal applications, it
  is vital that we expand our communications network infrastructure in
  such a way that it accommodates QKD deployment.}

The key objective of a QKD protocol is to exchange, in a secure way, a
secret cryptographic key between a pair of authenticated network
users, historically referred to as Alice and Bob. In the so-called
{\em prepare-and-measure schemes}, Alice generates a random key on her
side, and then maps the corresponding bits to quantum states and sends
them to Bob according to a certain protocol. Bob would then observe or
measure the potentially error-infested received signals, and by
exchanging some classical information with Alice via an authenticated
channel, they attempt to agree about the specific choice of an
identical secret key with each other. The level of security is
typically characterized by a security parameter, which specifies the
distance between an ideal randomly generated secret key and the one
that can be obtained by the QKD protocol, accounting for the fact that
Alice and Bob have the option to abort the protocol if the estimated
error rate os higher than a certain threshold, because it is deemed to
be tampered with by the eavesdropper (Eve)~\cite{Renner_ComposSec}.
Indeed, one could argue that if E attempts to observe the confidential
quantum-domain signal, it collapses back in the classical domain.

A unique property of QKD is that we can detect eavesdropping attempts,
which is not possible in the classical domain.  This allow us to limit
the amount of information leakage concerning the secret key to a third
party in a point-to-point system. In order for QKD to be accessible to
a wide range of customers it is essential to cover long distances, in
conjunction with a variety of classical services that next-generation
(NG) wireless systems aim to offer. However, the challenge is that
given a 0.2~dB/km fiber attenuation, at a distance of 100~km the
quantum signal is attenuated by 20~dB, ie. to 1~\% of its original
power and it cannot be amplified without first observing/measuring the
signal, which returns it to the classical domain. Therefore it is
plausible that the achievable rate is a direct function of the SNR in
dB, which is in turn determined by the distance. This physically
tangible relationship results in the formulation of the salient QKD
performance metric, namely the {\em key-rate vs. distance
  relationship}. Hence sophisticated terrestrial relaying techniques
are required for longer distances and there is also a clear {\em
  trade-off between the key-rate vs. the number of relays}, when
covering a fixed distance. More explicitly, the key-rate may be
increased by shortening the distance of relays, but at the current
state-of-the-art the so-called {\em trusted relays} must be hosted in
secure customer premises for preventing tampering by Eve.  For
example, the Eurasian QKD link spanning from Vladivostok to Helsinke
has a length in excess of 15 000~km and relies on numerous expensive
relays. As a design alternative, free-space optical (FSO) satellite
links have to be harnessed for lomng-haul transmission. Furthermore,
as alluded to above, QKD systems also require a classical channel in
support of the key negotiations between Alice and Bob, which may be
accommodated by the same fiber as the quantum channel by using
wave-length division multiplexing (WDM).  However, utter care must be
excersized to avoid that the weak quantum signal is overwhelmed by the
out-of-band interference of the high-power classical signal.

We continue with a brief critical appraisal of the state-of-the-art in
Sec.~\ref{Sec:QKD-SOTA}, followed by challenges we are facing, when
aiming for providing end-to-end security in networking scenarios, as
detailed in Sec.~\ref{Sec:QKD-Chall}. We conclude by a vision for the
roll-out of a long-distance QKD network in Sec.~\ref{Sec:QKD-Roadmap}.

\subsection{State-of-the-Art}
\label{Sec:QKD-SOTA}
\subsubsection{Key Principles}
\begin{figure}[t]
\centering
\includegraphics[width=\columnwidth]{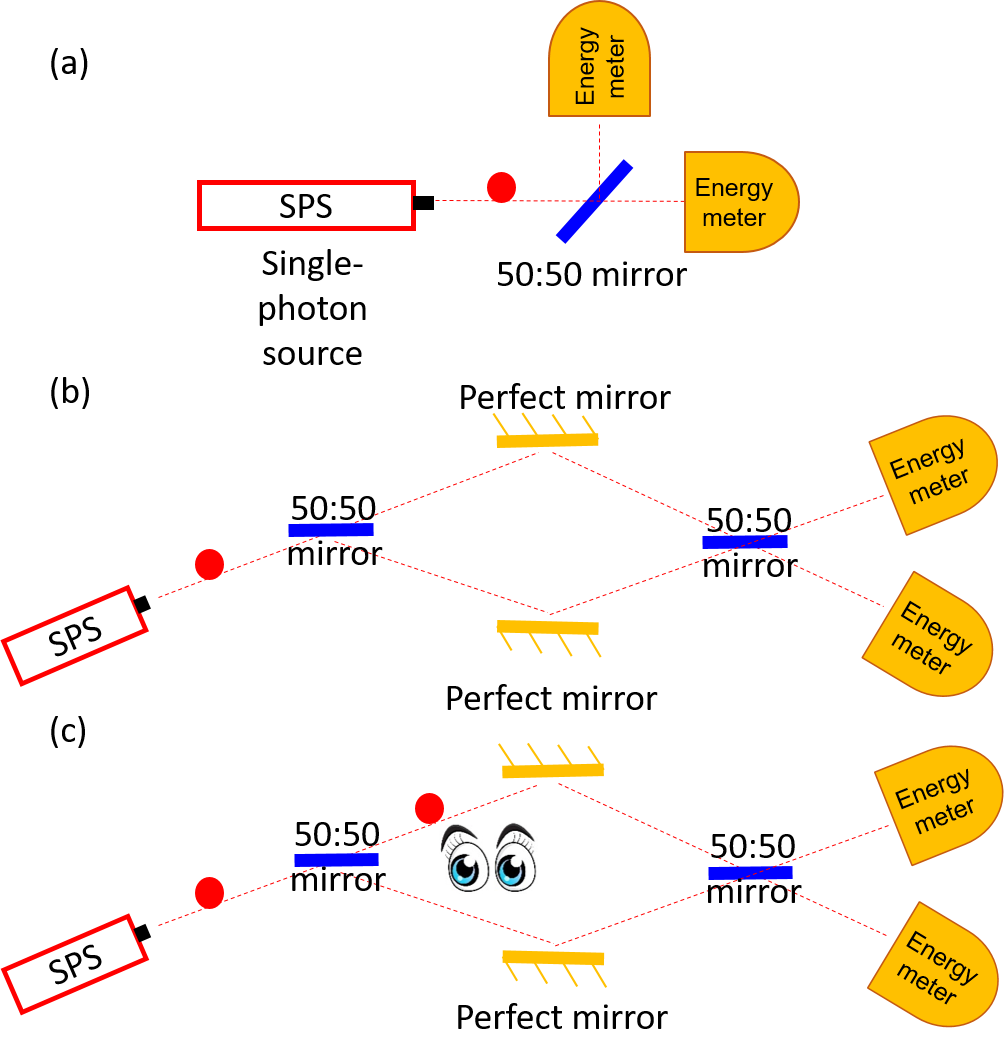}
\caption{Three experiments at the heart of BB84: (a) A single photon
  does not split, but randomly chooses a path; (b) Interference holds
  even at the single-photon level; and (c) Making observations can
  change the superposition state. } \label{fig:QKD-gedanken}
	\end{figure}

The basic idea behind the original QKD protocol, known as BB84 after
its inventors Bennett and Brassard \cite{bennett1984brassard}, can be
explained using the experiments shown in
Fig.~\ref{fig:QKD-gedanken}. Assume that we have a source that can
generate a single photon. Let this photon propagate through a 50:50
mirror, known as a {\em balanced beam splitter}. Then the question
arises - which one of the two ideal energy meters in
Fig.~\ref{fig:QKD-gedanken}(a) would click? Because quantum mechanics
only allows discrete energy levels for each mode of light, this single
photon cannot be split into a pair of smaller packets of energy. Hence
only one of the single-photon detectors would click - each with a
probability of 1/2.

The second experiment in Fig.~\ref{fig:QKD-gedanken}(b) uses two
balanced beam splitters, which jointly constitute an
interferometer. We know from classical optics that for a laser source
at the input, the two output ports of the second beam splitter exhibit
constructive and destructive interference, respectively, hence
resulting in only one of the detectors registering some
energy. Interestingly the same thing happens at the quantum mechanical
level for the single-photon input. This is due to the superposition
principle where, after the first beam splitter the state of the system
is characterized by the single photon being in the superposition of
the upper and lower arms of the interferometer. Finally, let us assume
that a curious third party observer measures which arm the photon in
Fig.~\ref{fig:QKD-gedanken}(b) has actually taken. Given this
knowledge, the experiment in Fig.~\ref{fig:QKD-gedanken}(c) becomes
identical to that of Fig.~\ref{fig:QKD-gedanken}(a), where -- in
contrast to the setup of Fig.~\ref{fig:QKD-gedanken}(b) -- either of
the detectors may actually click.

\begin{figure}[b]
  \centering
\includegraphics[width=\columnwidth]{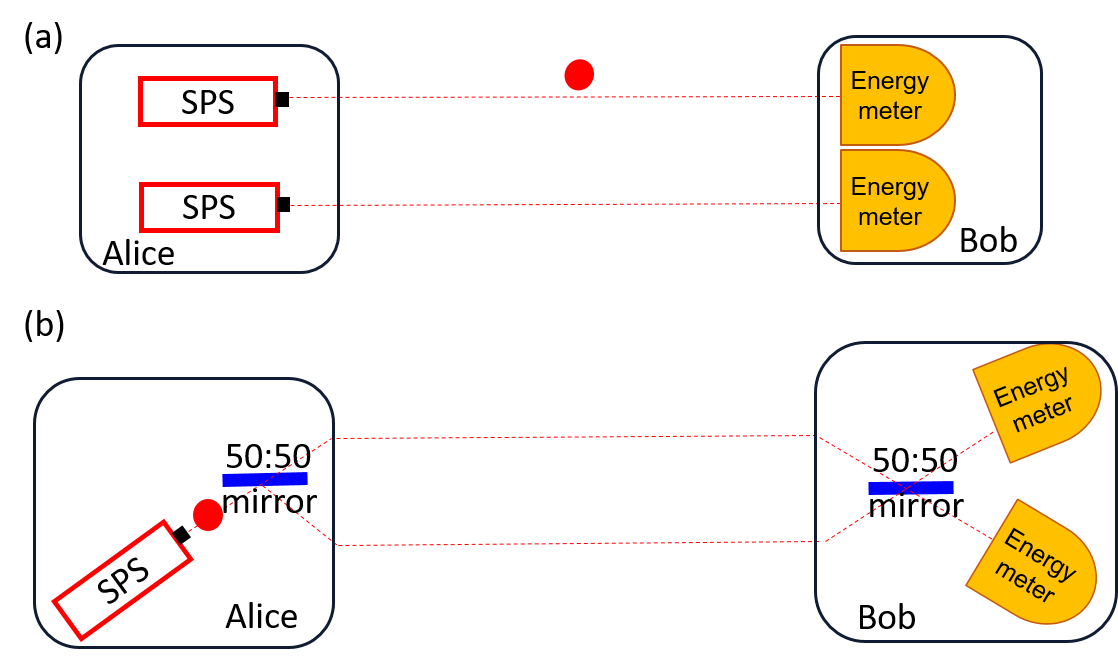}
\caption{The conjugate encodings used in our toy model QKD
  protocol. (a) In key transmission rounds, Alice sends a single
  photon on the upper (lower) arm to encode bit 1 (0). (b) In test
  rounds, Alice sends a superposition state to
  Bob. } \label{fig:QKD-toymodel}
	\end{figure}	

\begin{table*}
\begin{center}
\begin{tabular}{|c|l|c|c|c|c|c|c|c|c|c|c|c|c|}
\hline 
\multicolumn{14}{|l|}{\textbf{Alice}}\\ \hline
1 &Raw key & $1$  & \cellcolor{red}$0$ & $0$ & $1$ & \cellcolor{red}$1$ & $1$ & $0$ & $1$ & $0$ & \cellcolor{red}$0$ & $0$ & $1$\\ 
2 &Preparation (or encoding) basis & \cellcolor{green}$+$ &  $+$ & \cellcolor{green}$\times$ & $\times$ & $+$ & \cellcolor{green}$\times$ &  \cellcolor{green}$+$ & $+$ & \cellcolor{green}$\times$ & $\times$ & \cellcolor{green}$+$ & $\times$\\
3 &Quantum state preparation & $\rightarrow$ & $\uparrow$ & $\nearrow$ & $\searrow$ & $\rightarrow$ &$\searrow$ & $\uparrow$ & $\rightarrow$ & $\nearrow$ &$\nearrow$ & $\uparrow$ & $\searrow$\\ \hline 
\multicolumn{14}{|l|}{\textbf{Bob}}\\ \hline
4 &Measurement basis & \cellcolor{green}$+$ &  $\times$ & \cellcolor{green}$\times$ & $+$ & $\times$ & \cellcolor{green}$\times$ &  \cellcolor{green}$+$ & $\times$ & \cellcolor{green}$\times$ & $+$ & \cellcolor{green}$+$ & $+$ \\
5 &Quantum state detected & $\rightarrow$ & $\searrow$ & $\nearrow$ & $\rightarrow$ & $\nearrow$ & $\searrow$ & $\uparrow$ & $\searrow$ & $\nearrow$ & $\rightarrow$ & $\uparrow$ & $\rightarrow$ \\
6 &Detected key & $1$ & \cellcolor{red}$1$ & $0$ & $1$ & \cellcolor{red}$0$& $1$ & $0$ & $1$ & $0$ & \cellcolor{red}$1$ & $0$ & $1$\\ \hline
\multicolumn{14}{|l|}{\textbf{Classical post-processing}}\\ \hline
7 & Sifted key & $1$ &&  $0$ &&& $1$&$0$&&$0$&&$0$&\\ \hline
8 & \multicolumn{13}{l|}{Parameter (or error) estimation} \\
9 &  \multicolumn{13}{l|}{Information reconciliation} \\
10 &  \multicolumn{13}{l|}{Privacy amplification} \\ \hline
\end{tabular}
 \end{center}
 \caption{Prepare-and-measure discrete variable BB84 QKD example (in
   the absence of Eve and noise) \copyright Hosseinidehaj {\em et al.}
   IEEE~\cite{Hosseinidehaj_SatQKD2017}}
\label{tab:BB84}
\end{table*}

By appropriately combining the above ideas one can design a simple QKD
protocol as follows. In this protocol, Alice uses two types of
encoding, chosen at random, for transmitting her key bits to Bob. In
the first type, used for key generation, she uses a dual rail system,
as shown in Fig.~\ref{fig:QKD-toymodel}(a). To send a bit 1 (0), she
sends a single photon via the upper (lower) channel. In the second
encoding type seen at the bottom of the figure, Alice generates a
superimposed state from the photon in the upper and lower arms, as
shown in Fig.~\ref{fig:QKD-toymodel}(b). Let us assume that at the
receiver, Bob happens to opt for the matching decoder. Naturally, in a
real experiment, Bob does not know which encoder Alice has used. But
he can randomly choose one of his two decoders and later check with
Alice via their authenticated classical channel, whether the encoder
and decoder match for a particular bit interval. In the absence of
eavesdropping attempts -- an ideal scenario -- we would expect that
the decoder of Fig.~\ref{fig:QKD-toymodel}(a) can register clicks on
either of the detectors, whereas the decoder in
Fig.~\ref{fig:QKD-toymodel}(b) should observe clicks in only one of
the detectors. In this setting, if an Eve attempts to check, which
channel the photon is travelling through, she would then perturb the
statistics of the test rounds, which results in an increased bit error
rate. For this scenario it is possible to formulate bounds for
quantifying, how much information might have been leaked to Eve. In
practical QKD protocols so-called {\em reconciliation} schemes relying
on sophisticated classical error correction codes may be used for
mitigating the error rate experienced. Finally, some of the secret key
bits generated may be dropped by relying on the technique of {\em
  privacy amplification} for further confusing Eve.  By contrast, if
the error rate is excessive, the legitimate users may decide to abort
the protocol and re-commence the key-negotiation process.

To elaborate a little further on the Bennett and
Brassard~\cite{bennett1984brassard} protocol, we briefly refer to
Table~\ref{tab:BB84} abridged from~\cite{Hosseinidehaj_SatQKD2017},
where further details may be found.
\begin{enumerate}
\item A random binary key is generated by Alice in the classical
  domain, which is referred to as the raw key;

\item Then she randomly selects either a rectilinear or diagonal
  polarization basis represented by a + and x character in order to
  convey the 0/1 bits of the raw key; This is the so-called
  preparation basis;

\item As seen in Table~\ref{tab:BB84}, the quantum state is prepared
  by mapping the binary key of Step~(1) according to the specific
  polarizations of Step~(2), as indicated by the arrows;

\item As for Bob, he randomly selects a so-called measurement basis
  and the instances where the preparation as well as measurement bases
  match are marked in green in Table~\ref{tab:BB84};
  
\item The received quantum states are then measured by Bob using the
  random measurement basis coined in Step~(4), where nothing is
  output, where the measurement and preparation bases are different;

\item At this stage Bob's detected states are mapped onto 1/0
  classical bits. The instances where the detected and raw key bits
  differ are marked in red in Table~\ref{tab:BB84}; These classical
  bits are then post-processed as follows;

 \item Only those bits are retained, which have the same preparation
    and measurement basis, resulting in the {\em sifted key};

 \item The error rate is estimated for detecting the presence/absence
   of Eve;

 \item The information reconciliation scheme corrects errors in the
   sifted key;

 \item Finally, the {\em reconciled key} is shortened by the {\em
   privacy amplification} scheme, hence reducing Eve's chances of
   guessing the agreed key.
\end{enumerate}


\subsubsection{The evolution of QKD protocols}
Historically, the BB84 protocol was inspired by some of the ideas in
an earlier paper by Wiesner~\cite{Wiesner}, but in recent years the
field of QKD has evolved quite rapidly in different directions. Some
developments have focused on simplifying the hardware requirements,
while others have been related to security proofs that support these
new or modified protocols. They belong to the family of
discrete-variable (DV) and continuous-variable (CV) QKD protocols,
which are detailed in~\cite{Hosseinidehaj_SatQKD2017,Hanzo:QInternet}.

\begin{figure*}[!th]
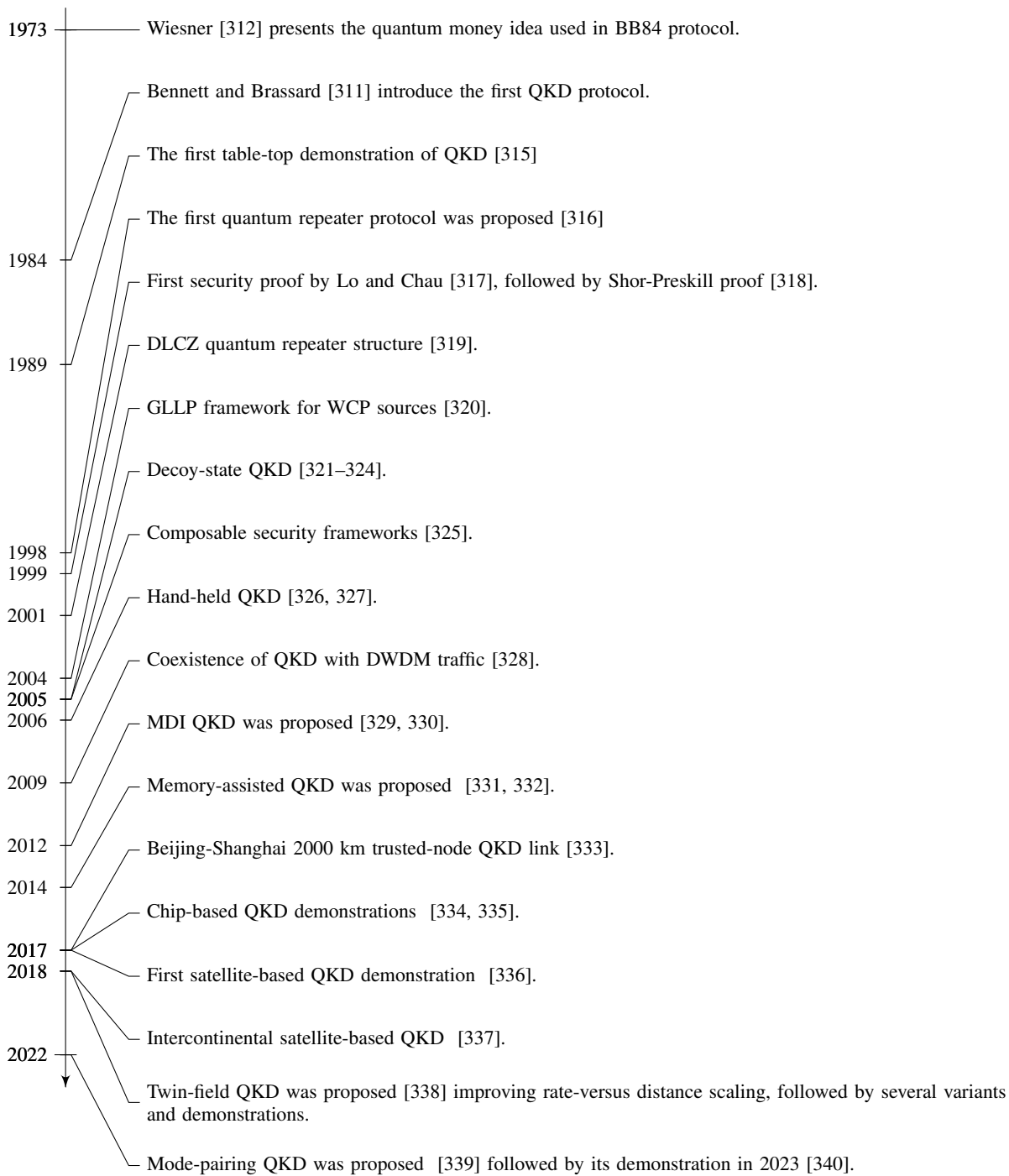

\begin{small}
\begin{timeline}{1973}{2022}{1cm}{1cm}{14cm}{0.7\textheight}
\entry{1973}{Wiesner~\cite{Wiesner} presents the quantum money idea used in BB84 protocol.}
\entry{1984}{Bennett and Brassard~\cite{bennett1984brassard} introduce the first QKD protocol.}
\entry{1989}{The first table-top demonstration of QKD \cite{tabletopQKD}} 
\entry{1998}{The first quantum repeater protocol was proposed \cite{briegel1998}} 
\entry{1999}{First security proof by Lo and Chau \cite{LoChauSecProof}, followed by Shor-Preskill proof \cite{ShorPreskill_00}.}
\entry{2001}{DLCZ quantum repeater structure \cite{duan2001}.}
\entry{2004}{GLLP framework for WCP sources \cite{gottesman2004security}.}
\entry{2005}{Decoy-state QKD \cite{Hwang:Decoy:2003, Wang:Decoy:2005, Wang:Decoy2:2005, ma2005practical}.}
\entry{2005}{Composable security frameworks \cite{Renner_Thesis_05}.}
\entry{2006}{Hand-held QKD \cite{HP_HandheldQKD, chun2017handheld}.}
\entry{2009}{Coexistence of QKD with DWDM traffic \cite{Telcordia_1550_1550}.}
\entry{2012}{MDI QKD was proposed \cite{lo2012measurement, Braunstein:MIQKD:2012}.}
\entry{2014}{Memory-assisted QKD was proposed ~\cite{panayi2014memory, MAQKD-Bruss}.}
\entry{2017}{Beijing-Shanghai 2000~km trusted-node QKD link \cite{Chen2021}.}
\entry{2017}{Chip-based QKD demonstrations ~\cite{ChipBasedQKD, ChipBasedQKDRev}.}
\entry{2017}{First satellite-based QKD demonstration ~\cite{Liao_Nat_2017}.}
\entry{2018}{Intercontinental satellite-based QKD ~\cite{Liao:ChinaAustria_PRL2018}.}
\entry{2018}{Twin-field QKD was proposed \cite{Lucamarini2018} improving rate-versus distance scaling, followed by several variants and demonstrations.}
\entry{2022}{Mode-pairing QKD was proposed ~\cite{ModePairingQKD} followed by its demonstration in 2023 \cite{MPQKDdemo}.}
\end{timeline}
\end{small}
\caption{Timeline of QKD milestones.} 
\label{fig:QKDtimeline}
\end{figure*}

In both cases, the security level attained effectively boils down to
being able to exploit the correlation between Alice and Bob in terms
of an entangled state. In the DV case, for instance, it can be shown
that if Alice and Bob share a maximally entangled state, they can
readily agree upon a shared secret key by measuring their share of the
entangled state. In fact, the second major QKD protocol following the
BB84 -- which was proposed by Ekert in 1991~\cite{ekert1991quantum} --
relied on verifying Bell inequality violations for a shared entangled
state between Alice and Bob. Then in 1992 a simplified
entanglement-based QKD protocol was conceived by Bennett, Brassard and
Mermin~\cite{BBM92}, which is termed as the BBM92 protocol. This
required similar measurement actions to those in the BB84 protocol. Since
then a large number of QKD protocols have been devised and
Fig.~\ref{fig:QKDtimeline} captures these evolutionary developments.

To expound a little further, the first decade of the new millennium
witnessed the development of rigorous security proofs for QKD
protocols. Shor and Preskill offered a simple security proof for ideal
BB84 based on entanglement distillation~\cite{ShorPreskill_00}. Later
in 2004, the security proofs were extended to the case of using weak
coherent pulses (WCPs) generated by lasers, instead of ideal single
photons. The Gottesman, Lo, L\"utkenhaus and Preskill (GLLP)
framework~\cite{GLLP_04} predicted that the key rate would not scale
desirably {\em vs.} the distance when we consider the extra photons
that can exist in a coherent state. This problem was fixed by the
development of the decoy-state
idea~\cite{Hwang:Decoy:2003,Wang:Decoy:2005,
  Wang:Decoy2:2005,ma2005practical}, where having multiple intensities
of light would facilitate for her and Bob to better estimate the key
bits generated by true single-photon states. New or improved security
proofs were introduced by Renner~\cite{Renner_Thesis_05} and
Koashi~\cite{Koashi:Comp:09}, which have been the basis for many later
protocols conceived in the next decade.

Still referring to Figure~\ref{fig:QKDtimeline}, the decade of
2010-2020 started with the conception of several experimental attacks
on some QKD system
implementations~\cite{Wiechers:AftergateAttack:2011,
  Weier:DeadtimeAttack:2011}. Interestingly, most these attacks
capitalised on some imperfections of the measurement devices in the
QKD systems. As a remedy, measurement-device-independent (MDI) QKD
systems~\cite{lo2012measurement, Braunstein:MIQKD:2012} emerged as a
practical solution against this type of attacks. Naturally, both Alice
and Bob have a transmitter, but all the measurements can be carried by
an untrusted party somewhere in the middle of the link. This idea led
to numerous breakthroughs in the ensuing era, resulting in protocols
capable of improving the key-rate versus distance scaling.  These are
exemplified by the twin-field~\cite{Lucamarini2018, send-or-not-send,
  PhaseMatchingQKD} and the mode-pairing~\cite{ModePairingQKD} QKD
protocols, and to a range of others that lend themselves to
integration into future generations of quantum communications
networks~\cite{Trust-freeQKD}.

\subsection{Knowledge Gaps and Challenges}
\label{Sec:QKD-Chall}
\subsubsection{The issue of distance}
Again, owing to the transmission of weak signals, point-to-point QKD
suffers from a channel attenuation of about 0.2~dB/km in fiber and
from hostile atmospheric propagation phenomena in FSO satellite
scenarios.  This may result in complete loss of signal in DV-QKD, or
in an excessively low signal-to-noise ratio in CV-QKD. As a result,
the total distance over which we can exchange a secret key without
using repeater or relay nodes is limited. Current terrestrial records
are at around 1000~km~\cite{QKD1000km}, albeit at very low secret key
generation rates. In Sec.~\ref{Sec:QKD-Roadmap}, we will hypothesize
about the roadmap of extending the reach of QKD networks to
arbitrarily long distances. In our formulation, we will show how we
can achieve end-to-end security in QKD networks. This may require
scientific advances in several technical areas, including quantum
devices, e.g. quantum memories, and even our quantum processing
capabilities.
\subsubsection{Cost and network-wide deployment}
For QKD to become ubiquitously available as a technology, we have to
reduce the cost of deployment by sharing it amongst many users. This
includes the cost of individual devices that have to be produced on a
commercial scale as well as the infrastructure costs of running
quantum applications. There are a number of promising directions to be
pursued for facilitating wide-scale deployment. This includes using
photonic integrated circuits to design QKD transmitter and receiver
modules~\cite{ChipBasedQKD, ChipBasedQKDRev,
  ChipBasedQKDToshiba}. This is particularly important for the
transmitter side, which is expected to be the main terminal that all
the end users would need. In terms of the infrastructure, it is very
important to use the existing fiber-optic networks laid out across the
globe for both quantum and classical applications. A lot of efforts
has therefore been directed at integrating QKD links with WDM
channels. Again, the key challenge that we face when sending quantum
and classical signals over the same optical fiber is that the
crosstalk generated by the high-power classical channels may overwhelm
the weak signals travelling through the quantum
channels. Sophisticated filtering techniques have been used to make
this possible, and several demonstrations have confirmed the
feasibility of this~\cite{Telcordia_1550_1550, Telcordia_1550_1310,
  Townsend_QI_home_2011, Shields.PRX.coexist, patel2014quantum,
  kumar2015coexistence, Coexist800Gbps}. Resource allocation in such
networks is another challenging issue, requiring the holistic
optimization of the overall performance~\cite{crosstalk2016,
  Bahrani2018, Bahrani:19}. A further largely unxplore area is the
development of the network stack for quantum
applications~\cite{QKDStack, Hanzo_MultiProt}, and ensuring its
compatibility with the evolving classical communications solutions,
such as software-defined networking~\cite{SDNQKD}.
%
\subsubsection{QKD in wireless terrestrial and satellite settings}
\begin{figure}[t]
\centering
\includegraphics[width=.95\linewidth]{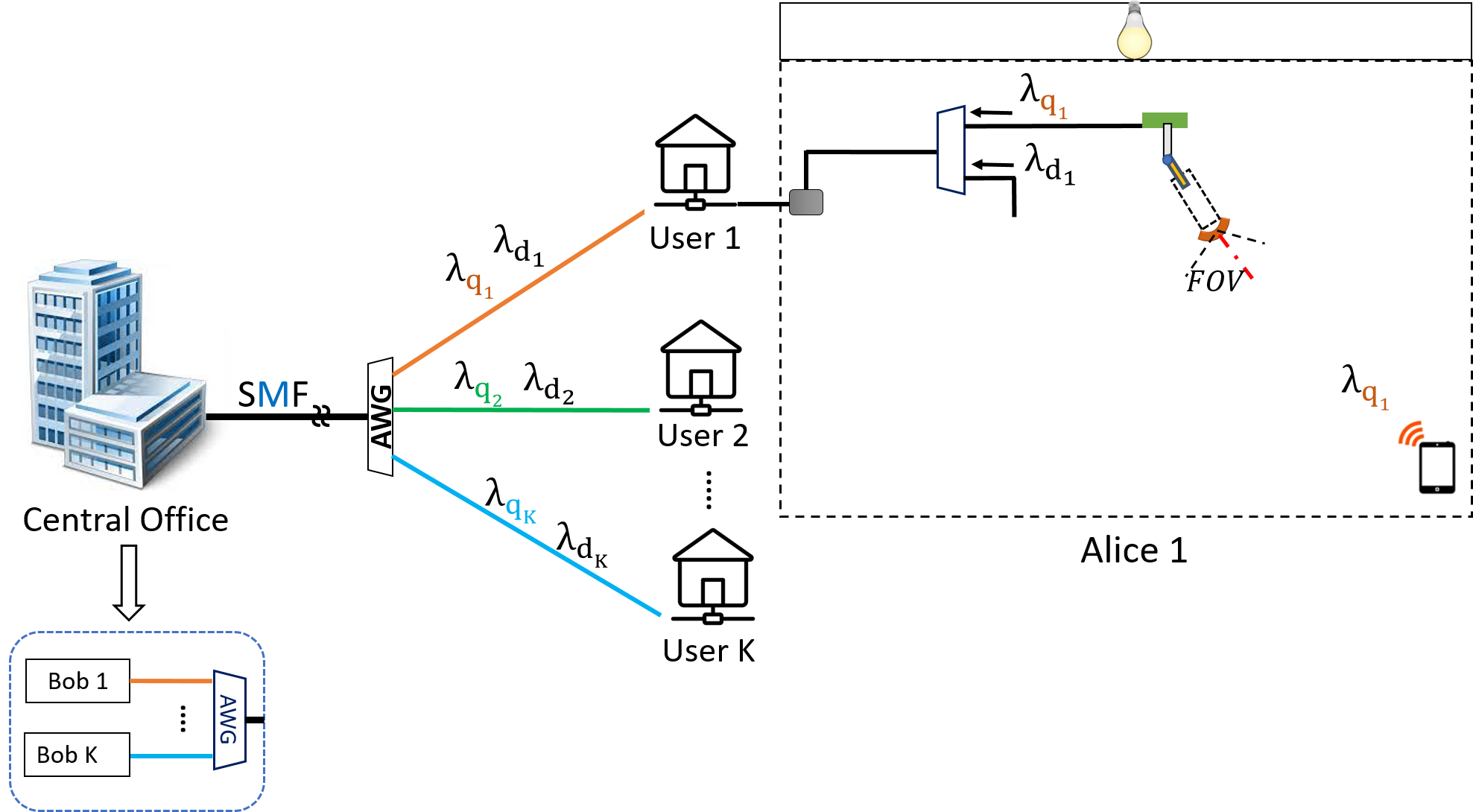}
\caption{Schematic view of exchanging secret keys between an indoor
  wireless user, Alice, and the central office (CO) within a DWDM
  passive optical network. SMF: Single-Mode Fiber, AWG: Arrayed
  Waveguide Grating.}
		\label{fig:QKDwireless}
	\end{figure}

Another important issue, especially with regard to compatibility with
emerging 6G technologies, is the adaptability of QKD to wireless
solutions. This has been of interest from early on, and in 2006 a
group of scientists at Bristol University in the UK along with their
collaborators at HP labs developed the first demonstration of a
handheld QKD device exchanging a key with an ATM-like
receiver~\cite{HP_HandheldQKD}. This was further enhanced recently and
demonstrated in indoor settings~\cite{chun2017handheld,
  HandheldQKD-folllowup}. The next step along this line is to enable
wireless access to QKD networks; see Fig.~\ref{fig:QKDwireless}. The
related theoretical analysis suggests that this is within reach even
with the aid of off-the-shelf technology in benign indoor settings
under mild lighting conditions~\cite{Wireless_indoor_QKD,
  Quantum_Access_to_quantum_networks, bahrani2018finite,
  bahrani2019wavelength}, although extending it to outdoor
applications is still an open challenge~\cite{Liao_NatPhoton_2017}.

On a completely different distance-scale, wireless settings are
capable of substantially improving the reach of QKD systems by
employing satellite-based quantum communications techniques. The key
advantage of relying on satellites is that the attenuation in free
space scales with the square of the distance, rather than attenuating
it exponentially as in optical fibers. Hence the total pathloss in the
satellite-to-ground link of a low-earth-orbit (LEO) satellite may be
as low as 30-40~dB, when the satellite passes over a ground
station. This allows us to connect nodes on different continents of
the world via orbiting satellites~\cite{Liao:ChinaAustria_PRL2018},
and eventually via a constellation of such satellites. Early
demonstrations by the in-orbit Chinese satellite referred to as Micius
has already paved the way in this regard~\cite{Liao_Nat_2017,
  Ren_Nat_2017}. A lot of global investment has since been allocated
to further this line of technology, since it is envisaged that a
satellite-based infrastructure could be part of our future generations
of quantum networks~\cite{liorni2020quantum}.

Satellite-based QKD may also allow for new security frameworks, where
some of the assumptions concerning the eavesdroppers are
relaxed~\cite{Ling-SatQKD-Man, Saikat_RestricedEve_PRApplied,
  Hugo_RestricedEve_PRApplied, Razavi_restricted_PRX}. This is
primarily because it is hard for Eve to intercept an optical
line-of-sight satellite link without being visually
detected~\cite{Razavi_restricted_PRX}.

Another interesting matter when it comes to wireless QKD systems is
the range of frequencies over which QKD can operate. Conventionally,
QKD has only been demonstrated in the optical regime. The key
advantage of the optical band over the radio frequency (RF) bands is
that the effect of quantum-domain noise is less dramatic than in the
RF band. This is because there is only a negligible amount of
thermally generated optical noise compared to the optical shot noise.
By cooling down our detectors, QKD can operate in a shot-noise-limited
regime, with the thermal noise effects imposed by the dark current
being orders of magnitude lower. However, this balance would gradually
shift, as we move to lower and lower frequencies, and once we reach
the THz regime, we should deal with the inherent thermal noise in our
devices and channels~\cite{THzQKD1}. That said, theoretically it might
be possible to reduce the frequency even down to 0.1~THz, and be able
to exchange a secret key over a few meters in indoor QKD scenarios.
At larger distances, THz could be an option in inter-satellite quantum
communications, where the free-space channel has low
temperature~\cite{Inter-SatTHzQKD}. Finally, regardless whether
considering the THz or the optical regime, wireless QKD systems would
substantially benefit from sophisticated multiple-input multiple-out
(MIMO) techniques, especially in hostile channel
conditions~\cite{MIMITHzQKD, MIMOQKD_Sahu,10094014}.
 
\subsubsection{Implementation security}
As QKD becomes more and more practical, it is important to
appropriately adjust the corresponding security proofs to match the
reality of the system implemented. For instance, in a typical QKD
protocol, we may have to generate certain states for the protocol to
rely on. However, once we implement the protocol, the states generated
by the devices employed may deviate from the required
one. Implementation-oriented security deals with such
issues~\cite{Pereira2019}, and aims for offering rigorous security
proofs that can be used in commercial settings. Other issues that
become relevant in such settings is the use of a finite number of data
points to bound the amount of information leaked to Eve, as discussed
in~\cite{zhang2017improved, NumericalFinteKey, TFQKD-finite-Guillermo}
for example. The assumptions we make about the probability of certain
attacks constitute another area of QKD research. For instance, in the
so-called Trojan horse attack, an eavesdropper can shed strong light
on Alice's transmitter in order to glean information about the
settings employed. In the field of implementation security, we account
for possible leakage of information under practical constraints and
derive the achievable secret key rate~\cite{TrojanHorseFiniteKey}.

Another important factor in the deployment of QKD is the need for
characterizing the devices employed. Different QKD protocols set
different requirements in this regard and as expected, typically the
best performers require detailed knowledge of the transmitter and
receiver specifications, while others such as MDI-QKD and device
independent QKD~\cite{Arnon-Friedman2018} aim for alleviating this
requirement. Naturally, it might be very challenging for an end user
to engage with the characterization process. To circumvent this issue,
standardization bodies are putting together certification processes
that enable the QKD industry and customers to do business with each
other in a convenient and reliable way. The entire list of
QKD-oriented standards formulated across the globe can be found
in~\cite{Hanzo:QInternet}.

\subsection{Research Roadmap}
\label{Sec:QKD-Roadmap}
\begin{figure*}
	\centering
        \includegraphics[width=\linewidth]{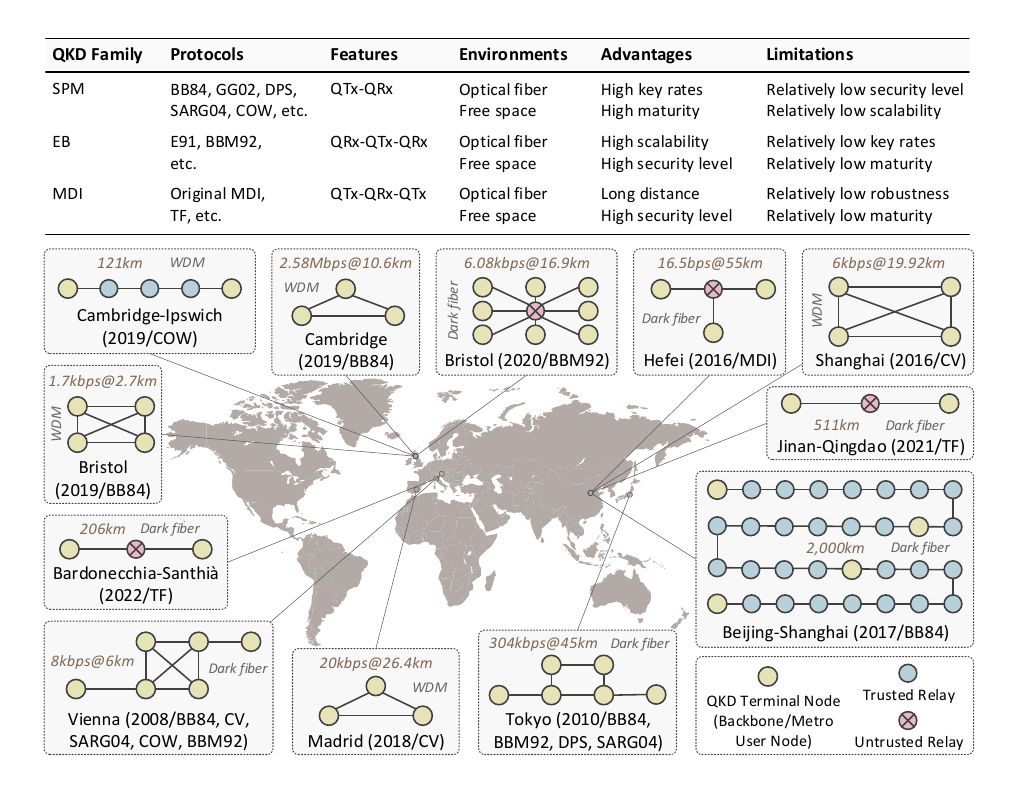}
	\caption{Diverse families of QKDPs and typical field trials of quantum networks around the world. {\bf SPM:} Single-Prepare-and-Measure; {\bf EB:} Entanglement-Based; {\bf MDI:} Measurement-Device-Independent; {\bf QTx:} QKD Transmitter; {\bf QRx:} QKD Receiver; {\bf BB84:} Bennett-Brassard-1984; {\bf GG02:} Grosshans-Grangier-2002; {\bf DPS:} Differential-Phase-Shift; {\bf SARG04:} Scarani-Ac\'{i}n-Ribordy-Gisin-2004; {\bf COW:} Coherent-One-Way; {\bf E91:} Ekert-91; {\bf BBM92:} Bennett-Brassard-Mermin-1992; {\bf CV:} Continuous-Variable; {\bf TF:} Twin-Field \copyright IEEE Cao {\em et al.}~\cite{Hanzo_MultiProt}}
        \label{yuan-fig1}
\end{figure*}

The prospect of deploying wide-scale quantum communications networks
has received a considerable boost over the past decade. This is partly
driven by several scientific breakthroughs facilitated by substantial
funding from governmental and industrial bodies. As seen at a glance
in Figure~\ref{yuan-fig1} to be detailed later in this section,
different countries across the globe -- including the UK, Germany,
Netherlands, and France, among others in Europe, as well as China,
Japan, and the US -- have invested on the order of several billion
dollars overall. European Union has also initiated an EU wide 10-year
flagship programme of the same scale to lead the second quantum
revolution.

In the context of the above developments it is important to have a
realistic view of how quantum technologies will evolve in the future,
and in which fields we as a community have to invest research
efforts. The research roadmap envisioned in this section is going to
address this subject from the perspective of deploying QKD systems
across our communications networks. There have been several other
recent roadmap documents that approach the question of 'quantum
futures' from different angles. Notably, Wehner {\em et
  al.}~\cite{Wehner_Roadmap} adopted an application-centric approach
for predicting the future evolution of the quantum Internet
(Qinternet) relying on six milestones along their predicted
roadmap. By contrast, the authors of~\cite{PRXQuantum_Roadmap} focused
more on the hardware required. As a further advance, Long {\em et
  al.}~\cite{long2022evolutionary} added an extra stage to the
six-stage roadmap of~\cite{Wehner_Roadmap} for the development of the
Qinternet by introducing the concept of {\em secure repeater networks
  (SRNs)}, which is compatible with the existing classical Internet
and relies on the popular philosophy of requiring excessively complex
operations for breaking security.  However, their SRN concept relying
on the QSDC concept of Section~\ref{qsdc-dl04}~\cite{deng2004secure}
-- rather than QKD -- incorporates eavesdropping detection, which is a
valuable extra feature.

Again, at this stage we use Figure~\ref{yuan-fig1}
of~\cite{Hanzo_MultiProt} by Cao {\em et al.}  for connecting the
state-of-the-art Subsection~\ref{Sec:QKD-SOTA}, the knowledge-gap
Subsection~\ref{Sec:QKD-Chall} and the future roadmap
Subsection~\ref{Sec:QKD-Roadmap}.  Observe in the figure that the
architecture of all existing QKD networks is rather
simple. Furthermore, for covering large distances by optical fiber
requires numerous repeaters, where the quantum signal is observed and
the resultant classical signal is amplified before preparing the
quantume states for transmission to the next relay. A similar
procedure has to be applied also for relaying-aided FSO-based
satellite systems, but naturally, they are capable of covering larger
distances. Observe furthermore in the figure that these networks tend
to rely on rather diverse QKD protocols. The acronyms identifying the
protocols are listed in the caption of the figure and their specific
features are detailed in~\cite{Hanzo:QInternet} for readers, who would
like to probe further. Suffice to say that the key performance metric
of QKD networks is the secret key-rate {\em vs.}  distance. This is
because the quantum signal must not be aplified and hence it is
gradually attenuated throughout its propagation. The resultant
attenuated signal can only support a reduced key-rate.

Observe for example at the bottom right corner of
Figure~\ref{yuan-fig1} that the Beijing-Shanghai link using the BB84
protocol has 31 hops between the source and destination and it relies
on so-called {\em trusted relays}, which must be hosted in secure
premises to avoid eavesdropping on the classical signal to be
amplified.  By contrast, some of these networks rely on so-called {\em
  untrusted relays}, which rely on more secure protocols than their
trusted counterparts.' They may also be referred to {\em trust-free
  relays/nodes}, which are capable of resisting eavesdropping owing to
their sophisticated entangelement-based protocol design. Hence they do
not have to be in secure premises. Clearly, a wide variety of
protocols having different levels of security have been harnessed
across the globe, and the issues of {\em trusted vs. trust-free} relay
nodes will be further detailed as part of our evolutionary roadmap of
Phase~I -- Phase~III.  Suffice to say that for constructing the global
Qinternet of the future, QKD protocol converters are required, as
detailed in~\cite{Hanzo_MultiProt}. This also underlines the
importance of global standardization for connecting the
network-segments relying on different protocols, as discussed
in~\cite{Hanzo:QInternet} for readers, who might like to explore
further.

We believe that the roadmap envisioned here complements the above
efforts and altogether offers a tangible serve-oriented perspective on
how quantum communications technologies may evolve.

Our solution envisaged for long-distance QKD evolves through multiple
developmental phases, which would naturally define relevant milestones in our
roadmap seen in Table~\ref{tab:roadmap}. Below, we highlight what might be
delivered in each developmental phase, and what would be required
to achieve it, with speculative timescales for the relevant
milestones.

\subsubsection{Phase I: Trusted Node QKD Networks}
\begin{figure}
\centering
				\includegraphics[width=.8\columnwidth]{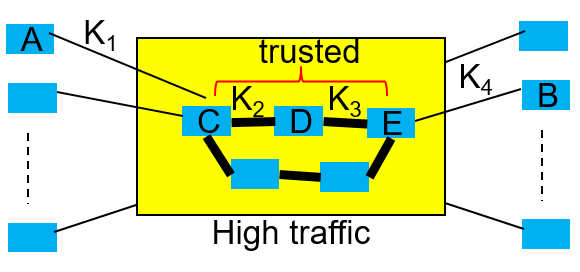}
				\caption{The schematic of a trusted node QKD network. In order for users A and B to exchange a secret key, we need to create a secret key between any two adjacent nodes that connect A to B (e.g., C, D, and E). We can then use these keys to securely relay a key from A to B. If there are multiple paths between A and B, we can generate separate keys using each path and then combine them in the end. }	\label{fig:TN}
	\end{figure}

The first phase of deployment, which is already in service, relies on
trusted node based QKD. Briefly, in trusted node QKD, secret key
exchange between parties A and B is carried out via multiple
intermediate nodes located at sufficiently short distances from each
other. Hence efficient point-to-point QKD is feasible between the
adjacent nodes without excessively reducing the key-rate, as
illustrated in Fig.~\ref{fig:TN}. If these intermediate relay nodes
can be trusted by parties A and B, then the {\em local key} exchanged
between the adjacent nodes can be readily used for relaying an {\em
  end-to-end key} between A and B. If multiple independent paths
exist between A and B, then the requirement on trusting the middle
nodes can be alleviated~\cite{Sanders:TrustedNode}.

\begin{table*}
    \centering
    	\includegraphics[width=\linewidth]{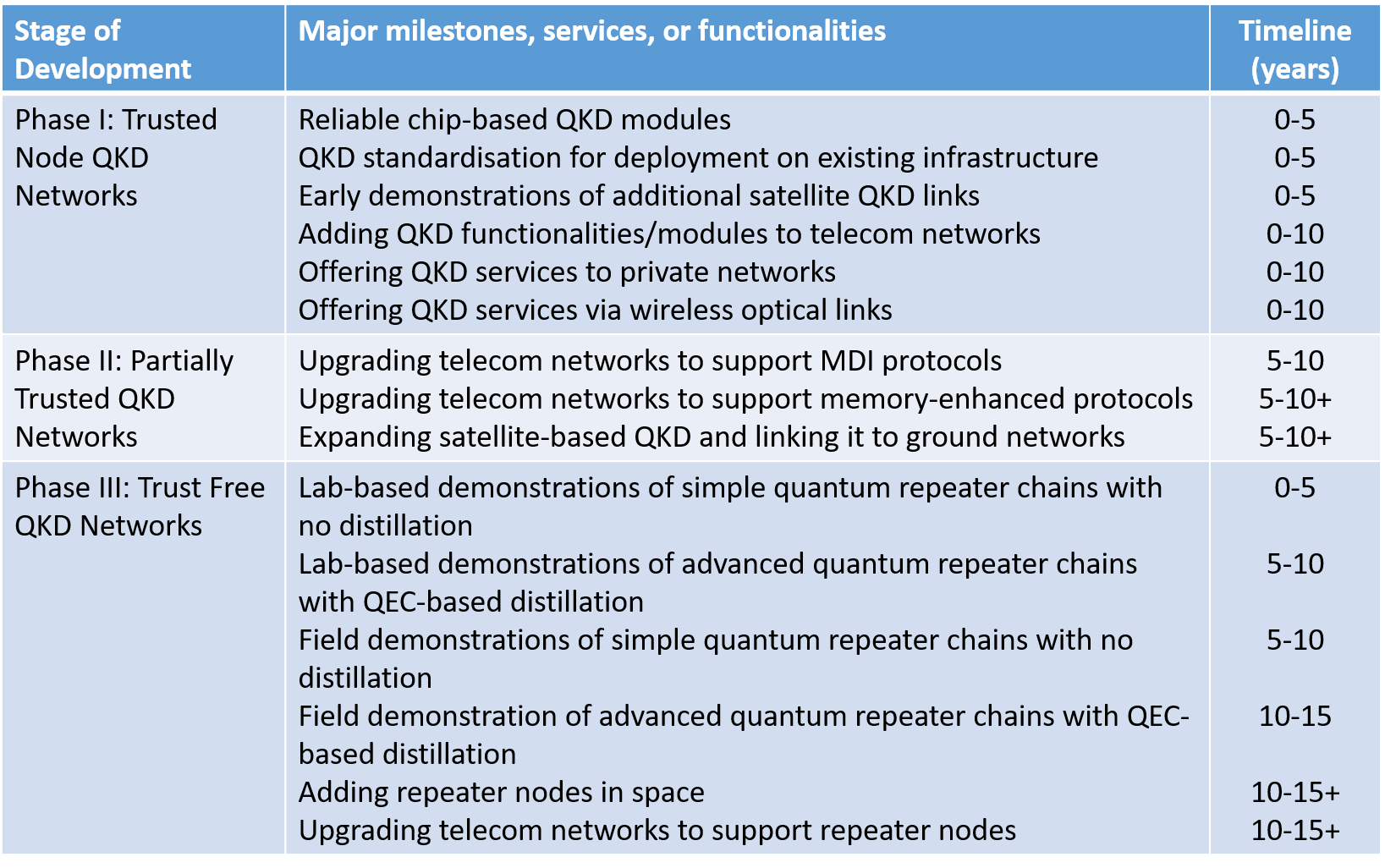}
	    \caption{A possible roadmap, with speculative timelines, for deploying QKD technologies in our existing and developing infrastructure. The deployment can take place in three phases, where in each phase, by employing more advanced technologies, the trust requirement on the service provider nodes is reduced. }
    \label{tab:roadmap}
\end{table*}

From a technology development perspective, trusted node based QKD
would certainly serve as a significant stepping stone toward future
phases of deployment. This is why at the time of writing trusted node
QKD research is at the core of almost all network demonstrations. This
includes the Chinese backbone network, the EU OpenQKD networks, and
the UK Quantum network, {\em inter alia}. This structure is expected
to have certain niche markets among high-security sectors, including
military and government bases as well as the financial and health
sector.  While the assumption of all nodes being trusted may be
acceptable in certain use cases, this is not necessarily acceptable in
high-security scenarios.  Gazing into the future, the community has to
construct chip-based QKD~\cite{Enzo_Chip_2019}, efficient detectors
and reliable sources. Radical frontier research is required also on
how to manage the resources of a hybrid communications network that
supports both quantum and classical applications.

Hence we recognise that the expansion of trusted node based QKD
constitutes an essential part of the roadmap leading to large-scale
QKD networks. We might speculate that the next decade would be
dedicated to improving the performance of all the different
components, as well as expanding the market within its relevant target
sectors.

\subsubsection{Phase II: Partially Trusted QKD Networks}

\begin{figure}[htbp]
\centering
				\includegraphics[width=\columnwidth]{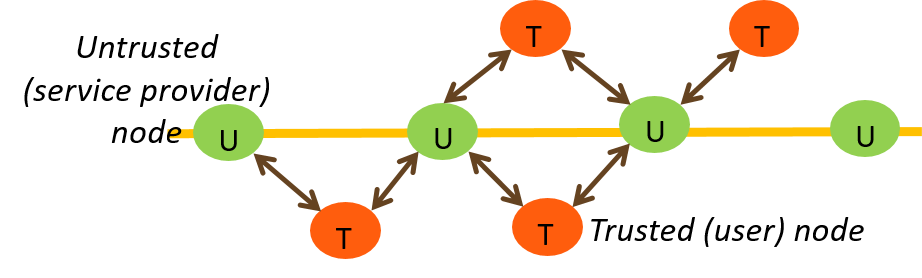}
				\caption{The schematic of a partially trusted QKD network. We exchange a key, using, e.g., MDI techniques, between two adjacent trusted (T) nodes via the untrusted (U) node that connects them together. The key exchange between the two trusted nodes not directly linked via an untrusted node can then be done similar to that of Fig.~\ref{fig:TN}. }	\label{fig:PT}
	\end{figure}

The next evolutionary phase following a trusted node QKD network is an
upgraded network in which the trust requirement on relay nodes in the
middle have been reduced, so that a larger groups of customers may opt
for harnessing QKD services. There are several promising technologies
that facilitate this transition:

\begin{itemize}
\item {\bf Measurement-device independent (MDI) QKD:} MDI QKD allows a
  pair of users to exchange a secret key via an untrusted node. This
  may only sound like a small adjustment to the trust issue, but in
  practice this will allow a larger number of enterprises to use the
  service, since they can use the service provider nodes to connect
  two of their trusted nodes, as seen in Fig.~\ref{fig:PT}. This way
  the need for having a fully private network would be
  alleviated. Moreover, with the advent of the new twin-field QKD
  protocols~\cite{TFQKD-Lucamarini2018, PRL509km, Pittaluga2021} the
  MDI structure can be used for improving the rate-versus-distance
  scaling as well. MDI protocols have been around for a while, but
  have not been widely used in commercial settings. They pave the way
  for future phases of deployment.

\item {\bf Memory-assisted QKD:} An alternative technique of improving
  the rate versus distance scaling is to harness quantum memories in
  the MDI setup~\cite{MAQKD-Bruss, panayi2014memory}. This will
  constitute a rudimentary repeater system that relies on quantum
  memories, and will be the stepping stone to the solutions that have
  to be developed in the third phase. The first demonstrations of such
  systems have just been reported in literature~\cite{MAQKD-Harvard,
    MAQKD-Rempe}, paving the way for their introduction into
  realistic/commercial settings.

\item {\bf Satellite-based QKD:} one of the emerging routes to
  long-distance quantum communications is via satellites, possibly in
  different orbits. Tehy may be harnessed as intermediate relay nodes
  between two ground stations. Prototype experiments of this nature
  have already been carried out using the Micius satellite, for
  example for exchanging a secret key between China and Austria by
  only trusting the satellite
  node~\cite{Liao:ChinaAustria_PRL2018}. This structure can also be
  expanded by using a constellation of satellites to serve a large
  number of users~\cite{liorni2020quantum}. It can also be employed in
  the near future for further expanding quantum communications
  networks during the developmental phase III to the space
  domain~\cite{liorni2020quantum}.
\end{itemize}

It is envisaged that the above technologies can make substantial
advances in the next decade or so, and will prepare us for the final
phase of deployment when no trust concerning the intermediate relay
nodes of the network is needed.

\subsubsection{Phase III: Trust-Free QKD Networks}
The key enabling idea behind trust-free QKD networks is to distribute
entanglement between a pair of remote nodes in an efficient way. The
users can then run an entanglement-based QKD protocol~\cite{BBM92} to
share a secret key while still being able to limit the amount of
information that might become leaked to any potential eavesdropper. In
effect, how the network provides the users with the entangled state
does not matter from a security-assurance perspective. Hence no trust
is required concerning intermediate relay nodes. An entanglement-based
network can also accommodate many other quantum applications, such as
distributed quantum computing, since reliable quantum data transfer
can be achieved via quantum teleportation~\cite{Wehner_Roadmap}.

The creation of long-distance entanglement distribution requires
fully-fledged quantum repeaters~\cite{briegel1998, duan2001,
  jiang2009quantum, munro2012quantum, RazaviRepeaterCh}. Quantum
repeaters extend the single-hop entanglement over a short distance to
longer distances by employing certain joint measurements, while
relying on quantum memories. The entangled state generated in this way
may have to be distilled for obtaining a higher-quality entangled
state. Based on the specific stage of development, the joint
measurement and/or entanglement distillation process may be carried
out either in a probabilistic~\cite{dur1999quantum, duan2001} or
deterministic~\cite{jiang2009quantum} manner. The probabilistic
solutions often offer lower key rates and require longer storage
times. By contrast, the deterministic solutions require reliable
quantum processing capabilities. Depending on the quality of quantum
processing operations carried out in the quantum repeater, we can
specify what security level may be expected from our repeater-based
network. In the long run, when high-performing quantum computers are
available, we can in principle use quantum repeaters that do not
require long storage times, but, rather they map the quantum-domain
data to large clusters of photons and send them from one node to
another, where each node can mitigate the errors along the way and
regenerate the encoded state~\cite{munro2012quantum}.

The exact timing of commercial quantum repeaters may be hard to
predict, but we envisage their appearance in 15+ years, as indicated
by the early demonstrations relying on probabilistic
measurements~\cite{yu2020entanglement, Delft-Pompili259}.  Quantum
repeaters that rely on the QEM techniques of Section~\ref{qem-lh} and
on the quantum error correction codes~\cite{jing2020quantum,
  jing2020simple} similar to those discussed in
Sections~\ref{quantum-coding} are expected to speed up this
evolutionary process by eliminating the propagation of errors across
consecutive hops. But again, advances are also required in the field
of quantum memory units and their interaction with
light~\cite{Ortu2018}.


\begin{figure*}
\centering
\includegraphics[width=7.5 in]{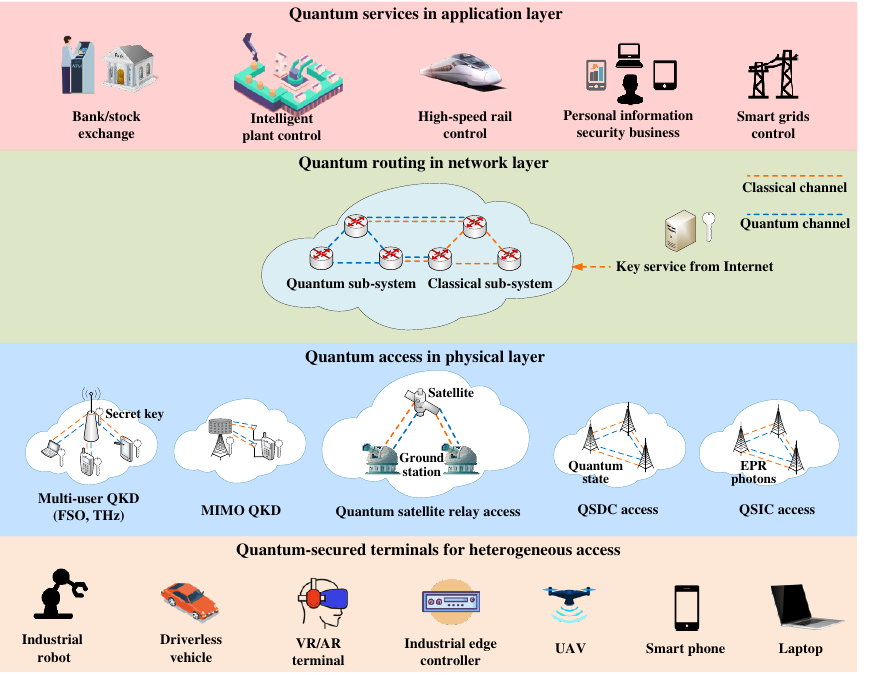}
\caption{A next-generation quantum-secured wireless system vision. The integration of QECC os Section~\ref{quantum-coding}, QEM of Section~\ref{qem-lh}, QML of Section~\ref{qml-osvaldo}, quantum radar of Sections~\ref{quantum-radar} and QKD of Section~\ref{qkd-mohsen} under the next-generation wireless vision will be a long-term development process. In the future, quantum-secured terminals are expected to appear in support of heterogeneous access.  Quantum-secured technologies, such as QKD and QSDC may be harnessed for quantum access in the physical layer. In the network layer, QML and quantum computing may be employed for quantum routing and at the application layer compelling quantum-secured services will become available, which also facilitate eavesdropping detection \copyright Zhou {\em et al.} \cite{xiaolin-cst}}
\label{Quantum system structure}
\end{figure*}

\section{A Next-Generation Quantum-Aided Wireless Roadmap}
\label{final}
\subsection{The Holistic Roadmap}
In this system-oriented section we further develop the initial vision of the
fiber-oriented Qinternet highlighted in Figure~\ref{fig:internet} and
extend it to the even broader quantum-native next-generation wireless
roadmap of Figure~\ref{Quantum system structure} relying on a
physical-, network- and application-layer vision. These holistic
system-oriented aspects are detailed in great depth by Zhou {\em et
  al.} in~\cite{xiaolin-cst}.

As seen at the bottom of Figure~\ref{Quantum system structure},
a diverse variety of industrial robots, autonomous vehicles,
virtual/augmented reality (VR/AR) services, industrial process
controllers, and even future smartphones and tablets are expected to
benefit.

Recall from Section~\ref{quantum-coding} that {\em the impairments
  imposed by quantum circuits are typically modelled by the
  depolarizing channel of Figure~\ref{fig:comp}}. In this context one
could draw a parallel with the classical Additive White Gaussian Noise
(AWGN) channel, which models the Brownian motion of electrons. {\em By
  contrast, when considering quantum communications over quantum
  channels, the physical properties of the FSO quantum transmission
  medium or of the fiber have to be considered.}

In this practical transmission context much of the quantum
communications research has been based on either optical fiber or
FSO-based satellite communications. For a detailed discourse on
satellite-based QKD channels please refer
to~\cite{Hosseinidehaj_SatQKD2017}. Indeed, non-terrestrial networking
(NTN) based on satellites constitutes one of the most topical subjects
in the 3GPP study schedule, as seen in Figure~6 of~\cite{kai-ten-acm},
which is a study item in Release~19 of the 5G/6G standardization.
This subject area is pictured in the center of Figure~\ref{Quantum
  system structure} and it is expected to attract continued research
atteantion.  The network layer echoes the architecture of
Figure~\ref{fig:internet} and the terminals portrayed at the bottom of
Figure~\ref{Quantum system structure} support the services of the
application layer seen at the top of the figure.  {\em In the rest of
  this section we will hypothesize as to what if any lessons of
  classical wireless communications may be relied upon in the design
  of future quantum communications systems.}

\subsection{Spectrum Harmonization in the Peta-Hz-Band}
\begin{figure}
	\centering
	\includegraphics[width=1.0\linewidth]{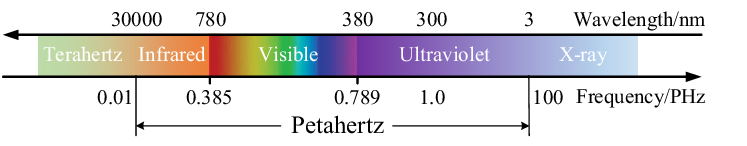}
	\caption{Petahertz band \copyright Xu {\em et al.} \cite{daniel-elsevier}}
	\label{fig1}
\end{figure}

In~\cite{daniel-elsevier} the concept of spectrum harmonization was
conceived, which relies on a sophisticated cognitive spectrum sensing
scheme capable of identifying the most suitable wave-length or
frequency domains for supporting a specific service.  Briefly, the
authors aim for unifying the existing infrared, visible light, and
ultraviolet subbands while also exploring the potential of the
Petahertz (PHz) band to support secure bandwidth-thirsty
telepresence/VR-style applications.  A hitherto scarcely-used
unlicensed spectral band is the Petahertz (PHz) band, which is defined
as the frequency range spanning from $0.01~PHz$ to $100~PHz$, where
$1~PHz=10^{15}~Hz$. The corresponding optical wireless wavelength
domain stretches from $30~\mu m$ to $3~nm$, as seen in
Fig. \ref{fig1}, which encompasses the THz, the Infrared (IR), the
Visible Light (VL) and the Ultraviolet (UV) subbands.  Their
propagation properties are detailed by Xu {\em et
  al.}~\cite{daniel-elsevier}, but these bands deserve further
exploration in the context of quantum communications. The most mature
solutions can be found in the realms of visible light communications.
Xu {\em et al.}~\cite{daniel-elsevier} also survey their modulation
schemes, system performance, multiple access techniques, and
networking. They conclude with a range of PetaCom challenges and open
research issues.

\subsection{Quantum Communications in the THz Band}
The feasibility of short-range quantum communication in the THz-band
has been critically appraised by the authors of~\cite{8976167}.  In
this context the authors of~\cite{e23091223} conceived a
continuous-valued satellite-to-satellite secret sharing
scheme~\cite{9674910} and a CV-QKD arrangement relying on THz-band
multi-carrier transmissions~\cite{9492803}. A range of THz-band
hardware aspects were discussed in recent years
in~\cite{8976167,7393447,7735237,9397381,nano10122436}, with special
attention dedicated to the conception of sensitive
detectors~\cite{doi:10.1021/acs.nanolett.9b05207,9566893,doi:10.1021/acs.nanolett.1c02022,Liu_2019,Yachmenev_2022,2022103,shao2021research,
  Thermoelectricgraphenephotodetectors}. The next few years may be
expected to unveil further advances in the THz band, which may be able
to usher in an era of radio-frequency based quantum communications.

\subsection{MIMO Techniques for Quantum Communications}
The field of classical wireless systems has been revolutionalized by
the conception of Multipla-Input Multiple Output (MIMO) techniques,
which rely on a combination of mutiple antennas and sophisticated
signal processing, as detailed in~\cite{5893900}. The four basic MIMO
types routinely used in the classical domain are~\cite{5742779}
space-time codes (STC), spatial multiplexing (SM) schemes, spatial
division multiple access (SDMA) arrangements and beamformers
(BF)~\cite{5742779}. Laser-based narrow beams naturally lend
themselves to angularly selective quantum communications, but a whole
suite of other MIMO solutions have been conceived either for improving
the reception quality or for increasing the throughput
attained~\cite{9099546,MIMITHzQKD,10094014,9739032,2005Quantum,zhou2019shot}.
As a representative result, the authors of~\cite{zhou2019shot}
benchmarked the performance of a multi-user MIMO scheme against that
of a classical scheme relying on a $N \times M = 4 \times
16$-dimensional array, which attained a substantial gain for
transmission over a Gamma-Gamma fading link.

\subsection{Multiple-Access for Quantum Communications}
In order to support multiple quantum communications users,
multiple-access methods are required. Hence the cardinal
question arises, as to which of the numerous classical multiple access
techniques might lend themselves to amalgamation with quantum
communications. In this context, a Code Division Multiple
Access (CDMA) based multi-user system was designed by the authors
of~\cite{6253218,rezai2021quantum,sharma2020quantum}, where unique
user-specific spreading codes were employed for distinguishing the
users.  By contrast, optical Orthogonal Frequency Division
Multiplexing (OFDM) was employed
in~\cite{6545781,bahrani2015orthogonal} for QKD transmission.


\subsection{Network-Coding for Quantum Communications}
In the simplest form of classical-domain network coding the
intermediate relay nodes between a pair of communicating parties use a
modulo-two gate for combining the data streams received from both
destination.  This allows us to detect both the identical and
different bit-positions. This capability can be exploited for reducing
the tele-traffic of the network.  Hence network coding is capable of
increasing the overall throughput, despite reducing the amount of
energy required per packet as well as the latency of
packets~\cite{Network_Coding_tutorial_Fraouli}.

Due to the inherent nature of quantum communications - namely that
copying of quantum information is precluded by the no-cloning theorem
- the question arises again as to whether the quantum domain
counterpart of classical network coding might be created. This dilemma
was raised in
\cite{RN1640,Ieee_Transactions_on_Information_Theory_Leung_2010}. However,
provided that preshared entanglement may be made available
\cite{Journal_of_Superconductivity_Mahdian_2015,Com_Letter_Jing_2015,PhysRevA.93.032302,Com_Letter_Shang_2014,Physical_Review_A_Satoh_2012,Journal_of_Mathematical_Physics_Jain_2011,Nature_Photonics_Munro_2010,Physical_Review_A_Hayashi_2007}
or a high-rate classical channel may be harnessed
\cite{Ieee_Transactions_on_Information_Theory_Leung_2010,Kobayashi_2011,Kobayashi_2010,Kobayashi_2009},
QNC is indeed realizable~\cite{8019800}.

\subsection{Quantum-Secured Direct Communications}
\label{qsdc-dl04}

Recall from Section~\ref{qkd-mohsen} that QKD solutions carry out
key-negotiation in the quantum-domain, but the associated encryption
process is reminiscent of its classical counterpart. By contrast,
quantum-secured direct communications (QSDC) constitutes a complete
quantum-domain communications solution~\cite{deng2004secure}. However,
the original QSDC protocol of Long and
Liu~\cite{long2002theoretically} relies on a block-based
communications philosophy, which necessitates the employment of
quantum memory. But quantum memory having an adequate
retention-duration is unavailable at the time of writing.  This
impediment has then been remedied in~\cite{long2002theoretically} and
sophisticated solutions circumventing this problem were proposed
in~\cite{sun2020toward}. At the time of writing QSDC is developing in
strides, to a great extent thanks to the pioneering frontier research
at Tsinghua University, as detailed in the work of Long and his
team~\cite{long2022evolutionary,WOS:000976372200001,WOS:000976372200002,WOS:000949858900001,WOS:001037124000012,guilu-cst}.

\subsection{Quantum-Search Aided Solution of Large-Scale Classical Wireless Search Problems}
When large-scale quantum computing becomes a commercial reality,
numerous demanding search-problems routinely found in science and
engineering may be solved more efficiently than by harnessing
classical search algorithms. The field of wireless communications also
has numerous large-scale search problems, as exemplified by the
following non-exhaustive list~\cite{6515077,8540839}:
\begin{itemize}
\item Multi-User Detection (MUD)~\cite{botsinis14tcom,botsinis15tcom}:
  The complexity of classical maximum-likelihood MUD in the uplink
  increases exponentially with the number of users, hence is not
  practical with classical algorithms, specifically for
  large-dimensional systems.
\item Non-Coherent Multiple Symbol Differential Detection in
  High-Doppler Wireless Systems~\cite{botsinis15acc}: Differential
  modulation relying on non-coherent detection over multiple symbols
  is an attractive alternative for coherent detection and provided
  performance gains over conventional non-coherent detection. However,
  it is computationally intensive.
\item Large-Scale Beam-Alignment Problems of mm-Wave Systems:
  Beamforming for large-scale Multi-User MIMO (MU-MIMO) systems is
  another computationally intensive search problem whose complexity
  increases with the number of users and antenna dimensions.
\item Joint Channel Estimation and Data Detection~\cite{7515148}:
  Joint channel estimation and multi-user detection is imperative for
  improving the performance of iterative receivers. However, such
  joint processing incurs high complexity, especially for
  high-dimensional wireless systems.
\item Multi-User Transmission (MUT)~\cite{7515210}: Optimizing MUT on
  the downlink is another computationally intensive search problem,
  particularly for rank deficient-systems with limited channel state
  feeback from the users.
\item Cooperative Multicell Processing (COMP): COMP is being used in
  LTE and 5G for enhancing network coverage and capacity. Resource
  allocation for COMP is another challenging search problem, as it
  requires joint optimization of data rate, interference and network
  capacity.
\item Localization Problems~\cite{7997701}: High localization accuracy
  and infrastructure/scenario constraints increase the computational
  complexity of optimal full-search based localization problems.
\item Multi-Objective System Optimization~\cite{7745885,7268791}:
  Wireless systems also have other multi-objective optimization
  functions, for example joint delay, energy and load optimization for
  routing in multi-hop networks having large number of nodes.

\item Recall that the quantum signal must not be amplified, because it
  would collapse back to the classical domain, relaying plays a
  pivotal role in extending the length of quantum networks, as
  detailed in~\cite{Hanzo:QInternet}. As a design alternative, LEO
  satellites may be employed for covering large
  distances~\cite{9463774} via the satellite-to-ground downlink and
  the ground-to-satellite uplink. Since the dimensions of the
  satellite are severely limited, its receiver aperture is typically
  much smaller than that of the ground station. Hence the uplink
  reception at the satellite requires substantial further research, as
  detailed in~\cite{9463774}.

 \end{itemize}
{\em Valued Colleague, join this community-effort, which is dedicated to
  solving the suite of open problems touched upon in this treatise!}

\bibliographystyle{IEEEtran} 
\footnotesize
\bibliography{mybibliography,xiaolin-ref,Hung_reference}

\begin{thebibliography}{100}
\providecommand{\url}[1]{#1}
\csname url@samestyle\endcsname
\providecommand{\newblock}{\relax}
\providecommand{\bibinfo}[2]{#2}
\providecommand{\BIBentrySTDinterwordspacing}{\spaceskip=0pt\relax}
\providecommand{\BIBentryALTinterwordstretchfactor}{4}
\providecommand{\BIBentryALTinterwordspacing}{\spaceskip=\fontdimen2\font plus
\BIBentryALTinterwordstretchfactor\fontdimen3\font minus
  \fontdimen4\font\relax}
\providecommand{\BIBforeignlanguage}[2]{{%
\expandafter\ifx\csname l@#1\endcsname\relax
\typeout{** WARNING: IEEEtran.bst: No hyphenation pattern has been}%
\typeout{** loaded for the language `#1'. Using the pattern for}%
\typeout{** the default language instead.}%
\else
\language=\csname l@#1\endcsname
\fi
#2}}
\providecommand{\BIBdecl}{\relax}
\BIBdecl

\bibitem{babar2018duality}
Z.~Babar, D.~Chandra, H.~V. Nguyen, P.~Botsinis, D.~Alanis, S.~X. Ng, and
  L.~Hanzo, ``{Duality of quantum and classical error correction codes: Design
  principles and examples},'' \emph{IEEE Communications Surveys \& Tutorials},
  vol.~21, no.~1, pp. 970--1010, 2018.

\bibitem{babar2019polar}
Z.~Babar, Z.~B.~K. Egilmez, L.~Xiang, D.~Chandra, R.~G. Maunder, S.~X. Ng, and
  L.~Hanzo, ``{Polar codes and their quantum-domain counterparts},'' \emph{IEEE
  Communications Surveys \& Tutorials}, vol.~22, no.~1, pp. 123--155, 2019.

\bibitem{chandra2019near}
D.~Chandra, Z.~Babar, S.-X. Ng, and L.~Hanzo, ``{Near-hashing-bound
  multiple-rate quantum turbo short-block codes},'' \emph{IEEE Access}, vol.~7,
  pp. 52\,712--52\,730, 2019.

\bibitem{chandra2023exit}
D.~Chandra, S.-X. Ng, and L.~Hanzo, ``{EXIT-chart aided design of irregular
  multiple-rate quantum turbo block codes},'' \emph{IEEE Access}, vol.~11, pp.
  96\,177--96\,195, 2019.

\bibitem{botsinis14tcom}
{Botsinis, P., Ng, S.X. and Hanzo, L.}, ``{Fixed-Complexity Quantum-Assisted
  Multi-User Detection for CDMA and SDMA},'' \emph{IEEE Transactions on
  Communications}, vol.~62, no.~3, pp. 990--1000, March 2014.

\bibitem{botsinis15tcom}
{P. Botsinis, D. Alanis, Z. Babar, S.X. Ng, and L. Hanzo}, ``{Iterative
  Quantum-Assisted Multi-User Detection for Multi-Carrier Interleave Division
  Multiple Access Systems},'' \emph{IEEE Transactions on Communications},
  vol.~pp, p.~pp, 2015.

\bibitem{botsinis15acc}
------, ``{Non-Coherent Quantum Multiple Symbol Differential Detection for
  Wireless Systems},'' \emph{IEEE Access}, vol.~pp, p.~pp, 2015.

\bibitem{8540839}
P.~{Botsinis}, D.~{Alanis}, Z.~{Babar}, H.~V. {Nguyen}, D.~{Chandra}, S.~X.
  {Ng}, and L.~{Hanzo}, ``Quantum search algorithms for wireless
  communications,'' \emph{IEEE Communications Surveys Tutorials}, vol.~21,
  no.~2, pp. 1209--1242, Secondquarter 2019.

\bibitem{7268791}
D.~Alanis, P.~Botsinis, Z.~Babar, S.~X. Ng, and L.~Hanzo, ``Non-dominated
  quantum iterative routing optimization for wireless multihop networks,''
  \emph{IEEE Access}, vol.~3, pp. 1704--1728, 2015.

\bibitem{7745885}
D.~Alanis, J.~Hu, P.~Botsinis, Z.~Babar, S.~X. Ng, and L.~Hanzo,
  ``Quantum-assisted joint multi-objective routing and load balancing for
  socially-aware networks,'' \emph{IEEE Access}, vol.~4, pp. 9993--10\,028,
  2016.

\bibitem{7997701}
P.~Botsinis, D.~Alanis, S.~Feng, Z.~Babar, H.~V. Nguyen, D.~Chandra, S.~X. Ng,
  R.~Zhang, and L.~Hanzo, ``Quantum-assisted indoor localization for uplink
  mm-wave and downlink visible light communication systems,'' \emph{IEEE
  Access}, vol.~5, pp. 23\,327--23\,351, 2017.

\bibitem{8019800}
H.~V. Nguyen, Z.~Babar, D.~Alanis, P.~Botsinis, D.~Chandra, M.~A.~M. Izhar,
  S.~X. Ng, and L.~Hanzo, ``Towards the quantum internet: Generalised quantum
  network coding for large-scale quantum communication networks,'' \emph{IEEE
  Access}, vol.~5, pp. 17\,288--17\,308, 2017.

\bibitem{hanzo-hte}
D.~Chandra, P.~Botsinis, D.~Alanis, Z.~Babar, S.-X. Ng, and L.~Hanzo, ``On the
  road to quantum communications,'' \emph{Infocommunications Journal}, vol.
  XIV, no.~3, pp. 2--8, Sept. 2022.

\bibitem{Hosseinidehaj_SatQKD2017}
N.~{Hosseinidehaj}, Z.~{Babar}, R.~{Malaney}, S.~X. {Ng}, and L.~{Hanzo},
  ``Satellite-based continuous-variable quantum communications:
  State-of-the-art and a predictive outlook,'' \emph{IEEE Communications
  Surveys Tutorials}, pp. 1--1, 2018.

\bibitem{Hanzo:QInternet}
Y.~Cao, Y.~Zhao, Q.~Wang, J.~Zhang, S.~X. Ng, and L.~Hanzo, ``The evolution of
  quantum key distribution networks: On the road to the qinternet,'' \emph{IEEE
  Communications Surveys \& Tutorials}, vol.~24, no.~2, pp. 839--894, 2022.

\bibitem{caleffi2020rise}
M.~Caleffi, D.~Chandra, D.~Cuomo, S.~Hassanpour, and A.~S. Cacciapuoti, ``{The
  Rise of the Quantum Internet},'' \emph{Computer}, vol.~53, no.~6, pp. 67--72,
  2020.

\bibitem{van2016path}
R.~Van~Meter and S.~J. Devitt, ``{The Path to Scalable Distributed Quantum
  Computing},'' \emph{Computer}, vol.~49, no.~9, pp. 31--42, 2016.

\bibitem{fitzsimons2017private}
J.~F. Fitzsimons, ``{Private Quantum Computation: An Introduction to Blind
  Quantum Computing and Related Protocols},'' \emph{npj Quantum Information},
  vol.~3, no.~1, pp. 1--11, 2017.

\bibitem{chuang1997bosonic}
I.~L. Chuang, D.~W. Leung, and Y.~Yamamoto, ``{Bosonic quantum codes for
  amplitude damping},'' \emph{Physical Review A}, vol.~56, no.~2, pp.
  1114--1125, 1997.

\bibitem{sarvepalli2009asymmetric}
P.~K. Sarvepalli, A.~Klappenecker, and M.~R{\"o}tteler, ``{Asymmetric quantum
  codes: Constructions, bounds and performance},'' \emph{Proceedings of the
  Royal Society A: Mathematical, Physical and Engineering Sciences}, vol. 465,
  no. 2105, pp. 1645--1672, 2009.

\bibitem{wootters1982single}
W.~K. Wootters and W.~H. Zurek, ``{A single quantum cannot be cloned},''
  \emph{Nature}, vol. 299, no. 5886, pp. 802--803, 1982.

\bibitem{shor1995scheme}
P.~W. Shor, ``{Scheme for reducing decoherence in quantum computer memory},''
  \emph{Physical Review A}, vol.~52, no.~4, pp. 2493--2496, 1995.

\bibitem{babar2015road}
Z.~Babar, P.~Botsinis, D.~Alanis, S.~X. Ng, and L.~Hanzo, ``{The road from
  classical to quantum codes: A hashing bound approaching design procedure},''
  \emph{IEEE Access}, vol.~3, pp. 146--176, 2015.

\bibitem{calderbank1996good}
A.~R. Calderbank and P.~W. Shor, ``{Good quantum error-correcting codes
  exist},'' \emph{Physical Review A}, vol.~54, no.~2, pp. 1098--1105, 1996.

\bibitem{steane1996multiple}
A.~M. Steane, ``{Multiple-particle interference and quantum error
  correction},'' \emph{Proceedings of the Royal Society A: Mathematical,
  Physical and Engineering Sciences}, vol. 452, no. 1954, pp. 2551--2577, 1996.

\bibitem{steane1996error}
------, ``{Error correcting codes in quantum theory},'' \emph{Physical Review
  Letters}, vol.~77, no.~5, pp. 793--797, 1996.

\bibitem{laflamme1996perfect}
R.~Laflamme, C.~Miquel, J.~P. Paz, and W.~H. Zurek, ``{Perfect quantum error
  correcting code},'' \emph{Physical Review Letters}, vol.~77, no.~1, pp.
  198--201, 1996.

\bibitem{bennett1996mixed}
C.~H. Bennett, D.~P. DiVincenzo, J.~A. Smolin, and W.~K. Wootters,
  ``{Mixed-state entanglement and quantum error correction},'' \emph{Physical
  Review A}, vol.~54, no.~5, pp. 3824--3851, 1996.

\bibitem{gottesman1997stabilizer}
D.~Gottesman, ``Stabilizer codes and quantum error correction,'' Ph.D.
  dissertation, California Institute of Technology, 1997.

\bibitem{gottesman1996class}
------, ``{Class of quantum error-correcting codes saturating the quantum
  Hamming bound},'' \emph{Physical Review A}, vol.~54, no.~3, pp. 1862--1868,
  1996.

\bibitem{cleve1997quantum}
R.~Cleve, ``{Quantum stabilizer codes and classical linear codes},''
  \emph{Physical Review A}, vol.~55, no.~6, pp. 4054--4059, 1997.

\bibitem{mackay2004sparse}
D.~J. MacKay, G.~Mitchison, and P.~L. McFadden, ``{Sparse-graph codes for
  quantum error correction},'' \emph{IEEE Transactions on Information Theory},
  vol.~50, no.~10, pp. 2315--2330, 2004.

\bibitem{steane1996simple}
A.~M. Steane, ``{Simple quantum error-correcting codes},'' \emph{Physical
  Review A}, vol.~54, no.~6, pp. 4741--4751, 1996.

\bibitem{grassl1997codes}
M.~Grassl, T.~Beth, and T.~Pellizzari, ``{Codes for the quantum erasure
  channel},'' \emph{Physical Review A}, vol.~56, no.~1, pp. 33--38, 1997.

\bibitem{calderbank1998quantum}
A.~R. Calderbank, E.~M. Rains, P.~W. Shor, and N.~J. Sloane, ``{Quantum error
  correction via codes over GF(4)},'' \emph{IEEE Transactions on Information
  Theory}, vol.~44, no.~4, pp. 1369--1387, 1998.

\bibitem{grassl1999quantum1}
M.~Grassl and T.~Beth, ``Quantum bch codes,'' in \emph{Proceedings of the X
  International Symposium on Theoretical Electrical Engineering (ISTET)}, 1999,
  pp. 207--212.

\bibitem{steane1999enlargement}
A.~M. Steane, ``{Enlargement of Calderbank-Shor-Steane quantum codes},''
  \emph{IEEE Transactions on Information Theory}, vol.~45, no.~7, pp.
  2492--2495, 1999.

\bibitem{xiaoyan2004quantum}
L.~Xiaoyan, ``{Quantum cyclic and constacyclic codes},'' \emph{IEEE
  Transactions on Information Theory}, vol.~50, no.~3, pp. 547--549, 2004.

\bibitem{steane1999quantum}
A.~M. Steane, ``{Quantum Reed-Muller codes},'' \emph{IEEE Transactions on
  Information Theory}, vol.~45, no.~5, pp. 1701--1703, 1999.

\bibitem{grassl1999quantum2}
M.~Grassl, W.~Geiselmann, and T.~Beth, ``{Quantum Reed—Solomon codes},'' in
  \emph{Proceedings of the 13th International Symposium of Applied Algebra,
  Algebraic Algorithms and Error-Correcting Codes (AAECC)}.\hskip 1em plus
  0.5em minus 0.4em\relax Springer, 1999, pp. 231--244.

\bibitem{kitaev1997quantum}
A.~Y. Kitaev, ``{Quantum computations: Algorithms and error correction},''
  \emph{Russian Mathematical Surveys}, vol.~52, no.~6, pp. 1191--1249, 1997.

\bibitem{kitaev2003fault}
------, ``{Fault-tolerant quantum computation by anyons},'' \emph{Annals of
  Physics}, vol. 303, no.~1, pp. 2--30, 2003.

\bibitem{fujii2015quantum}
K.~Fujii, \emph{{Quantum computation with topological codes: From qubit to
  topological fault-tolerance}}.\hskip 1em plus 0.5em minus 0.4em\relax
  Springer, 2015, vol.~8.

\bibitem{bravyi1998quantum}
S.~B. Bravyi and A.~Y. Kitaev, ``{Quantum codes on a lattice with boundary},''
  \emph{arXiv preprint quant-ph/9811052}, 1998.

\bibitem{horsman2012surface}
C.~Horsman, A.~G. Fowler, S.~Devitt, and R.~Van~Meter, ``{Surface code quantum
  computing by lattice surgery},'' \emph{New Journal of Physics}, vol.~14,
  no.~12, 2012.

\bibitem{bombin2006topological}
H.~Bombin and M.~A. Martin-Delgado, ``{Topological quantum distillation},''
  \emph{Physical Review Letters}, vol.~97, no.~18, 2006.

\bibitem{haah2011local}
J.~Haah, ``{Local stabilizer codes in three dimensions without string logical
  operators},'' \emph{Physical Review A}, vol.~83, no.~4, 2011.

\bibitem{zemor2009cayley}
G.~Z{\'e}mor, ``{On Cayley graphs, surface codes, and the limits of homological
  coding for quantum error correction},'' in \emph{Proceedings of the
  International Conference on Coding and Cryptology}.\hskip 1em plus 0.5em
  minus 0.4em\relax Springer, 2009, pp. 259--273.

\bibitem{delfosse2013tradeoffs}
N.~Delfosse, ``{Tradeoffs for reliable quantum information storage in surface
  codes and color codes},'' in \emph{Proceedings of the IEEE International
  Symposium on Information Theory (ISIT)}.\hskip 1em plus 0.5em minus
  0.4em\relax IEEE, 2013, pp. 917--921.

\bibitem{bravyi2014homological}
S.~Bravyi and M.~B. Hastings, ``{Homological product codes},'' in
  \emph{Proceedings of the 46th Annual ACM Symposium on Theory of Computing
  (STOC)}.\hskip 1em plus 0.5em minus 0.4em\relax ACM, 2014, pp. 273--282.

\bibitem{shannon1948mathematical}
C.~E. Shannon, ``{A mathematical theory of communication},'' \emph{The Bell
  System Technical Journal}, vol.~27, no.~3, pp. 379--423, 1948.

\bibitem{lloyd1997capacity}
S.~Lloyd, ``{Capacity of the noisy quantum channel},'' \emph{Physical Review
  A}, vol.~55, no.~3, pp. 1613--1622, 1997.

\bibitem{shor2003capacities}
P.~W. Shor, ``{Capacities of quantum channels and how to find them},''
  \emph{Mathematical Programming}, vol.~97, pp. 311--335, 2003.

\bibitem{devetak2005private}
I.~Devetak, ``{The private classical capacity and quantum capacity of a quantum
  channel},'' \emph{IEEE Transactions on Information Theory}, vol.~51, no.~1,
  pp. 44--55, 2005.

\bibitem{wilde2013entanglement}
M.~M. Wilde, M.-H. Hsieh, and Z.~Babar, ``{Entanglement-assisted quantum turbo
  codes},'' \emph{IEEE Transactions on Information Theory}, vol.~60, no.~2, pp.
  1203--1222, 2013.

\bibitem{divincenzo1998quantum}
D.~P. DiVincenzo, P.~W. Shor, and J.~A. Smolin, ``{Quantum-channel capacity of
  very noisy channels},'' \emph{Physical Review A}, vol.~57, no.~2, pp.
  830--839, 1998.

\bibitem{smith2007degenerate}
G.~Smith and J.~A. Smolin, ``{Degenerate quantum codes for Pauli channels},''
  \emph{Physical Review Letters}, vol.~98, no.~3, 2007.

\bibitem{fuentes2021degeneracy}
P.~Fuentes, J.~E. Martinez, P.~M. Crespo, and J.~Garcia-Fr{\'\i}as,
  ``{Degeneracy and its impact on the decoding of sparse quantum codes},''
  \emph{IEEE Access}, vol.~9, pp. 89\,093--89\,119, 2021.

\bibitem{postol2001proposed}
M.~S. Postol, ``{A proposed quantum low density parity check code},''
  \emph{arXiv preprint quant-ph/0108131}, 2001.

\bibitem{camara2005constructions}
T.~Camara, H.~Ollivier, and J.-P. Tillich, ``{Constructions and performance of
  classes of quantum LDPC codes},'' \emph{arXiv preprint quant-ph/0502086},
  2005.

\bibitem{camara2007class}
------, ``{A class of quantum LDPC codes: Construction and performances under
  iterative decoding},'' in \emph{Proceedings of the IEEE International
  Symposium on Information Theory (ISIT)}.\hskip 1em plus 0.5em minus
  0.4em\relax IEEE, 2007, pp. 811--815.

\bibitem{ollivier2003description}
H.~Ollivier and J.-P. Tillich, ``{Description of a quantum convolutional
  code},'' \emph{Physical Review Letters}, vol.~91, no.~17, 2003.

\bibitem{ollivier2004quantum}
------, ``{Quantum convolutional codes: Fundamentals},'' \emph{arXiv preprint
  quant-ph/0401134}, 2004.

\bibitem{forney2005simple}
G.~D. Forney and S.~Guha, ``{Simple rate-$1/3$ convolutional and tail-biting
  quantum error-correcting codes},'' in \emph{Proceedings of the International
  Symposium on Information Theory (ISIT)}.\hskip 1em plus 0.5em minus
  0.4em\relax IEEE, 2005, pp. 1028--1032.

\bibitem{forney2007convolutional}
G.~D. Forney, M.~Grassl, and S.~Guha, ``{Convolutional and tail-biting quantum
  error-correcting codes},'' \emph{IEEE Transactions on Information Theory},
  vol.~53, no.~3, pp. 865--880, 2007.

\bibitem{poulin2008quantum}
D.~Poulin, J.-P. Tillich, and H.~Ollivier, ``{Quantum serial turbo codes},'' in
  \emph{Proceedings of the IEEE International Symposium on Information Theory
  (ISIT)}.\hskip 1em plus 0.5em minus 0.4em\relax IEEE, 2008, pp. 310--314.

\bibitem{poulin2009quantum}
------, ``{Quantum serial turbo codes},'' \emph{IEEE Transactions on
  Information Theory}, vol.~55, no.~6, pp. 2776--2798, 2009.

\bibitem{renes2015efficient}
J.~M. Renes, D.~Sutter, F.~Dupuis, and R.~Renner, ``{Efficient quantum polar
  codes requiring no preshared entanglement},'' \emph{IEEE Transactions on
  Information Theory}, vol.~61, no.~11, pp. 6395--6414, 2015.

\bibitem{babar2016serially}
Z.~Babar, H.~V. Nguyen, P.~Botsinis, D.~Alanis, D.~Chandra, S.~X. Ng, and
  L.~Hanzo, ``{Serially concatenated unity-rate codes improve quantum codes
  without coding-rate reduction},'' \emph{IEEE Communications Letters},
  vol.~20, no.~10, pp. 1916--1919, 2016.

\bibitem{hagiwara2011spatially}
M.~Hagiwara, K.~Kasai, H.~Imai, and K.~Sakaniwa, ``{Spatially coupled
  quasi-cyclic quantum LDPC codes},'' in \emph{Proceedings of the IEEE
  International Symposium on Information Theory (ISIT)}.\hskip 1em plus 0.5em
  minus 0.4em\relax IEEE, 2011, pp. 638--642.

\bibitem{kasai2011nonbinary}
K.~Kasai, M.~Hagiwara, H.~Imai, and K.~Sakaniwa, ``{Non-binary quasi-cyclic
  quantum LDPC codes},'' in \emph{Proceedings of the IEEE International
  Symposium on Information Theory (IST)}.\hskip 1em plus 0.5em minus
  0.4em\relax IEEE, 2011, pp. 653--657.

\bibitem{kasai2011quantum}
------, ``{Quantum error correction beyond the bounded distance decoding
  limit},'' \emph{IEEE Transactions on Information Theory}, vol.~58, no.~2, pp.
  1223--1230, 2011.

\bibitem{andriyanova2012spatially}
I.~Andriyanova, D.~Maurice, and J.-P. Tillich, ``{Spatially coupled quantum
  LDPC codes},'' in \emph{Proceedings of the IEEE Information Theory Workshop
  (ITW)}.\hskip 1em plus 0.5em minus 0.4em\relax IEEE, 2012, pp. 327--331.

\bibitem{maurice2013family}
D.~Maurice, J.-P. Tillich, and I.~Andriyanova, ``{A family of quantum codes
  with performances close to the hashing bound under iterative decoding},'' in
  \emph{Proceedings of the IEEE International Symposium on Information Theory
  (ISIT)}.\hskip 1em plus 0.5em minus 0.4em\relax IEEE, 2013, pp. 907--911.

\bibitem{bowen2002entanglement}
G.~Bowen, ``{Entanglement required in achieving entanglement-assisted channel
  capacities},'' \emph{Physical Review A}, vol.~66, no.~5, 2002.

\bibitem{brun2006correcting}
T.~Brun, I.~Devetak, and M.-H. Hsieh, ``{Correcting quantum errors with
  entanglement},'' \emph{Science}, vol. 314, no. 5798, pp. 436--439, 2006.

\bibitem{brun2007general}
T.~A. Brun, I.~Devetak, and M.-H. Hsieh, ``General entanglement-assisted
  quantum error-correcting codes,'' in \emph{Proceedings of the IEEE
  International Symposium on Information Theory (ISIT)}.\hskip 1em plus 0.5em
  minus 0.4em\relax IEEE, 2007, pp. 2101--2105.

\bibitem{hsieh2007general}
M.-H. Hsieh, I.~Devetak, and T.~Brun, ``{General entanglement-assisted quantum
  error-correcting codes},'' \emph{Physical Review A}, vol.~76, no.~6, 2007.

\bibitem{hsieh2009entanglement}
M.-H. Hsieh, T.~A. Brun, and I.~Devetak, ``{Entanglement-assisted quantum
  quasi-cyclic low-density parity-check codes},'' \emph{Physical Review A},
  vol.~79, no.~3, 2009.

\bibitem{wilde2010entanglement}
M.~M. Wilde and T.~A. Brun, ``{Entanglement-assisted quantum convolutional
  coding},'' \emph{Physical Review A}, vol.~81, no.~4, 2010.

\bibitem{wilde2011entanglement}
M.~M. Wilde and M.-H. Hsieh, ``{Entanglement boosts quantum turbo codes},'' in
  \emph{Proceedings of the IEEE International Symposium on Information Theory
  (ISIT)}.\hskip 1em plus 0.5em minus 0.4em\relax IEEE, 2011, pp. 445--449.

\bibitem{wilde2012polar}
M.~M. Wilde and S.~Guha, ``{Polar codes for classical-quantum channels},''
  \emph{IEEE Transactions on Information Theory}, vol.~59, no.~2, pp.
  1175--1187, 2012.

\bibitem{wilde2013polar}
------, ``{Polar codes for degradable quantum channels},'' \emph{IEEE
  Transactions on Information Theory}, vol.~59, no.~7, pp. 4718--4729, 2013.

\bibitem{renes2012efficient}
J.~M. Renes, F.~Dupuis, and R.~Renner, ``Efficient polar coding of quantum
  information,'' \emph{Physical Review Letters}, vol. 109, no.~5, 2012.

\bibitem{chandra2017quantum}
D.~Chandra, Z.~Babar, H.~V. Nguyen, D.~Alanis, P.~Botsinis, S.~X. Ng, and
  L.~Hanzo, ``{Quantum coding bounds and a closed-form approximation of the
  minimum distance versus quantum coding rate},'' \emph{IEEE Access}, vol.~5,
  pp. 11\,557--11\,581, 2017.

\bibitem{BIBD2008}
I.~B. Djordjevic, ``{Quantum LDPC Codes from Balanced Incomplete Block
  Designs},'' \emph{IEEE Comm. Lett.}, vol.~12, no.~5, pp. 389--391, 2008.

\bibitem{Photonic_QLDPC2009}
------, ``{Photonic Quantum Dual-Containing LDPC Encoders and Decoders},''
  \emph{IEEE Photon. Technol. Lett.}, vol.~21, no.~13, pp. 842--844, 2009.

\bibitem{Photonic_EA_QLDPC2010}
------, ``{Photonic entanglement-assisted quantum low-density parity-check
  encoders and decoders},'' \emph{Opt. Lett.}, vol.~35, no.~9, pp. 1464--1466,
  2010.

\bibitem{pelchat2013degenerate}
E.~Pelchat and D.~Poulin, ``{Degenerate Viterbi decoding},'' \emph{IEEE
  Transactions on Information Theory}, vol.~59, no.~6, pp. 3915--3921, 2013.

\bibitem{babar2013near}
Z.~Babar, S.-X. Ng, and L.~Hanzo, ``{Near-capacity code design for
  entanglement-assisted classical communication over quantum depolarizing
  channels},'' \emph{IEEE Transactions on Communications}, vol.~61, no.~12, pp.
  4801--4807, 2013.

\bibitem{tillich2013quantum}
J.-P. Tillich and G.~Z{\'e}mor, ``{Quantum LDPC codes with positive rate and
  minimum distance proportional to the square root of the blocklength},''
  \emph{IEEE Transactions on Information Theory}, vol.~60, no.~2, pp.
  1193--1202, 2013.

\bibitem{babar2014exit}
Z.~Babar, S.-X. Ng, and L.~Hanzo, ``{EXIT-chart-aided near-capacity quantum
  turbo code design},'' \emph{IEEE Transactions on Vehicular Technology},
  vol.~64, no.~3, pp. 866--875, 2014.

\bibitem{leverrier2015quantum}
A.~Leverrier, J.-P. Tillich, and G.~Z{\'e}mor, ``{Quantum expander codes},'' in
  \emph{Proceedings of the 56th IEEE Annual Symposium on Foundations of
  Computer Science (FOCS)}.\hskip 1em plus 0.5em minus 0.4em\relax IEEE, 2015,
  pp. 810--824.

\bibitem{babar2016fully}
Z.~Babar, H.~V. Nguyen, P.~Botsinis, D.~Alanis, D.~Chandra, S.-X. Ng, R.~G.
  Maunder, and L.~Hanzo, ``{Fully-parallel quantum turbo decoder},'' \emph{IEEE
  Access}, vol.~4, pp. 6073--6085, 2016.

\bibitem{nguyen2016exit}
H.~V. Nguyen, Z.~Babar, D.~Alanis, P.~Botsinis, D.~Chandra, S.-X. Ng, and
  L.~Hanzo, ``{EXIT-chart aided quantum code design improves the normalised
  throughput of realistic quantum devices},'' \emph{IEEE Access}, vol.~4, pp.
  10\,194--10\,209, 2016.

\bibitem{panteleev2021degenerate}
P.~Panteleev and G.~Kalachev, ``{Degenerate quantum LDPC codes with good finite
  length performance},'' \emph{Quantum}, vol.~5, p. 585, 2021.

\bibitem{panteleev2021quantum}
------, ``{Quantum LDPC codes with almost linear minimum distance},''
  \emph{IEEE Transactions on Information Theory}, vol.~68, no.~1, pp. 213--229,
  2021.

\bibitem{leverrier2022quantum}
A.~Leverrier and G.~Z{\'e}mor, ``{Quantum Tanner codes},'' in \emph{Proceedings
  of the 63rd IEEE Annual Symposium on Foundations of Computer Science
  (FOCS)}.\hskip 1em plus 0.5em minus 0.4em\relax IEEE, 2022, pp. 872--883.

\bibitem{vuillot2022quantum}
C.~Vuillot and N.~P. Breuckmann, ``{Quantum pin codes},'' \emph{IEEE
  Transactions on Information Theory}, vol.~68, no.~9, pp. 5955--5974, 2022.

\bibitem{roffe2023bias}
J.~Roffe, L.~Z. Cohen, A.~O. Quintavalle, D.~Chandra, and E.~T. Campbell,
  ``Bias-tailored quantum ldpc codes,'' \emph{Quantum}, vol.~7, p. 1005, 2023.

\bibitem{christandl2022fault}
M.~Christandl and A.~M{\"u}ller-Hermes, ``{Fault-tolerant coding for quantum
  communication},'' \emph{IEEE Transactions on Information Theory}, vol. Early
  Access, pp. 1--38, 2022.

\bibitem{chandra2023universal}
D.~Chandra, Z.~B.~K. Egilmez, Y.~Xiong, S.-X. Ng, R.~G. Maunder, and L.~Hanzo,
  ``{Universal decoding of quantum stabilizer codes via classical guesswork},''
  \emph{IEEE Access}, vol.~11, pp. 19\,059--19\,072, 2023.

\bibitem{cruz2023quantum}
D.~Cruz, F.~A. Monteiro, and B.~C. Coutinho, ``{Quantum error correction via
  noise guessing decoding},'' \emph{IEEE Access}, vol.~11, pp.
  119\,446--119\,461, 2023.

\bibitem{yang2023quantum}
S.~Yang and A.~R. Calderbank, ``{Spatially-coupled QDLPC codes},'' \emph{arXiv
  preprint arXiv:2305.00137}, 2023.

\bibitem{hagiwara2007quantum}
M.~Hagiwara and H.~Imai, ``{Quantum quasi-cyclic LDPC codes},'' in
  \emph{Proceedings of the IEEE International Symposium on Information Theory
  (ISIT)}.\hskip 1em plus 0.5em minus 0.4em\relax IEEE, 2007, pp. 806--810.

\bibitem{yi2022quantum}
Z.~Yi, Z.~Liang, and X.~Wang, ``{Quantum polar stabilizer codes based on
  polarization of pure quantum channel are bad stabilizer codes for quantum
  computing},'' \emph{arXiv preprint arXiv:2204.11655}, 2022.

\bibitem{cleve1997efficient}
R.~Cleve and D.~Gottesman, ``{Efficient computations of encodings for quantum
  error correction},'' \emph{Physical Review A}, vol.~56, no.~1, pp. 76--82,
  1997.

\bibitem{fowler2012surface}
A.~G. Fowler, M.~Mariantoni, J.~M. Martinis, and A.~N. Cleland, ``{Surface
  codes: Towards practical large-scale quantum computation},'' \emph{Physical
  Review A}, vol.~86, no.~3, 2012.

\bibitem{shor1996fault}
P.~W. Shor, ``{Fault-tolerant quantum computation},'' in \emph{Proceedings of
  the 37th IEEE Annual Symposium on Foundations of Computer Science
  (FOCS}.\hskip 1em plus 0.5em minus 0.4em\relax IEEE, 1996, pp. 56--65.

\bibitem{steane1997active}
A.~M. Steane, ``{Active stabilization, quantum computation, and quantum state
  synthesis},'' \emph{Physical Review Letters}, vol.~78, no.~11, pp.
  2252--2255, 1997.

\bibitem{steane2002fast}
------, ``{Fast fault-tolerant filtering of quantum codewords},'' \emph{arXiv
  preprint quant-ph/0202036}, 2002.

\bibitem{chao2018quantum}
R.~Chao and B.~W. Reichardt, ``{Quantum error correction with only two extra
  qubits},'' \emph{Physical Review Letters}, vol. 121, no.~5, 2018.

\bibitem{huang2021between}
S.~Huang and K.~R. Brown, ``{Between Shor and Steane: A unifying construction
  for measuring error syndromes},'' \emph{Physical Review Letters}, vol. 127,
  no.~9, 2021.

\bibitem{dennis2002topological}
E.~Dennis, A.~Kitaev, A.~Landahl, and J.~Preskill, ``{Topological quantum
  memory},'' \emph{Journal of Mathematical Physics}, vol.~43, no.~9, pp.
  4452--4505, 2002.

\bibitem{bombin2015single}
H.~Bomb{\'\i}n, ``{Single-shot fault-tolerant quantum error correction},''
  \emph{Physical Review X}, vol.~5, no.~3, 2015.

\bibitem{campbell2019theory}
E.~T. Campbell, ``{A theory of single-shot error correction for adversarial
  noise},'' \emph{Quantum Science and Technology}, vol.~4, no.~2, 2019.

\bibitem{breuckmann2021single}
N.~P. Breuckmann and V.~Londe, ``{Single-shot decoding of linear rate LDPC
  quantum codes with high performance},'' \emph{IEEE Transactions on
  Information Theory}, vol.~68, no.~1, pp. 272--286, 2021.

\bibitem{quintavalle2021single}
A.~O. Quintavalle, M.~Vasmer, J.~Roffe, and E.~T. Campbell, ``{Single-shot
  error correction of three-dimensional homological product codes},'' \emph{PRX
  Quantum}, vol.~2, no.~2, 2021.

\bibitem{grospellier2021combining}
A.~Grospellier, L.~Grou{\`e}s, A.~Krishna, and A.~Leverrier, ``{Combining hard
  and soft decoders for hypergraph product codes},'' \emph{Quantum}, vol.~5, p.
  432, 2021.

\bibitem{higgott2023improved}
O.~Higgott and N.~P. Breuckmann, ``{Improved single-shot decoding of
  higher-dimensional hypergraph-product codes},'' \emph{PRX Quantum}, vol.~4,
  no.~2, 2023.

\bibitem{gu2023single}
S.~Gu, E.~Tang, L.~Caha, S.~H. Choe, Z.~He, and A.~Kubica, ``{Single-shot
  decoding of good quantum LDPC codes},'' \emph{arXiv preprint
  arXiv:2306.12470}, 2023.

\bibitem{zeng2019quantum}
W.~Zeng, A.~Ashikhmin, M.~Woolls, and L.~P. Pryadko, ``{Quantum convolutional
  data-syndrome codes},'' in \emph{Proceedings of the 20th IEEE International
  Workshop on Signal Processing Advances in Wireless Communications
  (SPAWC)}.\hskip 1em plus 0.5em minus 0.4em\relax IEEE, 2019, pp. 1--5.

\bibitem{etxezarreta2021time}
J.~Etxezarreta~Martinez, P.~Fuentes, P.~Crespo, and J.~Garcia-Frias,
  ``{Time-varying quantum channel models for superconducting qubits},''
  \emph{npj Quantum Information}, vol.~7, no.~1, p. 115, 2021.

\bibitem{wang2022construction}
Y.-J. Wang, Z.-Y. Xiao, Y.~Zhang, X.-Y. Xiong, and S.~Shi, ``{Construction of
  multiple-rate quantum LDPC codes sharing one scalable stabilizer circuit},''
  \emph{IEEE Transactions on Communications}, vol.~71, no.~2, pp. 1071--1082,
  2022.

\bibitem{skoric2023parallel}
L.~Skoric, D.~E. Browne, K.~M. Barnes, N.~I. Gillespie, and E.~T. Campbell,
  ``{Parallel window decoding enables scalable fault-tolerant quantum
  computation},'' \emph{Nature Communications}, vol.~14, no.~1, p. 7040, 2023.

\bibitem{battistel2023real}
F.~Battistel, C.~Chamberland, K.~Johar, R.~W. Overwater, F.~Sebastiano,
  L.~Skoric, Y.~Ueno, and M.~Usman, ``{Real-time decoding for fault-tolerant
  quantum computing: Progress, challenges and outlook},'' \emph{Nano Futures},
  vol.~7, no.~3, 2023.

\bibitem{yue2023efficient}
C.~Yue, V.~Miloslavskaya, M.~Shirvanimoghaddam, B.~Vucetic, and Y.~Li,
  ``{Efficient decoders for short block length codes in 6G URLLC},'' \emph{IEEE
  Communications Magazine}, vol.~61, no.~4, pp. 84--90, 2023.

\bibitem{fossorier1995soft}
M.~P. Fossorier and S.~Lin, ``{Soft-decision decoding of linear block codes
  based on ordered statistics},'' \emph{IEEE Transactions on Information
  Theory}, vol.~41, no.~5, pp. 1379--1396, 1995.

\bibitem{yue2021probability}
C.~Yue, M.~Shirvanimoghaddam, G.~Park, O.-S. Park, B.~Vucetic, and Y.~Li,
  ``{Probability-based ordered-statistics decoding for short block codes},''
  \emph{IEEE Communications Letters}, vol.~25, no.~6, pp. 1791--1795, 2021.

\bibitem{duffy2019capacity}
K.~R. Duffy, J.~Li, and M.~M{\'e}dard, ``{Capacity-achieving guessing random
  additive noise decoding},'' \emph{IEEE Transactions on Information Theory},
  vol.~65, no.~7, pp. 4023--4040, 2019.

\bibitem{roque2023efficient}
A.~Roque, D.~Cruz, F.~A. Monteiro, and B.~C. Coutinho, ``{Efficient
  entanglement purification based on noise guessing decoding},'' \emph{arXiv
  preprint arXiv:2310.19914}, 2023.

\bibitem{4675715}
A.~S. Fletcher, P.~W. Shor, and M.~Z. Win, ``Channel-adapted quantum error
  correction for the amplitude damping channel,'' \emph{IEEE Transactions on
  Information Theory}, vol.~54, no.~12, pp. 5705--5718, Dec 2008.

\bibitem{liEfficientVariationalQuantum2017}
Y.~Li and S.~C. Benjamin, ``Efficient {{Variational Quantum Simulator
  Incorporating Active Error Minimization}},'' \emph{Physical Review X},
  vol.~7, no.~2, p. 021050, Jun. 2017.

\bibitem{temmeErrorMitigationShortDepth2017}
K.~Temme, S.~Bravyi, and J.~M. Gambetta, ``Error {{Mitigation}} for
  {{Short-Depth Quantum Circuits}},'' \emph{Physical Review Letters}, vol. 119,
  no.~18, p. 180509, Nov. 2017.

\bibitem{mccleanHybridQuantumclassicalHierarchy2017}
J.~R. McClean, M.~E. {Kimchi-Schwartz}, J.~Carter, and W.~A. {de Jong},
  ``Hybrid quantum-classical hierarchy for mitigation of decoherence and
  determination of excited states,'' \emph{Physical Review A}, vol.~95, no.~4,
  p. 042308, Apr. 2017.

\bibitem{endoPracticalQuantumError2018}
S.~Endo, S.~C. Benjamin, and Y.~Li, ``Practical {{Quantum Error Mitigation}}
  for {{Near-Future Applications}},'' \emph{Physical Review X}, vol.~8, no.~3,
  p. 031027, Jul. 2018.

\bibitem{mcardleErrorMitigatedDigitalQuantum2019}
S.~McArdle, X.~Yuan, and S.~Benjamin, ``Error-{{Mitigated Digital Quantum
  Simulation}},'' \emph{Physical Review Letters}, vol. 122, no.~18, p. 180501,
  May 2019.

\bibitem{bonet-monroigLowcostErrorMitigation2018}
X.~{Bonet-Monroig}, R.~Sagastizabal, M.~Singh, and T.~E. O'Brien, ``Low-cost
  error mitigation by symmetry verification,'' \emph{Physical Review A},
  vol.~98, no.~6, p. 062339, Dec. 2018.

\bibitem{dumitrescuCloudQuantumComputing2018}
E.~F. Dumitrescu, A.~J. McCaskey, G.~Hagen, G.~R. Jansen, T.~D. Morris,
  T.~Papenbrock, R.~C. Pooser, D.~J. Dean, and P.~Lougovski, ``Cloud {{Quantum
  Computing}} of an {{Atomic Nucleus}},'' \emph{Physical Review Letters}, vol.
  120, no.~21, p. 210501, May 2018.

\bibitem{kandalaErrorMitigationExtends2019}
A.~Kandala, K.~Temme, A.~D. C{\'o}rcoles, A.~Mezzacapo, J.~M. Chow, and J.~M.
  Gambetta, ``Error mitigation extends the computational reach of a noisy
  quantum processor,'' \emph{Nature}, vol. 567, no. 7749, pp. 491--495, Mar.
  2019.

\bibitem{sagastizabalExperimentalErrorMitigation2019}
R.~Sagastizabal, X.~{Bonet-Monroig}, M.~Singh, M.~A. Rol, C.~C. Bultink, X.~Fu,
  C.~H. Price, V.~P. Ostroukh, N.~Muthusubramanian, A.~Bruno, M.~Beekman,
  N.~Haider, T.~E. O'Brien, and L.~DiCarlo, ``Experimental error mitigation via
  symmetry verification in a variational quantum eigensolver,'' \emph{Physical
  Review A}, vol. 100, no.~1, p. 010302, Jul. 2019.

\bibitem{czarnikErrorMitigationClifford2021}
P.~Czarnik, A.~Arrasmith, P.~J. Coles, and L.~Cincio, ``Error mitigation with
  {{Clifford}} quantum-circuit data,'' \emph{Quantum}, vol.~5, p. 592, Nov.
  2021.

\bibitem{strikis2021learning}
A.~Strikis, D.~Qin, Y.~Chen, S.~C. Benjamin, and Y.~Li, ``{Learning-based
  quantum error mitigation},'' \emph{PRX Quantum}, vol.~2, no.~4, 2021.

\bibitem{koczorExponentialErrorSuppression2021}
B.~Koczor, ``Exponential {{Error Suppression}} for {{Near-Term Quantum
  Devices}},'' \emph{Physical Review X}, vol.~11, no.~3, p. 031057, Sep. 2021.

\bibitem{hugginsVirtualDistillationQuantum2021}
W.~J. Huggins, S.~McArdle, T.~E. O'Brien, J.~Lee, N.~C. Rubin, S.~Boixo, K.~B.
  Whaley, R.~Babbush, and J.~R. McClean, ``Virtual {{Distillation}} for
  {{Quantum Error Mitigation}},'' \emph{Physical Review X}, vol.~11, no.~4, p.
  041036, Nov. 2021.

\bibitem{obrienPurificationbasedQuantumError2023}
T.~E. O'Brien, G.~Anselmetti, F.~Gkritsis, V.~E. Elfving, S.~Polla, W.~J.
  Huggins, O.~Oumarou, K.~Kechedzhi, D.~Abanin, R.~Acharya, I.~Aleiner,
  R.~Allen, T.~I. Andersen, K.~Anderson, M.~Ansmann, F.~Arute, K.~Arya,
  A.~Asfaw, J.~Atalaya, J.~C. Bardin, A.~Bengtsson, G.~Bortoli, A.~Bourassa,
  J.~Bovaird, L.~Brill, M.~Broughton, B.~Buckley, D.~A. Buell, T.~Burger,
  B.~Burkett, N.~Bushnell, J.~Campero, Z.~Chen, B.~Chiaro, D.~Chik, J.~Cogan,
  R.~Collins, P.~Conner, W.~Courtney, A.~L. Crook, B.~Curtin, D.~M. Debroy,
  S.~Demura, I.~Drozdov, A.~Dunsworth, C.~Erickson, L.~Faoro, E.~Farhi,
  R.~Fatemi, V.~S. Ferreira, L.~Flores~Burgos, E.~Forati, A.~G. Fowler,
  B.~Foxen, W.~Giang, C.~Gidney, D.~Gilboa, M.~Giustina, R.~Gosula,
  A.~Grajales~Dau, J.~A. Gross, S.~Habegger, M.~C. Hamilton, M.~Hansen, M.~P.
  Harrigan, S.~D. Harrington, P.~Heu, M.~R. Hoffmann, S.~Hong, T.~Huang,
  A.~Huff, L.~B. Ioffe, S.~V. Isakov, J.~Iveland, E.~Jeffrey, Z.~Jiang,
  C.~Jones, P.~Juhas, D.~Kafri, T.~Khattar, M.~Khezri, M.~Kieferov{\'a},
  S.~Kim, P.~V. Klimov, A.~R. Klots, A.~N. Korotkov, F.~Kostritsa, J.~M.
  Kreikebaum, D.~Landhuis, P.~Laptev, K.-M. Lau, L.~Laws, J.~Lee, K.~Lee, B.~J.
  Lester, A.~T. Lill, W.~Liu, W.~P. Livingston, A.~Locharla, F.~D. Malone,
  S.~Mandr{\`a}, O.~Martin, S.~Martin, J.~R. McClean, T.~McCourt, M.~McEwen,
  X.~Mi, A.~Mieszala, K.~C. Miao, M.~Mohseni, S.~Montazeri, A.~Morvan,
  R.~Movassagh, W.~Mruczkiewicz, O.~Naaman, M.~Neeley, C.~Neill, A.~Nersisyan,
  M.~Newman, J.~H. Ng, A.~Nguyen, M.~Nguyen, M.~Y. Niu, S.~Omonije,
  A.~Opremcak, A.~Petukhov, R.~Potter, L.~P. Pryadko, C.~Quintana, C.~Rocque,
  P.~Roushan, N.~Saei, D.~Sank, K.~Sankaragomathi, K.~J. Satzinger, H.~F.
  Schurkus, C.~Schuster, M.~J. Shearn, A.~Shorter, N.~Shutty, V.~Shvarts,
  J.~Skruzny, W.~C. Smith, R.~D. Somma, G.~Sterling, D.~Strain, M.~Szalay,
  D.~Thor, A.~Torres, G.~Vidal, B.~Villalonga, C.~Vollgraff~Heidweiller,
  T.~White, B.~W.~K. Woo, C.~Xing, Z.~J. Yao, P.~Yeh, J.~Yoo, G.~Young,
  A.~Zalcman, Y.~Zhang, N.~Zhu, N.~Zobrist, D.~Bacon, S.~Boixo, Y.~Chen,
  J.~Hilton, J.~Kelly, E.~Lucero, A.~Megrant, H.~Neven, V.~Smelyanskiy,
  C.~Gogolin, R.~Babbush, and N.~C. Rubin, ``Purification-based quantum error
  mitigation of pair-correlated electron simulations,'' \emph{Nature Physics},
  pp. 1--6, Oct. 2023.

\bibitem{kimEvidenceUtilityQuantum2023}
Y.~Kim, A.~Eddins, S.~Anand, K.~X. Wei, E.~{van den Berg}, S.~Rosenblatt,
  H.~Nayfeh, Y.~Wu, M.~Zaletel, K.~Temme, and A.~Kandala, ``Evidence for the
  utility of quantum computing before fault tolerance,'' \emph{Nature}, vol.
  618, no. 7965, pp. 500--505, Jun. 2023.

\bibitem{caiMultiexponentialErrorExtrapolation2021}
Z.~Cai, ``Multi-exponential error extrapolation and combining error mitigation
  techniques for {{NISQ}} applications,'' \emph{npj Quantum Information},
  vol.~7, p.~80, May 2021.

\bibitem{chowDetectingHighlyEntangled2010}
J.~M. Chow, L.~DiCarlo, J.~M. Gambetta, A.~Nunnenkamp, L.~S. Bishop,
  L.~Frunzio, M.~H. Devoret, S.~M. Girvin, and R.~J. Schoelkopf, ``Detecting
  highly entangled states with a joint qubit readout,'' \emph{Physical Review
  A}, vol.~81, no.~6, p. 062325, Jun. 2010.

\bibitem{kandalaHardwareefficientVariationalQuantum2017}
A.~Kandala, A.~Mezzacapo, K.~Temme, M.~Takita, M.~Brink, J.~M. Chow, and J.~M.
  Gambetta, ``Hardware-efficient variational quantum eigensolver for small
  molecules and quantum magnets,'' \emph{Nature}, vol. 549, no. 7671, pp.
  242--246, Sep. 2017.

\bibitem{PhysRevApplied.18.044064}
\BIBentryALTinterwordspacing
H.~Jnane, B.~Undseth, Z.~Cai, S.~C. Benjamin, and B.~Koczor, ``Multicore
  quantum computing,'' \emph{Phys. Rev. Appl.}, vol.~18, p. 044064, Oct 2022.
  [Online]. Available:
  \url{https://link.aps.org/doi/10.1103/PhysRevApplied.18.044064}
\BIBentrySTDinterwordspacing

\bibitem{koczor2021dominant}
B.~Koczor, ``The dominant eigenvector of a noisy quantum state,'' \emph{New
  Journal of Physics}, vol.~23, no.~12, p. 123047, 2021.

\bibitem{9638483}
Y.~Xiong, S.~X. Ng, and L.~Hanzo, ``Quantum error mitigation relying on
  permutation filtering,'' \emph{IEEE Transactions on Communications}, vol.~70,
  no.~3, pp. 1927--1942, 2022.

\bibitem{huoDualstatePurificationPractical2022}
M.~Huo and Y.~Li, ``Dual-state purification for practical quantum error
  mitigation,'' \emph{Physical Review A}, vol. 105, no.~2, p. 022427, Feb.
  2022.

\bibitem{caiResourceefficientPurificationbasedQuantum2021}
\BIBentryALTinterwordspacing
Z.~Cai, ``Resource-efficient {{Purification-based Quantum Error Mitigation}},''
  arXiv:2107.07279 [quant-ph], Jul. 2021. [Online]. Available:
  \url{https://doi.org/10.48550/arXiv.2107.07279}
\BIBentrySTDinterwordspacing

\bibitem{PhysRevLett.124.110501}
\BIBentryALTinterwordspacing
L.~J. Stephenson, D.~P. Nadlinger, B.~C. Nichol, S.~An, P.~Drmota, T.~G.
  Ballance, K.~Thirumalai, J.~F. Goodwin, D.~M. Lucas, and C.~J. Ballance,
  ``High-rate, high-fidelity entanglement of qubits across an elementary
  quantum network,'' \emph{Phys. Rev. Lett.}, vol. 124, p. 110501, Mar 2020.
  [Online]. Available:
  \url{https://link.aps.org/doi/10.1103/PhysRevLett.124.110501}
\BIBentrySTDinterwordspacing

\bibitem{caiLoopedPipelinesEnabling2023}
Z.~Cai, A.~Siegel, and S.~Benjamin, ``Looped {{Pipelines Enabling Effective 3D
  Qubit Lattices}} in a {{Strictly 2D Device}},'' \emph{PRX Quantum}, vol.~4,
  no.~2, p. 020345, Jun. 2023.

\bibitem{10048485}
Y.~Xiong, D.~Chandra, S.~X. Ng, and L.~Hanzo, ``Circuit symmetry verification
  mitigates quantum-domain impairments,'' \emph{IEEE Transactions on Signal
  Processing}, vol.~71, pp. 477--493, 2023.

\bibitem{hugginsEfficientNoiseResilient2021}
W.~J. Huggins, J.~R. McClean, N.~C. Rubin, Z.~Jiang, N.~Wiebe, K.~B. Whaley,
  and R.~Babbush, ``Efficient and noise resilient measurements for quantum
  chemistry on near-term quantum computers,'' \emph{npj Quantum Information},
  vol.~7, p.~23, Feb. 2021.

\bibitem{caiQuantumErrorMitigation2021}
Z.~Cai, ``Quantum {{Error Mitigation}} using {{Symmetry Expansion}},''
  \emph{Quantum}, vol.~5, p. 548, Sep. 2021.

\bibitem{bravyiFermionicQuantumComputation2002}
S.~Bravyi and A.~Kitaev, ``Fermionic quantum computation,'' \emph{Annals of
  Physics}, vol. 298, no.~1, pp. 210--226, May 2002.

\bibitem{setiaBravyiKitaevSuperfastSimulation2018}
K.~Setia and J.~D. Whitfield, ``Bravyi-{{Kitaev Superfast}} simulation of
  fermions on a quantum computer,'' \emph{The Journal of Chemical Physics},
  vol. 148, no.~16, p. 164104, Apr. 2018.

\bibitem{derbyCompactFermionQubit2021}
C.~Derby, J.~Klassen, J.~Bausch, and T.~Cubitt, ``Compact fermion to qubit
  mappings,'' \emph{Physical Review B}, vol. 104, no.~3, p. 035118, Jul. 2021.

\bibitem{jiangMajoranaLoopStabilizer2019}
Z.~Jiang, J.~McClean, R.~Babbush, and H.~Neven, ``Majorana {{Loop Stabilizer
  Codes}} for {{Error Mitigation}} in {{Fermionic Quantum Simulations}},''
  \emph{Physical Review Applied}, vol.~12, no.~6, p. 064041, Dec. 2019.

\bibitem{mccleanDecodingQuantumErrors2020}
J.~R. McClean, Z.~Jiang, N.~C. Rubin, R.~Babbush, and H.~Neven, ``Decoding
  quantum errors with subspace expansions,'' \emph{Nature Communications},
  vol.~11, p. 636, Jan. 2020.

\bibitem{collessComputationMolecularSpectra2018}
J.~I. Colless, V.~V. Ramasesh, D.~Dahlen, M.~S. Blok, M.~E. {Kimchi-Schwartz},
  J.~R. McClean, J.~Carter, W.~A. {de Jong}, and I.~Siddiqi, ``Computation of
  {{Molecular Spectra}} on a {{Quantum Processor}} with an {{Error-Resilient
  Algorithm}},'' \emph{Physical Review X}, vol.~8, no.~1, p. 011021, Feb. 2018.

\bibitem{urbanekChemistryQuantumComputers2020}
M.~Urbanek, D.~Camps, R.~Van~Beeumen, and W.~A. {de Jong}, ``Chemistry on
  {{Quantum Computers}} with {{Virtual Quantum Subspace Expansion}},''
  \emph{Journal of Chemical Theory and Computation}, vol.~16, no.~9, pp.
  5425--5431, Sep. 2020.

\bibitem{mottaDeterminingEigenstatesThermal2020}
M.~Motta, C.~Sun, A.~T.~K. Tan, M.~J. O'Rourke, E.~Ye, A.~J. Minnich, F.~G.
  S.~L. Brand{\~a}o, and G.~K.-L. Chan, ``Determining eigenstates and thermal
  states on a quantum computer using quantum imaginary time evolution,''
  \emph{Nature Physics}, vol.~16, no.~2, pp. 205--210, Feb. 2020.

\bibitem{yoshiokaGeneralizedQuantumSubspace2022}
N.~Yoshioka, H.~Hakoshima, Y.~Matsuzaki, Y.~Tokunaga, Y.~Suzuki, and S.~Endo,
  ``Generalized {{Quantum Subspace Expansion}},'' \emph{Physical Review
  Letters}, vol. 129, no.~2, p. 020502, Jul. 2022.

\bibitem{foldager2023can}
J.~Foldager and B.~Koczor, ``Can shallow quantum circuits scramble local noise
  into global white noise?'' \emph{J Phys A, \emph{accepted manuscript}}, 2023.

\bibitem{montanaro2021error}
A.~Montanaro and S.~Stanisic, ``Error mitigation by training with fermionic
  linear optics,'' \emph{arXiv preprint arXiv:2102.02120}, 2021.

\bibitem{caiPracticalFrameworkQuantum2021}
\BIBentryALTinterwordspacing
Z.~Cai, ``A {{Practical Framework}} for {{Quantum Error Mitigation}},''
  arXiv:2110.05389 [quant-ph], Oct. 2021. [Online]. Available:
  \url{http://arxiv.org/abs/2110.05389}
\BIBentrySTDinterwordspacing

\bibitem{quek2022exponentially}
Y.~Quek, D.~S. Fran{\c{c}}a, S.~Khatri, J.~J. Meyer, and J.~Eisert,
  ``Exponentially tighter bounds on limitations of quantum error mitigation,''
  \emph{arXiv preprint arXiv:2210.11505}, 2022.

\bibitem{9294106}
Y.~Xiong, D.~Chandra, S.~X. Ng, and L.~Hanzo, ``Sampling overhead analysis of
  quantum error mitigation: Uncoded vs. coded systems,'' \emph{IEEE Access},
  vol.~8, pp. 228\,967--228\,991, 2020.

\bibitem{9684862}
Y.~Xiong, S.~X. Ng, and L.~Hanzo, ``The accuracy vs. sampling overhead
  trade-off in quantum error mitigation using monte carlo-based channel
  inversion,'' \emph{IEEE Transactions on Communications}, vol.~70, no.~3, pp.
  1943--1956, 2022.

\bibitem{koczor2022quantum}
B.~Koczor and S.~C. Benjamin, ``Quantum natural gradient generalized to noisy
  and nonunitary circuits,'' \emph{Physical Review A}, vol. 106, no.~6, p.
  062416, 2022.

\bibitem{stokes2020quantum}
J.~Stokes, J.~Izaac, N.~Killoran, and G.~Carleo, ``Quantum natural gradient,''
  \emph{Quantum}, vol.~4, p. 269, 2020.

\bibitem{berg2022probabilistic}
E.~Van Den~Berg, Z.~K. Minev, A.~Kandala, and K.~Temme, ``Probabilistic error
  cancellation with sparse pauli--lindblad models on noisy quantum
  processors,'' \emph{Nature Physics}, pp. 1--6, 2023.

\bibitem{elben2023randomized}
A.~Elben, S.~T. Flammia, H.-Y. Huang, R.~Kueng, J.~Preskill, B.~Vermersch, and
  P.~Zoller, ``The randomized measurement toolbox,'' \emph{Nature Reviews
  Physics}, vol.~5, no.~1, pp. 9--24, 2023.

\bibitem{jnaneQuantumErrorMitigated2023}
\BIBentryALTinterwordspacing
H.~Jnane, J.~Steinberg, Z.~Cai, H.~C. Nguyen, and B.~Koczor, ``Quantum {{Error
  Mitigated Classical Shadows}},'' arXiv:2305.04956 [quant-ph], May 2023.
  [Online]. Available: \url{https://doi.org/10.48550/arXiv.2305.04956}
\BIBentrySTDinterwordspacing

\bibitem{seifShadowDistillationQuantum2023}
A.~Seif, Z.-P. Cian, S.~Zhou, S.~Chen, and L.~Jiang, ``Shadow {{Distillation}}:
  {{Quantum Error Mitigation}} with {{Classical Shadows}} for {{Near-Term
  Quantum Processors}},'' \emph{PRX Quantum}, vol.~4, no.~1, p. 010303, Jan.
  2023.

\bibitem{PhysRevX.12.041022}
\BIBentryALTinterwordspacing
G.~Boyd and B.~Koczor, ``Training variational quantum circuits with covar:
  Covariance root finding with classical shadows,'' \emph{Phys. Rev. X},
  vol.~12, p. 041022, Nov 2022. [Online]. Available:
  \url{https://link.aps.org/doi/10.1103/PhysRevX.12.041022}
\BIBentrySTDinterwordspacing

\bibitem{aruteQuantumSupremacyUsing2019}
F.~Arute, K.~Arya, R.~Babbush, D.~Bacon, J.~C. Bardin, R.~Barends, R.~Biswas,
  S.~Boixo, F.~G. S.~L. Brandao, D.~A. Buell, B.~Burkett, Y.~Chen, Z.~Chen,
  B.~Chiaro, R.~Collins, W.~Courtney, A.~Dunsworth, E.~Farhi, B.~Foxen,
  A.~Fowler, C.~Gidney, M.~Giustina, R.~Graff, K.~Guerin, S.~Habegger, M.~P.
  Harrigan, M.~J. Hartmann, A.~Ho, M.~Hoffmann, T.~Huang, T.~S. Humble, S.~V.
  Isakov, E.~Jeffrey, Z.~Jiang, D.~Kafri, K.~Kechedzhi, J.~Kelly, P.~V. Klimov,
  S.~Knysh, A.~Korotkov, F.~Kostritsa, D.~Landhuis, M.~Lindmark, E.~Lucero,
  D.~Lyakh, S.~Mandr{\`a}, J.~R. McClean, M.~McEwen, A.~Megrant, X.~Mi,
  K.~Michielsen, M.~Mohseni, J.~Mutus, O.~Naaman, M.~Neeley, C.~Neill, M.~Y.
  Niu, E.~Ostby, A.~Petukhov, J.~C. Platt, C.~Quintana, E.~G. Rieffel,
  P.~Roushan, N.~C. Rubin, D.~Sank, K.~J. Satzinger, V.~Smelyanskiy, K.~J.
  Sung, M.~D. Trevithick, A.~Vainsencher, B.~Villalonga, T.~White, Z.~J. Yao,
  P.~Yeh, A.~Zalcman, H.~Neven, and J.~M. Martinis, ``Quantum supremacy using a
  programmable superconducting processor,'' \emph{Nature}, vol. 574, no. 7779,
  pp. 505--510, Oct. 2019.

\bibitem{chan2022algorithmic}
H.~H.~S. Chan, R.~Meister, M.~L. Goh, and B.~Koczor, ``Algorithmic shadow
  spectroscopy,'' \emph{arXiv preprint arXiv:2212.11036}, 2022.

\bibitem{loweUnifiedApproachDatadriven2021}
A.~Lowe, M.~H. Gordon, P.~Czarnik, A.~Arrasmith, P.~J. Coles, and L.~Cincio,
  ``Unified approach to data-driven quantum error mitigation,'' \emph{Physical
  Review Research}, vol.~3, no.~3, p. 033098, Jul. 2021.

\bibitem{bultriniUnifyingBenchmarkingStateoftheart2023}
D.~Bultrini, M.~H. Gordon, P.~Czarnik, A.~Arrasmith, M.~Cerezo, P.~J. Coles,
  and L.~Cincio, ``Unifying and benchmarking state-of-the-art quantum error
  mitigation techniques,'' \emph{Quantum}, vol.~7, p. 1034, Jun. 2023.

\bibitem{suzukiQuantumErrorMitigation2022}
Y.~Suzuki, S.~Endo, K.~Fujii, and Y.~Tokunaga, ``Quantum {{Error Mitigation}}
  as a {{Universal Error Reduction Technique}}: {{Applications}} from the
  {{NISQ}} to the {{Fault-Tolerant Quantum Computing Eras}},'' \emph{PRX
  Quantum}, vol.~3, no.~1, p. 010345, Mar. 2022.

\bibitem{piveteauErrorMitigationUniversal2021}
C.~Piveteau, D.~Sutter, S.~Bravyi, J.~M. Gambetta, and K.~Temme, ``Error
  {{Mitigation}} for {{Universal Gates}} on {{Encoded Qubits}},''
  \emph{Physical Review Letters}, vol. 127, no.~20, p. 200505, Nov. 2021.

\bibitem{schuld2021machine}
M.~Schuld and F.~Petruccione, \emph{{Machine learning with quantum
  computers}}.\hskip 1em plus 0.5em minus 0.4em\relax Springer, 2021.

\bibitem{simeone2022introduction}
O.~Simeone \emph{et~al.}, ``{An introduction to quantum machine learning for
  engineers},'' \emph{Foundations and Trends{\textregistered} in Signal
  Processing}, vol.~16, no. 1-2, pp. 1--223, 2022.

\bibitem{valenti2019hamiltonian}
A.~Valenti, E.~van Nieuwenburg, S.~Huber, and E.~Greplova, ``{Hamiltonian
  learning for quantum error correction},'' \emph{Physical Review Research},
  vol.~1, no.~3, 2019.

\bibitem{nautrup2019optimizing}
H.~P. Nautrup, N.~Delfosse, V.~Dunjko, H.~J. Briegel, and N.~Friis,
  ``{Optimizing quantum error correction codes with reinforcement learning},''
  \emph{Quantum}, vol.~3, p. 215, 2019.

\bibitem{kim2020quantum}
C.~Kim, K.~D. Park, and J.-K. Rhee, ``{Quantum error mitigation with artificial
  neural network},'' \emph{IEEE Access}, vol.~8, pp. 188\,853--188\,860, 2020.

\bibitem{locher2023quantum}
D.~F. Locher, L.~Cardarelli, and M.~M{\"u}ller, ``{Quantum error correction
  with quantum autoencoders},'' \emph{Quantum}, vol.~7, p. 942, 2023.

\bibitem{nawaz2019quantum}
S.~J. Nawaz, S.~K. Sharma, S.~Wyne, M.~N. Patwary, and M.~Asaduzzaman,
  ``{Quantum machine learning for 6G communication networks: State-of-the-art
  and vision for the future},'' \emph{IEEE Access}, vol.~7, pp.
  46\,317--46\,350, 2019.

\bibitem{tabi2021evaluation}
Z.~I. Tabi, {\'A}.~Marosits, Z.~Kallus, P.~Vaderna, I.~G{\'o}dor, and
  Z.~Zimbor{\'a}s, ``{Evaluation of quantum annealer performance via the
  massive MIMO problem},'' \emph{IEEE Access}, vol.~9, pp. 131\,658--131\,671,
  2021.

\bibitem{cui2022quantum}
J.~Cui, Y.~Xiong, S.-X. Ng, and L.~Hanzo, ``{Quantum approximate optimization
  algorithm based maximum likelihood detection},'' \emph{IEEE Transactions on
  Communications}, vol.~70, no.~8, pp. 5386--5400, 2022.

\bibitem{chittoor2023quantum}
H.~H.~S. Chittoor and O.~Simeone, ``{Quantum machine learning for distributed
  quantum protocols with local operations and noisy classical
  communications},'' \emph{Entropy}, vol.~25, no.~2, 2023.

\bibitem{cerezo2022challenges}
M.~Cerezo, G.~Verdon, H.-Y. Huang, L.~Cincio, and P.~J. Coles, ``{Challenges
  and opportunities in quantum machine learning},'' \emph{Nature Computational
  Science}, vol.~2, no.~9, pp. 567--576, 2022.

\bibitem{banchi2023statistical}
L.~Banchi, J.~L. Pereira, S.~T. Jose, and O.~Simeone, ``{Statistical complexity
  of quantum learning},'' \emph{arXiv preprint arXiv:2309.11617}, 2023.

\bibitem{romero2017quantum}
J.~Romero, J.~P. Olson, and A.~Aspuru-Guzik, ``{Quantum autoencoders for
  efficient compression of quantum data},'' \emph{Quantum Science and
  Technology}, vol.~2, no.~4, 2017.

\bibitem{dallaire2018quantum}
P.-L. Dallaire-Demers and N.~Killoran, ``{Quantum generative adversarial
  networks},'' \emph{Physical Review A}, vol.~98, no.~1, 2018.

\bibitem{coyle2020born}
B.~Coyle, D.~Mills, V.~Danos, and E.~Kashefi, ``{The Born supremacy: Quantum
  advantage and training of an Ising Born machine},'' \emph{npj Quantum
  Information}, vol.~6, no.~1, p.~60, 2020.

\bibitem{gao2022enhancing}
X.~Gao, E.~R. Anschuetz, S.-T. Wang, J.~I. Cirac, and M.~D. Lukin, ``{Enhancing
  generative models via quantum correlations},'' \emph{Physical Review X},
  vol.~12, no.~2, 2022.

\bibitem{pirnay2023superpolynomial}
N.~Pirnay, R.~Sweke, J.~Eisert, and J.-P. Seifert, ``{Superpolynomial
  quantum-classical separation for density modeling},'' \emph{Physical Review
  A}, vol. 107, no.~4, 2023.

\bibitem{hinsche2023one}
M.~Hinsche, M.~Ioannou, A.~Nietner, J.~Haferkamp, Y.~Quek, D.~Hangleiter, J.-P.
  Seifert, J.~Eisert, and R.~Sweke, ``{One $T$ gate makes distribution learning
  hard},'' \emph{Physical Review Letters}, vol. 130, no.~24, 2023.

\bibitem{du2022theory}
Y.~Du, Z.~Tu, B.~Wu, X.~Yuan, and D.~Tao, ``{Theory of quantum generative
  learning models with maximum mean discrepancy},'' \emph{arXiv preprint
  arXiv:2205.04730}, 2022.

\bibitem{abbas2023quantum}
A.~Abbas, R.~King, H.-Y. Huang, W.~J. Huggins, R.~Movassagh, D.~Gilboa, and
  J.~R. McClean, ``{On quantum backpropagation, information reuse, and cheating
  measurement collapse},'' \emph{arXiv preprint arXiv:2305.13362}, 2023.

\bibitem{kak1995quantum}
S.~C. Kak, ``{Quantum neural computing},'' \emph{Advances in Imaging and
  Electron Physics}, vol.~94, pp. 259--313, 1995.

\bibitem{farhi2014quantum}
E.~Farhi, J.~Goldstone, and S.~Gutmann, ``{A quantum approximate optimization
  algorithm},'' \emph{arXiv preprint arXiv:1411.4028}, 2014.

\bibitem{wittek2014quantum}
P.~Wittek, \emph{{Quantum machine learning: What quantum computing means to
  data mining}}.\hskip 1em plus 0.5em minus 0.4em\relax Academic Press, 2014.

\bibitem{mcclean2016theory}
J.~R. McClean, J.~Romero, R.~Babbush, and A.~Aspuru-Guzik, ``{The theory of
  variational hybrid quantum-classical algorithms},'' \emph{New Journal of
  Physics}, vol.~18, no.~2, 2016.

\bibitem{liu2018differentiable}
J.-G. Liu and L.~Wang, ``{Differentiable learning of quantum circuit Born
  machines},'' \emph{Physical Review A}, vol.~98, no.~6, 2018.

\bibitem{farhi2018classification}
E.~Farhi and H.~Neven, ``{Classification with quantum neural networks on near
  term processors},'' \emph{arXiv preprint arXiv:1802.06002}, 2018.

\bibitem{arunachalam2018optimal}
S.~Arunachalam and R.~De~Wolf, ``{Optimal quantum sample complexity of learning
  algorithms},'' \emph{The Journal of Machine Learning Research}, vol.~19,
  no.~1, pp. 2879--2878, 2018.

\bibitem{schuld2018supervised}
M.~Schuld and F.~Petruccione, \emph{{Supervised learning with quantum
  computers}}.\hskip 1em plus 0.5em minus 0.4em\relax Springer, 2018, vol.~17.

\bibitem{mcclean2018barren}
J.~R. McClean, S.~Boixo, V.~N. Smelyanskiy, R.~Babbush, and H.~Neven, ``{Barren
  plateaus in quantum neural network training landscapes},'' \emph{Nature
  Communications}, vol.~9, no.~1, p. 4812, 2018.

\bibitem{cong2019quantum}
I.~Cong, S.~Choi, and M.~D. Lukin, ``{Quantum convolutional neural networks},''
  \emph{Nature Physics}, vol.~15, no.~12, pp. 1273--1278, 2019.

\bibitem{zoufal2019quantum}
C.~Zoufal, A.~Lucchi, and S.~Woerner, ``{Quantum generative adversarial
  networks for learning and loading random distributions},'' \emph{npj Quantum
  Information}, vol.~5, no.~1, p. 103, 2019.

\bibitem{tacchino2019artificial}
F.~Tacchino, C.~Macchiavello, D.~Gerace, and D.~Bajoni, ``{An artificial neuron
  implemented on an actual quantum processor},'' \emph{npj Quantum
  Information}, vol.~5, no.~1, p.~26, 2019.

\bibitem{broughton2020tensorflow}
M.~Broughton, G.~Verdon, T.~McCourt, A.~J. Martinez, J.~H. Yoo, S.~V. Isakov,
  P.~Massey, R.~Halavati, M.~Y. Niu, A.~Zlokapa \emph{et~al.}, ``{TensorFlow
  Quantum: A software framework for quantum machine learning},'' \emph{arXiv
  preprint arXiv:2003.02989}, 2020.

\bibitem{banchi2021generalization}
L.~Banchi, J.~Pereira, and S.~Pirandola, ``{Generalization in quantum machine
  learning: A quantum information standpoint},'' \emph{PRX Quantum}, vol.~2,
  no.~4, 2021.

\bibitem{sharma2022trainability}
K.~Sharma, M.~Cerezo, L.~Cincio, and P.~J. Coles, ``{Trainability of
  dissipative perceptron-based quantum neural networks},'' \emph{Physical
  Review Letters}, vol. 128, no.~18, 2022.

\bibitem{caro2022generalization}
M.~C. Caro, H.-Y. Huang, M.~Cerezo, K.~Sharma, A.~Sornborger, L.~Cincio, and
  P.~J. Coles, ``{Generalization in quantum machine learning from few training
  data},'' \emph{Nature Communications}, vol.~13, no.~1, p. 4919, 2022.

\bibitem{nietner2023average}
A.~Nietner, M.~Ioannou, R.~Sweke, R.~Kueng, J.~Eisert, M.~Hinsche, and
  J.~Haferkamp, ``{On the average-case complexity of learning output
  distributions of quantum circuits},'' \emph{arXiv preprint arXiv:2305.05765},
  2023.

\bibitem{liu2021rigorous}
Y.~Liu, S.~Arunachalam, and K.~Temme, ``{A rigorous and robust quantum speed-up
  in supervised machine learning},'' \emph{Nature Physics}, vol.~17, no.~9, pp.
  1013--1017, 2021.

\bibitem{ragone2023unified}
M.~Ragone, B.~N. Bakalov, F.~Sauvage, A.~F. Kemper, C.~O. Marrero, M.~Larocca,
  and M.~Cerezo, ``{A unified theory of barren plateaus for deep parametrized
  quantum circuits},'' \emph{arXiv preprint arXiv:2309.09342}, 2023.

\bibitem{pesah2021absence}
A.~Pesah, M.~Cerezo, S.~Wang, T.~Volkoff, A.~T. Sornborger, and P.~J. Coles,
  ``{Absence of barren plateaus in quantum convolutional neural networks},''
  \emph{Physical Review X}, vol.~11, no.~4, 2021.

\bibitem{peters2022generalization}
E.~Peters and M.~Schuld, ``{Generalization despite overfitting in quantum
  machine learning models},'' \emph{arXiv preprint arXiv:2209.05523}, 2022.

\bibitem{bronstein2021geometric}
M.~M. Bronstein, J.~Bruna, T.~Cohen, and P.~Veli{\v{c}}kovi{\'c}, ``{Geometric
  deep learning: Grids, groups, graphs, geodesics, and gauges},'' \emph{arXiv
  preprint arXiv:2104.13478}, 2021.

\bibitem{perrier2020quantum}
E.~Perrier, D.~Tao, and C.~Ferrie, ``{Quantum geometric machine learning for
  quantum circuits and control},'' \emph{New Journal of Physics}, vol.~22,
  no.~10, 2020.

\bibitem{nikoloska2023time}
I.~Nikoloska, O.~Simeone, L.~Banchi, and P.~Veli{\v{c}}kovi{\'c},
  ``{Time-warping invariant quantum recurrent neural networks via
  quantum-classical adaptive gating},'' \emph{Machine Learning: Science and
  Technology}, 2023.

\bibitem{nikoloska2022quantum}
I.~Nikoloska and O.~Simeone, ``{Quantum-aided meta-learning for Bayesian binary
  neural networks via Born machines},'' in \emph{Proceedings of the 32nd IEEE
  International Workshop on Machine Learning for Signal Processing
  (MLSP)}.\hskip 1em plus 0.5em minus 0.4em\relax IEEE, 2022, pp. 1--6.

\bibitem{carrasquilla2023quantum}
J.~Carrasquilla, M.~Hibat-Allah, E.~Inack, A.~Makhzani, K.~Neklyudov, G.~W.
  Taylor, and G.~Torlai, ``{Quantum hypernetworks: Training binary neural
  networks in quantum superposition},'' \emph{arXiv preprint arXiv:2301.08292},
  2023.

\bibitem{raussendorf2003measurement}
R.~Raussendorf, D.~E. Browne, and H.~J. Briegel, ``{Measurement-based quantum
  computation on cluster states},'' \emph{Physical Review A}, vol.~68, no.~2,
  2003.

\bibitem{majumder2023variational}
A.~Majumder, M.~Krumm, T.~Radkohl, H.~P. Nautrup, S.~Jerbi, and H.~J. Briegel,
  ``{Variational measurement-based quantum computation for generative
  modeling},'' \emph{arXiv preprint arXiv:2310.13524}, 2023.

\bibitem{tibaldi2023bayesian}
S.~Tibaldi, D.~Vodola, E.~Tignone, and E.~Ercolessi, ``{Bayesian optimization
  for QAOA},'' \emph{IEEE Transactions on Quantum Engineering}, vol. Early
  Access, pp. 1--12, 2023.

\bibitem{jose2022error}
S.~T. Jose and O.~Simeone, ``{Error-mitigation-aided optimization of
  parameterized quantum circuits: Convergence analysis},'' \emph{IEEE
  Transactions on Quantum Engineering}, vol.~3, pp. 1--19, 2022.

\bibitem{park2023quantum}
S.~Park and O.~Simeone, ``{Quantum conformal prediction for reliable
  uncertainty quantification in quantum machine learning},'' \emph{arXiv
  preprint arXiv:2304.03398}, 2023.

\bibitem{1}
M.~Lanzagorta, \emph{{Quantum Radar}}.\hskip 1em plus 0.5em minus 0.4em\relax
  Morgan and Claypool Publishers, 2012.

\bibitem{2}
I.~B. Djordjevic, \emph{{Quantum Communication, Quantum Networks, and Quantum
  Sensing}}.\hskip 1em plus 0.5em minus 0.4em\relax Elsevier/Academic Press,
  2022.

\bibitem{3}
------, ``{Entanglement assisted radars with transmitter side optical phase
  conjugation and classical coherent detection},'' \emph{IEEE Access}, vol.~10,
  pp. 49\,095--49\,100, 2022.

\bibitem{4}
------, ``{Entanglement-Assisted Joint Monostatic-Bistatic Radars},''
  \emph{Entropy}, vol.~24, no.~6, 2022.

\bibitem{5}
------, ``{On Entanglement-Assisted Multistatic Radar Techniques},''
  \emph{Entropy}, vol.~24, no.~7, 2022.

\bibitem{6}
S.~Lloyd, ``{Enhanced Sensitivity of Photodetection via Quantum
  Illumination},'' \emph{Science}, vol. 321, no. 5895, pp. 1463--1465, 2008.

\bibitem{7}
J.~H. Shapiro, ``{The Quantum Illumination Story},'' \emph{IEEE Aerospace and
  Electronic Systems Magazine}, vol.~35, p. 8–20, 2020.

\bibitem{8}
N.~B. B.-W. R.~G.~Torromé and P.~Knott, ``{Introduction to quantum radar},''
  \emph{arXiv preprint arXiv:2006.14238v3}, 2020.

\bibitem{9}
A.~Karsa, G.~Spedalieri, Q.~Zhuang, and S.~Pirandola, ``{Quantum illumination
  with a generic Gaussian source},'' \emph{Phys. Rev. Research}, vol.~2, no.~2,
  2020.

\bibitem{10}
S.~Barzanjeh and et~al., ``{Microwave quantum illumination},'' \emph{Phys. Rev.
  Lett.}, vol. 114, 2015.

\bibitem{11}
S.~Barzanjeh, S.~Pirandola, D.~Vitali, and J.~M. Fink, ``{Microwave quantum
  illumination using a digital receiver},'' \emph{Sci. Adv.}, vol.~6, 2020.

\bibitem{12}
G.~Sorelli, N.~Treps, F.~Grosshans, and F.~Boust, ``{Detecting a target with
  quantum entanglement},'' \emph{arXiv preprint arXiv:2005.07116}, 2021.

\bibitem{13}
C.~Noh, C.~Lee, and S.-Y. Lee, ``{Quantum illumination with definite
  photon-number entangled states},'' \emph{J. Opt. Soc. Am. B}, vol.~39, pp.
  1316--1322, 2022.

\bibitem{14}
H.~Corporation, ``{Quantum Sensors Program},'' \emph{Final Technical Report,
  AFRL-RI-RS-TR-2009-208}, 2009.

\bibitem{15}
P.~Kumar, V.~Grigoryan, and M.~Vasilyev, ``{Noise-Free Amplification: Towards
  Quantum Laser Radar},'' in \emph{Proc. 14th Coherent Laser Radar Conference},
  2007.

\bibitem{16}
Z.~Dutton, J.~H. Shapiro, and S.~Guha, ``{LADAR Resolution Improvement using
  Receivers Enhanced with Squeezed-Vacuum Injection and Phase-Sensitive
  Amplification},'' \emph{J. Opt. Soc. Am. B}, vol.~27, 2010.

\bibitem{17}
J.~Shapiro, ``{Quantum Pulse Compression Laser Radar},'' in \emph{Proc. SPIE
  6603, Noise and Fluctuations in Photonics, Quantum Optics, and
  Communications}, 2007.

\bibitem{18}
M.~Lanzagorta, ``{Quantum Radar Cross Sections},'' in \emph{Proc. Quantum
  Optics Conference, SPIE Photonics Europe}, 2010.

\bibitem{19}
L.~C. Andrews and R.~L. Philips, \emph{{Laser Beam Propagation through Random
  Media}}.\hskip 1em plus 0.5em minus 0.4em\relax SPIE Press, 2005.

\bibitem{20}
I.~B. Djordjevic, \emph{{Advanced Optical and Wireless Communications Systems,
  2nd Ed.}}\hskip 1em plus 0.5em minus 0.4em\relax Springer Nature Switzerland,
  2022.

\bibitem{21}
J.~F. Smith, ``{Quantum entangled radar theory and a correction method for the
  effects of the atmosphere on entanglement},'' in \emph{Proc. SPIE Defense,
  Security, and Sensing, SPIE Quantum Information and Computation VII
  conference}, vol. 7342, 2009.

\bibitem{22}
R.~K. Tyson, \emph{{Principles of Adaptive Optics}}.\hskip 1em plus 0.5em minus
  0.4em\relax CRC Press, Boca Raton, FL, 2015.

\bibitem{23}
V.~Nafria and I.~B. Djordjevic, ``{Entanglement Assisted Communication over the
  Free-Space Optical Link with Azimuthal Phase Correction for Atmospheric
  Turbulence by Adaptive Optics},'' \emph{Optics Express}, vol.~31, no.~24, pp.
  39\,906--39\,916, 2023.

\bibitem{24}
I.~B. Djordjevic and V.~Nafria, ``{Entanglement Assisted Quantum Radar
  Demonstration over Turbulent Free-Space Optical Channels},'' in \emph{Asia
  Communications and Photonics Conference (ACP) / The International Photonics
  and OptoElectronics Meetings (POEM) (ACPPOEM) 2023}, 2023.

\bibitem{Tan2008}
S.-H. Tan and et~al., ``{Quantum illumination with Gaussian states},''
  \emph{Phys. Rev. Lett.}, vol. 101, 2008.

\bibitem{26}
K.~M.~R. Audenaert and et~al., ``{Discriminating states: the quantum Chernoff
  bound},'' \emph{Phys. Rev. Lett.}, vol.~98, 2007.

\bibitem{27}
Z.~Zhang and et~al., ``{Entanglement-Enhanced Sensing in a Lossy and Noisy
  Environment},'' \emph{Phys. Rev. Lett.}, vol. 114, no.~11, 2015.

\bibitem{28}
Q.~Zhuang, Z.~Zhang, and J.~H. Shapiro, ``{Optimum Mixed-State Discrimination
  for Noisy Entanglement-Enhanced Sensing},'' \emph{Phys. Rev. Lett.}, vol.
  118, no.~4, 2017.

\bibitem{32}
I.~B. Djordjevic, ``{Entanglement assisted MIMO quantum radars},'' in
  \emph{Proc. 23rd International Conference on Transparent Optical Networks
  ICTON 2023}, 2023.

\bibitem{29}
L.~Fan and et~al., ``{Superconducting cavity electro-optics: A platform for
  coherent photon conversion between superconducting and photonic circuits},''
  \emph{Sci. Adv.}, vol.~4, 2018.

\bibitem{30}
X.~Han and et~al., ``{Microwave-optical quantum frequency conversion},''
  \emph{Optica}, vol.~8, p. 1050–1064, 2021.

\bibitem{31}
R.~Assouly and et~al., ``{Quantum advantage in microwave quantum radar},''
  \emph{Nature Physics}, vol.~19, p. 1418–1422, 2023.

\bibitem{40}
A.~I. Lvovsky, B.~C. Sanders, and W.~Tittel, ``{Optical quantum memory},''
  \emph{Nature Photonics}, vol.~3, no.~12, p. 706–714, 2009.

\bibitem{41}
D.~D. Sukachev, A.~Sipahigil, C.~T. Nguyen, M.~K. Bhaskar, R.~E. Evans,
  F.~Jelezko, and M.~D. Lukin, ``{Silicon vacancy spin qubit in diamond: A
  quantum memory exceeding 10 ms with single-shot state readout},'' \emph{Phys.
  Rev. Lett.}, vol. 119, 2017.

\bibitem{42}
S.~Sun, H.~Kim, Z.~Luo, G.~S. Solomon, and E.Waks, ``{A single-photon switch
  and transistor enabled by a solid-state quantum memory},'' \emph{Science},
  vol. 361, p. 57–60, 2018.

\bibitem{43}
Y.~Wang, J.~Li, S.~Zhang, K.~Su, Y.~Zhou, K.~Liao, S.~Du, H.~Yan, and S.-L.
  Zhu, ``{Efficient quantum memory for single-photon polarization qubits},''
  \emph{Nat. Photonics}, vol.~13, p. 346–351, 2019.

\bibitem{44}
A.Wallucks, I.~Marinkovic, B.~Hensen, R.~Stockill, and S.~Groeblacher, ``{A
  quantum memory at telecom wavelengths},'' \emph{Nat. Phys.}, vol.~16, p.
  772–777, 2020.

\bibitem{45}
X.~Liu, J.~Hu, Z.-F. Li, X.~Li, P.-Y. Li, P.-J. Liang, Z.-Q. Zhou, C.-F. Li,
  and G.-C. Guo, ``{Heralded entanglement distribution between two absorptive
  quantum memories},'' \emph{Nature}, vol. 594, p. 41–45, 2021.

\bibitem{46}
P.-J. Stas and et~al., ``{Robust multi-qubit quantum network node with
  integrated error detection},'' \emph{Science}, vol. 378, pp. 557--560, 2022.

\bibitem{47}
J.-M. Mol and et~al., ``{Quantum memories for fundamental science in space},''
  \emph{Quantum Sci. Technol.}, vol.~8, 2023.

\bibitem{48}
D.~V. Reddy and et~al., ``{Superconducting nanowire single-photon detectors
  with 98\% system detection efficiency at 1550 nm},'' \emph{Optica}, vol.~7,
  p. 1649–1653, 2020.

\bibitem{49}
G.-Z. Xu and et~al., ``{Superconducting microstrip single-photon detector with
  system detection efficiency over 90\% at 1550 nm},'' \emph{Photon. Res.},
  vol.~9, pp. 958--967, 2021.

\bibitem{50}
Y.~Pan and et~al., ``{Mid-infrared Nb4N3-based superconducting nanowire single
  photon detectors for wavelengths up to 10 µm},'' \emph{Opt. Express},
  vol.~30, pp. 40\,044--40\,052, 2022.

\bibitem{51}
M.~Eaton and et~al., ``{Resolution of 100 photons and quantum generation of
  unbiased random numbers},'' \emph{Nat. Photonics}, vol.~17, p. 106–111,
  2023.

\bibitem{52}
R.~Cheng and et~al., ``{A 100-pixel photon-number-resolving detector unveiling
  photon statistics},'' \emph{Nat. Photonics}, vol.~17, p. 112–119, 2023.

\bibitem{53}
M.~Forsch and et~al., ``{Microwave-to-optics conversion using a mechanical
  oscillator in its quantum ground state},'' \emph{Nat. Phys.}, vol.~16, p.
  69–74, 2020.

\bibitem{54}
W.~Jiang and et~al., ``{Efficient bidirectional piezo-optomechanical
  transduction between microwave and optical frequency},'' \emph{Nat. Commun.},
  vol.~11, 2020.

\bibitem{55}
K.~Tsujino and et~al., ``{Quantum receiver beyond the standard quantum limit of
  coherent optical communication},'' \emph{Phys. Rev. Lett.}, vol. 106, 2011.

\bibitem{56}
S.~Izumi and et~al., ``{Experimental demonstration of a quantum receiver
  beating the standard quantum limit at telecom wavelength},'' \emph{Phys. Rev.
  Appl.}, vol.~13, 2020.

\bibitem{57}
C.~Reimer and et~al., ``{Generation of multiphoton entangled quantum states by
  means of integrated frequency combs},'' \emph{Science}, vol. 351, p.
  1176–1180, 2016.

\bibitem{58}
M.~Kues and et~al., ``{Quantum optical microcombs},'' \emph{Nat. Photonics},
  vol.~13, p. 170–179, 2019.

\bibitem{59}
Z.~Yang and et~al., ``{A squeezed quantum microcomb on a chip},'' \emph{Nat.
  Commun.}, vol.~12, 2021.

\bibitem{38}
N.~Lauk and et~al., ``{Perspectives on quantum transduction},'' \emph{Quantum
  Sci. Technol.}, vol.~5, 2020.

\bibitem{39}
D.~Awschalom and et~al., ``{Development of quantum interconnects (quics) for
  next-generation information technologies},'' \emph{PRX Quantum}, vol.~2,
  2021.

\bibitem{35}
S.-R. Zhao and et~al., ``{Field demonstration of distributed quantum sensing
  without post-selection},'' \emph{Phys. Rev. X}, vol.~11, 2021.

\bibitem{36}
I.~B. Djordjevic, ``{On entanglement assisted classical optical communication
  with transmitter side optical phase-conjugation},'' \emph{IEEE Access},
  vol.~9, p. 168930 – 168936, 2021.

\bibitem{37}
S.~Hao and et~al., ``{Demonstration of entanglement-enhanced covert sensing},''
  \emph{Phys. Rev. Lett.}, vol. 129, 2022.

\bibitem{60}
G.~G. Taylor and et~al., ``{2.3 $\mu$m wavelength single photon LIDAR with
  superconducting nanowire detectors},'' in \emph{Conference on Lasers and
  Electro-Optics, OSA Technical Digest (Optica Publishing Group)}, 2019.

\bibitem{61}
T.~Staffas and et~al., ``{3D scanning quantum LIDAR},'' in \emph{Conference on
  Lasers and Electro-Optics, Technical Digest Series (Optica Publishing
  Group)}, 2022.

\bibitem{62}
M.~Reichert and et~al., ``{Quantum Illumination with a Hetero-Homodyne Receiver
  and Sequential Detection},'' \emph{Phys. Rev. Applied}, vol.~20, 2023.

\bibitem{63}
F.~Kronowetter and et~al., ``{Quantum Microwave Parametric Interferometer},''
  \emph{Phys. Rev. Applied}, vol.~20, 2023.

\bibitem{64}
S.-Y. Lee and et~al., ``{Bound for Gaussian-state quantum illumination using a
  direct photon measurement},'' \emph{Opt. Express}, vol.~31, pp.
  38\,977--38\,988, 2023.

\bibitem{66}
Z.~Xue and et~al., ``{Photonics-assisted joint radar and communication system
  based on an optoelectronic oscillator},'' \emph{Opt. Express}, vol.~29, pp.
  22\,442--22\,454, 2021.

\bibitem{67}
T.~Huang and et~al., ``{A Dual-Function Radar Communication System Using Index
  Modulation},'' in \emph{Signal Processing Advances in Wireless Communications
  (SPAWC), Cannes, France}, 2019, pp. 1--5.

\bibitem{68}
F.~Liu and et~al., ``{Joint Radar and Communication Design: Applications,
  State-of-the-Art, and the Road Ahead},'' \emph{IEEE Transactions on
  Communications}, vol.~68, no.~6, pp. 3834--3862, 2020.

\bibitem{69}
J.~Wang and et~al., ``{Integrated Sensing and Communication: Enabling
  Techniques, Applications, Tools and Data Sets, Standardization, and Future
  Directions},'' \emph{IEEE Internet of Things Journal}, vol.~9, no.~23, pp.
  23\,416--23\,440, 2022.

\bibitem{65}
Q.~Zhuang and J.~H. Shapiro, ``{Ultimate accuracy limit of quantum
  pulse-compression ranging},'' \emph{Phys. Rev. Lett.}, vol. 128, 2022.

\bibitem{long2022evolutionary}
G.-L. Long, D.~Pan, Y.-B. Sheng, Q.~Xue, J.~Lu, and L.~Hanzo, ``An evolutionary
  pathway for the quantum internet relying on secure classical repeaters,''
  \emph{IEEE Netw.}, vol.~36, no.~3, pp. 82--88, Jul. 2022.

\bibitem{Pirandola:AQCrypt}
\BIBentryALTinterwordspacing
S.~Pirandola, U.~L. Andersen, L.~Banchi, M.~Berta, D.~Bunandar, R.~Colbeck,
  D.~Englund, T.~Gehring, C.~Lupo, C.~Ottaviani, J.~L. Pereira, M.~Razavi,
  J.~S. Shaari, M.~Tomamichel, V.~C. Usenko, G.~Vallone, P.~Villoresi, and
  P.~Wallden, ``Advances in quantum cryptography,'' \emph{Adv. Opt. Photon.},
  vol.~12, no.~4, pp. 1012--1236, Dec 2020. [Online]. Available:
  \url{http://aop.osa.org/abstract.cfm?URI=aop-12-4-1012}
\BIBentrySTDinterwordspacing

\bibitem{RazaviQCbook}
\BIBentryALTinterwordspacing
M.~Razavi, \emph{An Introduction to Quantum Communications Networks}, ser.
  2053-2571.\hskip 1em plus 0.5em minus 0.4em\relax Morgan \& Claypool
  Publishers, 2018. [Online]. Available:
  \url{https://dx.doi.org/10.1088/978-1-6817-4653-1}
\BIBentrySTDinterwordspacing

\bibitem{Renner_ComposSec}
R.~Renner and R.~K{\"o}nig, ``Universally composable privacy amplification
  against quantum adversaries,'' in \emph{Theory of Cryptography}, J.~Kilian,
  Ed.\hskip 1em plus 0.5em minus 0.4em\relax Berlin, Heidelberg: Springer
  Berlin Heidelberg, 2005, pp. 407--425.

\bibitem{bennett1984brassard}
C.~H. Bennett and G.~Brassard, ``Quantum cryptography: public-key distribution
  and coin tossing,'' in \emph{Proceedings of IEEE International Conference on
  Computers Systems and Signal Processing}, 1984, pp. 175--179.

\bibitem{Wiesner}
\BIBentryALTinterwordspacing
S.~Wiesner, ``Conjugate coding,'' \emph{SIGACT News}, vol.~15, no.~1, p.
  78–88, jan 1983. [Online]. Available:
  \url{https://doi.org/10.1145/1008908.1008920}
\BIBentrySTDinterwordspacing

\bibitem{ekert1991quantum}
A.~K. Ekert, ``Quantum cryptography based on bell’s theorem,'' \emph{Physical
  review letters}, vol.~67, no.~6, p. 661, 1991.

\bibitem{BBM92}
C.~H. Bennett, G.~Brassard, and N.~D. Mermin, ``Quantum cryptography without
  {B}ell's theorem,'' \emph{Phys. Rev. Lett.}, vol.~68, p. 557, 1992.

\bibitem{tabletopQKD}
\BIBentryALTinterwordspacing
C.~H. Bennett and G.~Brassard, ``Experimental quantum cryptography: The dawn of
  a new era for quantum cryptography: The experimental prototype is working],''
  \emph{SIGACT News}, vol.~20, no.~4, p. 78–80, nov 1989. [Online].
  Available: \url{https://doi.org/10.1145/74074.74087}
\BIBentrySTDinterwordspacing

\bibitem{briegel1998}
H.-J. Briegel, W.~D{\"u}r, J.~I. Cirac, and P.~Zoller, ``Quantum repeaters: the
  role of imperfect local operations in quantum communication,'' \emph{Physical
  Review Letters}, vol.~81, no.~26, p. 5932, 1998.

\bibitem{LoChauSecProof}
\BIBentryALTinterwordspacing
H.-K. Lo and H.~F. Chau, ``Unconditional security of quantum key distribution
  over arbitrarily long distances,'' \emph{Science}, vol. 283, no. 5410, pp.
  2050--2056, 1999. [Online]. Available:
  \url{https://www.science.org/doi/abs/10.1126/science.283.5410.2050}
\BIBentrySTDinterwordspacing

\bibitem{ShorPreskill_00}
P.~W. Shor and J.~Preskill, ``Simple proof of security of the {BB84} quantum
  key distribution protocol,'' \emph{Phys.~Rev.~Lett.~}, vol.~85, no.~2, p.
  441, July 2000.

\bibitem{duan2001}
L.-M. Duan, M.~Lukin, J.~I. Cirac, and P.~Zoller, ``Long-distance quantum
  communication with atomic ensembles and linear optics,'' \emph{Nature}, vol.
  414, no. 6862, p. 413, 2001.

\bibitem{gottesman2004security}
D.~Gottesman, H.-K. Lo, N.~Lutkenhaus, and J.~Preskill, ``Security of quantum
  key distribution with imperfect devices,'' in \emph{International Symposium
  on Information Theory, 2004. ISIT 2004. Proceedings.}\hskip 1em plus 0.5em
  minus 0.4em\relax IEEE, 2004, p. 136.

\bibitem{Hwang:Decoy:2003}
W.-Y. Hwang, ``Quantum key distribution with high loss: Toward global secure
  communication,'' \emph{Phys.~Rev.~Lett.~}, vol.~91, p. 057901, August 2003.

\bibitem{Wang:Decoy:2005}
X.-B. Wang, ``Beating the $pns$ attack in practical quantum cryptography,''
  \emph{Phys.~Rev.~Lett.~}, vol.~94, p. 230503, 2005.

\bibitem{Wang:Decoy2:2005}
------, ``A decoy-state protocol for quantum cryptography with 4 intensities of
  coherent states,'' \emph{Phys.~Rev.~A}, vol.~72, p. 012322, 2005.

\bibitem{ma2005practical}
X.~Ma, B.~Qi, Y.~Zhao, and H.-K. Lo, ``Practical decoy state for quantum key
  distribution,'' \emph{Physical Review A}, vol.~72, no.~1, p. 012326, 2005.

\bibitem{Renner_Thesis_05}
R.~Renner, ``Security of quantum key distribution,'' Ph.D. dissertation, Swiss
  Federal Institute of Technology, 2005, also available in Int. J. Quant. Inf.
  {\bf 6}, 1 (2008).

\bibitem{HP_HandheldQKD}
J.~L. Duligall, M.~S. Godfrey, K.~A. Harrison, W.~J. Munro, and J.~G. Rarity,
  ``Low cost and compact quantum key distribution,'' \emph{New Journal of
  Physics}, vol.~8, no.~10, p. 249, 2006.

\bibitem{chun2017handheld}
H.~Chun, I.~Choi, G.~Faulkner, L.~Clarke, B.~Barber, G.~George, C.~Capon,
  A.~Niskanen, J.~Wabnig, D.~O'Brien, and D.~Bitauld, ``Handheld free space
  quantum key distribution with dynamic motion compensation,'' \emph{Optics
  Express}, vol.~25, no.~6, pp. 6784--6795, 2017.

\bibitem{Telcordia_1550_1550}
N.~A. Peters, P.~Toliver, T.~E. Chapuran, R.~J. Runser, S.~R. McNown, C.~G.
  Peterson, D.~Rosenberg, N.~Dallmann, R.~J. Hughes, K.~P. McCabe, J.~E.
  Nordholt, and K.~T. Tyagi, ``Dense wavelength multiplexing of 1550nm {QKD}
  with strong classical channels in reconfigurable networking environments,''
  \emph{New J. Phys.}, vol.~11, p. 045012, April 2009.

\bibitem{lo2012measurement}
H.-K. Lo, M.~Curty, and B.~Qi, ``Measurement-device-independent quantum key
  distribution,'' \emph{Physical review letters}, vol. 108, no.~13, p. 130503,
  2012.

\bibitem{Braunstein:MIQKD:2012}
S.~L. Braunstein and S.~Pirandola, ``Side-channel-free quantum key
  distribution,'' \emph{Phys. Rev. Lett.}, vol. 108, p. 130502, Mar 2012.

\bibitem{panayi2014memory}
C.~Panayi, M.~Razavi, X.~Ma, and N.~L{\"u}tkenhaus, ``Memory-assisted
  measurement-device-independent quantum key distribution,'' \emph{New Journal
  of Physics}, vol.~16, no.~4, p. 043005, 2014.

\bibitem{MAQKD-Bruss}
\BIBentryALTinterwordspacing
S.~Abruzzo, H.~Kampermann, and D.~Bru\ss{}, ``Measurement-device-independent
  quantum key distribution with quantum memories,'' \emph{Phys. Rev. A},
  vol.~89, p. 012301, Jan 2014. [Online]. Available:
  \url{https://link.aps.org/doi/10.1103/PhysRevA.89.012301}
\BIBentrySTDinterwordspacing

\bibitem{Chen2021}
\BIBentryALTinterwordspacing
Y.-A. Chen, Q.~Zhang, T.-Y. Chen, W.-Q. Cai, S.-K. Liao, J.~Zhang, K.~Chen,
  J.~Yin, J.-G. Ren, Z.~Chen, S.-L. Han, Q.~Yu, K.~Liang, F.~Zhou, X.~Yuan,
  M.-S. Zhao, T.-Y. Wang, X.~Jiang, L.~Zhang, W.-Y. Liu, Y.~Li, Q.~Shen,
  Y.~Cao, C.-Y. Lu, R.~Shu, J.-Y. Wang, L.~Li, N.-L. Liu, F.~Xu, X.-B. Wang,
  C.-Z. Peng, and J.-W. Pan, ``An integrated space-to-ground quantum
  communication network over 4,600 kilometres,'' \emph{Nature}, vol. 589, no.
  7841, pp. 214--219, Jan 2021. [Online]. Available:
  \url{https://doi.org/10.1038/s41586-020-03093-8}
\BIBentrySTDinterwordspacing

\bibitem{ChipBasedQKD}
\BIBentryALTinterwordspacing
P.~Sibson, C.~Erven, M.~Godfrey, S.~Miki, T.~Yamashita, M.~Fujiwara, M.~Sasaki,
  H.~Terai, M.~G. Tanner, C.~M. Natarajan, R.~H. Hadfield, J.~L. O'Brien, and
  M.~G. Thompson, ``Chip-based quantum key distribution,'' \emph{Nature
  Communications}, vol.~8, no.~1, p. 13984, Feb 2017. [Online]. Available:
  \url{https://doi.org/10.1038/ncomms13984}
\BIBentrySTDinterwordspacing

\bibitem{ChipBasedQKDRev}
\BIBentryALTinterwordspacing
L.-C. Kwek, L.~Cao, W.~Luo, Y.~Wang, S.~Sun, X.~Wang, and A.~Q. Liu,
  ``Chip-based quantum key distribution,'' \emph{AAPPS Bulletin}, vol.~31,
  no.~1, p.~15, Jun 2021. [Online]. Available:
  \url{https://doi.org/10.1007/s43673-021-00017-0}
\BIBentrySTDinterwordspacing

\bibitem{Liao_Nat_2017}
\BIBentryALTinterwordspacing
S.-K. Liao, W.-Q. Cai, W.-Y. Liu, L.~Zhang, Y.~Li, J.-G. Ren, J.~Yin, Q.~Shen,
  Y.~Cao, Z.-P. Li, F.-Z. Li, X.-W. Chen, L.-H. Sun, J.-J. Jia, J.-C. Wu, X.-J.
  Jiang, J.-F. Wang, Y.-M. Huang, Q.~Wang, Y.-L. Zhou, L.~Deng, T.~Xi, L.~Ma,
  T.~Hu, Q.~Zhang, Y.-A. Chen, N.-L. Liu, X.-B. Wang, Z.-C. Zhu, C.-Y. Lu,
  R.~Shu, C.-Z. Peng, J.-Y. Wang, and J.-W. Pan, ``Satellite-to-ground quantum
  key distribution,'' \emph{Nature}, vol. 549, p.~43, 2017. [Online].
  Available: \url{https://doi.org/10.1038/nature23655}
\BIBentrySTDinterwordspacing

\bibitem{Liao:ChinaAustria_PRL2018}
\BIBentryALTinterwordspacing
S.-K. Liao, W.-Q. Cai, J.~Handsteiner, B.~Liu, J.~Yin, L.~Zhang, D.~Rauch,
  M.~Fink, J.-G. Ren, W.-Y. Liu, Y.~Li, Q.~Shen, Y.~Cao, F.-Z. Li, J.-F. Wang,
  Y.-M. Huang, L.~Deng, T.~Xi, L.~Ma, T.~Hu, L.~Li, N.-L. Liu, F.~Koidl,
  P.~Wang, Y.-A. Chen, X.-B. Wang, M.~Steindorfer, G.~Kirchner, C.-Y. Lu,
  R.~Shu, R.~Ursin, T.~Scheidl, C.-Z. Peng, J.-Y. Wang, A.~Zeilinger, and J.-W.
  Pan, ``{Satellite-Relayed Intercontinental Quantum Network},'' \emph{Phys.
  Rev. Lett.}, vol. 120, p. 030501, Jan 2018. [Online]. Available:
  \url{https://link.aps.org/doi/10.1103/PhysRevLett.120.030501}
\BIBentrySTDinterwordspacing

\bibitem{Lucamarini2018}
\BIBentryALTinterwordspacing
M.~Lucamarini, Z.~L. Yuan, J.~F. Dynes, and A.~J. Shields, ``Overcoming the
  rate--distance limit of quantum key distribution without quantum repeaters,''
  \emph{Nature}, vol. 557, no. 7705, pp. 400--403, May 2018. [Online].
  Available: \url{https://doi.org/10.1038/s41586-018-0066-6}
\BIBentrySTDinterwordspacing

\bibitem{ModePairingQKD}
\BIBentryALTinterwordspacing
P.~Zeng, H.~Zhou, W.~Wu, and X.~Ma, ``Mode-pairing quantum key distribution,''
  \emph{Nature Communications}, vol.~13, no.~1, p. 3903, Jul 2022. [Online].
  Available: \url{https://doi.org/10.1038/s41467-022-31534-7}
\BIBentrySTDinterwordspacing

\bibitem{MPQKDdemo}
\BIBentryALTinterwordspacing
L.~Zhou, J.~Lin, Y.-M. Xie, Y.-S. Lu, Y.~Jing, H.-L. Yin, and Z.~Yuan,
  ``Experimental quantum communication overcomes the rate-loss limit without
  global phase tracking,'' \emph{Phys. Rev. Lett.}, vol. 130, p. 250801, Jun
  2023. [Online]. Available:
  \url{https://link.aps.org/doi/10.1103/PhysRevLett.130.250801}
\BIBentrySTDinterwordspacing

\bibitem{GLLP_04}
D.~Gottesman, H.-K. Lo, N.~L\"utkenhaus, and J.~Preskill, ``Security of quantum
  key distribution with imperfect devices,'' \emph{Quant.~Inf.~Comput.},
  vol.~4, p. 325, 2004.

\bibitem{Koashi:Comp:09}
M.~Koashi, ``Simple security proof of quantum key distribution based on
  complementarity,'' \emph{New Journal of Physics}, vol.~11, no.~4, p. 045018
  (12pp), 2009.

\bibitem{Wiechers:AftergateAttack:2011}
C.~Wiechers, L.~Lydersen, C.~Wittmann, D.~Elser, J.~Skaar, C.~Marquardt,
  V.~Makarov, and G.~Leuchs, ``After-gate attack on a quantum cryptosystem,''
  \emph{New Journal of Physics}, vol.~13, no.~1, p. 013043, 2011.

\bibitem{Weier:DeadtimeAttack:2011}
H.~Weier, H.~Krauss, M.~Rau, M.~F\"urst, S.~Nauerth, and H.~Weinfurter,
  ``Quantum eavesdropping without interception: an attack exploiting the dead
  time of single-photon detectors,'' \emph{New Journal of Physics}, vol.~13,
  no.~7, p. 073024, 2011.

\bibitem{send-or-not-send}
\BIBentryALTinterwordspacing
X.-B. Wang, Z.-W. Yu, and X.-L. Hu, ``Twin-field quantum key distribution with
  large misalignment error,'' \emph{Phys. Rev. A}, vol.~98, p. 062323, Dec
  2018. [Online]. Available:
  \url{https://link.aps.org/doi/10.1103/PhysRevA.98.062323}
\BIBentrySTDinterwordspacing

\bibitem{PhaseMatchingQKD}
\BIBentryALTinterwordspacing
X.~Ma, P.~Zeng, and H.~Zhou, ``Phase-matching quantum key distribution,''
  \emph{Phys. Rev. X}, vol.~8, p. 031043, Aug 2018. [Online]. Available:
  \url{https://link.aps.org/doi/10.1103/PhysRevX.8.031043}
\BIBentrySTDinterwordspacing

\bibitem{Trust-freeQKD}
N.~Lo~Piparo and M.~Razavi, ``{Long-Distance Trust-Free Quantum Key
  Distribution},'' \emph{IEEE Journal of Selected Topics in Quantum
  Electronics}, vol.~21, p. 6600508, 2015.

\bibitem{QKD1000km}
\BIBentryALTinterwordspacing
Y.~Liu, W.-J. Zhang, C.~Jiang, J.-P. Chen, C.~Zhang, W.-X. Pan, D.~Ma, H.~Dong,
  J.-M. Xiong, C.-J. Zhang, H.~Li, R.-C. Wang, J.~Wu, T.-Y. Chen, L.~You, X.-B.
  Wang, Q.~Zhang, and J.-W. Pan, ``Experimental twin-field quantum key
  distribution over 1000 km fiber distance,'' \emph{Phys. Rev. Lett.}, vol.
  130, p. 210801, May 2023. [Online]. Available:
  \url{https://link.aps.org/doi/10.1103/PhysRevLett.130.210801}
\BIBentrySTDinterwordspacing

\bibitem{ChipBasedQKDToshiba}
\BIBentryALTinterwordspacing
J.~A. Dolphin, T.~K. Para{\"i}so, H.~Du, R.~I. Woodward, D.~G. Marangon, and
  A.~J. Shields, ``A hybrid integrated quantum key distribution transceiver
  chip,'' \emph{npj Quantum Information}, vol.~9, no.~1, p.~84, Sep 2023.
  [Online]. Available: \url{https://doi.org/10.1038/s41534-023-00751-3}
\BIBentrySTDinterwordspacing

\bibitem{Telcordia_1550_1310}
T.~E. Chapuran, P.~Toliver, N.~A. Peters, J.~Jackel, M.~S. Goodman, R.~J.
  Runser, S.~R. McNown, N.~Dallmann, R.~J. Hughes, K.~P. McCabe, J.~E.
  Nordholt, C.~G. Peterson, K.~T. Tyagi, L.~Mercer, and H.~Dardy, ``Optical
  networking for quantum key distribution and quantum communications,''
  \emph{New J. Phys.}, vol.~11, p. 105001, Oct. 2009.

\bibitem{Townsend_QI_home_2011}
I.~Choi, R.~J. Young, and P.~D. Townsend, ``Quantum information to the home,''
  \emph{New J. Phys.}, vol.~13, p. 063039, June 2011.

\bibitem{Shields.PRX.coexist}
K.~A. {\rm Patel \em et al.}, ``Coexistence of high-bit-rate quantum key
  distribution and data on optical fiber,'' \emph{Phys. Rev. X}, vol.~2, p.
  041010, Nov. 2012.

\bibitem{patel2014quantum}
K.~Patel, J.~Dynes, M.~Lucamarini, I.~Choi, A.~Sharpe, Z.~Yuan, R.~Penty, and
  A.~Shields, ``Quantum key distribution for 10 {G}b/s dense wavelength
  division multiplexing networks,'' \emph{Applied Physics Letters}, vol. 104,
  no.~5, p. 051123, 2014.

\bibitem{kumar2015coexistence}
R.~Kumar, H.~Qin, and R.~All{\'e}aume, ``Coexistence of continuous variable
  {QKD} with intense {DWDM} classical channels,'' \emph{New Journal of
  Physics}, vol.~17, no.~4, p. 043027, 2015.

\bibitem{Coexist800Gbps}
\BIBentryALTinterwordspacing
M.~Pistoia, O.~Amer, M.~R. Behera, J.~A. Dolphin, J.~F. Dynes, B.~John, P.~A.
  Haigh, Y.~Kawakura, D.~H. Kramer, J.~Lyon, N.~Moazzami, T.~D. Movva,
  A.~Polychroniadou, S.~Shetty, G.~Sysak, F.~Toudeh-Fallah, S.~Upadhyay, R.~I.
  Woodward, and A.~J. Shields, ``Paving the way toward 800 gbps quantum-secured
  optical channel deployment in mission-critical environments,'' \emph{Quantum
  Science and Technology}, vol.~8, no.~3, p. 035015, may 2023. [Online].
  Available: \url{https://dx.doi.org/10.1088/2058-9565/acd1a8}
\BIBentrySTDinterwordspacing

\bibitem{crosstalk2016}
S.~Bahrani, M.~Razavi, and J.~A. Salehi, ``Crosstalk reduction in hybrid
  quantum-classical networks,'' \emph{Scientia Iranica D}, vol.~23, no.~6, pp.
  2898--2906, Dec. 2016.

\bibitem{Bahrani2018}
\BIBentryALTinterwordspacing
------, ``Wavelength assignment in hybrid quantum-classical networks,''
  \emph{Scientific Reports}, vol.~8, no.~1, p. 3456, Feb 2018. [Online].
  Available: \url{https://doi.org/10.1038/s41598-018-21418-6}
\BIBentrySTDinterwordspacing

\bibitem{Bahrani:19}
\BIBentryALTinterwordspacing
S.~Bahrani, O.~Elmabrok, G.~C. Lorenzo, and M.~Razavi, ``Wavelength assignment
  in quantum access networks with hybrid wireless-fiber links,'' \emph{J. Opt.
  Soc. Am. B}, vol.~36, no.~3, pp. B99--B108, Mar 2019. [Online]. Available:
  \url{https://opg.optica.org/josab/abstract.cfm?URI=josab-36-3-B99}
\BIBentrySTDinterwordspacing

\bibitem{QKDStack}
\BIBentryALTinterwordspacing
P.~K. Tysowski, X.~Ling, N.~Lütkenhaus, and M.~Mosca, ``The engineering of a
  scalable multi-site communications system utilizing quantum key distribution
  (qkd),'' \emph{Quantum Science and Technology}, vol.~3, no.~2, p. 024001, jan
  2018. [Online]. Available: \url{https://dx.doi.org/10.1088/2058-9565/aa9a5d}
\BIBentrySTDinterwordspacing

\bibitem{Hanzo_MultiProt}
Y.~Cao, Y.~Zhao, J.~Zhang, Q.~Wang, D.~Niyato, and L.~Hanzo, ``From
  single-protocol to large-scale multi-protocol quantum networks,'' \emph{IEEE
  Network}, vol.~36, no.~5, pp. 14--22, 2022.

\bibitem{SDNQKD}
A.~Aguado, V.~Lopez, D.~Lopez, M.~Peev, A.~Poppe, A.~Pastor, J.~Folgueira, and
  V.~Martin, ``The engineering of software-defined quantum key distribution
  networks,'' \emph{IEEE Communications Magazine}, vol.~57, no.~7, pp. 20--26,
  2019.

\bibitem{HandheldQKD-folllowup}
\BIBentryALTinterwordspacing
D.~Lowndes, S.~Frick, A.~Hart, and J.~Rarity, ``A low cost, short range quantum
  key distribution system,'' \emph{EPJ Quantum Technology}, vol.~8, no.~1,
  p.~15, May 2021. [Online]. Available:
  \url{https://doi.org/10.1140/epjqt/s40507-021-00101-2}
\BIBentrySTDinterwordspacing

\bibitem{Wireless_indoor_QKD}
O.~Elmabrok and M.~Razavi, ``Wireless quantum key distribution in indoor
  environments,'' \emph{JOSA B}, vol.~35, no.~2, pp. 197--207, 2018.

\bibitem{Quantum_Access_to_quantum_networks}
O.~Elmabrok, M.~Ghalaii, and M.~Razavi, ``Quantum-classical access networks
  with embedded optical wireless links,'' \emph{JOSA B}, vol.~35, no.~3, pp.
  487--499, 2018.

\bibitem{bahrani2018finite}
S.~Bahrani, O.~Elmabrok, G.~C. Lorenzo, and M.~Razavi, ``Finite-key effects in
  quantum access networks with wireless links,'' in \emph{2018 IEEE Globecom
  Workshops (GC Wkshps)}.\hskip 1em plus 0.5em minus 0.4em\relax IEEE, 2018,
  pp. 1--5.

\bibitem{bahrani2019wavelength}
------, ``Wavelength assignment in quantum access networks with hybrid
  wireless-fiber links,'' \emph{JOSA B}, vol.~36, no.~3, pp. B99--B108, 2019.

\bibitem{Liao_NatPhoton_2017}
\BIBentryALTinterwordspacing
S.-K. Liao, H.-L. Yong, C.~Liu, G.-L. Shentu, D.-D. Li, J.~Lin, H.~Dai, S.-Q.
  Zhao, B.~Li, J.-Y. Guan, W.~Chen, Y.-H. Gong, Y.~Li, Z.-H. Lin, G.-S. Pan,
  J.~S. Pelc, M.~M. Fejer, W.-Z. Zhang, W.-Y. Liu, J.~Yin, J.-G. Ren, X.-B.
  Wang, Q.~Zhang, C.-Z. Peng, and J.-W. Pan, ``Long-distance free-space quantum
  key distribution in daylight towards inter-satellite communication,''
  \emph{Nat. Photon.}, vol. 311, p. 509, 2017. [Online]. Available:
  \url{https://doi.org/10.1038/nphoton.2017.116}
\BIBentrySTDinterwordspacing

\bibitem{Ren_Nat_2017}
\BIBentryALTinterwordspacing
J.-G. Ren, P.~Xu, H.-L. Yong, L.~Zhang, S.-K. Liao, J.~Yin, W.-Y. Liu, W.-Q.
  Cai, M.~Yang, L.~Li, K.-X. Yang, X.~Han, Y.-Q. Yao, J.~Li, H.-Y. Wu, S.~Wan,
  L.~Liu, D.-Q. Liu, Y.-W. Kuang, Z.-P. He, P.~Shang, C.~Guo, R.-H. Zheng,
  K.~Tian, Z.-C. Zhu, N.-L. Liu, C.-Y. Lu, R.~Shu, Y.-A. Chen, C.-Z. Peng,
  J.-Y. Wang, and J.-W. Pan, ``Ground-to-satellite quantum teleportation,''
  \emph{Nature}, vol. 549, p.~70, 2017. [Online]. Available:
  \url{https://doi.org/10.1038/nature23675}
\BIBentrySTDinterwordspacing

\bibitem{liorni2020quantum}
C.~Liorni, H.~Kampermann, and D.~Bruss, ``Quantum repeaters in space,'' 2020.

\bibitem{Ling-SatQKD-Man}
\BIBentryALTinterwordspacing
T.~Vergoossen, R.~Bedington, J.~A. Grieve, and A.~Ling, ``Satellite quantum
  communications when man-in-the-middle attacks are excluded,'' 2019. [Online].
  Available: \url{https://www.mdpi.com/1099-4300/21/4/387}
\BIBentrySTDinterwordspacing

\bibitem{Saikat_RestricedEve_PRApplied}
\BIBentryALTinterwordspacing
Z.~Pan, K.~P. Seshadreesan, W.~Clark, M.~R. Adcock, I.~B. Djordjevic, J.~H.
  Shapiro, and S.~Guha, ``Secret-key distillation across a quantum wiretap
  channel under restricted eavesdropping,'' \emph{Phys. Rev. Applied}, vol.~14,
  p. 024044, Aug 2020. [Online]. Available:
  \url{https://link.aps.org/doi/10.1103/PhysRevApplied.14.024044}
\BIBentrySTDinterwordspacing

\bibitem{Hugo_RestricedEve_PRApplied}
\BIBentryALTinterwordspacing
A.~V\'azquez-Castro, D.~Rusca, and H.~Zbinden, ``Quantum keyless private
  communication versus quantum key distribution for space links,'' \emph{Phys.
  Rev. Applied}, vol.~16, p. 014006, Jul 2021. [Online]. Available:
  \url{https://link.aps.org/doi/10.1103/PhysRevApplied.16.014006}
\BIBentrySTDinterwordspacing

\bibitem{Razavi_restricted_PRX}
\BIBentryALTinterwordspacing
M.~Ghalaii, S.~Bahrani, C.~Liorni, F.~Grasselli, H.~Kampermann, L.~Wooltorton,
  R.~Kumar, S.~Pirandola, T.~P. Spiller, A.~Ling, B.~Huttner, and M.~Razavi,
  ``Satellite-based quantum key distribution in the presence of bypass
  channels,'' \emph{PRX Quantum}, vol.~4, p. 040320, Nov 2023. [Online].
  Available: \url{https://link.aps.org/doi/10.1103/PRXQuantum.4.040320}
\BIBentrySTDinterwordspacing

\bibitem{THzQKD1}
C.~Ottaviani, M.~J. Woolley, M.~Erementchouk, J.~F. Federici, P.~Mazumder,
  S.~Pirandola, and C.~Weedbrook, ``Terahertz quantum cryptography,''
  \emph{IEEE Journal on Selected Areas in Communications}, vol.~38, no.~3, pp.
  483--495, 2020.

\bibitem{Inter-SatTHzQKD}
Z.~Wang, R.~Malaney, and J.~Green, ``Inter-satellite quantum key distribution
  at terahertz frequencies,'' in \emph{ICC 2019 - 2019 IEEE International
  Conference on Communications (ICC)}, 2019, pp. 1--7.

\bibitem{MIMITHzQKD}
N.~K. Kundu, S.~P. Dash, M.~R. McKay, and R.~K. Mallik, ``Mimo terahertz
  quantum key distribution,'' \emph{IEEE Communications Letters}, vol.~25,
  no.~10, pp. 3345--3349, 2021.

\bibitem{MIMOQKD_Sahu}
S.~Sahu, A.~Lawey, and M.~Razavi, ``Continuous variable quantum key
  distribution in multiple-input multiple-output settings,'' 2023.

\bibitem{10094014}
N.~K. Kundu, M.~R. McKay, A.~Conti, R.~K. Mallik, and M.~Z. Win, ``Mimo
  terahertz quantum key distribution under restricted eavesdropping,''
  \emph{IEEE Transactions on Quantum Engineering}, vol.~4, pp. 1--15, 2023.

\bibitem{Pereira2019}
\BIBentryALTinterwordspacing
M.~Pereira, M.~Curty, and K.~Tamaki, ``Quantum key distribution with flawed and
  leaky sources,'' \emph{npj Quantum Information}, vol.~5, no.~1, p.~62, Jul
  2019. [Online]. Available: \url{https://doi.org/10.1038/s41534-019-0180-9}
\BIBentrySTDinterwordspacing

\bibitem{zhang2017improved}
Z.~Zhang, Q.~Zhao, M.~Razavi, and X.~Ma, ``Improved key-rate bounds for
  practical decoy-state quantum-key-distribution systems,'' \emph{Physical
  Review A}, vol.~95, no.~1, p. 012333, 2017.

\bibitem{NumericalFinteKey}
\BIBentryALTinterwordspacing
D.~Bunandar, L.~C.~G. Govia, H.~Krovi, and D.~Englund, ``Numerical finite-key
  analysis of quantum key distribution,'' \emph{npj Quantum Information},
  vol.~6, no.~1, p. 104, Dec 2020. [Online]. Available:
  \url{https://doi.org/10.1038/s41534-020-00322-w}
\BIBentrySTDinterwordspacing

\bibitem{TFQKD-finite-Guillermo}
\BIBentryALTinterwordspacing
G.~Curr{\'a}s-Lorenzo, {\'A}.~Navarrete, K.~Azuma, G.~Kato, M.~Curty, and
  M.~Razavi, ``Tight finite-key security for twin-field quantum key
  distribution,'' \emph{npj Quantum Information}, vol.~7, no.~1, p.~22, Feb
  2021. [Online]. Available: \url{https://doi.org/10.1038/s41534-020-00345-3}
\BIBentrySTDinterwordspacing

\bibitem{TrojanHorseFiniteKey}
\BIBentryALTinterwordspacing
Álvaro Navarrete and M.~Curty, ``Improved finite-key security analysis of
  quantum key distribution against trojan-horse attacks,'' \emph{Quantum
  Science and Technology}, vol.~7, no.~3, p. 035021, jun 2022. [Online].
  Available: \url{https://dx.doi.org/10.1088/2058-9565/ac74dc}
\BIBentrySTDinterwordspacing

\bibitem{Arnon-Friedman2018}
\BIBentryALTinterwordspacing
R.~Arnon-Friedman, F.~Dupuis, O.~Fawzi, R.~Renner, and T.~Vidick, ``Practical
  device-independent quantum cryptography via entropy accumulation,''
  \emph{Nature Communications}, vol.~9, no.~1, p. 459, Jan 2018. [Online].
  Available: \url{https://doi.org/10.1038/s41467-017-02307-4}
\BIBentrySTDinterwordspacing

\bibitem{Wehner_Roadmap}
\BIBentryALTinterwordspacing
S.~Wehner, D.~Elkouss, and R.~Hanson, ``Quantum internet: A vision for the road
  ahead,'' \emph{Science}, vol. 362, no. 6412, 2018. [Online]. Available:
  \url{https://science.sciencemag.org/content/362/6412/eaam9288}
\BIBentrySTDinterwordspacing

\bibitem{PRXQuantum_Roadmap}
\BIBentryALTinterwordspacing
D.~Awschalom, K.~K. Berggren, H.~Bernien, S.~Bhave, L.~D. Carr, P.~Davids,
  S.~E. Economou, D.~Englund, A.~Faraon, M.~Fejer, S.~Guha, M.~V. Gustafsson,
  E.~Hu, L.~Jiang, J.~Kim, B.~Korzh, P.~Kumar, P.~G. Kwiat,
  M.~Lon\ifmmode~\check{c}\else \v{c}\fi{}ar, M.~D. Lukin, D.~A. Miller,
  C.~Monroe, S.~W. Nam, P.~Narang, J.~S. Orcutt, M.~G. Raymer, A.~H.
  Safavi-Naeini, M.~Spiropulu, K.~Srinivasan, S.~Sun, J.~Vu\ifmmode
  \check{c}\else \v{c}\fi{}kovi\ifmmode~\acute{c}\else \'{c}\fi{}, E.~Waks,
  R.~Walsworth, A.~M. Weiner, and Z.~Zhang, ``Development of quantum
  interconnects (quics) for next-generation information technologies,''
  \emph{PRX Quantum}, vol.~2, p. 017002, Feb 2021. [Online]. Available:
  \url{https://link.aps.org/doi/10.1103/PRXQuantum.2.017002}
\BIBentrySTDinterwordspacing

\bibitem{deng2004secure}
F.-G. Deng and G.~L. Long, ``Secure direct communication with a quantum
  one-time pad,'' \emph{Phys. Rev. A}, vol.~69, no.~5, p. 052319, May 2004.

\bibitem{Sanders:TrustedNode}
T.~R. Beals and B.~C. Sanders, ``Distributed relay protocol for probabilistic
  information-theoretic security in a randomly-compromised network,'' in
  \emph{Information Theoretic Security}, R.~Safavi-Naini, Ed.\hskip 1em plus
  0.5em minus 0.4em\relax Berlin, Heidelberg: Springer Berlin Heidelberg, 2008,
  pp. 29--39.

\bibitem{Enzo_Chip_2019}
\BIBentryALTinterwordspacing
T.~K. Para{\"i}so, I.~De~Marco, T.~Roger, D.~G. Marangon, J.~F. Dynes,
  M.~Lucamarini, Z.~Yuan, and A.~J. Shields, ``A modulator-free quantum key
  distribution transmitter chip,'' \emph{npj Quantum Information}, vol.~5,
  no.~1, p.~42, May 2019. [Online]. Available:
  \url{https://doi.org/10.1038/s41534-019-0158-7}
\BIBentrySTDinterwordspacing

\bibitem{TFQKD-Lucamarini2018}
\BIBentryALTinterwordspacing
M.~Lucamarini, Z.~L. Yuan, J.~F. Dynes, and A.~J. Shields, ``Overcoming the
  rate--distance limit of quantum key distribution without quantum repeaters,''
  \emph{Nature}, vol. 557, no. 7705, pp. 400--403, May 2018. [Online].
  Available: \url{https://doi.org/10.1038/s41586-018-0066-6}
\BIBentrySTDinterwordspacing

\bibitem{PRL509km}
\BIBentryALTinterwordspacing
J.-P. Chen, C.~Zhang, Y.~Liu, C.~Jiang, W.~Zhang, X.-L. Hu, J.-Y. Guan, Z.-W.
  Yu, H.~Xu, J.~Lin, M.-J. Li, H.~Chen, H.~Li, L.~You, Z.~Wang, X.-B. Wang,
  Q.~Zhang, and J.-W. Pan, ``Sending-or-not-sending with independent lasers:
  Secure twin-field quantum key distribution over 509 km,'' \emph{Phys. Rev.
  Lett.}, vol. 124, p. 070501, Feb 2020. [Online]. Available:
  \url{https://link.aps.org/doi/10.1103/PhysRevLett.124.070501}
\BIBentrySTDinterwordspacing

\bibitem{Pittaluga2021}
\BIBentryALTinterwordspacing
M.~Pittaluga, M.~Minder, M.~Lucamarini, M.~Sanzaro, R.~I. Woodward, M.-J. Li,
  Z.~Yuan, and A.~J. Shields, ``600-km repeater-like quantum communications
  with dual-band stabilization,'' \emph{Nature Photonics}, vol.~15, no.~7, pp.
  530--535, Jul 2021. [Online]. Available:
  \url{https://doi.org/10.1038/s41566-021-00811-0}
\BIBentrySTDinterwordspacing

\bibitem{MAQKD-Harvard}
\BIBentryALTinterwordspacing
M.~K. Bhaskar, R.~Riedinger, B.~Machielse, D.~S. Levonian, C.~T. Nguyen, E.~N.
  Knall, H.~Park, D.~Englund, M.~Lon{\v{c}}ar, D.~D. Sukachev, and M.~D. Lukin,
  ``Experimental demonstration of memory-enhanced quantum communication,''
  \emph{Nature}, vol. 580, no. 7801, pp. 60--64, Apr 2020. [Online]. Available:
  \url{https://doi.org/10.1038/s41586-020-2103-5}
\BIBentrySTDinterwordspacing

\bibitem{MAQKD-Rempe}
\BIBentryALTinterwordspacing
S.~Langenfeld, P.~Thomas, O.~Morin, and G.~Rempe, ``Quantum repeater node
  demonstrating unconditionally secure key distribution,'' \emph{Phys. Rev.
  Lett.}, vol. 126, p. 230506, Jun 2021. [Online]. Available:
  \url{https://link.aps.org/doi/10.1103/PhysRevLett.126.230506}
\BIBentrySTDinterwordspacing

\bibitem{jiang2009quantum}
L.~Jiang, J.~M. Taylor, K.~Nemoto, W.~J. Munro, R.~Van~Meter, and M.~D. Lukin,
  ``Quantum repeater with encoding,'' \emph{Physical Review A}, vol.~79, no.~3,
  p. 032325, 2009.

\bibitem{munro2012quantum}
W.~J. Munro, A.~M. Stephens, S.~J. Devitt, K.~A. Harrison, and K.~Nemoto,
  ``Quantum communication without the necessity of quantum memories,''
  \emph{Nature Photonics}, vol.~6, no.~11, p. 777, 2012.

\bibitem{RazaviRepeaterCh}
\BIBentryALTinterwordspacing
M.~Razavi, \emph{Fiber-Based Quantum Repeaters}.\hskip 1em plus 0.5em minus
  0.4em\relax John Wiley \& Sons, Ltd, 2023, ch.~24, pp. 675--691. [Online].
  Available:
  \url{https://onlinelibrary.wiley.com/doi/abs/10.1002/9783527837427.ch24}
\BIBentrySTDinterwordspacing

\bibitem{dur1999quantum}
W.~D{\"u}r, H.-J. Briegel, J.~Cirac, and P.~Zoller, ``Quantum repeaters based
  on entanglement purification,'' \emph{Physical Review A}, vol.~59, no.~1, p.
  169, 1999.

\bibitem{yu2020entanglement}
Y.~Yu, F.~Ma, X.-Y. Luo, B.~Jing, P.-F. Sun, R.-Z. Fang, C.-W. Yang, H.~Liu,
  M.-Y. Zheng, X.-P. Xie \emph{et~al.}, ``Entanglement of two quantum memories
  via fibres over dozens of kilometres,'' \emph{Nature}, vol. 578, no. 7794,
  pp. 240--245, 2020.

\bibitem{Delft-Pompili259}
\BIBentryALTinterwordspacing
M.~Pompili, S.~L.~N. Hermans, S.~Baier, H.~K.~C. Beukers, P.~C. Humphreys,
  R.~N. Schouten, R.~F.~L. Vermeulen, M.~J. Tiggelman, L.~dos Santos~Martins,
  B.~Dirkse, S.~Wehner, and R.~Hanson, ``Realization of a multinode quantum
  network of remote solid-state qubits,'' \emph{Science}, vol. 372, no. 6539,
  pp. 259--264, 2021. [Online]. Available:
  \url{https://science.sciencemag.org/content/372/6539/259}
\BIBentrySTDinterwordspacing

\bibitem{jing2020quantum}
Y.~Jing, D.~Alsina, and M.~Razavi, ``Quantum key distribution over quantum
  repeaters with encoding: Using error detection as an effective postselection
  tool,'' \emph{Physical Review Applied}, vol.~14, no.~6, p. 064037, 2020.

\bibitem{jing2020simple}
Y.~Jing and M.~Razavi, ``Simple efficient decoders for quantum key distribution
  over quantum repeaters with encoding,'' \emph{Phys. Rev. Applied}, vol.~15,
  p. 044027, Apr 2021.

\bibitem{Ortu2018}
\BIBentryALTinterwordspacing
A.~Ortu, A.~Tiranov, S.~Welinski, F.~Fr{\"o}wis, N.~Gisin, A.~Ferrier,
  P.~Goldner, and M.~Afzelius, ``Simultaneous coherence enhancement of optical
  and microwave transitions in solid-state electronic spins,'' \emph{Nature
  Materials}, vol.~17, no.~8, pp. 671--675, Aug 2018. [Online]. Available:
  \url{https://doi.org/10.1038/s41563-018-0138-x}
\BIBentrySTDinterwordspacing

\bibitem{xiaolin-cst}
X.~Zhou, A.~Shen, S.~Hu, W.~Ni, X.~Wang, E.~Hossain, and L.~Hanzo, ``{Towards
  Quantum-Native Communication Systems: New Developments, Trends, and
  Challenges},'' \emph{https://arxiv.org/abs/2311.05239}, 2023.

\bibitem{kai-ten-acm}
L.-H. Shen, K.-T. Feng, and L.~Hanzo, ``{Five Facets of 6G: Research Challenges
  and Opportunities},'' \emph{ACM Computing Surveys}, vol.~55, no.~11, pp.
  1--39, Feb. 2023.

\bibitem{daniel-elsevier}
Z.~Xu, W.~Liu, Z.~Wang, and L.~Hanzo, ``Petahertz communication: Harmonizing
  optical spectra for wireless communications,'' \emph{Digital Communications
  and Networks}, vol.~7, no.~4, pp. 605--614, Nov. 2021.

\bibitem{8976167}
C.~Ottaviani, M.~J. Woolley, M.~Erementchouk, J.~F. Federici, P.~Mazumder,
  S.~Pirandola, and C.~Weedbrook, ``{T}erahertz quantum cryptography,''
  \emph{IEEE Journal on Selected Areas in Communications}, vol.~38, no.~3, pp.
  483--495, 2020.

\bibitem{e23091223}
C.~Liu, C.~Zhu, Z.~Li, M.~Nie, H.~Yang, and C.~Pei, ``Continuous-variable
  quantum secret sharing based on thermal {T}erahertz sources in
  inter-satellite wireless links,'' \emph{Entropy}, vol.~23, no.~9, 2021.

\bibitem{9674910}
K.~Senthoor and P.~K. Sarvepalli, ``Theory of communication efficient quantum
  secret sharing,'' \emph{IEEE Transactions on Information Theory}, vol.~68,
  no.~5, pp. 3164--3186, 2022.

\bibitem{9492803}
C.~Liu, C.~Zhu, X.~Liu, M.~Nie, H.~Yang, and C.~Pei, ``Multicarrier
  multiplexing continuous-variable quantum key distribution at {T}erahertz
  bands under indoor environment and in inter-satellite links communication,''
  \emph{IEEE Photonics Journal}, vol.~13, no.~4, pp. 1--13, 2021.

\bibitem{7393447}
I.~B. Djordjevic, ``Integrated optics modules based proposal for quantum
  information processing, teleportation, {QKD}, and quantum error correction
  employing photon angular momentum,'' \emph{IEEE Photonics Journal}, vol.~8,
  no.~1, pp. 1--12, 2016.

\bibitem{7735237}
X.~Cai, ``Photonic integrated devices for exploiting the orbital angular
  momentum of light in optical communications,'' in \emph{Progress in
  Electromagnetic Research Symposium (PIERS)}, 2016, pp. 3164--3164.

\bibitem{9397381}
Z.~Lin, X.~Pan, J.~Yao, Y.~Wu, Z.~Wang, D.~Zhang, C.~Ye, S.~Xu, F.~Yang, and
  X.~Wang, ``Characterization of orbital angular momentum applying
  single-sensor compressive imaging based on a microwave spatial wave
  modulator,'' \emph{IEEE Transactions on Antennas and Propagation}, vol.~69,
  no.~10, pp. 6870--6880, 2021.

\bibitem{nano10122436}
Z.~Wang, Q.~Tan, Y.~Liang, X.~Zhou, W.~Zhou, and X.~Huang, ``Active
  manipulation of the spin and orbital angular momentums in a {T}erahertz
  graphene-based hybrid plasmonic waveguide,'' \emph{Nanomaterials}, vol.~10,
  no.~12, p. 2436, 2020.

\bibitem{doi:10.1021/acs.nanolett.9b05207}
L.~Viti, D.~G. Purdie, A.~Lombardo, A.~C. Ferrari, and M.~S. Vitiello,
  ``{HBN}-encapsulated, graphene-based, room-temperature {T}erahertz receivers,
  with high speed and low noise,'' \emph{Nano Letters}, vol.~20, no.~5, pp.
  3169--3177, 2020.

\bibitem{9566893}
M.~Asgari, D.~Coquillat, G.~Menichetti, V.~Zannier, N.~Dyakonova, W.~Knap,
  L.~Sorba, L.~Viti, and M.~Serena~Vitiello, ``Highly sensitive photodetectors
  at 0.6 {TH}z based on quantum dot single electron transistors,'' in
  \emph{Proc. 46th International Conference on Infrared, Millimeter and
  {T}erahertz Waves (IRMMW-THz)}, 2021, pp. 1--2.

\bibitem{doi:10.1021/acs.nanolett.1c02022}
M.~Asgari, D.~Coquillat, G.~Menichetti, V.~Zannier, N.~Diakonova, W.~Knap,
  L.~Sorba, L.~Viti, and M.~S. Vitiello, ``Quantum-dot single-electron
  transistors as thermoelectric quantum detectors at {T}erahertz frequencies,''
  \emph{Nano Letters}, vol.~21, no.~20, pp. 8587--8594, 2021.

\bibitem{Liu_2019}
T.~Liu, Y.~Huang, Q.~Wei, K.~Liu, X.~Duan, and X.~Ren, ``Optimized
  uni-traveling carrier photodiode and mushroom-mesa structure for high-power
  and sub-{T}erahertz bandwidth under zero-and low-bias operation,''
  \emph{Journal of Physics Communications}, vol.~3, no.~9, p. 095004, 2019.

\bibitem{Yachmenev_2022}
A.~E. Yachmenev, R.~A. Khabibullin, and D.~S. Ponomarev, ``Recent advances in
  {TH}z detectors based on semiconductor structures with quantum confinement: A
  review,'' \emph{Journal of Physics D: Applied Physics}, vol.~55, no.~19, p.
  193001, 2022.

\bibitem{2022103}
Z.~Zhang, Z.~Fu, C.~Wang, and J.~Cao, ``Research on {T}erahertz quantum well
  photodetector,'' \emph{Journal of Infrared and Millimeter Waves}, vol.~41,
  no.~1, pp. 103--109, 2022.

\bibitem{shao2021research}
D.~Shao, Z.~Fu, Z.~Tan, C.~Wang, F.~Qiu, L.~Gu, W.~Wan, and J.~Cao, ``Research
  progress on terahertz quantum-well photodetector and its application,''
  \emph{Frontiers in Physics}, vol.~9, p. 581, 2021.

\bibitem{Thermoelectricgraphenephotodetectors}
L.~Viti, A.~R. Cadore, X.~Yang, A.~Vorobiev, J.~E. Muench, K.~Watanabe,
  T.~Taniguchi, J.~Stake, A.~C. Ferrari, and M.~S. Vitiello, ``Thermoelectric
  graphene photodetectors with sub-nanosecond response times at {T}erahertz
  frequencies,'' \emph{Nanophotonics}, vol.~10, no.~1, pp. 89--98, 2020.

\bibitem{5893900}
L.~Hanzo, M.~El-Hajjar, and O.~Alamri, ``Near-capacity wireless transceivers
  and cooperative communications in the mimo era: Evolution of standards,
  waveform design, and future perspectives,'' \emph{Proceedings of the IEEE},
  vol.~99, no.~8, pp. 1343--1385, Aug 2011.

\bibitem{5742779}
S.~Sugiura, S.~Chen, and L.~Hanzo, ``Mimo-aided near-capacity turbo
  transceivers: Taxonomy and performance versus complexity,'' \emph{IEEE
  Communications Surveys Tutorials}, vol.~14, no.~2, pp. 421--442, Second 2012.

\bibitem{9099546}
R.~Yuan and J.~Cheng, ``Free-space optical quantum communications in turbulent
  channels with receiver diversity,'' \emph{IEEE Transactions on
  Communications}, vol.~68, no.~9, pp. 5706--5717, 2020.

\bibitem{9739032}
N.~K. Kundu, S.~P. Dash, M.~R. McKay, and R.~K. Mallik, ``Channel estimation
  and secret key rate analysis of {MIMO} {T}erahertz quantum key
  distribution,'' \emph{IEEE Transactions on Communications}, vol.~70, no.~5,
  pp. 3350--3363, 2022.

\bibitem{2005Quantum}
M.~Gabay and S.~Arnon, ``Quantum key distribution by a free-space {MIMO}
  system,'' \emph{Journal of lightwave technology}, vol.~24, no.~8, p. 3114,
  2006.

\bibitem{zhou2019shot}
X.~Zhou, C.~Wei, D.~Shen, C.~Xu, L.~Wang, and X.~Yu, ``A shot noise limited
  quantum iterative massive {MIMO} system over {P}oisson atmospheric
  channels,'' in \emph{Proc. IEEE International Conference on Communications
  (ICC)}, 2019, pp. 1--6.

\bibitem{6253218}
M.~Razavi, ``Multiple-access quantum key distribution networks,'' \emph{IEEE
  Transactions on Communications}, vol.~60, no.~10, pp. 3071--3079, 2012.

\bibitem{rezai2021quantum}
M.~Rezai and J.~A. Salehi, ``Quantum {CDMA} communication systems,'' \emph{IEEE
  Transactions on Information Theory}, vol.~67, no.~8, pp. 5526--5547, 2021.

\bibitem{sharma2020quantum}
V.~Sharma and S.~Banerjee, ``Quantum communication using code division multiple
  access network,'' \emph{Optical and Quantum Electronics}, vol.~52, no.~8, pp.
  1--22, 2020.

\bibitem{6545781}
M.~Anandan, S.~Choudhary, and K.~Pradeep~Kumar, ``{OFDM} for frequency coded
  quantum key distribution,'' in \emph{Proc. International Conference on Fiber
  Optics and Photonics (PHOTONICS)}, 2012, pp. 1--3.

\bibitem{bahrani2015orthogonal}
S.~Bahrani, M.~Razavi, and J.~A. Salehi, ``Orthogonal frequency-division
  multiplexed quantum key distribution,'' \emph{Journal of Lightwave
  Technology}, vol.~33, no.~23, pp. 4687--4698, 2015.

\bibitem{Network_Coding_tutorial_Fraouli}
C.~Fragouli and E.~Soljanin, ``Network coding fundamentals,'' \emph{Foundation
  and Trends in Networking}, vol.~2, no.~1, pp. 1--133, 2007.

\bibitem{RN1640}
\BIBentryALTinterwordspacing
M.~Hayashi, K.~Wama, H.~Nishimura, R.~Raymond, and S.~Yamashita, \emph{Quantum
  network coding}, ser. Lecture Notes in Computer Science.\hskip 1em plus 0.5em
  minus 0.4em\relax Berlin: Springer-Verlag Berlin, 2007, vol. 4393, pp.
  610--621. [Online]. Available: \url{<Go to ISI>://WOS:000245503800052}
\BIBentrySTDinterwordspacing

\bibitem{Ieee_Transactions_on_Information_Theory_Leung_2010}
\BIBentryALTinterwordspacing
D.~Leung, J.~Oppenheim, and A.~Winter, ``Quantum network communication-the
  butterfly and beyond,'' \emph{IEEE Transactions on Information Theory},
  vol.~56, no.~7, pp. 3478--3490, 2010. [Online]. Available: \url{<Go to
  ISI>://WOS:000278812000035
  http://ieeexplore.ieee.org/ielx5/18/5484964/05485004.pdf?tp=&arnumber=5485004&isnumber=5484964}
\BIBentrySTDinterwordspacing

\bibitem{Journal_of_Superconductivity_Mahdian_2015}
M.~Mahdian and R.~Bayramzadeh, ``Perfect k-pair quantum network coding using
  superconducting qubits,'' \emph{Journal of Superconductivity and Novel
  Magnetism}, vol.~28, no.~2, pp. 345--348, 2015.

\bibitem{Com_Letter_Jing_2015}
\BIBentryALTinterwordspacing
L.~Jing, C.~Xiu-Bo, X.~Gang, Y.~Yi-Xian, and L.~Zong-Peng, ``Perfect quantum
  network coding independent of classical network solutions,''
  \emph{Communications Letters, IEEE}, vol.~19, no.~2, pp. 115--118, 2015.
  [Online]. Available:
  \url{http://ieeexplore.ieee.org/ielx7/4234/7033064/06980095.pdf?tp=&arnumber=6980095&isnumber=7033064}
\BIBentrySTDinterwordspacing

\bibitem{PhysRevA.93.032302}
\BIBentryALTinterwordspacing
T.~Satoh, K.~Ishizaki, S.~Nagayama, and R.~Van~Meter, ``Analysis of quantum
  network coding for realistic repeater networks,'' \emph{Phys. Rev. A},
  vol.~93, p. 032302, Mar 2016. [Online]. Available:
  \url{http://link.aps.org/doi/10.1103/PhysRevA.93.032302}
\BIBentrySTDinterwordspacing

\bibitem{Com_Letter_Shang_2014}
\BIBentryALTinterwordspacing
T.~Shang, X.-J. Zhao, and J.-W. Liu, ``Quantum network coding based on
  controlled teleportation,'' \emph{IEEE Communications Letters}, vol.~18,
  no.~5, pp. 865--868, 2014. [Online]. Available: \url{<Go to
  ISI>://WOS:000338114700037
  http://ieeexplore.ieee.org/ielx7/4234/6818356/06784159.pdf?tp=&arnumber=6784159&isnumber=6818356}
\BIBentrySTDinterwordspacing

\bibitem{Physical_Review_A_Satoh_2012}
\BIBentryALTinterwordspacing
T.~Satoh, F.~Le~Gall, and H.~Imai, ``Quantum network coding for quantum
  repeaters,'' \emph{Physical Review A}, vol.~86, no.~3, 2012. [Online].
  Available: \url{<Go to ISI>://WOS:000309101700003
  http://journals.aps.org/pra/abstract/10.1103/PhysRevA.86.032331}
\BIBentrySTDinterwordspacing

\bibitem{Journal_of_Mathematical_Physics_Jain_2011}
\BIBentryALTinterwordspacing
A.~Jain, M.~Franceschetti, and D.~A. Meyer, ``On quantum network coding,''
  \emph{Journal of Mathematical Physics}, vol.~52, no.~3, 2011. [Online].
  Available: \url{<Go to ISI>://WOS:000289152100008
  http://scitation.aip.org/docserver/fulltext/aip/journal/jmp/52/3/1.3555801.pdf?expires=1447779606&id=id&accname=2103930&checksum=932D2C468D049497975E80ED33E678FD}
\BIBentrySTDinterwordspacing

\bibitem{Nature_Photonics_Munro_2010}
\BIBentryALTinterwordspacing
W.~J. Munro, K.~A. Harrison, A.~M. Stephens, S.~J. Devitt, and K.~Nemoto,
  ``From quantum multiplexing to high-performance quantum networking,''
  \emph{Nature Photonics}, vol.~4, no.~11, pp. 792--796, 2010. [Online].
  Available: \url{<Go to ISI>://WOS:000283589200021
  http://www.nature.com/nphoton/journal/v4/n11/pdf/nphoton.2010.213.pdf}
\BIBentrySTDinterwordspacing

\bibitem{Physical_Review_A_Hayashi_2007}
\BIBentryALTinterwordspacing
M.~Hayashi, ``Prior entanglement between senders enables perfect quantum
  network coding with modification,'' \emph{Physical Review A}, vol.~76, no.~4,
  2007. [Online]. Available: \url{<Go to ISI>://WOS:000250619700001}
\BIBentrySTDinterwordspacing

\bibitem{Kobayashi_2011}
\BIBentryALTinterwordspacing
H.~Kobayashi, F.~Le~Gall, H.~Nishimura, and M.~Rotteler, ``Constructing quantum
  network coding schemes from classical nonlinear protocols,'' in
  \emph{Information Theory Proceedings (ISIT), 2011 IEEE International
  Symposium on}, Conference Proceedings, pp. 109--113. [Online]. Available:
  \url{http://ieeexplore.ieee.org/ielx5/6026198/6033677/06033701.pdf?tp=&arnumber=6033701&isnumber=6033677}
\BIBentrySTDinterwordspacing

\bibitem{Kobayashi_2010}
H.~Kobayashi, F.~L. Gall, H.~Nishimura, and M.~Rötteler, ``Perfect quantum
  network communication protocol based on classical network coding,'' in
  \emph{2010 IEEE International Symposium on Information Theory}, June 2010,
  pp. 2686--2690.

\bibitem{Kobayashi_2009}
\BIBentryALTinterwordspacing
H.~Kobayashi, F.~Le~Gall, H.~Nishimura, and M.~Roetteler, \emph{General Scheme
  for Perfect Quantum Network Coding with Free Classical Communication}, ser.
  Lecture Notes in Computer Science, 2009, vol. 5555, pp. 622--633. [Online].
  Available: \url{<Go to ISI>://WOS:000270963700051}
\BIBentrySTDinterwordspacing

\bibitem{long2002theoretically}
G.-L. Long and X.-S. Liu, ``Theoretically efficient high-capacity
  quantum-key-distribution scheme,'' \emph{Phys. Rev. A}, vol.~65, no.~3, p.
  032302, Feb. 2002.

\bibitem{sun2020toward}
Z.~Sun, L.~Song, Q.~Huang, L.~Yin, G.~Long, J.~Lu, and L.~Hanzo, ``Toward
  practical quantum secure direct communication: a quantum-memory-free protocol
  and code design,'' \emph{IEEE Trans. Commun.}, vol.~68, no.~9, pp.
  5778--5792, Sep. 2020.

\bibitem{WOS:000976372200001}
X.-J. Li, D.~Pan, G.-L. Long, and L.~Hanzo, ``Single-photon-memory
  measurement-device-independent quantum secure direct communication-part {I}:
  Its fundamentals and evolution,'' \emph{IEEE Communications Letters},
  vol.~27, no.~4, pp. 1055--1059, Apr. 2023.

\bibitem{WOS:000976372200002}
------, ``Single-photon-memory measurement-device-independent quantum secure
  direct communication-part {II}: A practical protocol and its secrecy
  capacity,'' \emph{IEEE Communications Letters}, vol.~27, no.~4, pp.
  1060--1064, Apr. 2023.

\bibitem{WOS:000949858900001}
F.~Li, X.~Zhang, J.~Li, J.~Wang, S.~Shi, L.~Tian, Y.~Wang, L.~Chen, and
  Y.~Zheng, ``Demonstration of fully-connected quantum communication network
  exploiting entangled sideband modes,'' \emph{Frontiers of Physics}, vol.~18,
  no.~4, Aug. 2023.

\bibitem{WOS:001037124000012}
Z.-Z. Sun, D.~Pan, D.~Ruan, and G.-L. Long, ``One-sided
  measurement-device-independent practical quantum secure direct
  communication,'' \emph{Journal of Lightwave Technology}, vol.~41, no.~14, pp.
  4680--4690, Jul. 2023.

\bibitem{guilu-cst}
D.~Pan, G.-L. Long, L.~Yin, Y.-B. Sheng, D.~Ruan, S.-X. Ng, J.~Lu, and
  L.~Hanzo, ``The evolution of quantum secure direct communication: On the road
  to the qinternet,'' \emph{IEEE Communications Surveys \& Tutorials,
  https://arxiv.org/abs/2311.13974}, 2024.

\bibitem{6515077}
P.~Botsinis, S.~X. Ng, and L.~Hanzo, ``Quantum search algorithms, quantum
  wireless, and a low-complexity maximum likelihood iterative quantum
  multi-user detector design,'' \emph{IEEE Access}, vol.~1, pp. 94--122, 2013.

\bibitem{7515148}
P.~Botsinis, D.~Alanis, Z.~Babar, S.~X. Ng, and L.~Hanzo, ``Joint
  quantum-assisted channel estimation and data detection,'' \emph{IEEE Access},
  vol.~4, pp. 7658--7681, 2016.

\bibitem{7515210}
P.~Botsinis, D.~Alanis, Z.~Babar, H.~V. Nguyen, D.~Chandra, S.~X. Ng, and
  L.~Hanzo, ``Quantum-aided multi-user transmission in non-orthogonal multiple
  access systems,'' \emph{IEEE Access}, vol.~4, pp. 7402--7424, 2016.

\bibitem{9463774}
E.~Villaseñor, M.~He, Z.~Wang, R.~Malaney, and M.~Z. Win, ``Enhanced uplink
  quantum communication with satellites via downlink channels,'' \emph{IEEE
  Transactions on Quantum Engineering}, vol.~2, pp. 1--18, 2021.

\end{thebibliography}

\begin{IEEEbiography}[{\includegraphics[width=1in,height=1.25in,clip,keepaspectratio]{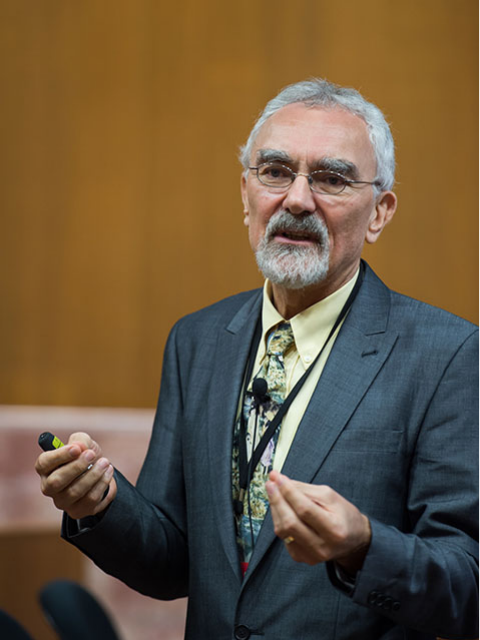}}]{\bf Lajos Hanzo} (http://www-mobile.ecs.soton.ac.uk, https://en.wikipedia.org/wiki/Lajos\_Hanzo) (FIEEE'04, Fellow of the Royal Academy of Engineering (FREng), of the IET and of EURASIP) received Honorary Doctorates  from the Technical University of Budapest (2009) and Edinburgh University (2015). He is a Foreign Member of the Hungarian Science-Academy, Fellow of the Royal Academy of Engineering (FREng), of the IET, of EURASIP and holds the IEEE Eric Sumner Technical Field Award.
\end{IEEEbiography}

\begin{IEEEbiography}
[
{
\includegraphics[width=1in,height=1.2in,clip,keepaspectratio]{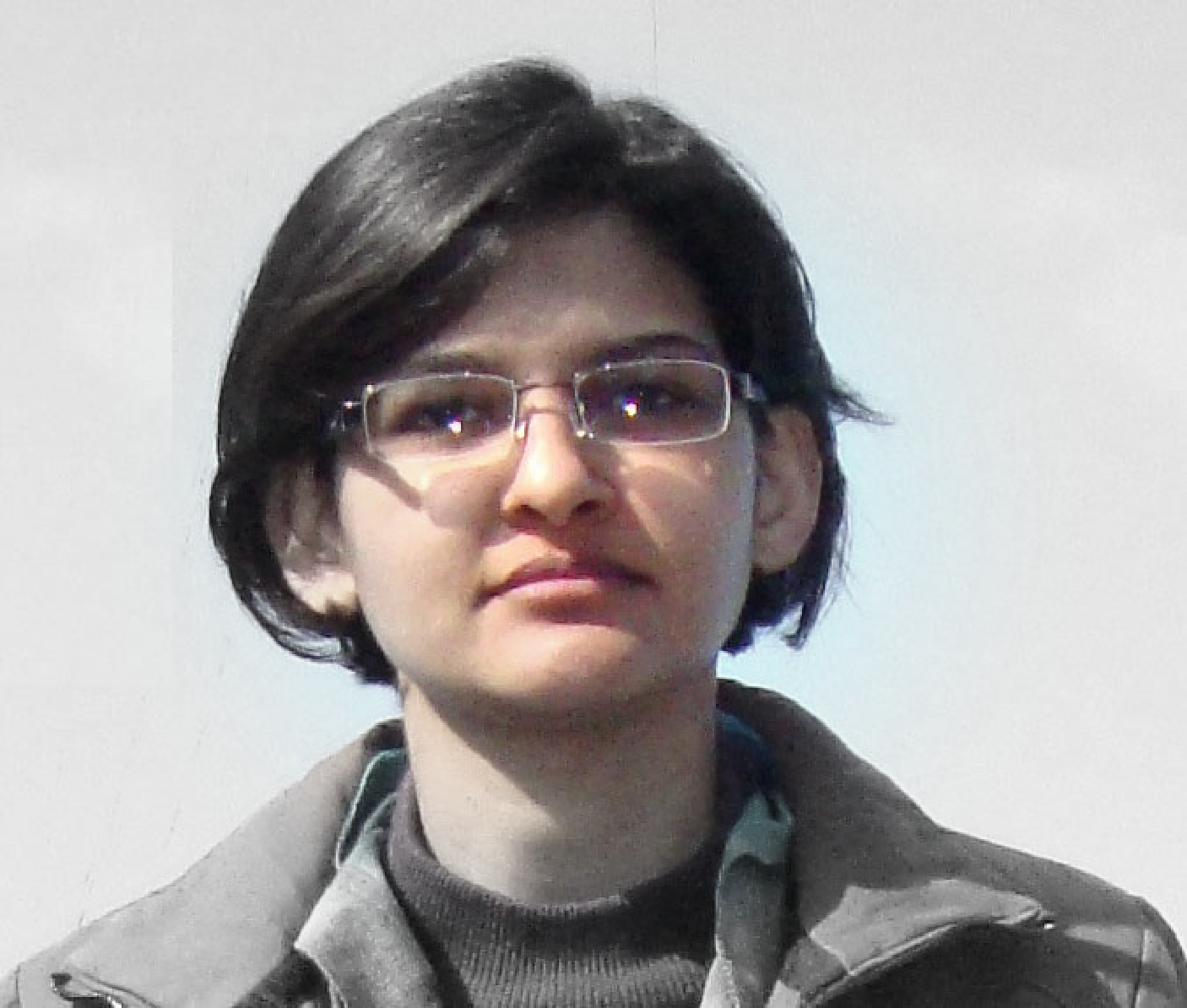}
}
]
  {\bf Zunaira Babar} (SMIEEE) received B.Eng. degree from the National University of Science and Technology (NUST), Pakistan, in 2008, and the M.Sc. (Hons) and Ph.D. degrees from the University of Southampton, UK,
in 2011 and 2015, respectively. Currently, she is a Staff Research Scientist at VIAVI Marconi Labs and also an Adjunct Fellow at the University of Southampton. {\bf She published about 40 journal papers on quantum communications.}
\end{IEEEbiography}

\begin{IEEEbiography}
[
  {\includegraphics[width=1in,height=1.25in,clip,keepaspectratio]{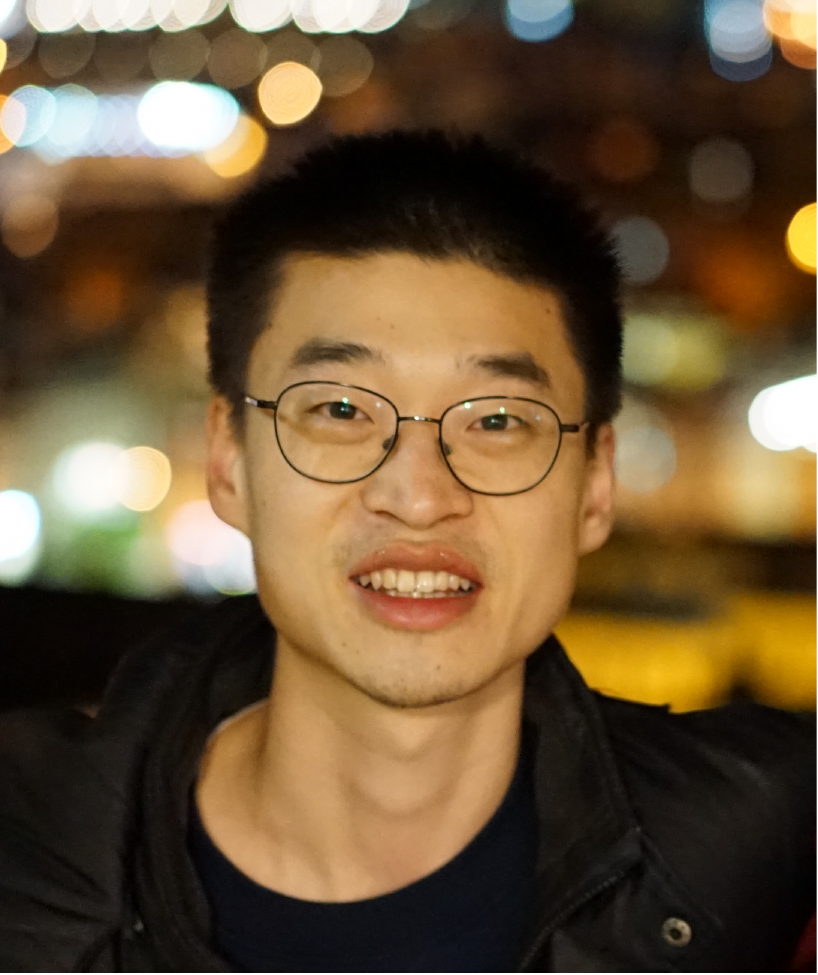}}
  ]
{\bf Zhenyu Cai} received his B.A. and M.Sc. from University of Cambridge in 2017 and his Ph.D. degree in quantum technologies from University of Oxford in 2020. Since then he has been a Junior Research Fellow in Physics at St John’s College, University of Oxford. {\bf His current research interests centre around quantum error correction, quantum error mitigation and their practical implementations in actual quantum hardware.}
\end{IEEEbiography}

\begin{IEEEbiography}
[
{
\includegraphics[width=1in,height=1.2in,clip,keepaspectratio]{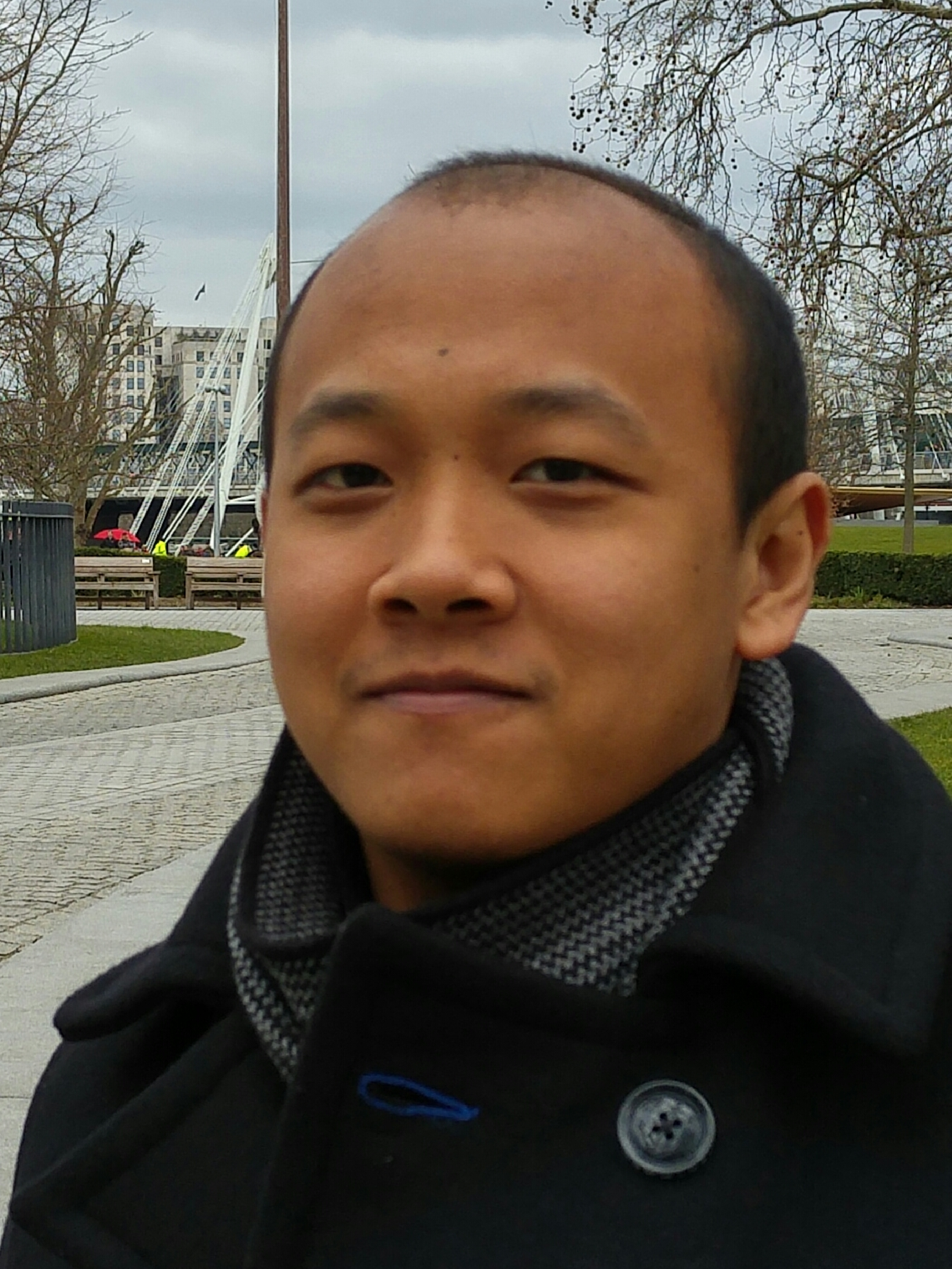}
}
]
  {\bf Daryus Chandra} received the B.Eng. and M.Eng. degrees from Universitas Gadjah Mada (UGM), Indonesia, in 2013 and 2014, respectively, and the Ph.D. degree from the University of Southampton, UK, in 2020. Currently, he is a quantum error correction researcher at Photonic Inc., Canada, and also a visiting research fellow at the University of Southampton. {\bf He published about 30 journal papers on quantum communications.}
\end{IEEEbiography}

\begin{IEEEbiography}
[
{
\includegraphics[width=1in,height=1.2in,clip,keepaspectratio]{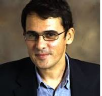}
}
]
  {\bf Ivan B. Djordjevic} (Fellow, IEEE) is a professor of ECE and optical sciences at the University of Arizona. He also serves as the director of the Optical Communications Systems Laboratory (OCSL) and Quantum Communications (QuCom) Lab. He is both IEEE Fellow and the Optica (formerly OSA) Fellow. Dr. Djordjevic received his PhD degree from the Faculty of Electronic Engineering at the University of Nis, Yugoslavia in 1999. His current research interests are on quantum communications, networking, and sensing. 

     Professor Djordjevic has authored/co-authored 11 books, 13 book chapters, more than 590 journal and conference publications, and he holds 56 US patents. He has been serving as an Editor for: IEEE Transactions on Communications, Optical and Quantum Electronics, and Frequenz. He was serving in the past as an associate editor for Journal of Optical Communications and Networking (2019 to 2023).  Further, he was serving as area editor/senior editor/editor of IEEE Communications Letters (2012 to 2021). Finally, he  was serving as associate editor/editorial board member for both IOP Journal of Optics and Elsevier Physical Communication Journal (2016 to 2021). 

     Before joining the University of Arizona, he held appointments at the University of Bristol and the University of the West of England in UK, Tyco Telecommunications in USA, National Technical University of Athens in Greece, and State Telecommunication Company in Nis, Yugoslavia.  
\end{IEEEbiography}

\begin{IEEEbiography}
[
{
\includegraphics[width=1in,height=1.2in,clip,keepaspectratio]{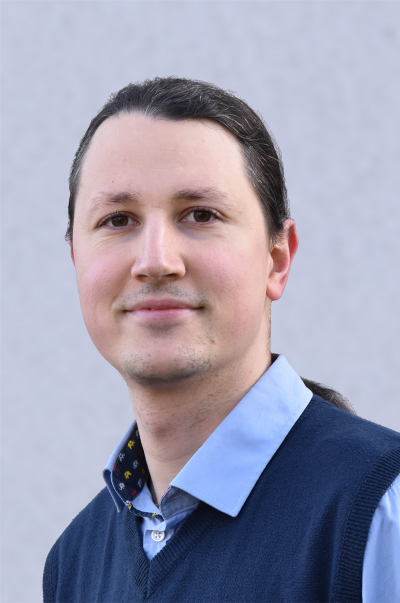}
}
]
  {\bf Balint Koczor} is an Associate Professor in Quantum Information Theory and Future Leaders Fellow in the Mathematical Institute in Oxford. He obtained his PhD degree from the Technical University of Munich where he worked on fundamental quantum theory and mathematical physics. He then joined the group of Prof. Simon Benjamin in Oxford to work on the theory of early quantum computers and held a Glasstone Research Fellowship. He is internationally recognized for his research on quantum error mitigation and near-term quantum computing. He works part-time as Lead Quantum Theorist for the company Quantum Motion. {\bf He published widely on quantum science and engineering.}
\end{IEEEbiography}

\begin{IEEEbiography}[{\includegraphics[width=1in,height=1.25in,clip,keepaspectratio]{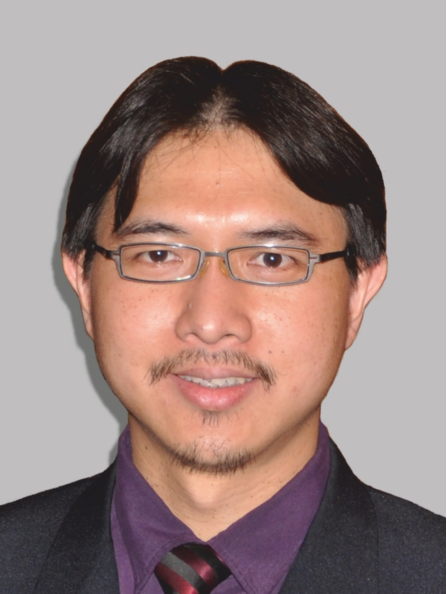}}]{\bf Soon Xin Ng} (S'99-M'03-SM'08) received the B.Eng. degree (First class) in electronic engineering and the Ph.D. degree in telecommunications from the University of Southampton, Southampton, U.K., in 1999 and 2002, respectively. From 2003 to 2006, he was a postdoctoral research fellow working on collaborative European research projects known as SCOUT, NEWCOM and PHOENIX. Since August 2006, he has been a member of academic staff in the School of Electronics and Computer Science, University of Southampton. He is involved in the OPTIMIX and CONCERTO European projects as well as the IU-ATC and UC4G projects. He is a full Professor in telecommunications at the University of Southampton.

His research interests include adaptive coded modulation, coded modulation, channel coding, space-time coding, joint source and channel coding, iterative detection, OFDM, MIMO, cooperative communications, distributed coding, quantum error correction codes and joint wireless-and-optical-fibre communications. He has published over 200 papers and co-authored two John Wiley/IEEE Press books in this field. He is a Senior Member of the IEEE, a Chartered Engineer and a Fellow of the Higher Education Academy in the UK. {\bf He is one of the four members of the IEEE Quantum Communications and information Technology (QCIT)
emerging technical committee's leadership team and published widely in quantum communications.}
\end{IEEEbiography}

\begin{IEEEbiography}
  [
{
\includegraphics[width=1in,height=1.2in,clip,keepaspectratio]{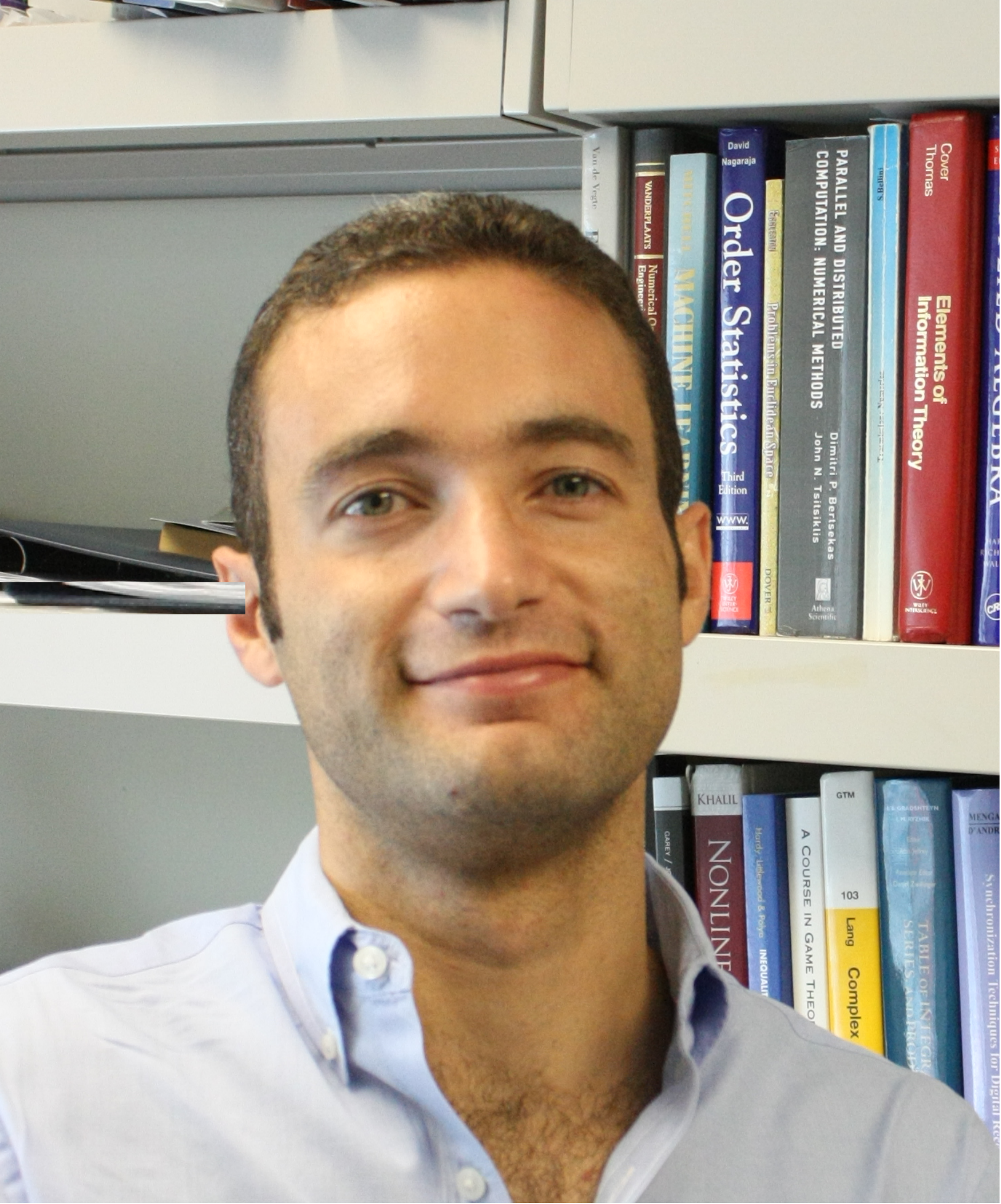}
}
]
{\bf Osvaldo Simeone} is a Professor of Information Engineering with the Centre for Telecommunications Research at the Department of Engineering of King's College London, where he directs the King's Communications, Learning and Information Processing lab. He received an M.Sc. degree (with honors) and a Ph.D. degree in information engineering from Politecnico di Milano, Milan, Italy, in 2001 and 2005, respectively. From 2006 to 2017, he was a faculty member of the Electrical and Computer Engineering (ECE) Department at New Jersey Institute of Technology (NJIT), where he was affiliated with the Center for Wireless Information Processing (CWiP). His research interests include information theory, machine learning, wireless communications, neuromorphic computing, and quantum machine learning. Dr Simeone is a co-recipient of the 2022 IEEE Communications Society Outstanding Paper Award, the 2021 IEEE Vehicular Technology Society Jack Neubauer Memorial Award, the 2019 IEEE Communication Society Best Tutorial Paper Award, the 2018 IEEE Signal Processing Best Paper Award, the 2017 JCN Best Paper Award, the 2015 IEEE Communication Society Best Tutorial Paper Award and of the Best Paper Awards of IEEE SPAWC 2007 and IEEE WRECOM 2007. He was awarded an Open Fellowship by the EPSRC in 2022 and a Consolidator grant by the European Research Council (ERC) in 2016. His research has been also supported by the U.S. National Science Foundation, the European Commission, the European Research Council, the Vienna Science and Technology Fund, the European Space Agency, as well as by a number of industrial collaborations including with Intel Labs and InterDigital. He is the Chair of the Signal Processing for Communications and Networking Technical Committee of the IEEE Signal Processing Society and of the UK \& Ireland Chapter of the IEEE Information Theory Society. He is currently a Distinguished Lecturer of the IEEE Communications Society, and he was a Distinguished Lecturer of the IEEE Information Theory Society in 2017 and 2018. {\bf Dr Simeone is the author of the textbook "Machine Learning for Engineers" published by Cambridge University Press, four monographs, including "An introduction to quantum machine learning"} on the Foundations and Trends in Signal Processing, and more than 180 research journal and magazine papers. He is a Fellow of the IET, EPSRC, and IEEE.
\end{IEEEbiography}

\begin{IEEEbiography}
  [
{
\includegraphics[width=1in,height=1.2in,clip,keepaspectratio]{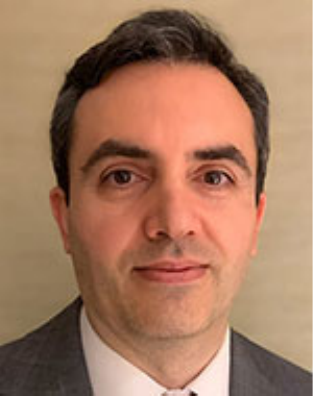}
}
]
{\bf Mohsen Razavi} received his B.Sc. and M.Sc. degrees in Electrical Engineering from Sharif University of Technology, respectively, in 1998 and 2000, and his PhD from MIT, in 2006. He was a postdoctoral fellow at the Institute for Quantum Computing at the University of Waterloo until September 2009, when he joined the School of Electronic and Electrical Engineering at the University of Leeds, where he is now a Professor of Quantum Communications. He is a recipient of the Marie-Curie International Reintegration Grant. He organized the first International Workshop on Quantum Communication Networks in 2014. He was the Coordinator of the European Innovative Training Network, QCALL, which aimed at providing quantum communications services to all users. {\bf He has authored a book on quantum communications networks} in IOP Concise Physics series. His research interests include a variety of topics in quantum optical communications, quantum optics, and quantum communications networks.
\end{IEEEbiography}

\end{document}